\documentclass[american,longbibliography]{book}
\usepackage[latin9]{inputenc}
\usepackage{geometry}
\usepackage{appendix}
\usepackage{float}
\usepackage{calc}
\usepackage{cite}
\usepackage{amsmath}
\usepackage{amssymb}
\usepackage{graphicx}
\usepackage{esint}
\PassOptionsToPackage{normalem}{ulem}
\usepackage{ulem}
\makeatletter

\usepackage[english]{babel}

\usepackage{booktabs}\usepackage{xcolor}\definecolor{light-gray}{gray}{0.9}
\usepackage{calc}\usepackage{units}\usepackage{url}\PassOptionsToPackage{normalem}{ulem}
\usepackage{fancyhdr}

\newcommand{\nocontentsline}[3]{}
\newcommand{\tocless}[2]{\bgroup\let\addcontentsline=\nocontentsline#1{#2}\egroup}

\fancypagestyle{basicstyle}{%
  \fancyhf{}
  \fancyhead[LO]{\nouppercase{\rightmark}}
  \fancyhead[RE]{\nouppercase{\leftmark}}
  \fancyhead[LE,RO]{\thepage}
  
  \fancyfoot[C]{}
  
  }

\fancypagestyle{otherstyle}{%
  \fancyhf{}
  \fancyhead[C]{header}
  }

\usepackage{titlesec}\titleformat{\chapter}[display]
  {\bfseries\Large}
  {\filright\MakeUppercase{\chaptertitlename} \Huge\thechapter}
  {1ex}
  {\titlerule\vspace{1ex}\filleft}
  [\vspace{1ex}\titlerule]

\fancypagestyle{constant}{%
  \fancyhf{}
  
  \fancyhead[RE]{\nouppercase{CONSTANTES, ORDRES DE GRANDEUR UTILES} }
  \fancyhead[LE,RO]{\thepage}
  
  }

  \fancypagestyle{avantpropos}{%
  \fancyhf{}
  
  \fancyhead[RE]{\nouppercase{AVANT-PROPOS} }
  \fancyhead[LE,RO]{\thepage}
  
  }

\fancypagestyle{plain}{%
  \fancyhf{}%
}

\usepackage{babel}
\usepackage{perpage}\MakePerPage{footnote}

\addto\extrasfrench{%
   \providecommand{\fg}{\ifdim\lastskip>\z@\unskip\fi~\frqq}%
}

\usepackage{listings}

\makeatother
\usepackage{makeidx}
\usepackage{babel}
\makeatletter
\addto\extrasfrench{%
   \providecommand{\fg}{\ifdim\lastskip>\z@\unskip\fi~\frqq}%
}
\makeatother
\makeindex

\begin{document}

%\begin{minipage}[b]{0.5\linewidth}
%\flushleft
%\includegraphics[scale=0.24]{Logo_Université_Paris-Saclay.png}\\
%~\\
%\end{minipage}
%\begin{minipage}[b]{0.5\linewidth}
%\flushright
%\includegraphics[scale=0.15]{X.jpg} 
%ladder\end{minipage}

%\begin{figure}[h!]
 %\includegraphics[scale=0.3]{X.png} 
%\end{figure}

%\begin{figure}[h!]
 %\includegraphics[scale=0.2]{CNRS.png} 
%\end{figure}

NNT : 2016SACLX036 
\begin{center}
 \begin{Large}
TH\`{E}SE DE DOCTORAT DE L'UNIVERSIT\'{E} PARIS-SACLAY PREPAR\'{E}E  \`{A} \\
~\\
 \begin{LARGE}
\'{E}COLE POLYTECHNIQUE
 \end{LARGE}
 \end{Large}
 \end{center}

%\maketitle

\begin{center}
 \begin{large}
\'{E}COLE DOCTORALE N. 564 PIF | Physique de l'Ile-de-France\\
Sp\'{e}cialit\'{e} de doctorat: Physique\\
par
 \end{large}
 \end{center}

\begin{center}
 \begin{large}
 \textbf{Lo\"{i}c Henriet}
 \end{large}
 \end{center}
 
 \begin{center}
 \begin{large}
Dynamique hors \'{e}quilibre de syst\`{e}mes quantiques \`{a} N-corps.
 \end{large}
 \end{center}

\noindent \textbf{Th\`{e}se pr\'{e}sent\'{e}e et soutenue \`{a} Palaiseau, le 08/09/2016}\\
 \begin{large}
  \begin{center}{Composition du jury:}  
  \end{center}
  
  \begin{minipage}[c]{0.6\linewidth}
\noindent \textbf{Jonathan Keeling}\\
 \textbf{Wilhelm Zwerger}\\
%  \textbf{Jacqueline Bloch}\\
  \textbf{Beno\^{i}t Dou\c{c}ot}\\
  \textbf{Philippe Lecheminant}\\
  \textbf{Laurent Sanchez-Palencia}\\
   \textbf{Karyn Le Hur}\\
   \textbf{Marco Schir\'{o}}\\
 \end{minipage}
   \begin{minipage}[c]{0.4\linewidth}
Rapporteur\\
Rapporteur\\
Pr\'{e}sident du jury\\
Examinateur\\
Examinateur\\
Directrice de th\`{e}se\\
Membre invit\'{e}\\
 \end{minipage}
 \end{large}

\thispagestyle{empty}
\newpage

\begin{small}
\noindent \textbf{Titre}: Dynamique hors \'{e}quilibre de syst\`{e}mes quantiques \`{a} N corps.\\

\noindent\textbf{Mots cl\'{e}s}: Dynamique, Hors \'{e}quilibre, Syst\`{e}mes ouverts\\

\noindent\textbf{R\'{e}sum\'{e}}: Cette th\`{e}se porte sur l'\'{e}tude de propri\'{e}t\'{e}s dynamiques de mod\`{e}les quantiques port\'{e}s hors \'{e}quilibre. Nous introduisons en particulier des mod\`{e}les g\'{e}n\'{e}raux de type spin-boson, qui d\'{e}crivent par exemple l'interaction lumi\`{e}re-mati\`{e}re ou certains ph\'{e}nom\`{e}nes de dissipation. Nous contribuons au d\'{e}veloppement d'une approche stochastique exacte permettant de d\'{e}crire la dynamique hors \'{e}quilibre du spin dans ces mod\`{e}les. Dans ce contexte, l'effet de l'environnement bosonique est pris en compte par l'interm\'{e}di\'{e}aire des degr\'{e}s de libert\'{e} al\'{e}atoires suppl\'{e}mentaires, dont les corr\'{e}lations temporelles d\'{e}pendent des propri\'{e}t\'{e}s spectrales de l'environnement bosonique. Nous appliquons cette approche \`{a} l'\'{e}tude de ph\'{e}nom\`{e}nes \`{a} N-corps, comme par exemple la transition de phase dissipative induite par un environnement bosonique de type ohmique. Des ph\'{e}nom\`{e}nes de synchronisation spontan\'{e}e, et de transition de phase topologique sont aussi identifi\'{e}s. Des progr\`{e}s sont aussi r\'{e}alis\'{e}s dans l'\'{e}tude de la dynamique dans les r\'{e}seaux de syst\`{e}mes lumi\`{e}re-mati\`{e}re coupl\'{e}s. Ces d\'{e}veloppements th\'{e}oriques sont motiv\'{e}s par les progr\`{e}s exp\'{e}rimentaux r\'{e}cents, qui permettent d'envisager une \'{e}tude approfondie de ces ph\'{e}nom\`{e}nes. Cela inclut notamment les syst\`{e}mes d'atomes ultra-froids, d'ions pi\'{e}g\'{e}s, et les plateformes d'\'{e}lectrodynamique en cavit\'{e} et en circuit. Nous int\'{e}ressons aussi \`{a} la physique des syst\`{e}mes hybrides comprenant des dispositifs \`{a} points quantiques m\'{e}soscopiques coupl\'{e}s \`{a} un r\'{e}sonateur \'{e}lectromagn\'{e}tique. L'av\`{e}nement de ces syst\`{e}\`{e}mes permet des mesures de la formation d'\'{e}tats \`{a} N-corps de type Kondo gr\^{a}ce au r\'{e}sonateur; et d'envisager de dispositifs thermo\'{e}lectriques.
\\

\noindent  \textbf{Title}: Non-equilibrium dynamics of many body quantum systems.\\


\noindent\textbf{Abstract}: This thesis deals with the study of dynamical properties of out-of-equilibrium quantum systems. We introduce in particular a general class of Spin-Boson models, which describe for example light-matter interaction or dissipative phenomena. We contribute to the development of a stochastic approach to describe the spin dynamics in these models. In this context, the effect of the bosonic environment is encapsulated into additional stochastic degrees of freedom whose time-correlations are determined by spectral properties of the bosonic environment. We use this approach to study many-body phenomena such as the dissipative quantum phase transition induced by an ohmic bosonic environment. Synchronization phenomena as well as dissipative topological transitions are identified. We also progress in the study of arrays of interacting light-matter systems. These theoretical developments follow recent experimental achievements, which could ensure a quantitative study of these phenomena. This notably includes ultra-cold atoms, trapped ions and cavity and circuit electrodynamics setups. We also investigate hybrid systems comprising electronic quantum dots coupled to electromagnetic resonators, which enable us to provide a spectroscopic analysis of many-body phenomena linked to the Kondo effect. We also introduce 
thermoelectric applications in these devices. 

%\thispagestyle{empty}
%\addtocontents{toc}{\protect\thispagestyle{empty}}

\end{small}

%\thispagestyle{empty}
%\addtocontents{toc}{\protect\thispagestyle{empty}}
\chapter*{Remerciements}

Je remercie tout d'abord les rapporteurs, professeurs J. Keeling et W. Zwerger, d'avoir accept\'{e}  de lire et de porter un oeil critique sur mon manuscrit. Je tiens aussi \`{a} remercier les autres membres du jury d'avoir accept\'{e} d'assister \`{a} ma soutenance.\\

La plus grande reconnaissance va \`{a} Karyn Le Hur, qui a constamment suivi avec soin mon parcours de th\`{e}se. Ses id\'{e}es toujours excellentes ont su guider mon effort, et sont \`{a} l'origine de beaucoup de ce qui est pr\'{e}sent\'{e}  dans ce manuscrit. Merci pour la confiance qu'elle m'a accord\'{e}e.\\

Merci \`{a} tous les membres du Centre de Physique Th\'{e}orique pour leur accueil et leur gentillesse. Je remercie particuli\`{e}rement mon parrain Christoph Kopper pour sa gentillesse et sa disponibilit\'{e}, ainsi que Bernard Pire qui nous avait soutenu Lucien et moi lors de notre tentative de mise en place d'un s\'{e}minaire pour les doctorants et post-doctorants du laboratoire. Un grand merci \`{a} Fadila, Florence, Jeannine, Malika ainsi qu'\`{a} Danh, David, et Jean-Luc pour leur aide. J'aimerais aussi souligner ma reconnaissance envers l'Ecole Polytechnique, et j'en profite aussi pour remercier tous les professeurs qui ont influenc\'{e} mon parcours et m'ont pouss\'{e} \`{a} poursuivre en th\`{e}se.\\

Merci \`{a} Peter Orth et Zoran Ristivojevic pour leur aide pr\'{e}cieuse au d\'{e}but de ma th\`{e}se et leurs conseils pour aborder un probl\`{e}me scientifique. Leur contribution au d\'{e}veloppement de l'approche stochastique pr\'{e}sent\'{e}e dans ce manuscrit est \`{a} souligner particuli\`{e}rement. Il est aussi important de mentionner que cette technique a \'{e}t\'{e} \'{e}labor\'{e}e suivant les id\'{e}es d'Adilet Imambekov, que je n'ai pas eu le plaisir de conna\^{i}tre. J'ai aussi eu l'opportunit\'{e} de pouvoir travailler en collaboration avec Andrew Jordan et je le remercie pour sa gentillesse et ses id\'{e}es \'{e}clairantes. J'ai aussi beaucoup appr\'{e}ci\'{e} travailler avec Guillaume Roux et Marco Schir\'{o}, et les remercie pour cette chance. Je salue aussi Guang-wei Deng et le remercie pour le dialogue que nous avons pu lier entre th\'{o}riciens et exp\'{e}rimentateurs lors de notre collaboration. Ces ann\'{e}es de th\`{e}se m'ont aussi permis d'int\'{e}ragir avec de nombreux groupes de recherche et je tiens \`{a} remercier les universit\'{e}s de B\^{a}le, Cologne, Heifei, Orsay, Princeton, Stony Brook, Ulm, ainsi que l'Ecole Normale Sup\'{e}rieure, le CEA-Saclay et l'ICFO --et plus particuli\`{e}rement Christoph Bruder, Darrick Chang, Sebastian Diehl, Jean-No\"{e}l Fuchs, Guo-Ping Guo, Susana Huelga, Christophe Mora, Olivier Parcollet, Fr\'{e}d\'{e}ric Pi\'{e}chon, Dominik Schneble, Pascal Simon, et Hakan T\"{u}reci. Je remercie aussi les organisateurs de l'\'{e}cole \textit{Physique m\'{e}soscopique} au Qu\'{e}bec et du GdR CNRS \`{a} Aussois, l'Institut Canadien de Recheche Avanc\'{e}e, et les organisateurs de \textit{Qlight 2016} en Cr\^{e}te.\\

 Sur un plan plus personnel, je remercie Camille, Michele, et Marco pour les discussions que j'ai pu avoir avec eux. Merci \`{a} tous les anciens et actuels membres du groupe: Peter, Zoran, Alex, Tianhan, Lo\"{i}c, Kirill, et Antonio avec qui j'ai eu le plaisir de collaborer pendant son stage de master.\\

Merci \`{a} tous mes amis, et \`{a} ma famille. \\

\newpage

\textit{\`{A} C\'{e}cile,}

\tableofcontents

\chapter*{Introduction}
\addcontentsline{toc}{chapter}{\protect\numberline{}Introduction}%
Modern experimental platforms now allow to control and monitor interacting quantum systems in a precise manner. Motivated by quantum information purposes, these platforms are also of great interest to explore many-body quantum physics. The unprecedented control over the system's parameters permits indeed to devise and tune desired Hamiltonians. This ability to emulate complex many-body Hamiltonian motivates the development of precise theoretical predictions to better understand the physical phenomena underlying the system's dynamics. Most experimental setups also involve periodic pumping of energy into the system, as well as dissipative effects, fostering the development of specific theoretical tools to tackle the dynamics in out-of-equilibrium conditions.\\

In Chapter I we introduce the class of Spin-Boson models and focus in particular on the Rabi model, the ohmic spinboson model, and their lattice versions. The Rabi model considers a two-level system coupled to a quantized harmonic oscillator and describes the simplest interaction between matter and light. Its lattice version describing a set of interacting light-matter systems opens the door to many-body physics with light. The ohmic spinboson model was first introduced to describe dissipative effects on a two-level system. In this description dissipation is modelled by a bath of quantized harmonic oscillators, and many-body effects of the bath notably induce a dissipative quantum phase transition at large spin-bath coupling (see ``Many-Body Quantum Electrodynamics Networks: Non-Equilibrium Condensed Matter Physics with Light").  \\

In Chapter II, we derive a Stochastic Schr\"{o}dinger Equation (SSE) describing the time evolution of the spin-reduced density matrix for Spin-boson problems. We test this framework and recover known results for the dynamics of the Rabi model (see ``Quantum dynamics of the driven and dissipative Rabi model") and the ohmic spinboson model. We also compare the SSE approach to other known methods and present the limitations related to the SSE.\\

In Chapter III, we extend the applicability of the SSE to the case of two spins coupled to a common ohmic bosonic environment. We investigate the dissipative quantum phase transition induced by the bath, and study the quenched dynamics in both phases. We also focus on bath-induced synchronization effects between two spins with different frequencies  (see ``Quantum sweeps, synchronization, and Kibble-Zurek physics in dissipative quantum spin systems").\\

In Chapter IV, we study the influence of an ohmic environment on the topological properties of a two-level system measured with a dynamical protocol. We show in particular that the bath induces a dissipative topological transition at strong coupling, and we investigate the properties of this transition. \\

In Chapter V, we focus on hybrid devices coupling quantum dots and lead electronic degrees of freedom to light. We first introduce and study theoretically a nano heat-engine setup consisting of a quantum dot coupled to fermionic (electronic) leads and different Left/Right bosonic environments (see ``Electrical Current from Quantum Vacuum Fluctuations in Nano-Engines"). We show how zero-point quantum fluctuations stemming from bosonic environments permit the rectification of the current. Then we present and interpert the results of a recent experiment studying a double quantum dot setup coupled to a resonator. We establish a link between the light measurements and the formation of a SU(4) Kondo bound state at low temperatures (see ``A Quantum Electrodynamics Kondo Circuit: Probing Orbital and Spin Entanglement").\\

\underline{Publication list}:
\begin{itemize}
\item ``Quantum dynamics of the driven and dissipative Rabi model", L. Henriet, Z. Ristivojevic, P.P. Orth, K. Le Hur, Physical Review A \textbf{90} (2), 023820 (2014).
\item ``Electrical current from quantum vacuum fluctuations in nanoengines", L. Henriet, A. N. Jordan, K. Le Hur, Physical Review B \textbf{92} (12), 125306 (2015).
\item ``Quantum sweeps, synchronization, and Kibble-Zurek physics in dissipative quantum spin systems", L. Henriet, K. L. Hur, Physical Review B \textbf{93} (6), 064411 (2016).
\item ``Many-Body Quantum Electrodynamics Networks: Non-Equilibrium Condensed Matter Physics with Light", K. Le Hur, L. Henriet, A. Petrescu, K. Plekhanov, G. Roux, M. Schir\'{o}, Comptes Rendus Physique, \textbf{17} (8), 808-835  (2016).
\item ``Topology of a dissipative spin: Dynamical Chern number, bath-induced nonadiabaticity, and a quantum dynamo effect",
L. Henriet, A. Sclocchi, P. P. Orth, and K. Le Hur, Phys. Rev. B \textbf{95}, 5, 054307 (2017).
\item ``A Quantum Electrodynamics Kondo Circuit: Probing Orbital and Spin Entanglement", G.-W. Deng, L. Henriet, D. Wei, S.-X. Li, H.-O. Li, G. Cao, M. Xiao, G.-C. Guo, M. Schir\'{o}, K. Le Hur, G.-P. Guo, arXiv preprint arXiv:1509.06141 (2015).
\end{itemize}

\chapter{Spin-Boson models}

In this chapter, we will intoduce the Rabi model and the ohmic Spinboson model as paradigmatic models to describe respectively light-matter interaction and decoherence. We will put a particular emphasis on the investigation of many-body physics phenomena related to these models. In the Rabi case, this notably includes the possible realization of arrays of interacting light-matter elements, giving access to exotic phases where light behaves as a quantum fluid. In the ohmic spinboson case, many-body effects play naturally an important role and trigger a dissipative quantum phase transition at large coupling. Realizing dissipative arrays may allow to facilitate the investigation of this transition.

\section{Rabi model and applications}
\label{presentation_rabi}
The Rabi model had been originally introduced to describe the effect of a magnetic field on an atom possessing a nuclear spin \cite{rabi1,rabi2}. Nowadays, this model is applied to a variety of quantum systems in relation with strong light-matter interaction. It follows indeed naturally from the general description of a set of spinless point charges interacting with the quantized electromagnetic field. The Rabi model turns out then to be a paradigmatic model to describe cavity quantum electrodynamics experiments (cavity QED), in which an atom is placed inside a microwave or optical cavity. The introduction of cavity QED experiments and the derivation of the Rabi model in this context will be exposed in section \ref{cavity_qed}, following the general description provided in Refs. \cite{Cohen_interactions_photons_atomes,Haroche_Raymond:exploring_the_quantum}.\\

In Sec. \ref{circuit_qed}, we focus on the description of so-called circuit quantum electrodynamics experiments (circuit QED), describing the interaction of an artificial atom made of superconducting materials with microwave photons. These systems permits to reproduce the physics of the Rabi model on chip, with a very large light-matter coupling. Recent experimental progress in this field allows now to envision the creation of larger circuits composed of a large number of coupled light-matter elements, which are of particular interest to investigate many-body phenomena with light.

\subsection{Rabi Hamiltonian in cavity quantum electrodynamics}
\label{cavity_qed}

\subsubsection{Charges in a box}
We consider a set of charges placed inside a three dimensional box of typical length $L$. The Hamiltonian describing these charges in interaction with the electromagnetic field reads in the Coulomb gauge \cite{Cohen_interactions_photons_atomes} $\mathcal{H}=\mathcal{H}_{kin}+\mathcal{H}_r+V_{coul}$, with
\begin{align}
\mathcal{H}_{kin}&=\sum_{\alpha} \frac{\left[\textbf{p}_{\alpha}-q_{\alpha}\textbf{A}(\textbf{r}_{\alpha})\right]^2}{2 m_{\alpha}}\\
\mathcal{H}_r&=\sum_{\textbf{k}} \sum_{n=1,2} \hbar \omega_\textbf{k} a^{\dagger}_{\textbf{k},n} a_{\textbf{k},n}.
\label{Hamiltonian_electrodynamics} 
\end{align}
In Eq. (\ref{Hamiltonian_electrodynamics}), $\textbf{r}_{\alpha}$ and $\textbf{p}_{\alpha}$ are the position and momentum operators of the particle $\alpha$ with charge $q_{\alpha}$ and mass $m_{\alpha}$.  The vector potential $\textbf{A}(r_{\alpha})$ is described by a set of bosonic operators $a^{\dagger}_{\textbf{k},n}$ and $a_{\textbf{k},n}$ corresponding respectively to the creation and the annihilation of a photon in the mode $\textbf{k}$ with energy $\hbar \omega_\textbf{k}$ and polarization $n\in(1,2)$. 
\begin{equation}
\textbf{A}(\textbf{r}_{\alpha})=\sum_{\textbf{k}} \sum_{n=1,2}  \sqrt{\frac{\hbar}{2\epsilon_0 \omega_\textbf{k} L^3}}\left(a_{\textbf{k},n} e^{i\textbf{k}. \textbf{r}_{\alpha}}  +a^{\dagger}_{\textbf{k},n} e^{-i\textbf{k}.\textbf{r}_{\alpha}} \right) \textbf{e}_{\textbf{k},n},
\label{Vector_potential}
\end{equation}
The frequencies $\omega_{\textbf{k},n}$ are quantized in relation with the boundary conditions imposed by the box, and the vectors $\textbf{e}_{\textbf{k},n}$ correspond to the two possible polarizations for the transverse vector potential. The kinetic term $\mathcal{H}_{kin}$ in Eq. (\ref{Hamiltonian_electrodynamics}) will describe the interaction between the charges and the electromagnetic field through the vector potential. We remark from Eq. (\ref{Vector_potential}) that the vector potential grows as the size of the box we consider decreases. In the case of a small box the vector potential exists only as a superposition of a discrete set of modes. This interesting case led to the the design of cavity quantum electrodynamics experiments, where an atom placed inside a cavity interacts resonantly with the confined electromagnetic field. 

\subsubsection{Cavity Quantum electrodynamics setup}

In a typical cavity QED setup, the atom is chosen such that the transition energy $\Delta$ between a reference state $|g\rangle$ and an excited state $|e\rangle$ is close to the energy $\hbar \omega_0$ of one of the first modes of the confined light, in the microwave \cite{Raimond:RMP} or optical regime \cite{Hood:science}. The joint evolution of the light-matter system inside the cavity will result from this resonant interaction, and the effect of higher harmonics will be neglected. The wavelength corresponding to the resonant mode under consideration is much greater than the atomic length-scale, so that one can safely replace the kinetic term $\mathcal{H}_{kin}$ in Eq. (\ref{Hamiltonian_electrodynamics}) by
\begin{equation}
\mathcal{H}_{kin}=\sum_{\alpha} \frac{\left[\textbf{p}_{\alpha}-q_{\alpha}\textbf{A}(\textbf{0})\right]^2}{2 m_{\alpha}},
\label{Long_wavelength_approximation} 
\end{equation}
where we define $\textbf{r}=\textbf{0}$ as the position of the atomic center of mass. This simplification is known as the long-wavelength approximation. Expanding Eq. (\ref{Long_wavelength_approximation}) then leads to a sum of three terms,
\begin{equation}
\mathcal{H}_{kin}=\sum_{\alpha} \frac{\textbf{p}_{\alpha}^2}{2 m_{\alpha}}- \left(\sum_{\alpha} \frac{q_{\alpha}\textbf{p}_{\alpha}}{2 m_{\alpha}}\right).\textbf{A}(\textbf{0})+\sum_{\alpha} \frac{\left[q_{\alpha}\textbf{A}(\textbf{0})\right]^2}{2 m_{\alpha}}.
\label{H_kin_expanded} 
\end{equation}
In most cavity-QED experiments, the last term on the right hand side of Eq. (\ref{H_kin_expanded}) can be neglected due to the smallness of the electromagnetic field \cite{Haroche_Raymond:exploring_the_quantum}. The atom-field coupling is captured by the second term of Eq. (\ref{H_kin_expanded}) which can be understood better in terms of the dipole of the atom. Following Ref. \cite{Keeling_course}, we write
\begin{equation}
\frac{\textbf{p}_{\alpha}}{m}=\frac{i}{\hbar} \left[ \mathcal{H}_{at},\textbf{r}_{\alpha}\right],
\label{Hamilton_eq_bare_atom} 
\end{equation}
where $H_{at}$ would be the Hamiltonian describing the atom without its interaction with the electromagnetic field. Introducing the dipole operator 
\begin{equation}
\textbf{d}=\sum_{\alpha} q_{\alpha}\textbf{r}_{\alpha},
\label{dipole} 
\end{equation}
one can then write
\begin{equation}
\mathcal{H}_{kin}=\sum_{\alpha} \frac{\textbf{p}_{\alpha}^2}{2 m_{\alpha}}- \frac{i}{2\hbar}\left[\mathcal{H}_{at}, \textbf{d}\right].\textbf{A}(\textbf{0})+\sum_{\alpha} \frac{\left[q_{\alpha}\textbf{A}(\textbf{0})\right]^2}{2 m_{\alpha}}.
\label{H_kin_expanded_bis} 
\end{equation}
As stated previously, the atomic degree of freedom can be effectively described by a two-level system composed of the states $|g\rangle$ and $|e\rangle$. The dipole operator is off-diagonal in this basis, as the atom is neutral. We call $\textbf{d}=\langle e |\textbf{d}|g\rangle$ the dipole matrix element between the two atomic states and we reach the Hamiltonian
\begin{equation}
\mathcal{H}=\left( \begin{array}{cc}
\frac{\Delta}{2}&0\\
0&-\frac{\Delta}{2}
\end{array} \right)-i \sum_n \frac{\Delta}{2} \frac{1}{\sqrt{2 \epsilon_0 \hbar \omega_0 L^3}} \left( \begin{array}{cc}
0&d_n\\
-d_n^*&0
\end{array} \right)\left(a_n+a_n^{\dagger} \right)+ \sum_n \hbar \omega_0 a_n^{\dagger}a_n.
\end{equation}
Considering only one light polarization and after a rotation in the atomic subspace, we reach the Rabi Hamiltonian,
\begin{equation}
\mathcal{H}_{Rabi}=\frac{\Delta}{2}\sigma^z+\frac{g}{2} \sigma^x (a+a^{\dagger})+\hbar \omega_0 \left(a^{\dagger}a+\frac{1}{2}\right)
\label{Rabi}
\end{equation}
where the light-matter coupling strength $g$ reads
\begin{equation}
g=\frac{d \Delta}{\sqrt{2 \epsilon_0 \hbar \omega_0 L^3}}.
\end{equation}
In Eq. (\ref{Rabi}), $\sigma^x$ and  $\sigma^z$ belong to the set of Pauli matrices,
\begin{equation}
\sigma^x=\left( \begin{array}{cc}
0&1\\
1&0
\end{array} \right)\qquad \sigma^y=\left( \begin{array}{cc}
0&-i\\
i&0
\end{array} \right)\qquad \sigma^z=\left( \begin{array}{cc}
1&0\\
0&-1
\end{array} \right).
\label{Pauli_matricess}
\end{equation}

\subsubsection{Weak-coupling regime and Jaynes-Cummings physics}
 The typical value of the light matter coupling in 3D optical \cite{Hood:science} or microwave \cite{Raimond:RMP} cavity QED setups is of the order of $g/(\hbar\omega_0) \sim 10^{-7}$. In this range of parameters and for low detuning $|\Delta- \hbar\omega_0|/(\hbar \omega_0) \ll1$, Eq. (\ref{Rabi}) can be simplified further with the application of the rotating wave approximation (RWA). This approximation consists in neglecting the counter-rotating terms, $(\sigma^+ a^{\dagger} + h.c.)$ where $\sigma^{\pm}=(\sigma^x\pm i\sigma^y)/2$, which do not conserve the number of excitations. This ensures a continuous $U(1)$ symmetry and an associated conserved quantity, the polariton number $N = a^{\dagger} a + \sigma^+ \sigma^-$ (which counts the total number of excitations, as the sum of light and matter excitations). The resulting Hamiltonian is the Jaynes-Cummings Hamiltonian $\mathcal{H}_{JC}$ \cite{JC:Proc_IEEE},
\begin{equation}
\mathcal{H}_{JC}=\frac{\Delta}{2}\sigma^z+\frac{g}{2} \left( \sigma^+ a+ \sigma^- a^{\dagger}\right)+\hbar \omega_0 \left( a^{\dagger}a+\frac{1}{2} \right).
\label{JC}
\end{equation}

\begin{figure}[h!]
\center
\includegraphics[scale=0.35]{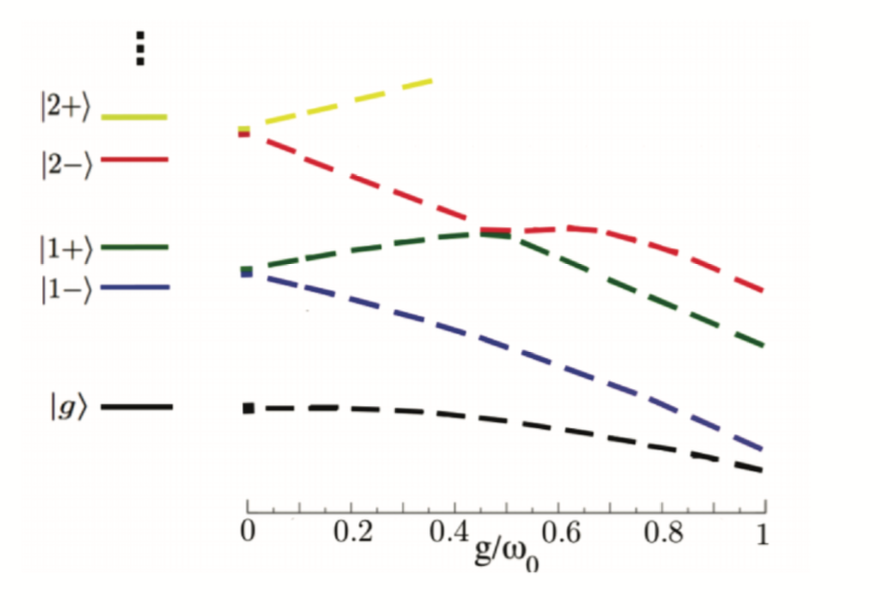}  
\caption{Jaynes-Cummings polariton ladder for
$g/\omega_0 = 0.2$, and evolution of the energy eigenstates of the Rabi model when approaching the strong coupling limit with $g/(\hbar\omega_0)$ (we take $\hbar=1$ for the label of the axis). The energy spacing between the two lowest energy states decreases as $e^{-g^2/(\hbar\omega_0)^2}$.}
\label{JC_ladder}
\end{figure}

The Jaynes-Cummings Hamiltonian is easily diagonalized in the so-called dressed basis \cite{JC:Proc_IEEE}. The ground state of the system consists of the two-level system in its lower state and vacuum for the photons, while the excited eigenstates $|n{\pm}\rangle$ are pairs of combined light-matter excitations (polaritons) characterized by their polariton number. One has $N|n_{\pm}\rangle=n|n_{\pm}\rangle$. These eigenstates form the well-known structure of the anharmonic JC ladder (Fig. \ref{JC_ladder}). More precisely, the light-matter eigenstates satisfy (with $n>0$): 
\begin{align}
 &|g \rangle= |\downarrow_z,0\rangle  \\ 
 &|n+\rangle=  \alpha_n  |\uparrow_z,n-1\rangle+\beta_n  |\downarrow_z,n\rangle \\          
 &|n-\rangle=     -\beta_n  |\uparrow_z,n-1\rangle+\alpha_n  |\downarrow_z,n\rangle,
\end{align}
where $|\downarrow_z\rangle$ and $|\uparrow_z\rangle$ are the two eigenstates of $\sigma^z$ with eigenvalues $-1$ and $+1$. The corresponding energies are:
\begin{align}
 &E_{|g \rangle}=\frac{\delta}{2}  \\ 
 &E_{|n+\rangle}=n \hbar \omega_0   +\frac{1}{2}\sqrt{\delta^2+ n g^2}   \label{Eplus}   \\         
 &E_{|n-\rangle}= n \hbar \omega_0 -\frac{1}{2}\sqrt{\delta^2+ n g^2}. \label{Emoins}
\end{align}
We have $\alpha_n=\sqrt{\left[A(n)-\delta\right]/2A(n)}$, $\beta_n=\sqrt{\left[A(n)+\delta\right]/2 A(n)}$, $A(n)=\sqrt{ng^2+\delta^2}$ and $\delta=\hbar\omega_0-\Delta$. 
 Let us consider the case of an atom initially excited in an empty cavity. Such a state is characterized by a wavefunction $|\psi\rangle$, which verifies $|\psi(t_0)\rangle=|\uparrow_z,0\rangle=\alpha_1 |1+\rangle-\beta_1 |1-\rangle$, where $t_0$ is the initial time. There is initially one excitation in the system, which will be conserved during the dynamics. As the initial state is a superposition of the first two polaritons, we will observe an oscillation phenomenon in terms of the atomic and photonic observables $\langle \sigma^z \rangle$ and $\langle a^{\dagger}a \rangle$. There is a periodic exchange of energy between the atom and the field, a phenomenon called Rabi oscillations. \\

We have seen that the Rabi model may be used to describe the interaction of the electromagnetic field with electronic energy levels in atoms. Recent progress in nanotechnology and electrical engineering have allowed to transpose this description of light-matter interaction from atomic physics to mesoscopic physics, with the development of the field of circuit quantum electrodynamics. In these setups, optical photons are replaced by microwave photons, while an association of superconducting elements effectively play the role of the atom. One important advantage of this class of devices is that they allow a great control over the system parameters, and one can reach higher values of the light-matter coupling $g/(\hbar\omega_0)$, opening the way to light-matter physics beyond the rotating wave approximation. These systems are moreover promising candidate to realize arrays of interacting light and matter elements, which would be useful both for quantum computation purposes and to simulate desired Hamiltonians. Let us now focus on how the Rabi model effectively describes the physical degrees of freedom of microwave light interacting with superconducting qubits

\subsection{Rabi Hamiltonian in circuit quantum electrodynamics}
\label{circuit_qed}

Superconductivity is a striking phenomenon occuring in certain materials below a characteristic temperature, notably associated with zero electric resistance and magnetic flux expulsion. It results from the condensation of pairs of electrons \cite{BCS:PR} (named Cooper pairs) into a common ground state below a certain temperature. In the superconducting regime, a nanoscale portion of a superconducting material can be described by a mesoscopic wavefunction with quantized energy levels, providing an ``artificial atom" of typical micrometer size.  \\

One key ingredient to describe effectively the physics of electronic levels in an atom is to provide a non-linear element (typically a Josepshon junction), so that the energy difference between successive energy levels is not the same. Several types of artificial atoms/qubits have been proposed in the past few years and we will focus on the example of the Cooper pair box \cite{Bouchiat:Physica_scripta}. We will focus on the setup proposed in Ref. \cite{Blais:PRB} where the role of the cavity is played by a section of a superconducting transmission line, acting like a resonator. Other superconducting circuits known as flux \cite{Lloyd:Science} or phase qubits \cite{martinis:PRL} provide equivalent systems.  \\

\subsubsection{Artificial atom: the Cooper pair box}
\begin{figure}[h!]
\center
\includegraphics[scale=0.45]{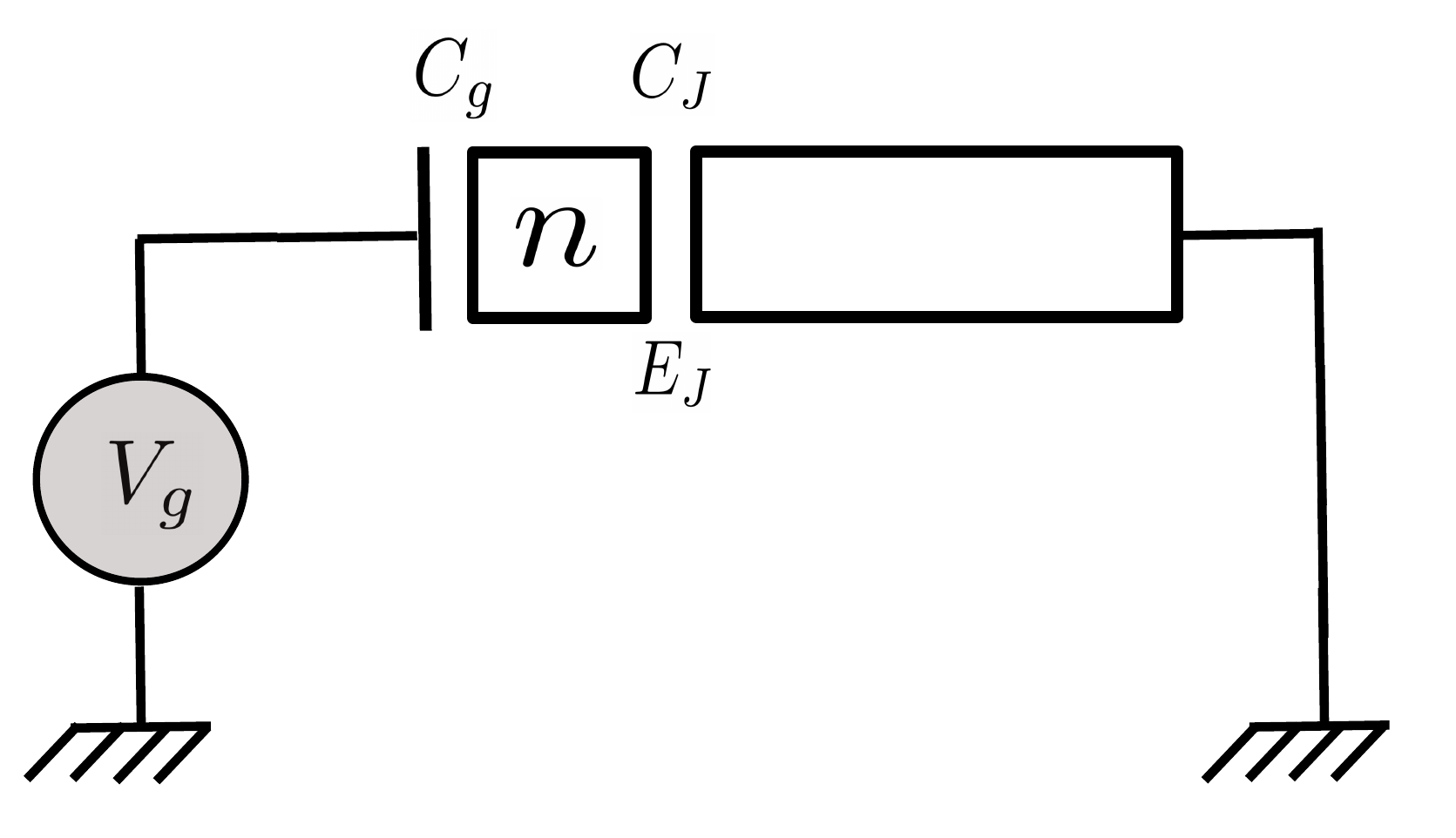}  
\caption{Cooper pair box.
}
\label{fig:Cooper_pair}
\end{figure}

The Cooper pair box, shown in Fig. (\ref{fig:Cooper_pair}), is a nanoscale superconducting circuit which was introduced in Ref. \cite{Bouchiat:Physica_scripta}. A mesoscopic superconducting island is connected on one side to a superconducting electron reservoir through a Josephson junction with energy $E_J$ and capacitance $C_J$, while the other end is connected to the gate voltage source through a capacitance $C_g$. In the superconducting regime, one can consider that all electrons in the island are paired. The superconductor can then be described by a single degree of freedom: the number $n$ of excess Cooper pairs. Using the natural basis associated with this observable one can write the Hamiltonian of the system as a sum of electrostatic and Josephson terms $\mathcal{H}=\mathcal{H}_{el}+\mathcal{H}_{J}$, with 
\begin{align}
\mathcal{H}_{el}&=E_C \sum_n (n-n_g)^2 |n\rangle \langle n |\\
\mathcal{H}_J &=\frac{E_J}{2} \sum_n \left(|n\rangle\langle n+1|+|n+1\rangle\langle n| \right).
\label{Hamiltonian_Cooper_pair_box} 
\end{align}
$E_C=(2e)^2/2(C_J+C_g)$ is the Coulomb energy of an extra Cooper pair on the island and $n_g=C_g V_g/(2e)$ is the dimensionless gate voltage. This Hamiltonian $\mathcal{H}$ leads to particularly simple behavior in the charge regime when the electrostatic energy dominates over the Josephson coupling.

\begin{figure}[h!]
\center
\includegraphics[scale=0.4]{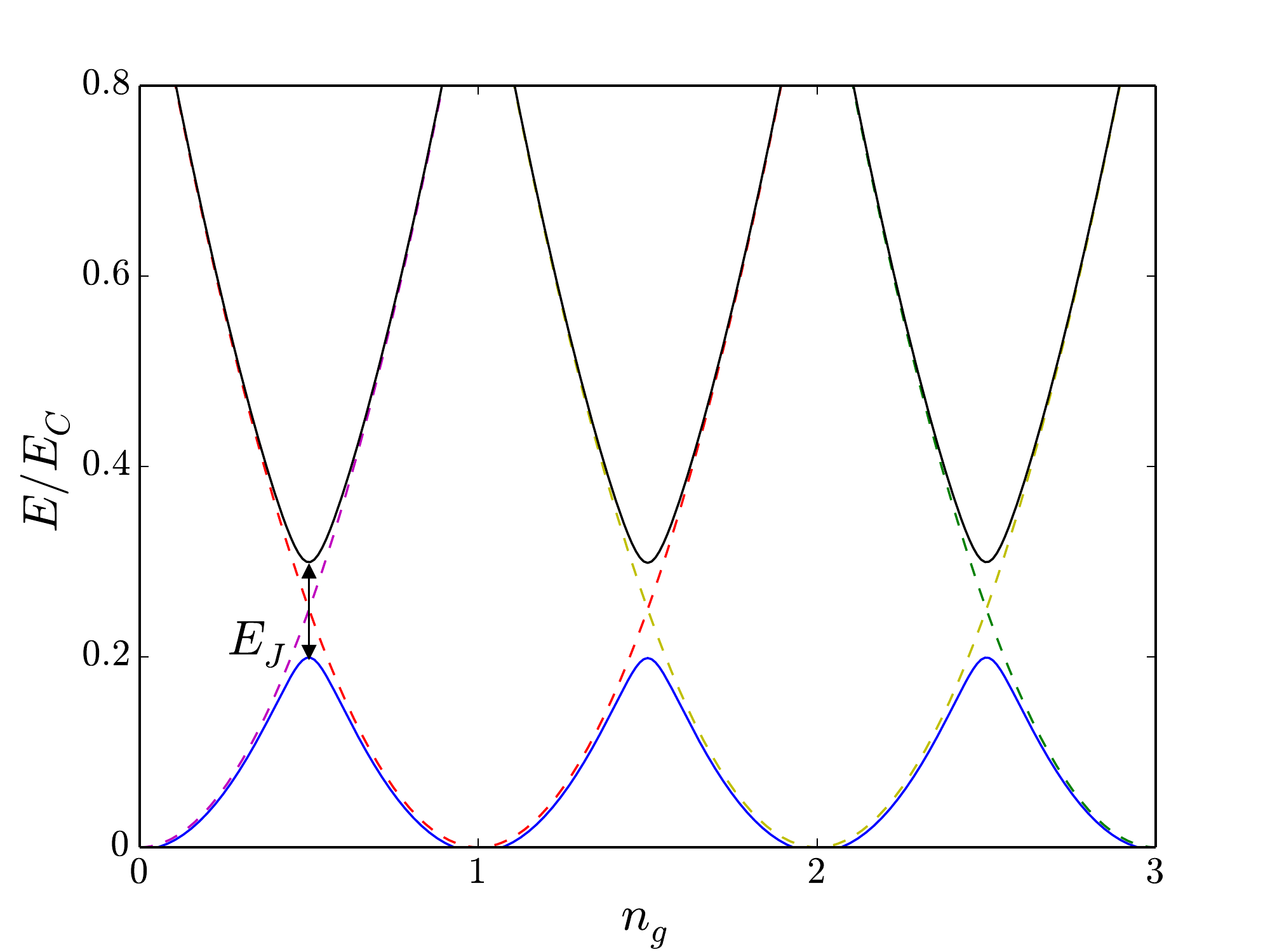}  
\caption{$E_{el}(n,n_g)/E_C$ as a function of $n_g$ for $n=0$ (dashed magenta), $n=1$ (dashed red), $n=2$ (dashed yellow) and $n=3$ (dashed green). The blue and black lines show the eigenvalues of the total Hamiltonian $\mathcal{H}_{CPB}$ (in units of $E_C$) as a function of $n_g$. Here $E_J=E_C/10$.
}
\label{fig:Cooper_pair_energies}
\end{figure}

When $n_g$ is a half-integer, the electromagnetic energies of the two states with $\lfloor n_g \rfloor$ and $\lfloor  n_g+1 \rfloor $ Cooper pairs are equal (where $\lfloor  a \rfloor $ denotes the integer part of $a$). The Josephson tunneling mixes these two states and opens up a gap in the spectrum (see Fig. \ref{fig:Cooper_pair_energies}). All other charge states have a much higher energy and will be neglected in the following. Near these degeneracy points, the Cooper pair box can be described as an effective two level system.
Let us consider for clarity that the dimensionless gate voltage $n_g$ is close to $1/2$, and we restrict the dynamics of the system to the two states with either $0$ or $1$ excess Cooper pairs. In this basis, one can write $\mathcal{H}_{CPB}=E_J \frac{\sigma^x}{2}-E_C (1-2n_g)\frac{\sigma^z}{2}$ up to a constant term, where $\sigma^x=|0\rangle\langle1|+|1\rangle\langle 0|$ and $\sigma^z=|0\rangle \langle0|-|1\rangle \langle 1|$.\\

The effective field along the $z$-direction can be controlled dynamically by the gate voltage $V_g$. A magnetic control of the field along the $x$-direction is also possible if one replaces the Josephson junction by a pair of junctions in parallel, each with energy $E_J/2$ (see \cite{Blais:PRB}). In this case the transverse field becomes $E_J \cos (\pi \phi/\phi_0)$, where $\phi$ is the magnetic flux in the loop formed by the two junctions and $\phi_0=h/2e$ is the flux quantum. \\

\label{cqed} 
%These quantum degrees of freedom can couple with an electromagnetic environment in different ways, allowing to reproduce on-chip the physics of quantum electrodynamics.
\subsubsection{Cavity: The superconducting resonator}
\label{resonator_section}
\begin{figure}[h!]
\center
\includegraphics[scale=0.2]{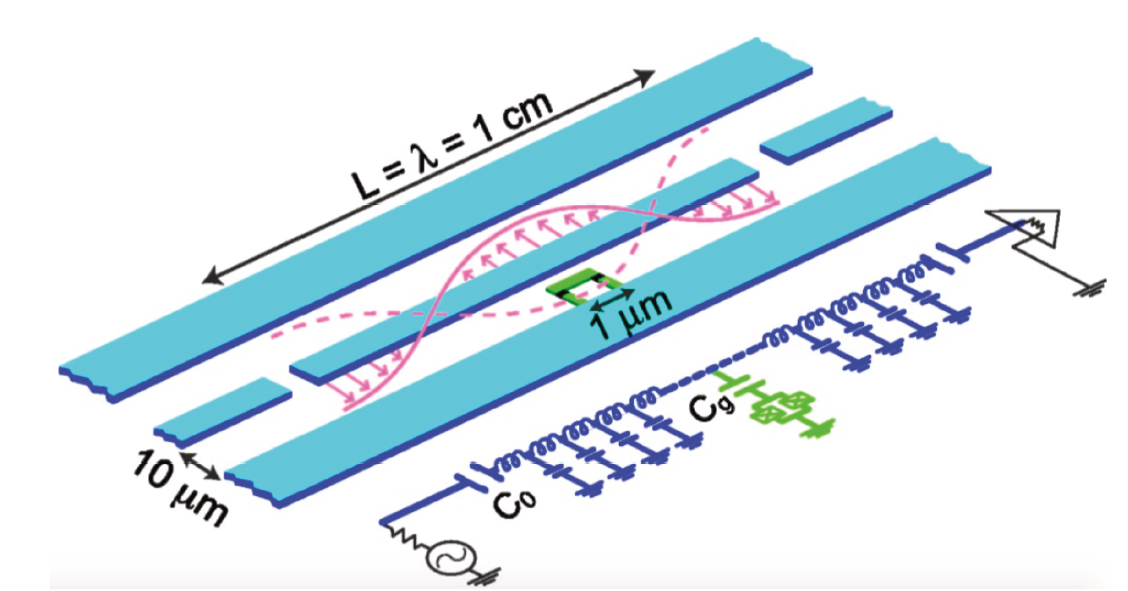}  
\caption{Schematic and circuit representation of a superconducting circuit quantum electrodynamics setup. A Cooper pair box with two Josephson elements in parallel (in green) is placed inside a 1D transmission line resonator (in blue) consisting of a superconducting coplanar waveguide. Figure from Ref. \cite{Blais:PRB}.
}
\label{resonator}
\end{figure}

The circuit-equivalent of the cavity consists in a section of a superconducting transmission line represented in Fig. \ref{resonator}. In the setup of Ref. \cite{Blais:PRB}, the artificial atom is directly placed at the center of the transmission line. The gate voltage acquires then a quantum part $v$, stemming from the fluctuations of the electromagnetic field inside the transmission line.
The electromagnetic Hamiltonian $\mathcal{H}_{el}$ in Eq. (\ref{Hamiltonian_Cooper_pair_box}) is replaced by
\begin{align}
\mathcal{H}_{el}&=E_C \sum_n \left(n-n^{dc}_g-\frac{C_g v}{2e}\right)^2 |n\rangle \langle n |,
\label{Hamiltonian_CPB_inside_resonator}
\end{align}
where $n^{dc}_g$ represents the direct current part of the gate voltage. When $n^{dc}_g$ is close to $1/2$, the restriction to the two relevant states with either $0$ or $1$ excess Cooper pairs gives
\begin{align}
\mathcal{H}_{el}&=-E_C (1-2n^{dc}_g)\frac{\sigma^z}{2}+e\frac{C_g}{C_J+C_g}v \frac{\sigma^z}{2}+E_C\left(\frac{C_g v}{2e}\right)^2.
\label{Hamiltonian_CPB_inside_resonator_2}
\end{align}
The second term on the right hand side of Eq. (\ref{Hamiltonian_CPB_inside_resonator_2}) couples the two-level system and the electromagnetic field in the transmission line. For a finite length $L$ (and a large quality factor), the transmission line acts as a resonator with resonant frequencies. To show this, we model the transmission line as an infinite series of inductors (see Fig. \ref{resonator}), where each node is capacitively coupled to the ground (see Fig. \ref{resonator}). The system is then well described by the charge and flux operators at node $n$, $Q_n$ and $\phi_n$ (see \cite{Devoret:les_houches}). These operators obey canonical commutation relations $[\phi_n,Q_n]=i \hbar \delta_{m,n}$. \\

The Hamiltonian $\mathcal{H}_{TL}$ of the isolated transmission line reads in the continuum limit
\begin{align}
\mathcal{H}_{TL}&=\int_{-L/2}^{L/2} dx \left( \frac{Q^2}{2c}+\frac{(\nabla \phi)^2}{l} \right),
\label{Hamiltonian_resonator}
\end{align}
where $l$ and $c$ are the inductance and capacitance per unit length. The charge neutrality imposes that $Q$ have anti-nodes at the boundaries $x=\pm L/2$. Diagonalization leads to \cite{Blais:PRB},
\begin{align}
\mathcal{H}_{TL}&=\sum_{p=1}^{\infty}\left[ \omega_{2p}\left( \hbar a^{\dagger}_{2p}a_{2p}+1/2\right)+\hbar \omega_{2p-1}\left(a^{\dagger}_{2p-1}a_{2p-1}+1/2\right)  \right],
\end{align}
where $\omega_j = j\pi/(L \sqrt{lc})$. Even modes have voltage anti-nodes at $x=0$, while odd modes have voltage nodes at $x=0$, so that only the even modes participate to the voltage $v$ coupled to the Cooper pair box at $x=0$. If the energy of the mode $p=2$ is close to $E_J $, one can neglect the modes of higher energy (note that the last term on the right hand side of Eq. \ref{Hamiltonian_CPB_inside_resonator_2} renormalizes $\omega_2$). After a $\pi/2$ rotation in the spin space, we recover the Rabi Hamiltonian $\mathcal{H}_{Rabi}$ from Eq. (\ref{Rabi}) for $n_g=1/2$, with the identification: 
\begin{align}
&\omega_0=\omega_2\\
&\Delta=E_J\\
&g=e\frac{C_g}{C_J+C_g}\sqrt{\frac{\hbar\omega_2}{cL}}.
\label{identification_Rabi}
\end{align}\\

These 1D circuit-QED systems allow to reach larger values of light matter coupling $g/(\hbar\omega)$ compared to their 3D cavity analogues \cite{Blais:PRB} and constitutes one the most promising architecture for quantum computation.

\subsubsection{Strong coupling regime beyond the RWA}
\label{strong_coupling}

Light-matter coupling in these circuit QED experiments can reach $g/(\hbar\omega_0) \simeq 10^{-1}$ \cite{Mooij}. In this regime, the rotating wave approximation breaks down and one must take into account the presence of the counter-rotating terms $(\sigma_+ a^{\dagger}+\sigma_- a)$. The effect of such terms can be understood in perturbation theory, and they give rise to a shift of the resonance frequency between the atom and photon, leading to an additional negative detuning in the energies plap (\ref{Eplus}) and (\ref{Emoins})
\begin{align}
\delta \to \delta -g^2/\left[2(\hbar\omega_0+\Delta)\right],
\end{align}
 when $\Delta<(\hbar\omega_0)$ (the levels repel each other). This so-called Bloch-Siegert shift \cite{cohen_2} has been observed in Ref. \cite{Mooij}.\\

\begin{figure}[h!]
\center
\includegraphics[scale=0.2]{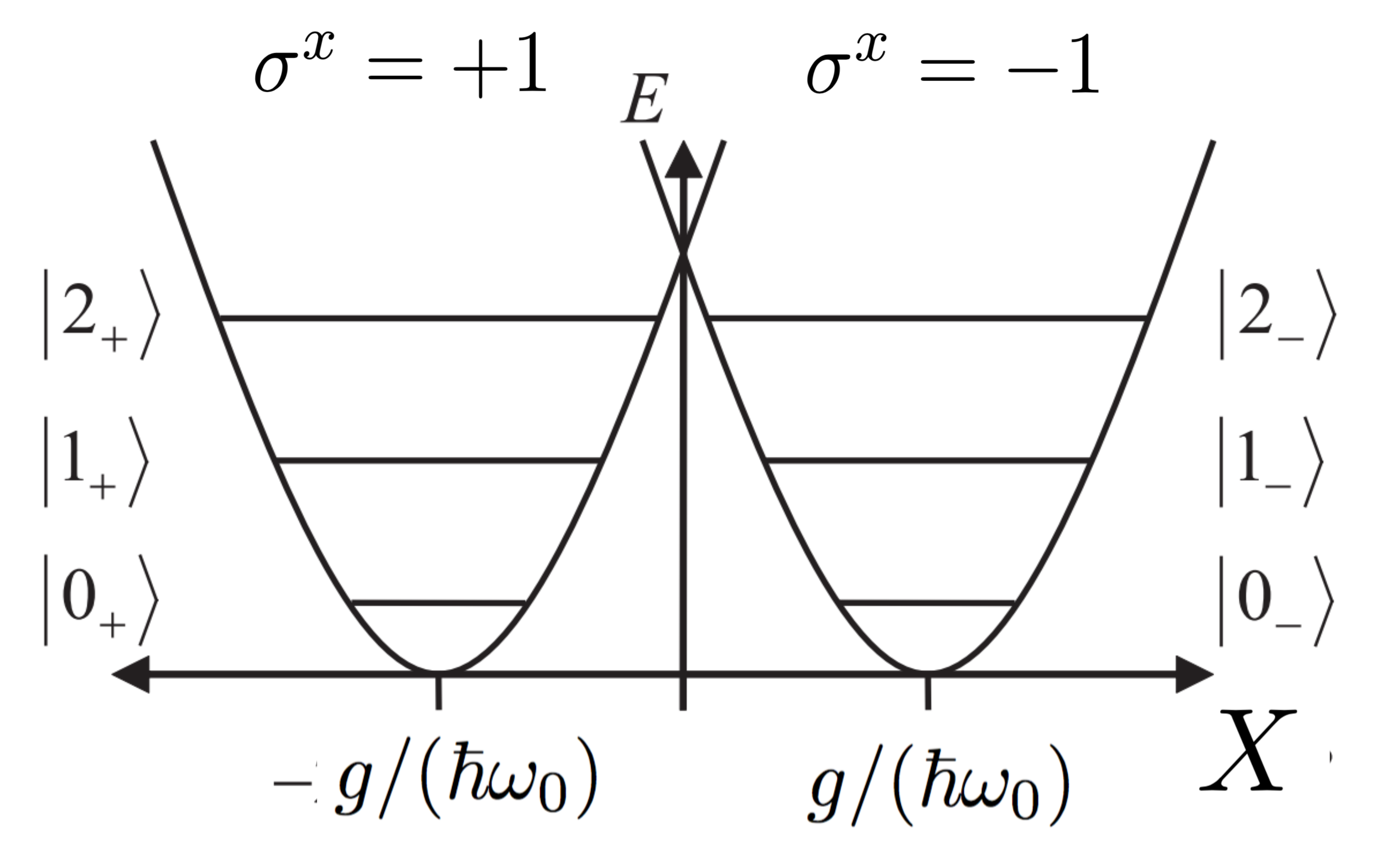}  
\caption{Two displaced wells characterized by the value of $\sigma^x$. This correspond to the limiting case $\Delta=0$. At first order in $\Delta/(\hbar \omega_0)$, one can consider that the term in $\sigma^z$ mixes only states corresponding to the same number of excitations of each well. The energy splitting resulting from this term will be proportional to $\Delta$ and to the overlap $\langle N_- | N_+\rangle=e^{-g^2/2(\hbar \omega_0)^2} L_N [g/(\hbar \omega_0)]$.
}
\label{fig_adiabatic}
\end{figure}

Another solvable limit, relevant to the strong coupling analysis of the Rabi model, is the so-called adiabatic regime \cite{adiabatic_0,adiabatic_1} or "quasi-degenerate limit" \cite{adiabatic_2,adiabatic_3,adiabatic_4}. This regime corresponds to a highly detuned system $\Delta/(\hbar \omega_0) \ll 1$ with arbitrary large light-matter coupling. One can visualize such a limit as a set of two displaced oscillator wells (characterized by the value of $\sigma^x$), whose degeneracy is lifted by the field along the $z$-direction (see Fig. \ref{fig_adiabatic}). At $\Delta=0$, one can diagonalize separately the bosonic Hamiltonian, for the two values of $\sigma^x$, by looking for eigenstates $|\phi_{\pm}\rangle$ with eigenenergies $E_{\pm}$:
\begin{align}
&\left[\hbar \omega_0 a^{\dagger} a \pm \frac{g}{2}(a+a^{\dagger})\right]|\phi_{\pm}\rangle=E_{\pm}|\phi_{\pm}\rangle.\notag \\
\Rightarrow &\left[ \left(a^{\dagger}\pm \frac{g}{2\omega_0} \right) \left( a \pm \frac{g}{2\omega_0}\right)\right]|\phi_{\pm}\rangle=\left[\frac{E_{\pm}}{\hbar \omega_0}+\left(\frac{g}{2\omega_0}\right)^2\right]|\phi_{\pm}\rangle.
\label{adiabatic_square}
\end{align}
The operator on the left-hand side of Equation (\ref{adiabatic_square}) can be re-expressed as\linebreak $D(\mp g/2\omega_0) a^{\dagger}a D^{\dagger}(\mp g/2\omega_0)$, where $D(\nu)=\exp [\nu (a^{\dagger}-a)]$ is a displacement operator \cite{Grynberg_Aspect_Fabre}. This permits to find that eigenstates $|\phi_{\pm}\rangle$ correspond to displaced number states
\begin{align}
&|\phi_{\pm,N}\rangle=e^{\mp \frac{g}{2\omega_0}(a^{\dagger}-a)}|N\rangle \equiv |N_{\pm}\rangle, \label{adiabatic_eigenstate}\\
&E_{\pm,N}=\hbar \omega_0 \left[N-\left(\frac{g}{2\omega_0}\right)^2\right].\label{adiabatic_eigenenergy}
\end{align}
In Eq. (\ref{adiabatic_eigenstate}), $|N\rangle $ is a standard number state. The case $N=0$ corresponds to well-known coherent states. \\

 The lowest order adiabatic approximation consists in considering that the term $\Delta/2 \sigma^z$ only couples states in opposite wells with the same number of excitations. The Hamiltonian can then be diagonalized as the number of displaced photons is a conserved quantity. A system initially prepared in a displaced state of one well would undergo coherent and complete oscillations between this state and its symmetric counterpart in the other well. The frequency of oscillations only depends on the overlap between these two states, and one can show that, starting from the $N$-th displaced state of one well, this frequency is $\Omega=\Delta e^{-g^2/2(\hbar \omega_0)^2} L_N [g/(\hbar \omega_0)]$, where $L_N$ is the $N$-th Laguerre Polynomial \cite{adiabatic_1}. \\

In these first two sections, we have explored two solvable limits of the Rabi model:
\begin{itemize}
\item  The Jaynes-Cummings regime corresponds to the limit $g/(\hbar\omega_0) \ll 1$ where the RWA holds. The corresponding Hamiltonian (\ref{JC}) can be diagonalized in terms of mixed light-matter excitations named polaritons.
\item	The adiabatic regime corresponds to the limit $\Delta/(\hbar \omega_0) \ll 1$. In this regime, the Hamiltonian can be diagonalized in terms of superpositions of displaced number states.
\end{itemize}

Between these two clear limits one can use standard numerical methods, which allow to explore the time-evolution of dynamical quantities. Recently, analytical solutions of the quantum Rabi model have been explored in Refs. \cite{braak,moroz,zhong,gritsev}, based on the underlying discrete $\mathbb{Z}_2$ (parity) symmetry. Some dynamical properties of the model have notably been addressed \cite{braak2} thanks to this exact solvability. The link between the exact solvability and integrability of the Rabi model with related models such as the Dicke \cite{Dicke} model have also been studied \cite{batchelor}. Dicke model, which is the $N$ spins version of the Rabi model, is notably known to exhibit a superradiant quantum phase transition in the limit $N\to\infty$ \cite{Emary_brandes}. By contrast to the Dicke model, the isolated quantum Rabi model does not exhibit a superradiant quantum phase transition when increasing the value of $g/(\hbar\omega_0)$. This is illustrated on the left side of Fig. \ref{JC_ladder} where we see that the ground state and the first excited state do not cross when one increases the coupling $g/(\hbar \omega_0)$, somehow illustrating the stability of an atom interacting with a quantum field\footnote{It is however important to note that the Rabi ground state contains a non-trivial photonic component at large coupling.}. This must be contrasted by what would be predicted by the Jaynes-Cummings model, for which crossings occur when $g/(\hbar \omega_0)$ increases, leading to unphysical polariton-like ground state.  \\
%General solutions of the quantum Rabi model.\\

\subsection{Many-body physics in large circuits}
In the past few years experimental progress was accomplished towards the realization of large arrays of circuit QED elements, where two neighbouring cavities $i$ and $j$ can be coupled using a capacitive or Josephson of the form $J(a_i+a_i^{\dagger})(a_j+a_j^{\dagger})$. Assuming that the coupling $J$ between cavities is much smaller than the internal frequencies of the cavities, the coupling turns into an effective hopping term from one cavity to the neighboring ones.  These architectures provide a robust architecture for solid-state quantum computation because of the unprecedented control over the system parameters and the low loss level. From a different viewpoint, they open a new way to explore many-body quantum systems in a precise and controllable manner.
\subsubsection{Arrays of cavities}
 From this perspective, one model that has been investigated a lot theoretically in the literature is the Jaynes-Cummings lattice model, which takes the form \cite{Angelakis_PRA,Hartmann,houck}:
\begin{equation}
\mathcal{H}_{array} = \sum_i \mathcal{H}_{JC} ^{i} - J\sum_{\langle i;j\rangle} \left(a^{\dagger}_i a_j +  a^{\dagger}_j a_i\right)-\mu_{eff}\sum_i \left(a^{\dagger}_i a_i + \sigma_i^+ \sigma_i^-\right),
\label{JC_array}
\end{equation}
where $\mathcal{H}_{JC} ^{i}$ describes the Jaynes-Cummings Hamiltonian in each cavity, introduced in Eq. (\ref{JC}), and the summation is over nearest neighbors. This model has attracted some attention, for example, in the light of realizing analogues of Mott insulators with polaritons \cite{Greentree,KochHur,SchmidtBlatter,SchmidtBlatter1}. More precisely, one observes two competing effects in these arrays. For large values of $J$, formally the system would tend to form a wave delocalized over the full lattice, by analogy to a polariton superfluid \cite{CiutiCarusotto}, whereas for very small values of $J$, the photon blockade in each cavity will play some role.  Theoretical works \cite{Greentree,SchmidtBlatter,SchmidtBlatter1,KochHur,KLHQPT}, have focused on solving the phase diagram at equilibrium in the presence of a tunable chemical potential $\mu_{eff}$.\\

 For $\mu_{eff}=0$, the ground state at small $J$ would correspond to the vacuum in each cavity. By changing $\mu_{eff}$ and keeping $J$ small, one could eventually turn the vacuum in each cavity into a polariton state. Simple energetic arguments predict that this change would occur when $E_{|1-\rangle} - \mu_{eff} = E_{|g\rangle}$. This result can be made more formal by using a mean-field theory and a strong-coupling expansion \cite{Greentree,KochHur,SchmidtBlatter,SchmidtBlatter1}. In the atomic limit where  $J$ is small, one then predicts the analogue of Mott-insulating incompressible phases, as observed in ultra-cold atoms \cite{Greiner}, where it costs a finite energy to change the polariton number \cite{Greentree,Hartmann,Fazio} (see Fig. \ref{fig_adiabatic_phase_diag}). By increasing the hopping $J$, one can build an equivalent of the $\psi^4$  theory, where $\psi\sim \langle a_i\rangle$, in order to describe the second-order quantum phase transition between the Mott region of polaritons and the superfluid limit \cite{KochHur}. \\

\subsubsection{Driven-dissipative problem}
Despite its great theoretical interest, this model is not readily implementable because no tunable chemical potential exists for photons. Similar to ultra-cold atoms \cite{coldatomreview,coldatom2}, it seems important to be able to engineer an effective chemical potential to develop further quantum simulation proposals and observe these Mott phase analogues. In Ref. \cite{Hafezi}, Hafezi and co-workers proposed to simulate an effective chemical potential for photons from a parametric coupling with a bath of the form $\lambda\cos(\omega_p t)H_{SB}$ where $H_{SB}=\sum_j(a_j+a^{\dagger}_j)B_j$ where $B_j$ is a bath operator. Considerable attention has been turned at a general level towards describing both driven and dissipative cavity arrays in order to realize novel steady states.\\

In realistic experimental conditions, photon leakage out of the system must indeed be taken into account. Each cavity is exposed to the vaccuum noise of the surrounding environment, and energy can leak out from the system. This effect can be addressed in a microscopic manner by considering a coupling of the inner photonic modes to an infinite number of external bosonic modes, so that one must add to the system Hamiltonian the term (within a rotating-wave approximation) \cite{Clerk2}:
\begin{align}
\sum_q \hbar \omega_q l^{\dagger}_q l_q-i \sum_q \left[ f_q a_i^{\dagger}l_q +f_q^* a_i l^{\dagger}_q  \right],
\end{align}
where $l^{\dagger}_q$ ($l_q$) is the creation (annihilation) operator of an external boson of frequency $\omega_q$. The use of the Heisenberg equations of motion in the Markov approximation, which assumes that the coupling strength $f=\sqrt{|f_q|}$ and the density of state $\rho=\sum_q \delta(\omega-\omega_q)$ are constant, allows to write the effect of the environment as a imaginary component $\Gamma=2 \pi f^2 \rho$ for the photon frequency \cite{Clerk2,Marco:PRB}. \\

In these conditions the system will eventually reach the ground state after a typical time of the order $1/\Gamma$. To compensate this relaxation mechanism and access non-trivial states of matter, energy is often pumped into the system through the intermediary of an external time-dependent coherent drive on the cavity $i$ \cite{bishop,bishop2}, of the form $V_i(t) (a_i+a_i^{\dagger})$. One important point would then be to understand how the interplay of drive and dissipation could play the role, or replace, the chemical potential $\mu_{eff}$ in Eq. (\ref{JC_array}).\\

Let us study the effect of this driving term on the non-dissipative Jaynes-Cummings model $(\ref{JC_drived})$.
\begin{equation}
\mathcal{H}_{JC}^{d}=\mathcal{H}_{JC}+\frac{V_0}{2} (a e^{i\omega_d t}+a^{\dagger } e^{-i\omega_d t}) .
\label{JC_drived}
\end{equation}
 We can get rid of the time-dependent part of the Hamiltonian through a unitary transformation $ | \tilde{\psi} \rangle=U(t) | \psi \rangle$ with $U(t)=\exp\left[ i \omega_d (a^{\dagger} a+\sigma_{+}\sigma_{-}) t\right]$. The evolution of $| \tilde{\psi} \rangle$ is governed by the time-independent Hamiltonian $\tilde{\mathcal{H}}_{sys}^{RWA}$:
\begin{equation}
\tilde{\mathcal{H}}_{sys}^{RWA}=\frac{\tilde{\Delta}}{2} \sigma^z+ \tilde{\omega}_0 a^{\dagger} a+ \frac{g}{2} (\sigma_{+} a+\sigma_{-} a^{\dagger})+\frac{V_0}{2} (a +a^{\dagger } ),
\label{drived_h}
\end{equation}
where $\tilde{\Delta}=\Delta-\omega_d$ and $\tilde{\omega}_0=\omega_0-\omega_d$. $\tilde{H}$ is the sum of a JC Hamiltonian with renormalized energies $\tilde{\Delta}$ and $\tilde{\omega}_0$, and a time independent driving term. It is convenient to express the last term of Eq. (\ref{drived_h}) in the dressed basis $\mathcal{B}=\{|g\rangle,|1,-\rangle,|1,+\rangle,|2,-\rangle,|2+\rangle,... \}$ of the coupled system \cite{driving_dressed_basis}

\begin{align}
a +a^{\dagger }=&\ \beta_1 |1,+\rangle\langle g |+\alpha_1 |1,-\rangle\langle g | \notag \\
+\sum_{n=1}^{\infty}& \left[\sqrt{n+1} \beta_n \beta_{n+1}+\sqrt{n} \alpha_n \alpha_{n+1}\right] |n+1,+\rangle\langle n,+|  \notag \\
+& \left[\sqrt{n} \beta_n \beta_{n+1}+\sqrt{n+1} \alpha_n \alpha_{n+1}\right]         |n+1,-\rangle\langle n, -|  \notag \\
+& \left[\sqrt{n+1} \alpha_{n+1} \beta_{n}-\sqrt{n}\alpha_n \beta_{n+1}\right]     |n+1,-\rangle\langle n,+| \notag \\
 +&\left[\sqrt{n+1} \beta_{n+1} \alpha_{n}-\sqrt{n} \alpha_{n+1} \beta_{n}\right]  |n+1,+\rangle\langle n,-|   \notag \\
 +&\textrm{h.c.}.
 \label{transition_dressed_states}
\end{align}
Driving the cavity induces transition between the dressed states, and changes the number $N$ of excitations by 
$\pm1$. As can be seen in Fig. (\ref{JC_ladder}), the Jaynes Cummings ladder is composed of two subladders of `minus' and `plus' polaritons. Eq. (\ref{transition_dressed_states}) illustrates the fact that the coupling between states of the same sub-ladder is  stronger than the coupling between states which belong to different sub-ladders. \\

The interplay of drive, dissipation, and lattice effects make difficult the realization of an effective chemical potential $\mu_{eff}$, motivating the development of numerical schemes allowing to tackle the real-time dynamics of such driven dissipative models. One may mention the development of a nonequilibrium extension of stochastic mean-field theory in Ref. \cite{Keeling_sto}, applicable to problems of coupled cavities with rather general forms of driving and dissipation. \\

\subsubsection{Mean field quantum phase transition in the adiabatic regime}
Alternatively, one could take advantage of the physical properties of the strong coupling regime to reach non-trivial phases with finite photon density. We have indeed remarked in Sec.  \ref{strong_coupling} that low energy states of the adiabatic regime correspond to a superposition of displaced number states for the photons. The non-triviality of the photon ground state results more generally from the presence of counter-rotating terms in the Hamiltonian. M. Schir\'{o} and coworkers followed this promising route in Refs. \cite{marco_rabi_1,marco_rabi_2} where they studied the effect of counter rotating terms on the ground state properties of Hamiltonian (\ref{JC_array}) without chemical potential. Interestingly, these counter-rotating terms drive the system accross a $Z_2$ parity breaking quantum phase transition, associated with the disparition of the multiple Mott lobes associated with Hamiltonian (\ref{JC_array}). We illustrate the mean-field transition exposed in \cite{marco_rabi_1,marco_rabi_2}, for which the low energy theory in the adiabatic regime corresponds to a transverse field anisotropic Ising model . We focus on the following Hamiltonian,
\begin{align}
\mathcal{H}=\sum_j H_{Rabi}^{j}- \sum_{\langle i,j \rangle} J(a_i+a_i^{\dagger})(a_j+a_j^{\dagger}),\\
\label{Model_adiabatic}
\end{align}
with $H_{Rabi}^{j}$ given by Eq. (\ref{Rabi}). We keep the general capacitive coupling of the form $(a_i+a_i^{\dagger})(a_j+a_j^{\dagger})$, which is valid for $J/\omega_0$ not too small. We consider that each site has $z$ neighbours and study the adiabatic regime characterized by $\Delta/(\hbar \omega_0) \ll 1$ (see Sec. \ref{strong_coupling} and references therein). We follow Refs. \cite{marco_rabi_1,marco_rabi_2} and decouple photon hopping at a mean field level, $(a_i+a_i^{\dagger})(a_j+a_j^{\dagger})=\langle a_i+a_i^{\dagger}\rangle(a_j+a_j^{\dagger})+( a_i+a_i^{\dagger}) \langle a_j+a_j^{\dagger}\rangle- \langle a_i+a_i^{\dagger}\rangle \langle a_j+a_j^{\dagger}\rangle$, which is exact in the limit $z\to \infty$. We are then left with the mean-field Hamiltonian,
\begin{align}
\mathcal{H}_{MF}^{j}=\frac{\Delta}{2} \sigma_j^z+ \left(\frac{g}{2}\sigma_j^x+J \psi \right) (a_j+a_j^{\dagger})+ \hbar \omega_0 a_j^{\dagger} a_j,
\label{Model_MF}
\end{align}
where $\psi=z \langle a +a^{\dagger}\rangle$. \\

\begin{figure}[h!]
\center
\includegraphics[scale=0.3]{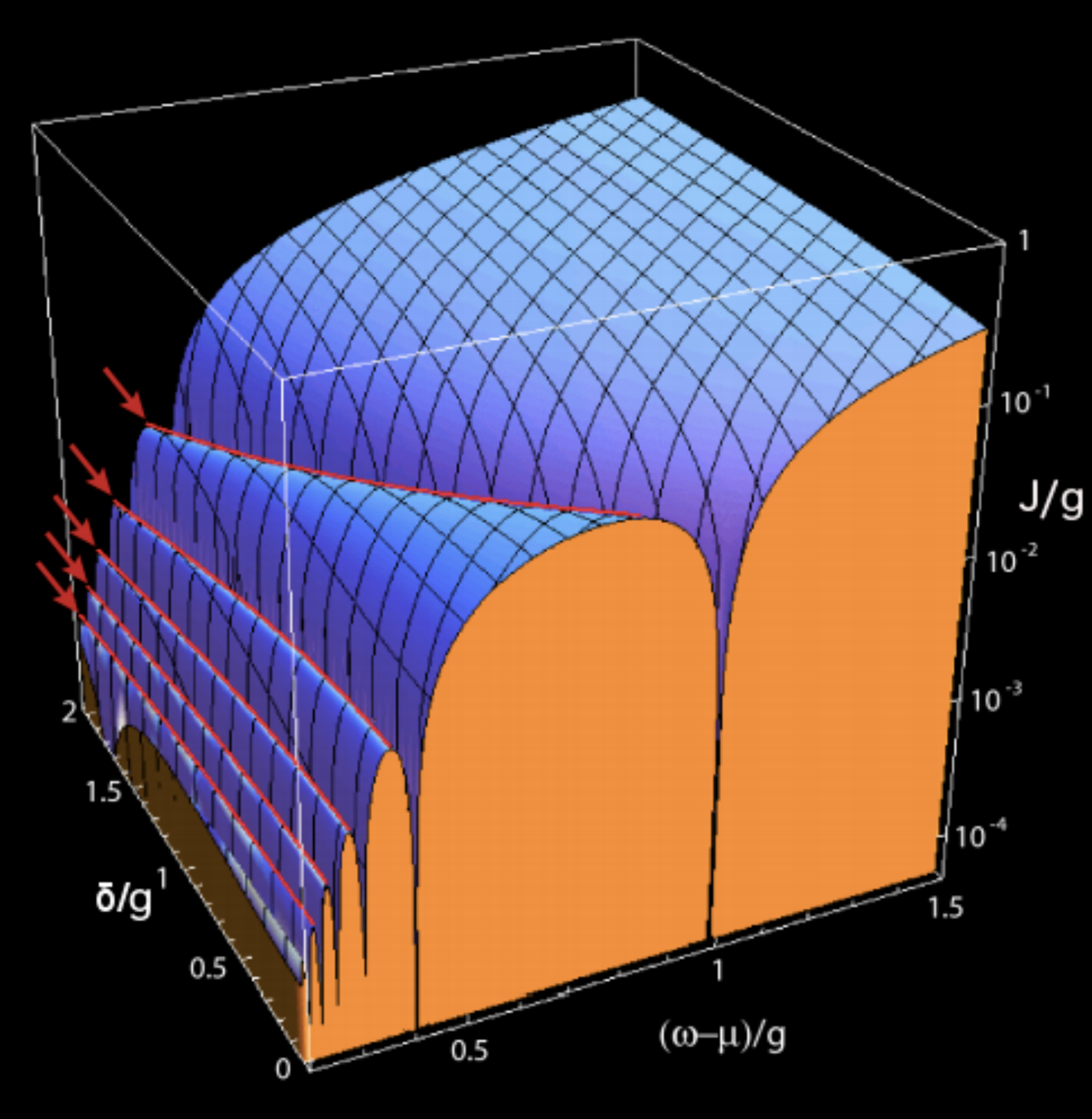}\includegraphics[scale=0.35]{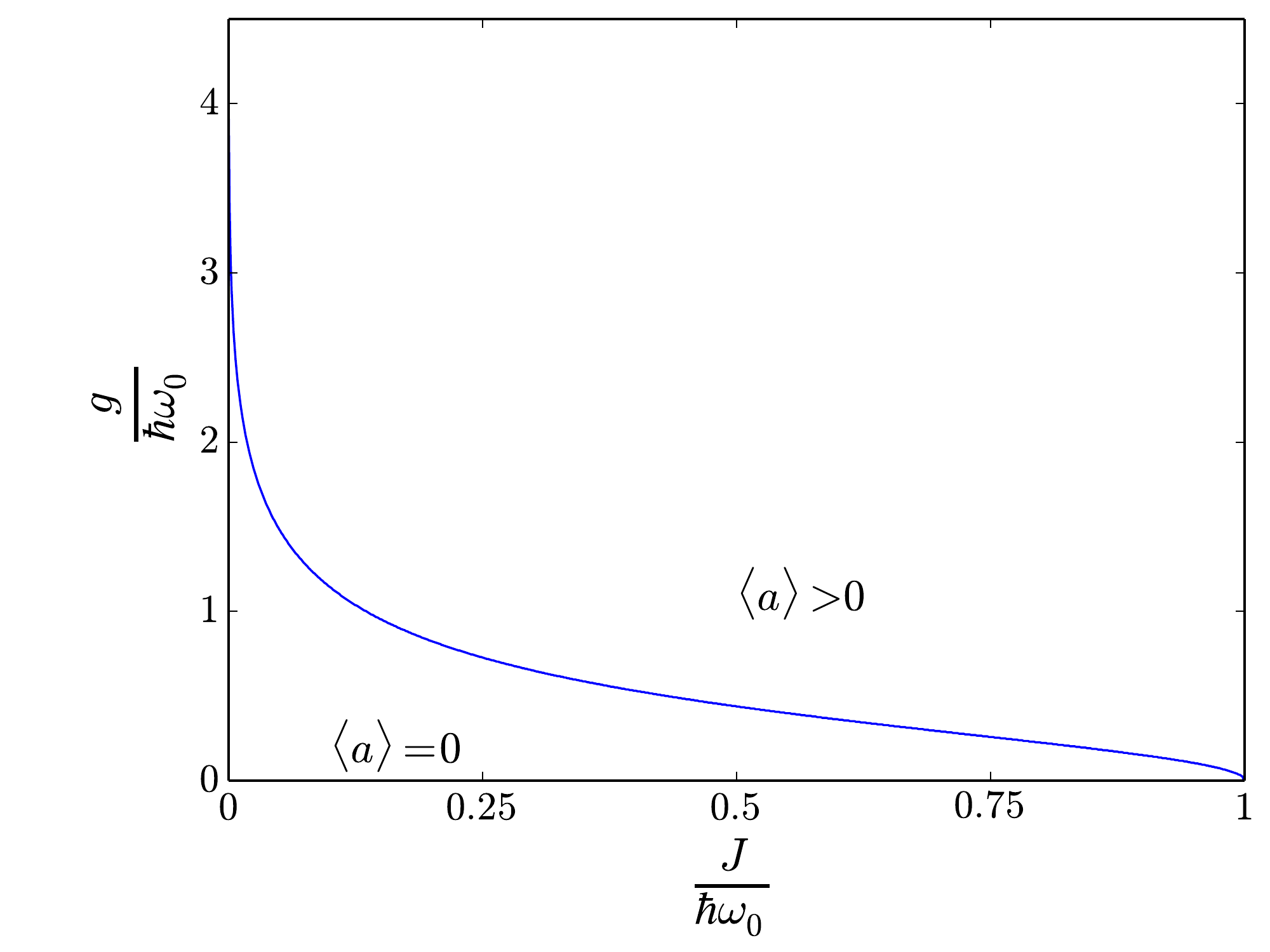}  
\caption{Left: Mean-field phase boundary of the Jaynes-Cummings lattice model as a function of
effective chemical potential $(\mu_{eff}-\hbar\omega_0)$ (we take $\hbar=1$ for the axis label) and atom-resonator detuning $\delta/g$. Figure from Ref. \cite{KochHur}. Right: Transition line for $\Delta/\omega_0=0.1$ in the case of the Rabi-Hubbard lattice. It reproduces the results of Ref. \cite{marco_rabi_1,marco_rabi_2}.
}
\label{fig_adiabatic_phase_diag}
\end{figure}

Studying the adiabatic regime characterized by $\Delta/(\hbar \omega_0) \ll 1$ allows now to precisely describe the phase transition. One can visualize the mean-field system as two displaced oscillator wells (characterized by the value of $\sigma^x$ and $\psi$). This permits to find that eigenstates $|\phi_{\pm}(\psi)\rangle$,
\begin{align}
&|\phi_{\pm,N}(\psi)\rangle=e^{\left(\mp\frac{g}{2 \hbar \omega_0}+\frac{J \psi}{ \hbar \omega_0}\right)(a^{\dagger}-a)}|N\rangle \equiv |N_{\pm}\rangle, \label{adiabatic_eigenstate_1}\\
&E_{\pm,N}=\hbar \omega_0 \left[N-\left(\pm\frac{g}{2\hbar \omega_0}-\frac{J\psi}{\hbar \omega_0}\right)^2\right].\label{adiabatic_eigenenergy_1}
\end{align}
As in Sec. \ref{strong_coupling}, we consider that the term $\Delta/2 \sigma^z$ only couples states in opposite wells with the same number of excitations, which is the lowest order of the adiabatic approximation. For all $\psi$, the minimal energy corresponds to (up to a constant term),
\begin{align}
&\langle\mathcal{H}_{MF}^{j}\rangle_{min}(\psi)=-\hbar \omega_0 \left[ \left(\frac{g}{2\hbar \omega_0}\right)^2+ \left(\frac{J \psi}{\hbar \omega_0}\right)^2 \right]-\left[\Delta^2 e^{-\frac{g^2}{2(\hbar\omega_0)^2}}+\frac{g^2}{(\hbar\omega_0)^2}J^2 \psi^2 \right]^{1/2}.
\end{align}
Minimization of $J \psi^2+\langle\mathcal{H}_{MF}^{j}\rangle_{min}(\psi)$ with respect to $\psi$ allows to determine the transition line of the mean-field transition (see Fig. \ref{fig_adiabatic_phase_diag}). One finds a critical line
\begin{align}
J_c \left(\frac{g}{\hbar\omega_0}\right)=\frac{\Delta \exp\left[-\frac{g^2}{2(\hbar\omega_0)^2}\right]}{\frac{\Delta \exp\left[-\frac{g^2}{2(\hbar\omega_0)^2}\right]}{\hbar\omega_0}+2 \left(\frac{g}{\hbar\omega_0} \right)^2}.
\end{align}
The order parameter $\psi$ is equal to zero below $J_c$ and we find for $J>J_c$
\begin{align}
|\psi|=\alpha (J-J_c)^{1/2}
\end{align}
with
\begin{align}
\alpha=\frac{\left\{2\left(\frac{g}{\hbar\omega_0}\right)^2 J+ \Delta \exp\left[-\frac{g^2}{2(\hbar\omega_0)^2}\right] \left( 1-\frac{J}{\hbar\omega_0}\right) \right\}^2}{2\frac{g}{\hbar\omega_0}J \left(1-\frac{J}{\hbar\omega_0} \right)\left\{\frac{\Delta \exp\left[-\frac{g^2}{2(\hbar\omega_0)^2}\right]}{\hbar\omega_0}+2 \left(\frac{g}{\hbar\omega_0} \right)^2 \right\}^{1/2}}.
\end{align}\\

This analysis was formulated in Refs. \cite{marco_rabi_1,marco_rabi_2} in terms of an effective transverse field anisotropic Ising model, where the Ising operators refer to the adiabatic states exposed above. The critical hopping $J_c$ at large $g/(\hbar \omega_0)$ is exponentially small, because the transverse field is renormalized by a Franck-Condon like exponential factor while the exchange not. In Ref. \cite{PRL_schiro_tureci}, the authors considered the effect of photon losses on the phase boundary. Interestingly, photon losses favours the disordered trivial phase and the phase boundary shows a tip structure above which the critical coupling $J_c$ grows with $g/(\hbar\omega_0)$. A similar effect happens in the opposite limit at large $J/(\hbar \omega_0)$ and small $g/(\hbar\omega_0)$. \\

We have seen in this section that circuit QED setup permits to emulate quantum many-body physics with light. In the same perspective, recent interest have turned to the realization of quantum impurity models, describing discrete local quantum degrees of freedom coupled to continuous baths of excitations. They were originally introduced for the description of magnetic impurities in metals. These models are also relevant for describing transport through quantum dots coupled to metallic leads. Some models of this kind are not integrable and necessitate complex many-body techniques understand their properties. Upon variation of a control parameter, such models may exhibit a quantum phase transition where order is destroyed only by quantum fluctuations. One paradigmatic model of this kind is the spinboson model, initially introduced to describe dissipation.

%We now turn to the spinboson model. \textbf{faire une transition}.

\section{Ohmic spinboson model}
Quantum mechanical problems of interest involving a (effective) two-level system are widespread in physics and chemistry. We have seen above the example of the Rabi Hamiltonian, describing transitions between two energy levels of an atom. Other examples include the description of a nucleus of spin 1/2 (with applications in NMR), or the inversion resonance of the ammonia molecule. Such two-level systems can be described by an Hamiltonian of the form, 
\begin{align}
\mathcal{H}_{TLS}&= \frac{\Delta}{2} \sigma^x +  \frac{\epsilon}{2} \sigma^z,
\label{Hamiltonian_two_level_isolated} 
\end{align}
where $\epsilon$ denotes the energy difference between the two levels and $\Delta$ is the tunneling amplitude between these states.\\

\subsection{Modelling of dissipation}

In many cases of interest, the description of the system solely in terms of a two-level system is inaccurate. In practise for a real experiment, the two-level system indeed interacts with its surrounding environment through a term of the form $\sum_{\nu \in \{x,y,z\} }\sigma^{\nu} \Omega_{\nu}$, where $\Omega_{\nu}$ for $\nu \in \{x,y,z\}$ are environment operators. In the following we will focus on cases where the coupling is non-zero only along one direction, let us say the $z$-direction. Provided that the coupling to the environment is sufficiently weak, it is relevant to describe it as a set of harmonic oscillators with a coupling linear in the oscillator coordinates \cite{Caldeira_Leggett,bookzwerger}. In the following and for the remaining of the manuscript, we will take the convention $\hbar=1$. We reach the spinboson Hamiltonian,

\begin{align}
\mathcal{H}_{SB}&= \frac{\Delta}{2} \sigma^x +  \frac{\epsilon}{2} \sigma^z+\sigma^z \sum_k \frac{\lambda_k}{2} (b_k +b_k^{\dagger})+ \sum_k  \omega_k \left(b_k^{\dagger}b_k+\frac{1}{2}\right),
\label{Hamiltonian_spin_boson} 
\end{align}
where $b_k$ ($b_k^{\dagger}$) is the annihilation (creation) operator of a boson in mode $k$ with frequency $\omega_k$. The spin-bath interaction is fully characterized by the spectral function $J(\omega)=\pi \sum_k \lambda_k^2 \delta(\omega-\omega_k)$. We will assume that $J$ is a smooth function of $\omega$, of the form
\begin{align}
J(\omega)=2 \pi \alpha \omega^s \omega_c^{1-s} \exp \left(-\frac{\omega}{\omega_c}\right),
\label{J_ohmic} 
\end{align}
where $\omega_c$ is a high frequency cutoff such that $\Delta\ll  \omega_c$ and the dimensionless parameter $\alpha$ quantifies the strength of the coupling. A case of particular interest corresponds to the so-called ohmic coupling which corresponds to $s=1$.\\ 

\subsection{Quantum phase transition and relation with Kondo model}
\label{quantum_phase_transition}
 The interaction with the bath plays an important role and affects both the equilibrium and the dynamical properties of the system. In this section, we focus on the ohmic spinboson model and investigate how the interaction with the bath triggers at high coupling a quantum phase transition from a delocalized to a localized phase, in relation to Kondo physics.
 
 \subsubsection{Polaron ansatz and quantum phase transition}
 \label{polaron}
 To understand better spin-bath effects, let us start by considering the limit $\epsilon=\Delta=0$. The model reduces then to a set of harmonic oscillators with finite displacements, exactly as considered in Sec. \ref{strong_coupling}. When $\Delta \neq 0$, the high frequency modes of the bath ($ \omega_k \gg \Delta$) could be treated in the adiabatic approximation fashion developped in Sec. \ref{strong_coupling}. For low frequency modes ($ \omega_k \leq \Delta$), the situation is however different and the term $\Delta/2 \sigma^x $ cannot be treated in a perturbative manner.  We therefore follow Ref. \cite{Silbey_Harris} (and references therein) and use a multi-mode coherent state ansatz for the ground state wavefunction,
 \begin{align}
|\psi_{var}\rangle=\frac{1}{\sqrt{2}}\left[ |\uparrow_z \rangle |f\rangle-|\downarrow_z \rangle |-f\rangle \rangle  \right],
\label{silbey_harris}
\end{align}
 where $ |f\rangle=\exp \left[ -\sum_k f_k (b_k^{\dagger}-b_k) \right] |0\rangle$ ($|0\rangle$ corresponds to the vacuum for all the oscillators). With this ansatz we do not specify the amplitude with which a given mode is displaced \textit{ab initio}, but these coefficients are determined by minimizing the mean energy of the system,
  \begin{align}
\langle \psi_{var}|\mathcal{H}_{SB}|\psi_{var}\rangle=-\frac{\Delta}{2}\exp\left[-2 \sum_k f_k^2 \right]+\sum_k \omega_k f_k^2-\sum_k \lambda_k f_k.
\label{silbey_harris_energy}
\end{align}
 We then minimize the variational energy $\partial_{f_k} \langle \psi_{var}|\mathcal{H}_{SB}|\psi_{var}\rangle=0$ and find the self-consistent displacements
  \begin{align}
f_k =\frac{\lambda_k/2}{ \omega_k+\Delta \exp\left[-2 \sum_k f_k^2 \right]}.
\label{Displacement}
\end{align}
We recover the adiabatic displacement $\lambda_k/(2\omega_k)$ for bath modes $k$ such as $\hbar \omega_k \gg \Delta$. On the other hand, the displacement tends to zero for low frequency modes if $\Delta >0$.\\

\label{polaron_ansatz}

Bath states now ``dress" the spin states, which leads to a renormalization of the tunneling element $\Delta$. Let us estimate this renormalized element $\Delta_r$.
\begin{align}
&\Delta_r =\Delta \exp \left[ - \frac{1}{2\hbar} \int_0^{\infty} d\omega \frac{J(\omega)}{(\omega+\Delta_r/\hbar)^2} \right]\notag \\
\Rightarrow &\Delta_r \simeq \Delta \exp \left[ - \alpha \int_{\Delta_r}^{\omega_c}   \frac{d\omega}{\omega} \right]\\
\Rightarrow &\Delta_r \simeq \Delta \left(\frac{\Delta}{\hbar \omega_c}\right)^{\frac{\alpha}{1-\alpha}}.
\label{deltar}
\end{align}
We see in particular that $\Delta_r \to 0$ when $\alpha \to 1$, indicating that the bath may forbid tunneling between spin states at sufficiently high coupling. The effect of the bath is then to polarize entirely the spins, by analogy to a ferromagnetic phase. The spin gets trapped in one of the states $|\uparrow_z\rangle$ or $|\downarrow_z\rangle$.\\

 This result sheds light on the mechanism at the origin of the dissipative quantum phase transition induced by the bath, which has been seen directly by applying the Numerical Renormalization Group (NRG) \cite{Bulla_vojta,Vojta_phil,Hur_Hofstetter}. The procedure involves the re-writing of the partition function using a kink gas representation \cite{Vojta_phil}, mapping the problem on an Ising chain with long range $1/r^2$ interactions. More precisely, the partition function at inverse temperature $\beta$ reads at $\epsilon=0$
 \begin{align}
Z=\sum_{m=0}^{\infty} \left(\frac{\Delta}{2}\right)^{2m} \int_0^{\beta} ds_{2m}  \int_0^{s_{2m}} ds_{2m-1}...  \int_0^{s_2} ds_{1} \mathcal{F}_m \left[\left\{s_j \right\} \right].
\label{partition_function}
\end{align}
$\mathcal{F}_m$ carries the environment influences in the form of ``kink" (or charge) interactions
  \begin{align}
\mathcal{F}_m \left[\left\{s_j \right\} \right]=\exp \left\{ \sum_{j=2}^{2m} \sum_{i=1}^{j-1} W(s_j-s_i) \right\},
\label{partition_function}
\end{align}
 with in the scaling limit $\Delta/\omega_c \ll 1$,
\begin{equation}
W(\tau)=2\alpha \ln \left[ \frac{\beta \omega_c}{\pi} \sin \left( \frac{\pi \tau}{\beta}\right) \right] 
\label{partition_function_charge}
\end{equation}
This representation allows to derive valid RG equations following Ref. \cite{Kosterlitz}, which are equivalent to the ones derived earlier in the case of the anisotropic Kondo model by Anderson Yuval and Hamann \cite{Anderson_Yuval_Hamann} and describe a Kosterlitz-Thouless transition. It is also important to note that expression (\ref{deltar}) for the renormalized tunneling element can also be found using an adiabatic renormalization procedure, developped in Refs. \cite{leggett:RMP,Weiss:QDS}.

%For one spin, the quantum phase transition belongs to the Kosterlitz-Thouless class, and the order parameter at equilibrium $\langle \sigma^z \rangle$ exhibits a jump\cite{KLH}.  \\
%\cite{Garst_Vojta,Peter_two_spins,Winter_Rieger}

\subsubsection{Relation with Kondo model}
\label{mapping_kondo_spinboson}

The dominant contribution to the electrical resistivity in metals comes from the scattering of the conduction electrons with lattice vibrations. This scattering grows with temperature as more and more lattice vibrations are excited. This results in a monotonical increase of electrical resistivity with temperature in most metals. A residual temperature-independent resistivity due to the scattering of the electrons with defects may subsist in the very low temperature range.\\

A resistance minimum as a function of temperature was however observed in a gold sample in 1934 \cite{resistance_gold_minimum}. The solution to this problem was formulated by Jun Kondo in 1964 \cite{Kondo}, when he described how certain scattering processes from magnetic impurities could give rise to a resistivity contribution increasing at low temperatures. We present here the \textit{anisotropic}\footnote{The anisotropy is essential to formulate the link with the ohmic spinboson model} Kondo model, which describes the exchange interaction between a band of non-interacting conduction electrons  with one magnetic impurity. The quantum impurity is represented by a spin 1/2 and it is coupled to the condution electrons by an antiferromagnetic exchange coupling. The corresponding Hamiltonian $\mathcal{H}_K$ is
\begin{align}
\mathcal{H}_K=& v_F \sum_{k,\sigma} k c_{k,\sigma}^{\dagger}c_{k,\sigma}+\frac{ J_{\perp}}{2}\sum_{k,k'}\left( c_{k\uparrow}^{\dagger} c_{k'\downarrow} S^{-}+c_{k\downarrow}^{\dagger} c_{k'\uparrow} S^{+}  \right)\notag \\
&+\frac{ J_{z}}{4} S^z\sum_{k,k'}\left( c_{k\uparrow}^{\dagger} c_{k'\uparrow}-c_{k\downarrow}^{\dagger} c_{k'\downarrow} \right),
\label{anisotropic_kondo}
\end{align}

where $c_{k\sigma}$ ($c_{k\sigma}^{\dagger}$) is the annihilation (creation) operator of a conduction electron in mode $k$ with spin $\sigma$. We assume a constant density of states $\rho=(2\pi v_F)^{-1}$, where $v_F$ is the Fermi velocity. The term in $J_z$ describes scattering of the electrons in which the spin is conserved while the term in $J_{\perp}$ describes spin-flip scattering. The equivalence between the two models can be shown through the kink gas representation  \cite{Bulla_vojta,Vojta_phil,Hur_Hofstetter,Kosterlitz,Anderson_Yuval_Hamann,Weiss:QDS,Hur}, as the partition function for the anisotropic Kondo model reads
\begin{align}
Z_K=\sum_{m=0}^{\infty} &\left(\frac{-\rho J_{\perp} \cos^2 \delta_e}{2 \tau_c} \right)^{2m} \int_0^{ \beta-\tau_c} ds_{2m}  \int_0^{s_{2m}-\tau_c} ds_{2m-1}...  \int_0^{s_2-\tau_c} ds_{1} \notag \\
&\times \exp\left\{2 \left(1+\frac{2\delta_e}{\pi} \right)^2\sum_{j>k=1}^{2m} \ln \left[\frac{ \beta}{\pi \tau_c} \sin \left(\frac{\pi(s_j-s_k)}{ \beta} \right) \right] \right\}.
\label{partition_function_anistropic_kondo}
\end{align}
where $\tau_c=1/\omega_c$ and $\delta_e =\tan^{-1}(-\pi \rho J_z/4)$. We have a correspondence between Eq. (\ref{partition_function}) and Eq. (\ref{partition_function_anistropic_kondo}) in the scaling regime, with the identification
\begin{align}
&\rho J_{\perp}\to \Delta/ \omega_c\\
&(1+2 \delta_e /\pi)^2\to \alpha.
\end{align}\\

It follows that the localized-delocalized quantum phase transition in the ohmic spin-boson model is equivalent to the ferromagnetic-antiferromagnetic transition in the anistropic Kondo model. The relationship between the ohmic spinboson model and the anisotropic Kondo model is due to the fact that the low-energy electron-hole excitations have a bosonic character and can be interpreted in terms of density fluctuations \cite{leggett:RMP}. The equivalence between these two models notably led to the original understanding of the localization phenomenon in the ohmic spinboson model \cite{equivalence_1,equivalence_2}.  The equivalence between these two models can also be shown through bosonization \cite{kondo_spinboson}. \\

%Both thermodynamical and dynamical properties are qualitatively the same for the two models \cite{Leggett}. 
The Kondo effect can be considered as an example of asymptotic freedom, i.e., the coupling of electrons and spin becomes weak at high temperatures or high energies. In this respect, it embodies a paradigmatic model of quantum many-body physics and represents the ``hydrogen atom'' of this field. Being able to engineer either the Kondo model or the ohmic spinboson model in a controllable manner would then provide a new perspective on these effects. We will come back in more details on related Kondo models in quantum dots in Chapter V.

%offer the opportunity to reproduce these many-body effects.

\subsection{Circuit quantum electrodynamics: semi-infinite transmission line}
\label{transmission_line}

%The description of our circuit is not complete yet, as we have not taken into account the natural presence of dissipative elements. 

\begin{figure}[h!]
\includegraphics[scale=0.4]{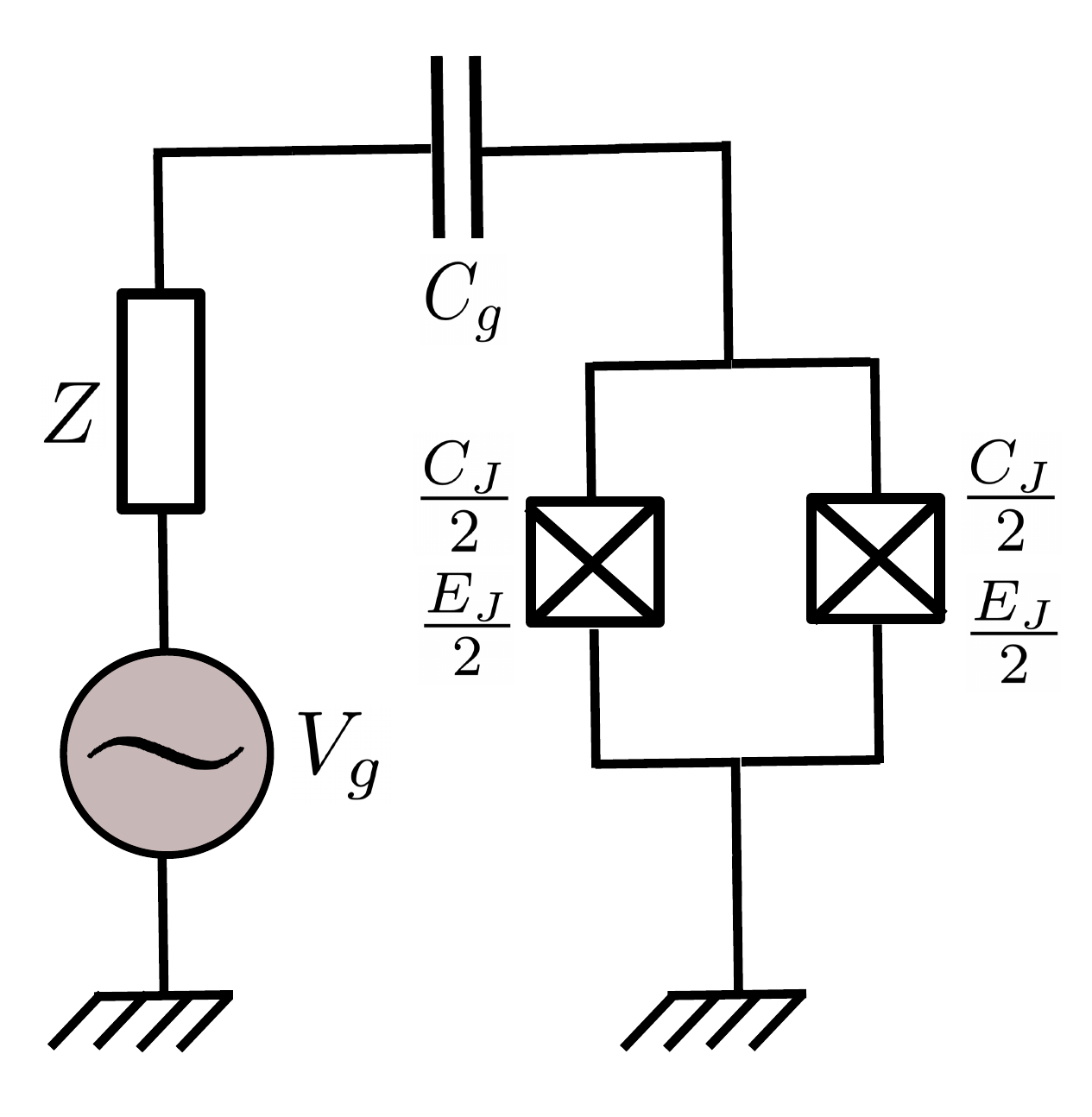}  \includegraphics[scale=0.2]{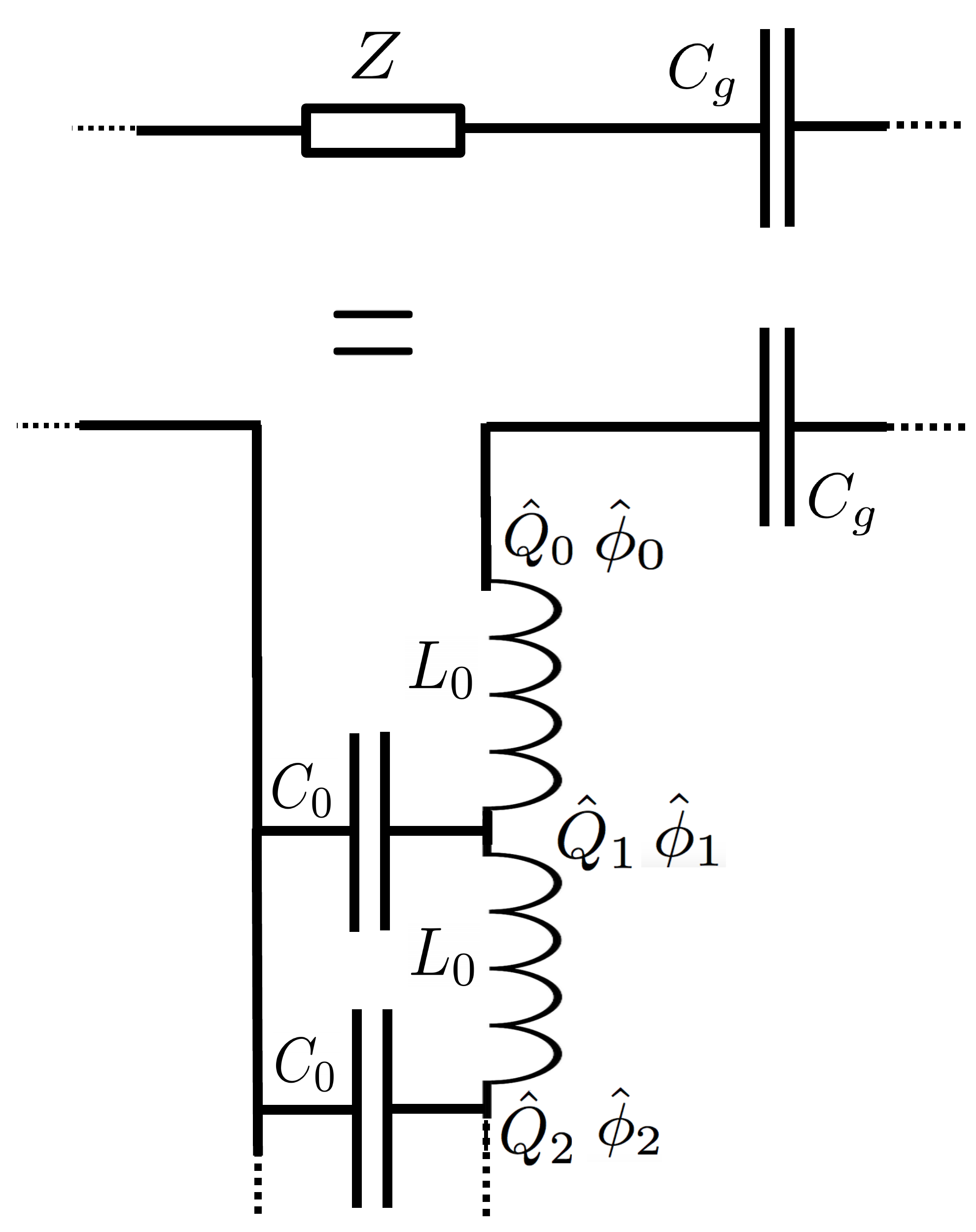}
\caption{Left panel: Circuit of the noisy Cooper pair box. The crosses indicate Josephson junctions. Right panel: Transmission line modelling of the resistive impedance.
}
\label{fig:Cooper_pair_box_final_transmission_line}
\end{figure}

Some proposals have focused on the implementation of the ohmic spinboson model in circuit QED setups. Let us consider the artificial atom introduced in \ref{cqed}. As shown in Refs. \cite{Cedraschi:Annals_of_physics,Cedraschi:PRL}, the coupling of the circuit to an external resistive environment naturally takes the form of Eq. (\ref{Hamiltonian_spin_boson}) with an ohmic spectral density (\ref{J_ohmic}). It is possible to show this by modelling the resistive environment of the circuit by a semi-infinite transmission line, as shown in Fig. \ref{fig:Cooper_pair_box_final_transmission_line}\footnote{For convenience, we take the capacitance of the transmission line to match with the effective capacitance of the mesoscopic structure $1/C_0=1/C_J+1/C_g$. The Ohmic resistance $R=\sqrt{L_0/C_0}$ can be then adjusted by controlling the inductance $L_0$.}. \\

The Hamiltonian of the transmission line can be diagonalized as in \ref{resonator_section} (See Refs. \cite{Cedraschi:Annals_of_physics,Cedraschi:PRL} for a detailed derivation), and one recovers Eq. (\ref{Hamiltonian_spin_boson})  with an ohmic spectral function with the identification
\begin{align}
&\Delta=E_J\\
&\alpha=\frac{R}{R_K}\\
&\omega_c=\frac{1}{\sqrt{L_0 C_0}},
\end{align}
where  $R_q = h/e^2$ is the resistance quantum. The spectrum is dense, as only one boundary condition holds in this semi-infinite case. The artificial atom is then coupled to a continuum of modes, with a dense spectrum at low frequencies.\\

Circuit QED setups may provide new ways to measure many-body effects. One way to experimentally measure the Kondo energy would be to have access to the renormalized parameter $\Delta_r$, for example through the observation of Rabi oscillations. Alternatively, the Kondo energy can be directly measured based on the resonant propagation of a photon inside the transmission line, as shown in Ref. \cite{Le_Hur:PRBR,kondo_microwave_houzet}. This proposal involves a periodic driving of the system, which can be treated through the input-output theory \cite{Clerk:RMP}. In the underdamped limit, the analysis of the microwave signal reveals a manybody Kondo resonance in the elastic power of a transmitted photon.\\

%The ohmic spinboson model can be also be realized in systems of trapped ions, as exposed for example in Ref. \cite{Cirac_spinboson}.

\subsection{Spinboson model in a cold-atomic setup}
\label{spinboson_cold_atoms}

\begin{figure}[h!]
\center
\includegraphics[scale=0.25]{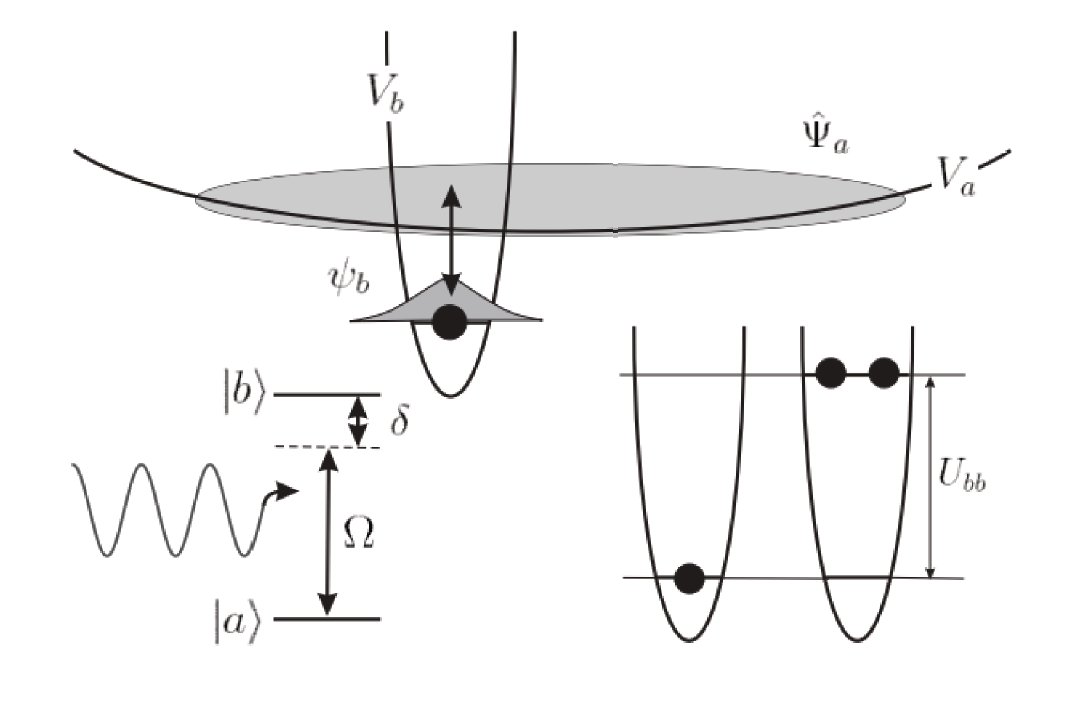}  
\caption{ Atomic quantum dot coupled
to a superfluid atomic reservoir. The Bose-liquid of atoms in state \textit{a} is confined in a shallow trap $V_a (x)$. The atom in state
textit{b} is localized in tightly confining potential $V_b (x)$. Atoms are coupled via a Raman transition. A large onsite interaction $U_{bb} > 0$ allows
only a single atom in the dot. Figure adapted from Ref. \cite{recati_fedichev}}
\label{fig:Recati_fedichev}
\end{figure}

The ohmic spinboson model can also be realized with cold atomic setup, as proposed in Refs. \cite{recati_fedichev}. This proposal rely on the use of cold bosons with two different ground states \textit{a} and \textit{b}, trapped by two types of potentials (see Fig. \ref{fig:Recati_fedichev}). 
\begin{itemize}
\item A rather shallow potential traps atoms in state \textit{a}. At sufficiently low temperatures, these atoms will form a Bose-Einstein Condensate (BEC).
\item A highly confining potential traps atoms in state \textit{b}. This latter potential can be produced for example by a deep optical lattice, which would be only seen by atoms in this state \textit{b} (see Refs. \cite{jaksch_briegel_cirac_gardiner_zoller,mandel_greiner} for details about the realization of such a state-selective potential). Only a discrete set of states are allowed for the atoms in state \textit{b} which are trapped in the highly confining potential, forming an atomic quantum dot.
\end{itemize}

The description of the collision interactions between the atoms can be done using a pseudopotential description and introducing effective coupling parameters $g_{\alpha \beta}$ between states $\alpha$ and $\beta$ $\in \{a,b\}$. The collisional interaction may result in a strong repulsion for atoms in state \textit{b}. Low energy states of the quantum dot can be coupled to the condensate reservoir by Raman transitions. Altogether, the Hamiltonian describing the system $\mathcal{H}_{sys}$ can be written,
\begin{align}
\mathcal{H}_{sys}=&\left(-\delta_0+g_{ab} \int dx |\psi_b(x)|^2 \rho_a(x) \right) b^{\dagger}b+\frac{U_{bb}}{2} b^{\dagger}b^{\dagger}bb \notag \\
&+ \int dx  \Omega \left( \Psi_a(x)\psi_b(x)b^{\dagger}+h.c.\right)+\mathcal{H}_{a},
\end{align}
where $\Psi_a(x)$ is the annihilation operator for an atom \textit{a} at point $x$, and $\rho_a(x)=\Psi_a^{\dagger}(x) \Psi_a(x)$ is the associated density operator. $b$ annihilates a boson in the atomic quantum dot, with the wavefunction $\psi_b^{\dagger}(x) $. $\delta_0$ is the Raman detuning. The term proportional to $g_{ab}$ describes collisional interactions between the atoms on the dot and the reservoir. $U_{bb}$ quantifies the on-site repulsion between different \textit{b} atoms on the dot. The last term describes the laser-induced coupling between the two types. \\

The low-energies excitations of the BEC consist of linear dispersion phonons with linear dispersion $\omega=v_s |q|$, where $v_s$ is the sound velocity and $q$ is the momentum of the excitation. For a large number of atoms in the condensate, one can then consider that the atoms \textit{b} are coupled to a coherent matter wave described in terms of a quantum hydrodynamic Hamiltonian. In the limit of large on-site repulsion $U_{bb} \gg  \Delta$, we can restrict the dynamics to the two lowest energy levels on the atomic quantum dot. After a unitary transformation $S=\exp[-i\sigma^z\phi(0)/2] $, we reach 

\begin{align}
\mathcal{H}_{sys}=-\frac{ \Delta}{2}\sigma^x+\sum_q \omega_q b_q^{\dagger} b_q +\left(-\delta +\sum_q \lambda_q (b_q +b_q^{\dagger}) \right) \frac{ \sigma^z}{2},
\end{align}
where the coupling coefficients $\lambda_q$ read
\begin{align}
\lambda_q=\left[ \frac{m |q|v_s^3}{2V\rho_a}\right]^{1/2}\left(\frac{g_{ab}\rho_a}{m v_s^2}-1 \right).
\end{align}\\

The dispersion relation above permits the realization on an ohmic spectral function. As shown in Ref. \cite{orth_stanic_lehur}, one could extend this proposal to an optical lattice and consider a lattice of well-separated tightly confining trapping potentials, so that atoms in state \textit{b} cannot hop from one site to the other. Following the same lines as for the one-site case, the interaction with the common bath of harmonic excitations leads to the following Hamiltonian, 
\begin{align}
\mathcal{H}_{SB}^M=&\frac{\Delta}{2}\sum_{p=1}^M  \sigma_p^x-\sum_{j\neq p} K_{|j-p|} \sigma^z_j \sigma^z_{p}+ \sum_{p=1}^M \sum_{k} \lambda_{k} e^{ik x_p} \left(b^{\dagger}_{-k}+b_k \right) \frac{\sigma_p^z}{2} +\sum_{k}  \omega_{k} b^{\dagger}_{k} b_{k}.
\label{ising_1}
\end{align}
where $M$ denotes the number of sites of the lattice. Eq. (\ref{ising_1}) constitutes the $M$ spin version of the ohmic spinboson model. $K_{|j-p|}$ notably depends on the characteristics of the bath and vanishes for $g_{aa}=2g_{ab}$ \cite{orth_stanic_lehur}. Let us now see what are the additional effects induced by the coupling to a common bath in this multi-spin case with respect to the single-mode case. This can be studied by applying an unitary transformation $\tilde{H}=V^{-1}HV$ on the Hamiltonian (\ref{ising_1}), with $V=\exp \left\{\frac{1}{2}\sum_{k}\sum_{j=1}^M\sigma_j^z e^{ikx_j}\frac{\lambda_{k}}{ \omega_{k}}   (b_k-b_{-k}^{\dagger}) \right\}$. The transformed Hamiltonian reads:
\begin{align}
\tilde{\mathcal{H}}_{SB}^M=&\ \sum_{j=1}^M \frac{\Delta}{2} \left( \sigma^{+}_j e^{i \Omega_j}+\sigma^{-}_j e^{-i \Omega_j} \right)- \sum_{j \neq p} K'_{|j-p|} \sigma^z_j \sigma^z_{p}+\sum_{k} \omega_{k} b^{\dagger}_{k} b_{k},
\label{N_spins}
\end{align}
where $\Omega_j=i\sum_{k} \frac{\lambda_{k}}{ \omega_{k}}  e^{ik x_j} (b_{k}-b_{-k}^{\dagger})$ and
\begin{align}
K'_{|j-p|}=K_{|j-p|}+\frac{\alpha \omega_c}{2} \frac{1}{1+\frac{\omega_c^2 (x_j-x_p)^2}{v_s^2}}.
\label{renormalized_coupling}
\end{align}
Note that we recover the renormalization of the tunneling element induced by the bath in this polaron-transformed rewriting. As can be seen from Eq. (\ref{N_spins}) the excitation of spin $j$ indeed comes with a simultaneous polarization of the neighboring bath into a coherent state $|\Omega_j\rangle=e^{i\Omega_j}|0\rangle$. The tunneling energy is thus renormalized due to this boson-dependent phase. On top of this effects, the bath also engenders in this multi-spin case a strong Ising-type {\it ferromagnetic} interaction $K'_{|j-p|}$ between the spins $j$ and $p$, which is mediated by an exchange of bosonic excitations at low wave vectors, as demonstrated in Ref. \cite{orth_stanic_lehur}. We saw in Sec. \ref{polaron_ansatz} that the critical value $\alpha_c$ of the coupling is $\alpha_c=1$. Due to the strong ferromagnetic interaction between the spins induced by the bath, this critical value decreases with the number $M$ of sites, as confirmed in Refs. \cite{Peter_two_spins,sougato,Winter_Rieger}.\\

We also note that the ohmic spinboson model can be realized in systems of trapped ions, as exposed for example in Ref. \cite{Cirac_spinboson}. Environment effects in relation with dissipative critical behaviour have also been studied in the transport properties of quantum dot systems of carbon nanotubes coupled to resistive environments \cite{Mebrahtu}, emulating tunnel coupled Luttinger liquids \cite{Kane_Fisher}. A prediction of this latter model is the existence of resonance peaks of perfect conductance which narrows when temperature decreases; which translates to environmental conductance suppression in the case of tunneling with dissipation \cite{IN} (see also Refs. \cite{Safi,Hur_li,NRG4,Chung} which explore the link between Luttinger liquid physics). We note related experimental progress studying the backaction of the environment on the conductance in a tunable GaAs/Ga(Al)As Quantum Point Contact setup \cite{Pierre:Nature_2}.

\subsection{Non-equilibrium dynamics and NIBA equation}
\label{NIBA_from_EOM}

Several methods were devised to tackle the spin dynamics in the spinboson model. Among them, the well-known Non-Interacting Blip Approximation (NIBA) allows to reach analytical results in the scaling regime characterized by $0\leq\alpha\leq 1/2$ and $\Delta/\omega_c \ll 1$. The derivation of NIBA was originally done in Ref. \cite{leggett:RMP} using a path integral formalism, that we will present in the next chapter. Interestingly, the NIBA equations can also be derived by using a weak-coupling decoupling over Hamiltonian after a unitary transformation as shown in Ref. \cite{Dekker}.\\

Let us focus on the non-equilibrium dynamics for one spin initially in the pure state $|\uparrow_z\rangle$, coupled to a bath at equilibrium at zero temperature at time $t_0$. We compute the Heisenberg equations of motion for the spin operators $\sigma^z$, $\sigma_+$ and $\sigma_{-}$ after having performed the one-spin version of the unitary transformation introduced in the previous subsection. Replacing the equations obtained for the transverse elements in the one obtained for $\sigma^z$, we reach
\begin{align}
\dot{\sigma^z}(t)=-\frac{\Delta^2}{2} \int_{t_0}^t ds \sigma^z(s)\left[ e^{i\Omega(t)}e^{-i\Omega(s)}+ e^{-i\Omega(t)}e^{+i\Omega(s)} \right].
\label{NIBA_EOM_0}
\end{align}
As $\sigma^z$ commutes with the unitary transformation, one can equally compute its evolution in the two frames. To recover equations of NIBA derived in Ref. \cite{leggett:RMP}, Dekker decoupled spin and bath expectation values and assumed that the time evolution of the bath operators was governed by the free bath Hamiltonian \cite{Dekker}. This leads to
\begin{align}
\langle\dot{\sigma^z}(t)\rangle+\int_{t_0}^t ds f(t-s)\langle\sigma^z(s)\rangle =0,
\label{NIBA_EOM}
\end{align}
where 
\begin{align}
f(t)=\Delta^2 \cos \left[\frac{1}{\pi}\int_0^{\infty} d\omega\frac{J(\omega)}{\omega^2}\sin \omega t\right] \exp\left[-\frac{1}{\pi}\int_0^{\infty} d\omega\frac{J(\omega)}{\omega^2}\left(1-\cos \omega t\right)\right].
\end{align} 
Eq. (\ref{NIBA_EOM}) can then be solved exactly using Laplace transformation. One finds that $\langle\dot{\sigma^z}(t)\rangle$ is the sum of a coherent term $p_{coh}(t)$ and an incoherent term $p_{inc}(t)$ where
\begin{align}
&p_{coh}(t)=\frac{1}{1-\alpha}e^{-\gamma t} cos \Gamma t.
\end{align} 
The oscillation frequency and the decay rate are characterized by $\Gamma/\gamma=\cot [\pi \alpha /(2-2\alpha)]$ while the incoherent behavior dominates the long-time dynamics as it behaves as $(\Delta_r t)^{2-2\alpha}$. This incoherent behavior is considered to be an incorrect prediction of NIBA \cite{Weiss:QDS}.\\

In this chapter, we introduced the Rabi model and the Spinboson model, and their relevance for modern experimental techniques. We have also seen the need to develop new techniques to tackle the non-equilibirum dynamics in these problems. A particular case of interest related to the Rabi case consists in the developement of a numerical/theoretical framework which would allow to take into account drive and dissipation effects. For the spinboson model, the free dynamics at strong coupling is already a challenge\footnote{An overview of the different techniques will be provided below}. \\

%We also demonstrated the need 

\chapter{SSE equation and applications}

In this chapter, we introduce the stochastic Schr\"{o}dinger equation applicable to Spin-boson models. We consider first a spin 1/2 interacting with a bosonic bath, described by the Hamiltonian
\begin{align}
\mathcal{H}=\frac{\Delta}{2}\sigma^x +\sigma^z \sum_k \frac{\lambda_k}{2} (b_k+b^{\dagger}_k) +\sum_k \omega_k b^{\dagger}_k b_k.
\label{Hamiltonien_general_spinboson_methode}
\end{align}
The coupling between the spin and the bosonic bath is characterized by the spectral function\footnote{We will keep the discussion general in this section and will not specify a particular form for $J$.} $J(\omega)=\pi \sum_k \lambda_k^2 \delta(\omega-\omega_k)$. \\

In the case of a continuous spectral function, computing the spin dynamics is generally a challenging task and a very large number of different methods were devised to this end. At very weak coupling, Markovian master equations permits to capture qualitatively relaxation and dephasing effects \cite{Breuer_Petruccione,Keeling_sto}. At higher spin-bath coupling, the influence of the environment on the dynamics becomes more subtle and memory effects have to be taken into account. \\

As Hamiltonian (\ref{Hamiltonien_general_spinboson_methode}) is quadratic in terms of bosonic operators, one can integrate out exactly these degrees of freedom in a path integral approach. This technique was pioneered by Feynman and Vernon \cite{FV}, and constitutes the starting point of the well-controlled Non Interacting Blip Approximation (NIBA) \cite{leggett:RMP,Weiss:QDS} and extensions to it \cite{NIBA_extension_1,NIBA_extension_2,NIBA_extension_3,NIBA_extension_4}. Despite great success in the delocalized phase for $\alpha<1/2$, this approximation is for example unable to describe the quantum phase transition occuring in the ohmic case (see Sec. \ref{quantum_phase_transition}). \\

Numerous analytical and numerical methods were built from the Feynman-Vernon influence functional and the ``Blip" and ``Sojourn" development at the origin of NIBA. This includes  stochastic Liouville equations \cite{Stockburger_Mac,Stockburger,Stockburger_2,Koch_morse}, Non-Markovian master equations \cite{Tu_Zhang,Zhang_Nori}, real-time Path Integral Monte Carlo methods \cite{QMC_1,QMC_2,QMC_3,QMC_4}, Quasi-Adiabatic Propagator Path Integral techniques \cite{QUAPI,QUAPI_recent} and the Stochastic Schr\"{o}dinger Equation under consideration.\\

Different approaches following a different path were also devised to tackle the real-time dynamics in this problem, with for example quantum jumps approaches on the wavefunction (or stochastic wavefunction approaches) \cite{stochastic_wavefunction_1,stochastic_wavefunction_2,stochastic_wavefunction_3}. More recently, various Numerical Renormalization Group (NRG) techniques \cite{NRG1,NRG2,NRG3,NRG4,NRG5,NRG6,NRG7,NRG8,NRG9,NRG10}, or Multilayer multiconfiguration time dependent Hartree method \cite{Wang_Thoss} were also developped. \\

%In the early years of the twentieth century, physicists understood that the influence of the external environment on a classical system could be effectively described by a random force \cite{Einstein,Langevin}. Similarly in the quantum case, one can imagine to describe the effect of a quantum environment by means of a stochastic force, and define a quantum analogue of Brownian motion \cite{Breuer_Petruccione,Tilloy}. 
% Real-time Monte Carlo methods \cite{QMC_1,QMC_2,QMC_3,QMC_4} were also devised to study the dynamics of such systems.\\

We present here the details of the Stochastic  Schr\"{o}dinger Equation method, following our Ref. \cite{Rabi}, and we will try to highlight the links with the other methods/approaches mentionned above along the derivation. 

%\section{Derivation of the Stochastic Schr\"{o}dinger Equation}

%{\color{red}
%Some authors did not express spin paths in the language of blips and sojourns, but rather reached an effective stochastic Liouville equation for the density matrix, see Refs. \cite{Stockburger_Mac,Stockburger,Stockburger_2}.This technique has notably been used to compute the dynamics for the Morse oscillator\cite{Koch_morse}. The main difference in this case is that the stochastic decoupling is done before the rewriting of the influence functionnal in terms of blips and sojourns.\\

% Non-Markovian master equations\cite{Tu_Zhang,Zhang_Nori} were derived thanks to the same Feynman-Vernon influence functional starting point. A review of the different path-integral methods developped to tackle the non-Markovian dynamics in spin-bath systems is provided in Ref.~\cite{de_Vega_review}.Keldysh: \cite{An_introduction_to_nonequilibrium_many_body_theory_Maciejko,Wagner_general_initial_condition}. Master equations.}

\section{SSE Equation in the case of one spin}

A state of the system is described by a wavefunction $|\psi \rangle$, which belongs to the Hilbert space $\epsilon=\epsilon_\mathcal{S} \otimes \epsilon_\mathcal{B}$, which is the tensor product of spin and bath spaces $\epsilon_\mathcal{S}$ and $\epsilon_\mathcal{B}$. This mixed spin-boson system is conveniently described in terms of density operators, or density matrices. Considering our quantum system, there is a unique operator $\rho$ such that 
\begin{align}
\langle A \rangle= \textrm{tr} \left(\rho A\right),
\label{def_density_matrix}
\end{align}
for all observable operator $A$. This operator $\rho$ is called the density operator, or density matrix, of the system. We are mainly interested in the dynamics of the spin observables, and the effect of the bath on their dynamics. It is then convenient to define reduced density operators $\rho_{\mathcal{S}}$ and $\rho_{\mathcal{B}}$ as partial traces of the total density matrix
\begin{align}
\rho_{\mathcal{S}}= \textrm{tr}_{\epsilon_{\mathcal{B}}} \left(\rho\right),\\
\rho_{\mathcal{B}}= \textrm{tr}_{\epsilon_{\mathcal{S}}} \left(\rho\right).
\label{def_reduced_density_matrix}
\end{align}\\

We are interested in the time-evolution of the spin-reduced density matrix $\rho_{\mathcal{S}}(t)$ for $t \geq t_0$, where $t_0$ denotes the initial time. We assume factorizing initial conditions $\rho (t_0)=\rho_B(t_0)\otimes \rho_S(t_0)$, with the bath in a thermal state at inverse temperature $\beta$. Under these assumptions, we can show the following result.\\

\noindent\begin{minipage}[c]{1\linewidth}%
\fcolorbox{gray}{light-gray}{
\begin{minipage}[c]{1.\linewidth}%
\underline{Stochastic Schr\"{o}dinger Equation for one spin}\\

The elements of the spin-reduced density matrix at time $t \geq t_0$ are given by 
\begin{equation} 
\left[\rho_S (t)\right]_{ij}=\overline{\langle \Sigma_{ij}| \Phi(t)\rangle},
\label{solution_density_matrix}
\end{equation}
where the overline denotes a stochastic average and $| \Phi\rangle$ is the four-dimensional vector solution of the Stochastic Schr\"{o}dinger-like differential equation (\ref{SSE}), 
\begin{equation} 
i \partial_t | \Phi \rangle = V (t) | \Phi \rangle,
\label{SSE}
\end{equation}
with $| \Phi (t_0) \rangle= \left(\left[\rho_S(t_0)\right]_{11} e^{k(t_0)},\left[\rho_S(t_0)\right]_{12} e^{h(t_0)}, \left[\rho_S(t_0)\right]_{21} e^{-h(t_0)},\left[\rho_S(t_0)\right]_{22} e^{-k(t_0)} \right)^T$.\\

In Eq. (\ref{SSE}), we have
\begin{equation}
V= \left( \begin{array}{cccc}
0&e^{-h+k }&-e^{h+k }&0 \\
e^{h-k }&0&0&-e^{h+k }\\
-e^{-h-k }&0&0&e^{-h+k }\\
0&-e^{-h-k }&e^{h-k }&0
\end{array} \right).
\label{eq:spin_hamiltonian}
\end{equation}
 $h$ and $k$ are two complex gaussian random fields with correlations
\begin{align}
 \overline{ h(t) h(s)} = & \frac{1}{\pi} Q_2(t-s) + l_1, \label{height_1} \\
 \overline{ k(t) k(s)} = &\  l_2,    \label{height_2}   \\
 \overline{ h(t) k(s) } = & \frac{i}{\pi}  Q_1(t-s) \theta(t-s) + l_3, \label{height_3}
\end{align}
where $l_j$ for $j\in \{1,2,3\}$ are arbitrary complex constants and
\begin{align}
 Q_1(t)&=\int_0^{\infty} d\omega\frac{J(\omega)}{\omega^2}\sin \omega t, \label{q1} \\
  Q_2(t)&=\int_0^{\infty} d\omega\frac{J(\omega)}{\omega^2}\left(1-\cos \omega t\right) \coth \frac{\beta\omega}{2} .\label{q2}
\end{align} 
Vectors $\langle \Sigma_{ij}|$ read $\langle \Sigma_{11}|=(e^{-k(t)},0,0,0)$; $\langle \Sigma_{12}|=(0,e^{-h(t)},0,0)$; $\langle \Sigma_{21}|=(0,0,e^{h(t)},0)$; $\langle \Sigma_{22}|=(0,0,0,e^{k(t)})$.

\end{minipage}}
\end{minipage}\\

The derivation of this result is based on different results related to Refs. \cite{FV,leggett:RMP,Weiss:QDS,2010stoch,Peter,Rabi}, which will be exposed below, and can be decomposed into three consecutive steps:
\begin{itemize}
\item Integration of the bosonic degrees of freedom in a path integral formalism \cite{FV}. This integration will induce spin-spin interactions, which are long range in time.
\item Rewriting of the spin path in the language of ``Blips'' and ``Sojourns'', following the work of Ref. \cite{leggett:RMP}.
\item  Stochastic unravelling of the bath-induced spin-spin interaction: we decouple this spin-spin interaction thanks to the introduction of stochastic degrees of freedom \cite{2010stoch,Peter,Rabi,Stockburger_Mac,Stockburger}. 
\end{itemize}

\label{result_method}

\subsection{Feynman-Vernon influence functional}

Hamiltonian (\ref{Hamiltonien_general_spinboson_methode}) is quadratic in terms of bosonic operators, which enables us to carry out an exact integration of these degrees of freedom in a path integral approach. This operation typically generates additional spin-spin interactions in time, whose kernel depends on the spectral properties of the bath. This technique was originally introduced by Feynman and Vernon in Ref. \cite{FV} with an integration over extended coordinates of the harmonic oscillators (see also Ref. \cite{Weiss:QDS}). One can also derive the result using a coherent-state path integral description. This derivation is done in Appendix \ref{appendeix_FV_derivation}, and we only reproduce the main steps here. Let $\{|u\rangle\}$ be the basis of coherent states of $\epsilon_B$, and $\{|\sigma_{k}\rangle\}=\{ |\uparrow_{z}\rangle,|\downarrow_{z}\rangle\}$ the canonical basis associated with the $z$-axis of $\epsilon_S$. The starting point is to express the density matrix $\rho_S$ in terms of the evolution operator of the whole system $U(t,t_0)$,
\begin{align} 
\rho_S(t)=\textrm{tr}_{\epsilon_{\mathcal{B}}} \left[U(t,t_0)\rho(t_0)U^{\dagger}(t,t_0)\right)]. 
\label{density_matrix_basic}
\end{align}
We insert then resolutions of the identity in terms of coherent-state and spin projectors
 \begin{align} 
\mathbb{I}=\sum_{k} \int d \mu (v)   |v,\sigma_k \rangle \langle v, \sigma_k |,
\label{eq_appendix:idnetity}
\end{align}
 on both sides of $\rho(t_0)$ in the expression (\ref{density_matrix_basic}). The coherent state measure is defined by
\begin{align} 
d \mu (u) = \frac{1}{\pi} d u_x du_y e^{-|u|^2},
\end{align}\\
with $u_x$ and $u_y$ respectively the real and imaginary part of $u$. The main idea is then to re-express the forward and backward propagators in terms of integrals over bosonic fields and real-valued spin fields, following the standard path integration procedure. The resulting action of each propagator can be expressed\footnote{up to terms coming from boundary conditions, see Appendix \ref{appendeix_FV_derivation}} in terms of a time-integral of a Lagrangian $\mathcal{L}$ which has only linear and square dependence on the bosonic trajectories $\psi_k$ and $\psi_k^*$. Each action can thus be evaluated in an exact manner by means of the stationary phase condition, 
\begin{align} 
&\frac{d}{d\tau} \frac{\partial \mathcal{L}}{\partial \dot{\psi_k}}=\frac{\partial \mathcal{L}}{\partial\psi_k},\label{Lagrange_equation_1}\\
&\frac{d}{d\tau} \frac{\partial \mathcal{L}}{\partial \dot{\psi_k^*}}=\frac{\partial \mathcal{L}}{\partial\psi_k^*}\label{Lagrange_equation_2},
\end{align}
with well-determined boundary conditions. A final integration over the endpoints of the trajectories give the final result that we summarize below.\\

At a given time $t \geq t_0$ and for any  $|\sigma_f\rangle, |\sigma_{f'}\rangle  \in \{ |\sigma_{1}\rangle \equiv |\uparrow_{z}\rangle, |\sigma_{2}\rangle\equiv|\downarrow_{z}\rangle\}$, the element of the spin-reduced density matrix between $| \sigma_f\rangle$ and $| \sigma_{f'}\rangle$  read
\begin{align}
\langle \sigma_f |\rho_S (t)|\sigma_{f'} \rangle=\sum_{k,k'}  [\rho_S (t_0)]_{k,k'} J_{k,k',f,f'},
\label{eq:densitymatrix_with_identity}
\end{align}
where $J_{k,k',f,f'}$ takes the form 
\begin{align}
J_{k,k',f,f'}=\int \mathcal{D}[\Sigma] \mathcal{D}[\Sigma'] \mathcal{A} [\Sigma]  \mathcal{A}^* [\Sigma'] \mathcal{F}[\Sigma, \Sigma'].
\label{eq:densitymatrixelement}
\end{align}
The integration in Eq. (\ref{eq:densitymatrixelement}) runs over all constant by parts paths $\Sigma$ and $\Sigma'$ taking values in $\{-1,1\}$ with endpoints verifying $\sigma^z|\sigma_k\rangle=\Sigma(t_0)|\sigma_k\rangle$ and $\sigma^z|\sigma_{k'}\rangle=\Sigma'(t_0)|\sigma_{k'}\rangle$, $\sigma^z|\sigma_f\rangle=\Sigma(t)|\sigma_f\rangle$ and $\sigma^z|\sigma_{f'}\rangle=\Sigma'(t)|\sigma_{f'}\rangle$. The term $\mathcal{A} [\Sigma]$ denotes the amplitude to follow one given path $\Sigma$ in the sole presence of the spin Hamiltonian. The effect of the environment is fully contained in the so-called Feynman-Vernon influence functional $\mathcal{F}[\Sigma, \Sigma']$ which reads \cite{FV} :
\begin{equation}
\mathcal{F}[\Sigma,\Sigma']=e^{ \left\{-\frac{1}{\pi} \int_{t_0}^t ds \int_{t_0}^s ds'\left[-i L_1(s-s')\frac{ \Sigma (s)-\Sigma '(s) }{2} \frac{ \Sigma (s')+\Sigma '(s') }{2} +L_2(s-s')\frac{\Sigma (s)-\Sigma'(s) }{2} \frac{ \Sigma (s')-\Sigma'(s')}{2}\right] \right\}}.
\label{eq:influence}
\end{equation}
 The functions $L_1$ and $L_2$ read
\begin{align} &L_1(t)=\int_0^{\infty} d \omega J(\omega) \sin \omega t ,  \notag \\
&L_2(t)=\int_0^{\infty} d \omega J(\omega) \cos \omega t \coth \frac{\beta \omega}{2}.
\label{Ls_2}
\end{align}\\

 From Eq. (\ref{eq:influence}), we see that the bosonic environment couples the symmetric and anti-symmetric spin paths 
\begin{align} 
&\eta(t)=\frac{1}{2}[\Sigma(t)+\Sigma'(t)],\label{eta} \\
&\xi(t)=\frac{1}{2}[\Sigma(t)-\Sigma'(t)],\label{xsi}
\end{align}
 at different times. These variables take values in $\{-1,0,+1\}$ and are the equivalent of the classical and quantum variables in the Schwinger-Keldysh representation. We have then integrated out the bosonic degrees of freedom, which no longer appear in the expression of the spin dynamics, but the prize to pay is the introduction of a spin-spin interaction term which is not local in time. Dealing with such terms is difficult at a general level. The spin dynamics at a given time $t$ depends on its state at previous times $s<t$: the dynamics is said to be non-Markovian. The effective action is reminiscent of the classical spin chains with long-range interaction, where time now replaces space. In particular for an ohmic spectral density given by (\ref{J_ohmic}), $L_2(t) \propto 1/(\omega_c t)^2$ at long times. When integrated twice, we recover the characteristic $\ln$ behavior found by Anderson, Yuval and Hamman \cite{Anderson_Yuval_Hamann} when studying the Kondo problem (see \ref{mapping_kondo_spinboson}). \\ 
 
 Non-Markovian master equations \cite{Tu_Zhang,Zhang_Nori} can be derived from this Feynman-Vernon influence functional in the case of excitation number conserving interaction terms (of the form $(\sigma_+ a +\sigma_- a^{\dagger})$), by integrating out the system variables trajectories. This point of view requires however to carry out the path integral using Grassman coherent states after a fermionization of the spin. An interesting open issue left for further work is to generalize this method in our case of spin-boson coupling with counter-rotating terms.\\

\subsection{ ``Blips'' and ``Sojourns''}
\label{Blips_sojourns}

The next step is the rewriting of the spin path in the language of ``Blips'' and ``Sojourns'', following the seminal work of Ref. \cite{leggett:RMP}. This technique is also well explained in Ref. \cite{Weiss:QDS}.\\

\begin{figure}[h!]
\center
\includegraphics[scale=0.30]{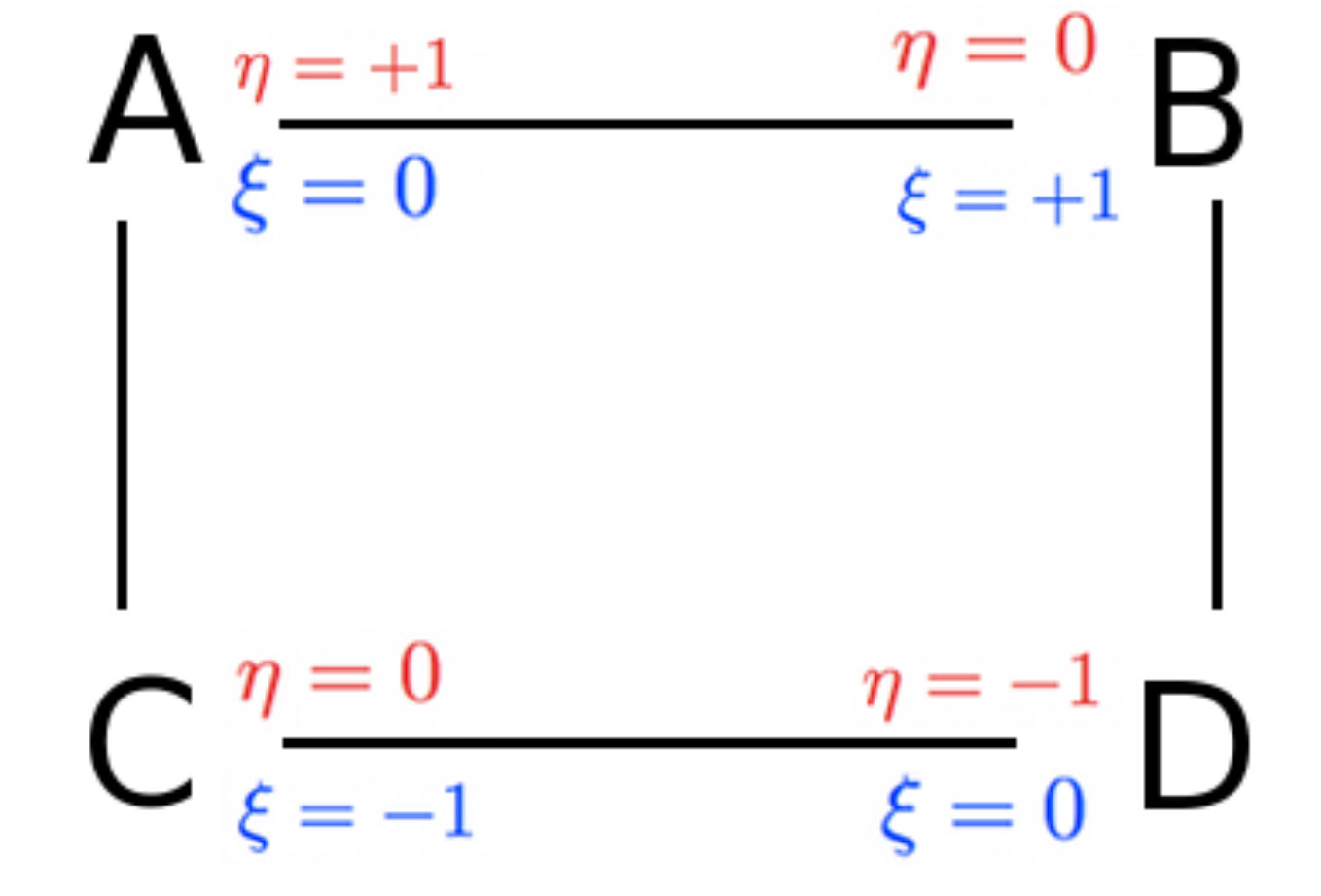}  
\caption{Spin states.}
\label{etats}
\end{figure}

 The double path integral along $\Sigma$ and $\Sigma'$ in Eq. (\ref{eq:densitymatrixelement}) can be viewed as a single path that visits the four states A (for which $\eta=1$ and $\xi=0$), B (for which $\eta=0$ and $\xi=1$), C (for which $\eta=0$ and $\xi=-1$) and D (for which $\eta=-1$ and $\xi=0$). States A and D correspond to the diagonal elements of the density matrix (also named `sojourn' states) whereas B and C correspond to the off-diagonal ones (also called `blip' states) \cite{leggett:RMP,Weiss:QDS}. The four states are depicted in Fig.~\ref{etats}. As stated below Eq. (\ref{eq:densitymatrixelement}), the initial state of these paths is characterized by the initial spin-reduced density matrix and the final state by $|\sigma_f\rangle$ and $|\sigma_f'\rangle$. \\
 
A careful examination of $J_{k,k',f,f'}$ in Eq. (\ref{eq:densitymatrix_with_identity}) is done in Ref. \cite{leggett:RMP} and in Chapter ``Two-state dynamics" of Ref. \cite{Weiss:QDS} in the general case, and presented in Appendix \ref{appendix_blips_sojourns}. It is however instructive to examine some details of this blips and sojourns rewriting, and we present below the computation of $\langle \uparrow_z |\rho_S(t)|\uparrow_z  \rangle$ for a spin starting initially in the state $\rho_S(t_0)=|\uparrow_z\rangle\langle \uparrow_z|$. \\

\begin{figure}[t!]
\center
\includegraphics[scale=1.05]{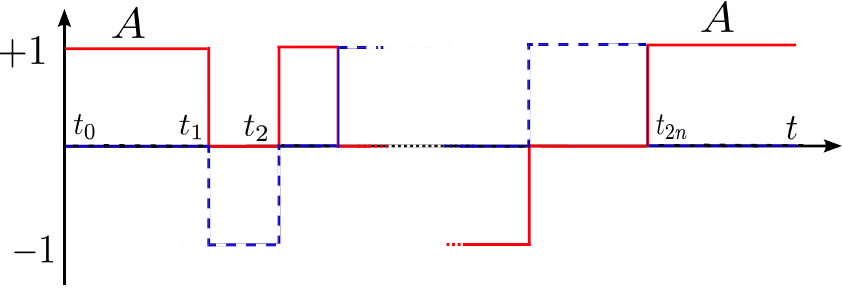}  
\caption{Spin path with initial and final state A - $\eta(t)=\sum_{j=0}^{2n} \Upsilon_j\theta(t-t_j)$ in red; $\xi(t)=\sum_{j=1}^{2n} \Xi_j\theta(t-t_j)$ in dashed blue.}
\label{spin_path_1}
\end{figure}

In this case the sum in Eq. (\ref{eq:densitymatrix_with_identity}) reduces to a single term as $k=k'=1$ from the initial condition and $f=f'=1$ for the term we seek to compute. To compute the corresponding $J_{1,1,1,1}$, we have to consider all the double spin paths which start and end in the sojourn state A. One path of this type makes $2n$ transitions along the way at times $t_i$, $i \in \{1,2,..,2n\}$ such that $t_0<t_1<t_2<...<t_{2n}$. We can write this spin path as $\xi(t)=\sum_{j=1}^{2n} \Xi_j\theta(t-t_j)$ and $\eta(t)=\sum_{j=0}^{2n} \Upsilon_j\theta(t-t_j)$ where the variables $\Xi_i$ and $\Upsilon_i$ take values in $\{-1,1\}$. Such a path is illustrated in Fig. \ref{spin_path_1}. The variables $\Xi$ (in blue) describe the blip parts, and the variables $\Upsilon$ (in red) on the other hand characterize the sojourn parts. \\

Using an explicit representation of the path measure (introduced in Eq. (\ref{spin_appendix_measure})), $J_{1,1,1,1}$ is given by an exact series in the tunneling coupling $\Delta^2$ :
\begin{equation}
\langle \uparrow_z |\rho_S(t)|\uparrow_z  \rangle=J_{1,1,1,1}=\sum_{n=0}^{\infty} \left(\frac{i\Delta}{2} \right)^{2n} \int_{t_0}^{t} dt_{2n} ... \int_{t_0}^{t_2} dt_{1} \sum_{\{\Xi_j\},\{\Upsilon_j\}' } \mathcal{F}_{n}.
\label{eq:p(t)}
\end{equation}
 The prime in $\{\Upsilon_j\}'$ in Eq. (\ref{eq:p(t)}) indicates that the initial and final sojourn states are fixed according to the initial and final conditions $\Upsilon_0=\Upsilon_{2n}=1$. $\mathcal{F}_{n}$ corresponds to the evaluation of $\mathcal{F}[\Sigma,\Sigma']$ for a given path with $2n$ spin flips, introduced above. Given this path, we can evaluate Eq. (\ref{eq:influence}). First we evaluate the contribution given by $L_1$.
\begin{align}
&\frac{i}{\pi}\int_{t_0}^t ds \int_{t_0}^s ds' L_1(s-s') \xi(s) \eta(s')=\frac{i}{\pi} \sum_{j>k=0}^n \xi_j \eta_k \int_{t_{2j-1}}^{t_{2j}} ds \int_{t_{2k}}^{t_{2k+1}} ds'L_1(s-s') \notag\\
&=\frac{i}{\pi} \sum_{j>k=0}^n \xi_j \eta_k [ Q_1(t_{2j-1}-t_{2k})+Q_1(t_{2j}-t_{2k+1})-Q_1(t_{2j}-t_{2k}) -Q_1(t_{2j-1}-t_{2k+1})] \notag\\
&=\frac{i}{\pi} \sum_{j>k=0}^{2n} \Xi_j \Upsilon_k  Q_1(t_j-t_k),
\label{eq:integration}
\end{align}
with $Q_1$ the opposite of the second integral of $L_1$ with $Q_1(0)=0$. In Eq. (\ref{eq:integration}), variables $\xi_j=\Xi_{2j-1} $ and $\eta_k=\Upsilon_{2k}$) denote the values of the $j$-th blip (starting from 1) and $k$-th sojourn (starting from 0). Then we evaluate the contribution given by $L_2$, which contains a self-interaction term,
\begin{align}
-\frac{1}{\pi}\int_{t_0}^t ds \int_{t_0}^s ds' L_2(s-s') \xi(s) \xi(s')=&-\frac{1}{\pi} \sum_{j>k=0}^n \xi_j \xi_k \int_{t_{2j-1}}^{t_{2j}} ds \int_{t_{2k-1}}^{t_{2k}} ds'L_2(s-s')\notag \\
&-\frac{1}{\pi} \xi_j \xi_j \int_{t_{2j-1}}^{t_{2j}} ds \int_{t_{2j-1}}^{s} ds'L_2(s-s')\notag\\
&=\frac{1}{\pi} \sum_{j>k=0}^{2n} \Xi_j \Xi_k  Q_2(t_j-t_k),
\end{align}
with $Q_2$ the second integral of $L_2$ with $Q_2(0)=0$. This leads to \cite{leggett:RMP}:
 \begin{align}
 &\mathcal{F}_{n}= \exp \left[ \frac{i}{\pi} \sum_{k=0}^{2n-1}\sum_{j=k+1}^{2n} \Xi_j \Upsilon_k  Q_1(t_j-t_k) \right] \exp \left[ \frac{1}{\pi} \sum_{k=1}^{2n-1}\sum_{j=k+1}^{2n} \Xi_j \Xi_k  Q_2(t_j-t_k) \right]\label{Q} . 
 \end{align}
 The functions $Q_1$ and $Q_2$, which describe the feedbacks of the dissipative environment, are directly obtained from the spectral function $J(\omega)$ ,
\begin{align}
 Q_1(t)&=\int_0^{\infty} d\omega\frac{J(\omega)}{\omega^2}\sin \omega t, \label{q1} \\
  Q_2(t)&=\int_0^{\infty} d\omega\frac{J(\omega)}{\omega^2}\left(1-\cos \omega t\right) \coth \frac{\beta \omega}{2}.\label{q2}
\end{align} 
From Eq.~(\ref{Q}), we see that the first term couples the blips to all the previous sojourns, while the second one couples the blips to all the previous blips (including self-interaction). Blips and sojourns do not have symmetric effects: the index for the $\Upsilon$ variables starts at $0$ and ends at $2n-1$ whereas the index for the $\Xi$ variables starts at $1$ and ends at $2n$. It is worth  noting that the last sojourn does not contribute and the latest coupling period is the blip which lasts from $t_{2n-1}$ to $t_{2n}$. We recall that we provide in Appendix \ref{appendix_blips_sojourns} the general rewriting in this language to compute the other elements of the spin-reduced density matrix, for any initial condition.\\
 
 Equations (\ref{Q},\ref{q1},\ref{q2}) constitute a more explicit rewriting of Eq. (\ref{eq:influence}), but we are still left with long range spin-spin interaction induced by the bath. Such terms can be evaluated in certain cases by analytical methods. For example, the authors of Ref. \cite{leggett:RMP} developped the so-called Non-Interacting-Blip-Approximation (NIBA) to tackle the dynamics of spinboson models. This approximation is notably justified in the Ohmic case at weak coupling and/or large temperatures. It assumes that the time spent by the system in the states B and C (blips) is very small compared to the time spent in states A and D. Consequently, the interblips correlations of Eq. (\ref{Q}) have little effect on the dynamics and may be ignored. In this limit, analytical results may be obtained (see \cite{leggett:RMP,Weiss:QDS}).\\
 
 Real-time path-integral simulation techniques combining Monte Carlo sampling over the quantum fluctuations $\Xi$ with an exact treatment of the quasi-classical degrees of freedom $\Upsilon$ were also developped from Eq. (\ref{Q}) \cite{QMC_1,QMC_2,QMC_3,QMC_4}.\\
 
 Here, we follow a different path and use additional stochastic degrees of freedom to unravel the spin-spin interactions.

\subsection{Stochastic unravelling}
\label{unravelling}
To decouple the spin-spin interaction(s) in time, we use Gaussian stochastic decoupling (or Hubbard-Stratonovitch transformation in function space). Some efforts in this direction were done previously in Refs. \cite{Stockburger_Mac,Stockburger,Stockburger_2} and we shall come back to these results at the end of the present subsection. This stochastic unravelling of the influence functional will allow us to write the dynamics of the spin-reduced density matrix as a solution of a stochastic differential equation \cite{2010stoch,Peter,Rabi}.\\

 We focus again on the computation of $\langle \uparrow_z | \rho_S(t) |\uparrow_z \rangle$ for a spin starting initially in the state $|\uparrow_z \rangle$. Let $h$ and $k$ be two complex gaussian random fields which verify \cite{Rabi}
\begin{align}
 \overline{ h(t) h(s)} = & \frac{1}{\pi} Q_2(t-s) + l_1, \label{height_1} \\
 \overline{ k(t) k(s)} = &\  l_2,    \label{height_2}   \\
 \overline{ h(t) k(s) } = & \frac{i}{\pi}  Q_1(t-s) \theta(t-s) + l_3. \label{height_3}
\end{align}
The overline denotes statistical average, $\theta(.)$ is the Heaviside step function and $l_1$, $l_2$ and $l_3$ are arbitrary complex constants. Making use of the identity $\overline{\exp(X)}=\exp(\overline{X^2}/2 )$, Eq. (\ref{Q}) can then be reexpressed as:
 \begin{align}
\mathcal{F}_{n}= \overline{  \prod_{j=1}^{2n} \exp\left[h(t_j) \Xi_j+k(t_{j-1}) \Upsilon_{j-1}    \right]}.
\label{functionnal_1}
\end{align}
The complex constants $l_p$ do not contribute to the average because $\sum_{k=0}^{2n-1} \Upsilon_k=\sum_{j=1}^{2n} \Xi_j=0$. This step was done in Refs. \cite{2010stoch,Peter} with the introduction of one stochastic field (which is valid in a certain limit, as we will see later), and with two fields in Ref. \cite{Rabi}. The summation over blips and sojourn variables $\{\Xi_j\}$ and $\{\Upsilon_j\}$ can be incorporated by considering a product of matrices of the form 
\begin{equation}
V= \left( \begin{array}{cccc}
0&e^{-h+k }&-e^{h+k }&0 \\
e^{h-k }&0&0&-e^{h+k }\\
-e^{-h-k }&0&0&e^{-h+k }\\
0&-e^{-h-k }&e^{h-k }&0
\end{array} \right),
\label{eq:spin_hamiltonian}
\end{equation}
in the four dimensional vector space of states $\{\textrm{A},\textrm{B},\textrm{C},\textrm{D}\}$. This rewriting was originally introduced in Ref. \cite{Lesovik}. Then, the computation of diagonal elements of the  we get
\begin{equation}
\langle \uparrow_z | \rho_S(t) |\uparrow_z \rangle=\overline{\sum_{n=0}^{\infty} \left(\frac{i\Delta}{2} \right)^{2n} \int_{t_0}^{t} dt_{2n} ... \int_{t_0}^{t_2} dt_{1} \prod_{j=1}^{2n} V (t_j)}.
\label{eq:p(t)_12}
\end{equation}\\

We remark that Eq. (\ref{eq:p(t)_12}) has the form of a time-ordered exponential, averaged over stochastic variables, so that we finally have:
\begin{equation} 
\langle \uparrow_z | \rho_S(t) |\uparrow_z \rangle=\overline{\langle \Phi_f | \Phi (t) \rangle},
\label{scalar_prod_SSE}
\end{equation}
where $\langle \Phi_f |=(e^{-k(t_{2n})},0,0,0)$ and  $| \Phi \rangle$ is the solution of the Stochastic Schr\"{o}dinger Equation (SSE),

\begin{equation} 
i \partial_t | \Phi \rangle = V(t) | \Phi \rangle
\label{SSE_bis}
\end{equation}
with initial condition $|\Phi_i \rangle=(e^{k(t_0)},0,0,0)^T$.\\

The vector $|\Phi(t) \rangle$ represents the double spin state which characterizes the spin density matrix. The vectors $|\Phi_i \rangle$ and $|\Phi_f \rangle$ are related to the initial and final conditions of the paths. As spin paths start and end in the sojourn state A, only the first component of these vectors is non-zero. The choice of the phases is linked to the asymmetry between blips and sojourns (see Eq.~(\ref{Q})). The contribution from the first sojourn is encoded in $|\Phi_i \rangle$, and we artificially suppress the contribution of the last sojourn via $|\Phi_f \rangle$. This final vector depends on an intermediate time, but we can notice that replacing $(e^{-k(t_{2n})},0,0,0)$ by $(e^{-k(t)},0,0,0)$ does not add any contribution on average. Generalization of the procedure to compute the other elements of the spin-reduced density matrix for an arbitrary spin initial state is straightforward, and leads to the general result presented at the beginning of the chapter.\\

The resolution protocol requires a large number of realizations of the fields $h$ and $k$. For each realization, we solve the stochastic equation and the spin density matrix is obtained by averaging over the results of all the realizations. The authors of Refs. \cite{Stockburger_Mac,Stockburger,Stockburger_2} used a similar stochastic decoupling directly on the expression (\ref{eq:influence}), before the blip and sojourn rewriting. The stochastic processes relevant in this case obey correlation relations similar to Eqs. (\ref{height_1},\ref{height_2},\ref{height_3}) where $L_1$ and $L_2$ replace $Q_1$ and $Q_2$. They reached an effective stochastic Liouville equation for the density matrix. This technique has notably been used to compute the dynamics for the Morse oscillator \cite{Koch_morse}. \\

\section{Properties of the SSE equation}

We characterize some properties of the SSE equation and verify its relevance to describe the dynamics of an open quantum system. 

\subsection{Non-unitarity and trace preservation}
As they verify complex-valued correlations given by Eqs. (\ref{height_1}-\ref{height_3}), the stochastic fields $h$ and $k$ are complex numbers and have both a real and an imaginary part. Thus, the effective Hamiltonian $V$ for the density matrix is not hermitian. As a consequence, the evolution is not unitary and the norm of the vector $ | \Phi \rangle$ (which gives the density matrix norm $||\rho(t)||=\left[1/2(1+\langle \sigma^x(t) \rangle^2+\langle \sigma^y(t) \rangle^2+\langle \sigma^z(t) \rangle^2)\right]^{1/2}$\linebreak when averaged) is not conserved over time. This is not surprising for the description of an open system subject to decoherence. The time evolution may bring the system from a pure quantum state with $||\rho||=1$ to a mixed state with $||\rho||<1$.\\

The trace of the density matrix at time $t$ is given by $\overline{e^{k(t)} \Phi_1 (t)+e^{-k(t)} \Phi_4 (t)}$. One can verify from equations (\ref{eq:spin_hamiltonian}) and (\ref{SSE_bis}) that $d\left\{\overline{e^{k(t)} \Phi_1 (t)+e^{-k(t)} \Phi_4 (t)}\right\}=0$ for all $t$ if $\overline{dk(t)}=0$ (or more simply in the discretized version of the process, if $\overline{k_j - k_{j-1}}=0$). In order to check that the SSE equation is trace-preserving on average, one needs to properly define the stochastic processes $h$ and $k$. This is done in Appendix \ref{appendix:sampling}, where we notably focus on the sampling procedure and present two options.
\begin{itemize}
\item  It seems first natural to take advantage of the translational invariance of equations (\ref{height_1}-\ref{height_3}) and use Fourier series decomposition of the functions $Q_1$ and $Q_2$. Fourier sampling constitues our first option and was already introduced in Refs. \cite{2010stoch,Peter}.
\item The will to give a precise definition of the stochastic processes $h$ and $k$ lead to a description in relation with Autoregressive models frequently used in statistics and signal processing. This constitutes our second sampling method.
\end{itemize}
We find that the SSE is indeed trace preserving on average, but the trace is however not constant for each realization, exluding then the interpretation of each trajectory in terms of a real physical process. \\

\subsection{Dynamical sign problem}

The non-unitarity of the SSE equation may lead to convergence problems when increasing the spin-environment coupling. The presence of a non-zero real part in the expression of the stochastic fields engenders generally an exponential slowing down of the convergence when increasing spin-bath coupling. This is the well-known dynamical sign problem occuring for real-time numerical methods. The SSE is not an exception. In particular cases however, this dynamical sign problem can be dealt with. We will show below that the SSE method gives reliable results for:
\begin{itemize}
\item  the Rabi model, even in the strong coupling regime $g/\omega_0 \simeq 1$.
\item ohmic systems in the scaling limit characterized by $0<\alpha<1/2$ and $\Delta/\omega_c \ll 1$. 
\end{itemize} 
 Generalizing its use in other cases such as the sub-ohmic and super-ohmic spinboson model remains however an open question. \\
 
 In the general case, one may try to minimize the real part of the fields in the sampling in order to improve convergence. In particular, the constants are chosen such that $\overline{\mathcal{R}h}(t)=\overline{\mathcal{R}k}(t)=0$. It is however not possible to simultaneously reduce the variance of the real parts at will, as they are contrained by Eq. (\ref{height_2}). In practice, we choose a naturally symetric allocation between the two fields $h$ and $k$ due to Fourier development.
%\subsection{Stochastic}

\subsection{Formulation in terms of a Stochastic Master equation}

The SSE equation (\ref{SSE}) can be formulated in terms of spin jump operators dressed by stochastic fields, by analogy to standard master equations. Considering the matrix version of $|\Phi \rangle$
\begin{equation} 
\Phi=\left( \begin{array}{cc}
\Phi_1&\Phi_2  \\
\Phi_3 &\Phi_4
\end{array} \right),
\end{equation}
we have more precisely
\begin{equation} 
i \partial_t  \Phi =\frac{\Delta}{2}\left( e^{h-k} \Phi \sigma_+ + e^{-h+k} \Phi \sigma_- -e^{h+k} \sigma_+ \Phi  -e^{-h-k} \sigma_- \Phi \right) .
\label{master_SSE}
\end{equation}
The density matrix $\rho$ is obtained from (\ref{master_SSE}) after a stochastic averaging.

\subsection{NIBA from the SSE}
We can recover NIBA equation by considering from Eq. (\ref{SSE}), using an independence hypothesis. The equation obtained with this derivation bears similarities with the one obtained in Sec. \ref{NIBA_from_EOM}. From Eq. (\ref{SSE}), one can indeed reach the closed form
\begin{align}
\partial_t \left[ \psi_1 -\psi_4  \right]=-\frac{\Delta^2}{2} \int_{t_0}^t ds &\Big\{ \psi_1(s)\left[ e^{-h(t)-k(t)+h(s)-k(s)}+e^{h(t)-k(t)-h(s)-k(s)} \right] \notag \\
 & -\psi_4(s)\left[ e^{-h(t)+k(t)+h(s)+k(s)}+e^{h(t)+k(t)-h(s)+k(s)} \right]  \Big\} .
\label{NIBA_from_SSE}
\end{align}
In this form, stochastic operators $h$ and $k$ play the role of the bath operators $\Omega$ of Eq.  (\ref{NIBA_EOM_0}). Taking the stochastic average with an independence hypothesis on the products appearing on the right hand side of (\ref{NIBA_from_SSE}) leads to NIBA equation (\ref{NIBA_EOM}). The independence hypothesis is equivalent to the decoupling in the expectation values used to derive NIBA with the Heisenberg equations of motion.
\newpage
\subsection{Relation to other methods}

\begin{figure}[h!]
\center
\includegraphics[scale=0.46]{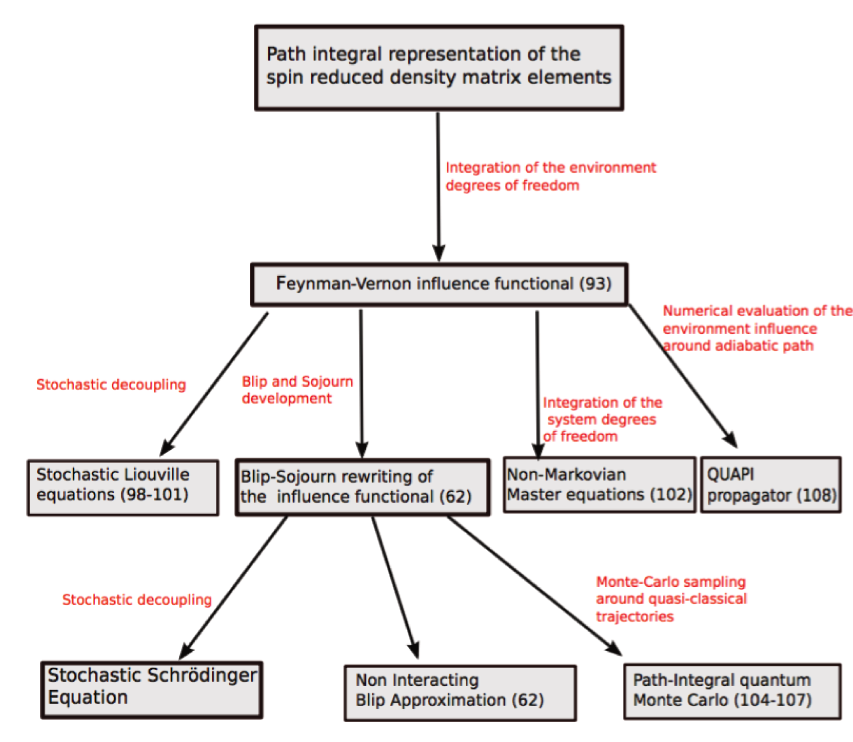}  
\caption{Relations to other methods.}
\label{methodes_relation}
\end{figure}

We show in Fig. \ref{methodes_relation} a table which aims to summarize the links that can be made between the SSE equations and other related methods based on the same grounds.

\section{Application : dynamics in the quantum Rabi model}

We test this framework on the well-known Rabi model studied in Section \ref{presentation_rabi}. We show that the SSE method gives reliable results in this simple case, even in the strong coupling regime where the RWA approximation breaks down. We use this simple example to illustrate that the SSE method also allows to take into account driving terms and particle losses in an exact manner. The Rabi model corresponds to a spectral function of the form
\begin{align}
J(\omega)=\pi g^2 \delta(\omega-\omega_0).
\end{align}

\subsection{Strong coupling regime}

Hamiltonian $\mathcal{H}_{Rabi}$ defined in Eq. (\ref{Rabi}) is obtained from Eq. (\ref{Hamiltonien_general_spinboson_methode}) by considering a coupling to one bosonic mode, after a $\pi/2$ rotation of axis $(Oy)$, generated by $\exp(-i \pi/4 \sigma^y)$. The identification is complete provided that we change the sign in front of the transverse field ($-\Delta \to \Delta$). \\

%We show here that the SSE framework gives reliable results concerning the spin dynamics in the strong coupling regime. We moreover show that this framework is quite flexible, as it allows to describe realistic experimental conditions with driving term and photon losses.

 %We also discuss non-trivial dynamical final states with one polariton achieved by driving the system, analogous to a standard $\pi$-pulse for polaritons. By increasing the drive amplitude we decrease the characteristic time to reach a pure state with one polariton under the typical coherence type of the system. This may find applications to realize a driven Mott state of polaritons, {\it i.e.}, dressed states (eigenstates) of light and matter \cite{houck}, in the weak-coupling limit between light and matter. 

\subsubsection{From the Jaynes-Cummings regime to the adiabatic regime}

\begin{figure}[h!]
\center
\begin{minipage}[b]{0.5\linewidth}
 \center
\includegraphics[scale=0.26]{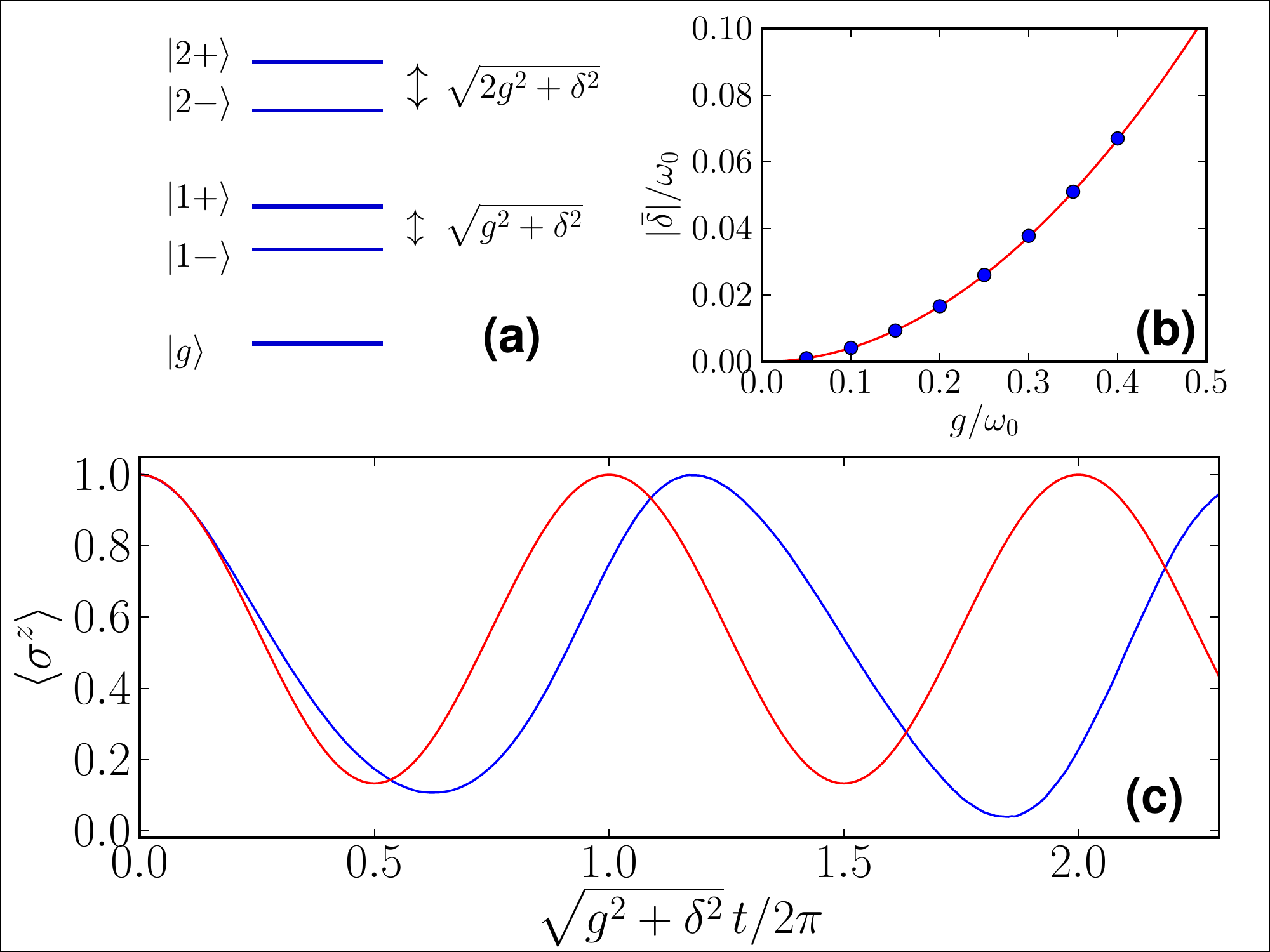}
 ~\\
  ~\\
\end{minipage}\begin{minipage}[b]{0.5\linewidth}
 \center
\includegraphics[scale=0.18]{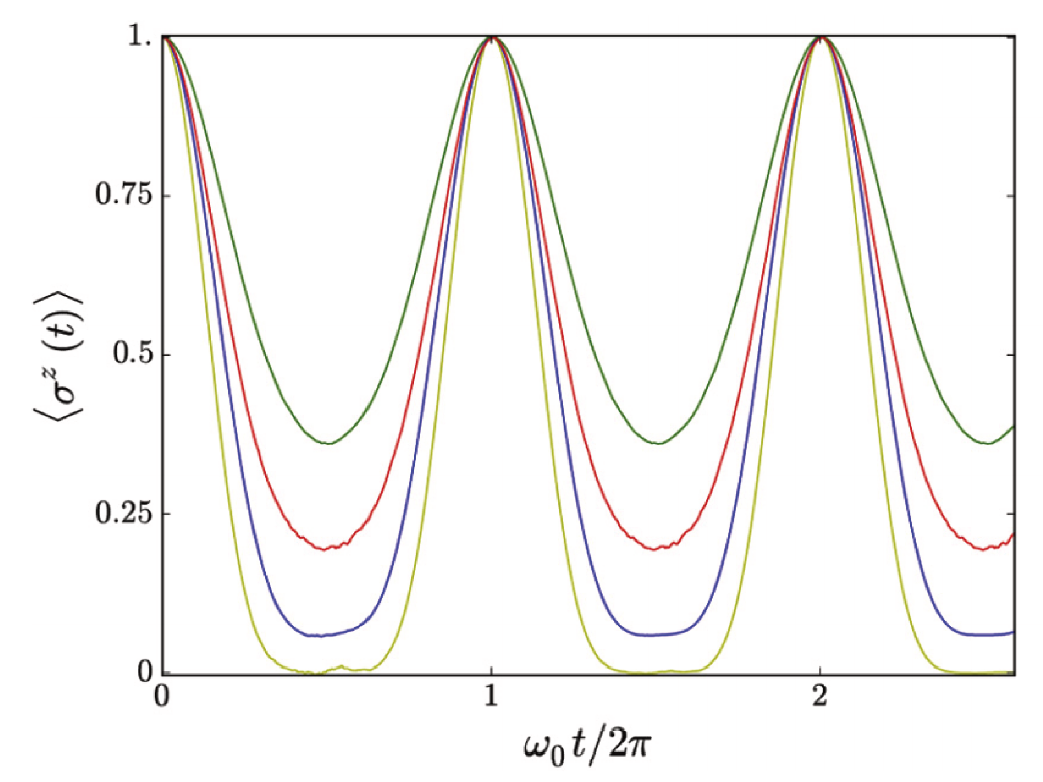}  
\end{minipage}
\caption{Left: (a) JC ladder of polaritons. (b) Absolute
value of the Bloch-Siegert shift $|\overline{\delta}|$ versus $g/\omega_0$, perturbation
theory in red \cite{BS_cohen}, and results from our method in blue (dots). (c) Example of dynamics of $\langle \sigma^z \rangle$ with the initial condition
$|\uparrow_z\rangle$, which is a linear superposition of $|1-\rangle$ and $|1+\rangle$, for quite strong couplings. Parameters are set to $g/\omega_0 = 0.7$, $\Delta/\omega_0 = 0.2$. The Rabi solution from our method is shown in dashed blue; within the RWA, the JC solution would rather
read $\langle \sigma^z (t) \rangle=1-2g^2 \sin^2 \left(\sqrt{g^2+\delta^2}t/2 \right)/(g^2+\delta^2)$ and is shown in red. Right: far detuned or adiabatic limit of the Rabi model: $g/\omega_0$ from 0.7 (green) to 1.5 (yellow) and $\Delta/\omega_0$ = 0.05. The system is prepared in the initial state $|\uparrow_z \rangle \otimes |0 \rangle $ and the revival probability of the initial state can be computed exactly in this limit \cite{Solano}. The positions of the maxima only depend on the cavity frequency $\omega_0$.}
\label{bloch_siegert}
\end{figure}

As studied in detail in Ref. \cite{Rabi}, the SSE gives correct results in the Jaynes-Cummings regime. Numerical results at higher coupling are also consistent with the analytical expression which quantify the effects of counter-rotating terms in perturbation theory, as can be seen on the left panel of Fig. \ref{bloch_siegert}. The presence of the counter-rotating terms in the quantum Rabi model gives rise at second order in perturbation theory to a shift of the resonance frequency between the atom and photon, leading to an additional negative detuning $\overline{\delta}=-g^2/[2(\omega_0+\Delta)]$ when $\Delta<\omega_0$ \cite{BS_cohen}. In Fig. \ref{bloch_siegert}, we also show numerical results in the adiabatic regime presented in Sec. \ref{strong_coupling}, characterized by $\Delta/\omega_0 \ll 1$. Starting from the initial state $|\uparrow_z\rangle\otimes |0\rangle$, it is possible to compute the revival probability of this state, which reads $p=\exp\left[- g^2/\omega_0^2 |e^{-i\omega_0t}-1|^2\right]$ \cite{Solano}. This evolution engenders periodic collapses and revivals of $\langle \sigma^z(t) \rangle$.

After this basic test, we use the single-mode case to introduce drive and dissipation in the SSE framework.

\subsection{Drive and photon losses}

It is actually possible to consider both photon leakage out of the cavity and driving. Photon losses can be incorporated by considering an imaginary part to the photon frequency. This leads to a change in the coupling functions  $Q_1$ and $Q_2$ \cite{Rabi}.\\

 The effect of a coherent semi-classical drive can be treated exactly by formally substituting $\Sigma (t)$ by $(\Sigma (t)+V(t))$ in the path integral approach. This is simply reflected by the appearance of a new coupling term. Assuming $V(t)$ to be of the form $V_0 \cos \omega_d t$ and beginning the procedure at time $t_0$, the functional $F[\Sigma, \Sigma']$ is changed into $F^d [\Sigma, \Sigma']$ which reads, for $t \geq t_0$:

\begin{equation}
 F^d[\Sigma, \Sigma']=e^{\left[ 2 i V_0 g \int_{t_0}^t ds \int_{t_0}^s ds' \sin \omega_0 (s-s') \xi(s)  \cos \omega_d s'    \right]} F[\Sigma, \Sigma']. 
 \end{equation}
The new contribution can be taken into account exactly. For example, if $\omega_d \ne \omega_0$, one subtitutes $h_d$ to $h$ with 
\begin{align}
\label{hxi_drived}
h_d(t)=h (t)+\frac{2iV_0 g \omega_0}{\omega_d^2-\omega_0^2} \Bigg\{&\frac{\sin \left[\omega_0 t+(\omega_0+\omega_d) t_0\right]}{\omega_0} +\frac{ \sin \omega_d t}{\omega_d}\Bigg\}.
\end{align}
It is also possible to consider the drive term with a RWA-type approximation\linebreak $V_0/2 \left(ae^{i\omega_d t}+a^{\dagger}e^{-i\omega_d t}\right)$, which only results for $\omega_d \ne \omega_0$ in the replacement of $2 V_0 g  \omega_0/(\omega_d^2-\omega_0^2)$ by $V_0 g/(\omega_d-\omega_0)$ in Eq. (\ref{hxi_drived}).

\subsubsection{Polariton $\pi$-pulse}

As an application, we present a driven protocol allowing to reach quantitatively a polariton state in the detuned regime $\Delta<\omega_0$. We set the drive frequency $\omega_d$ to match exactly the energy difference between the ground state and the first polariton (which has a greater ``atomic'' component, due to the negative detuning). In the limit of infinitely small drive $V_0/g\ll1$ the dynamics shows complete semi-classical Bloch oscillations of frequency $\alpha_1 V_0/2$ between these two levels; with $\alpha_1= [(A -\delta_r)/2A]^{1/2}$ and $A=\sqrt{g^2+\delta^2}$. This is due to the anharmonicity of the Jaynes-Cummings ladder: $E_{|1-\rangle}-E_{|g\rangle} \neq E_{|2-\rangle}-E_{|1-\rangle}$. Driving with frequency $\omega_d=E_{|1-\rangle}-E_{|g\rangle}$ triggers then predominantly one-photon excitations, resulting in the photon-blockade phenomenon (or polariton blockade). As the drive frequency does not match the energy difference between $|1-\rangle$ and $|2-\rangle$, no additional excitations occur at low drive strength. This blockade is often quantified by the measure of the quantity $g^{(2)}(\tau)=\langle a^{\dagger}(t)a^{\dagger}(t+\tau) a (t+\tau)a (t) \rangle/(\langle a^{\dagger}(t)a^{\dagger}(t+\tau)\rangle \langle a (t+\tau)a (t) \rangle)$. For a classical field, one has $g^{(2)}(0) \geq 1$ while $g^{(2)}(0)<1$ for a \textit{non-classical} state (with in particular $g^{(2)}(0)=0$ for a single photon-state). However, the switch-off time $t_s$ necessary to bring the system into the state $|1-\rangle$ with this weak drive protocol is typically longer than the decoherence time. This forbids then such an operation.

\begin{figure}[h!]
%\vskip -0.2cm
\center
\includegraphics[scale=0.3]{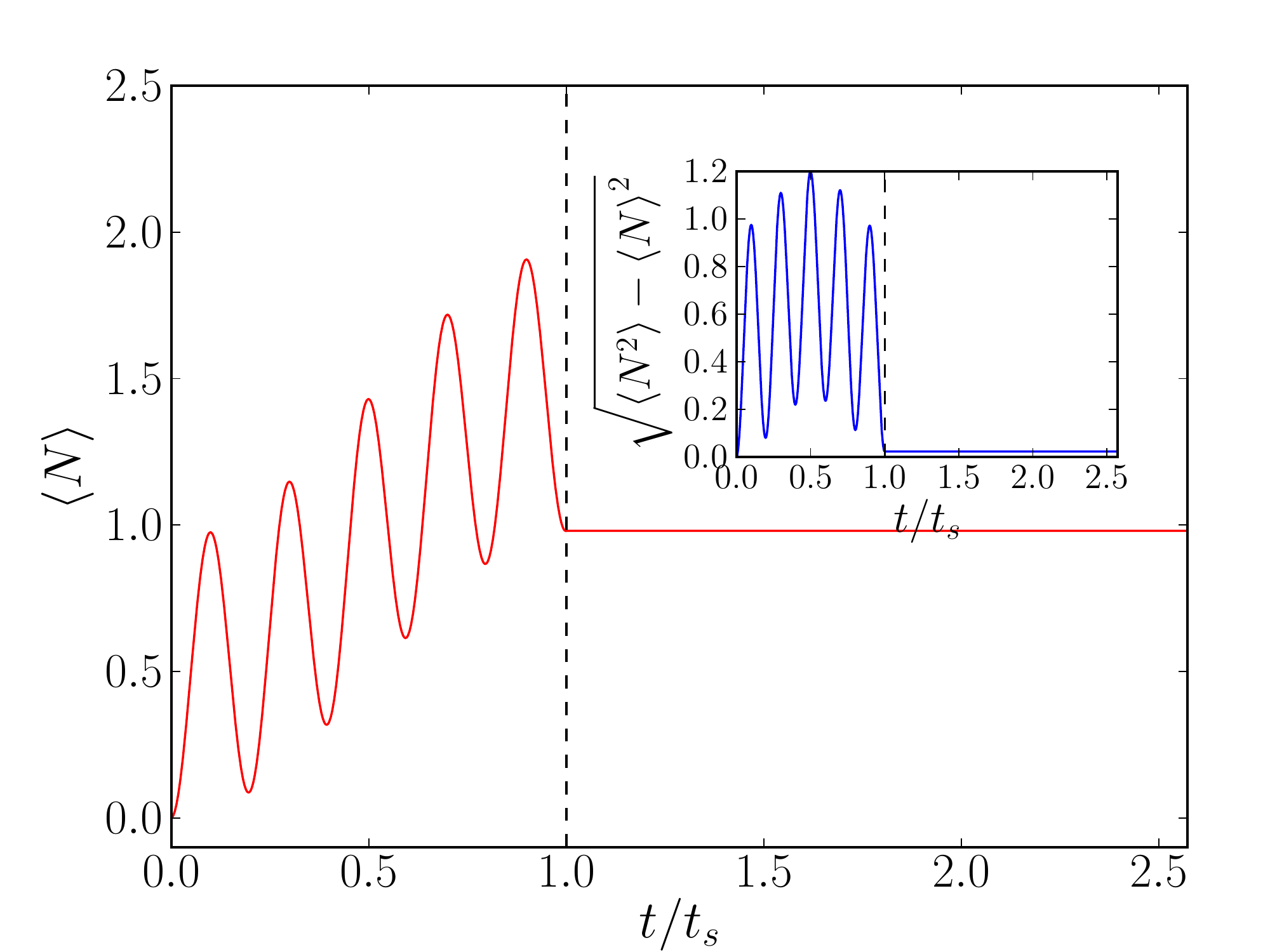} \includegraphics[scale=0.3]{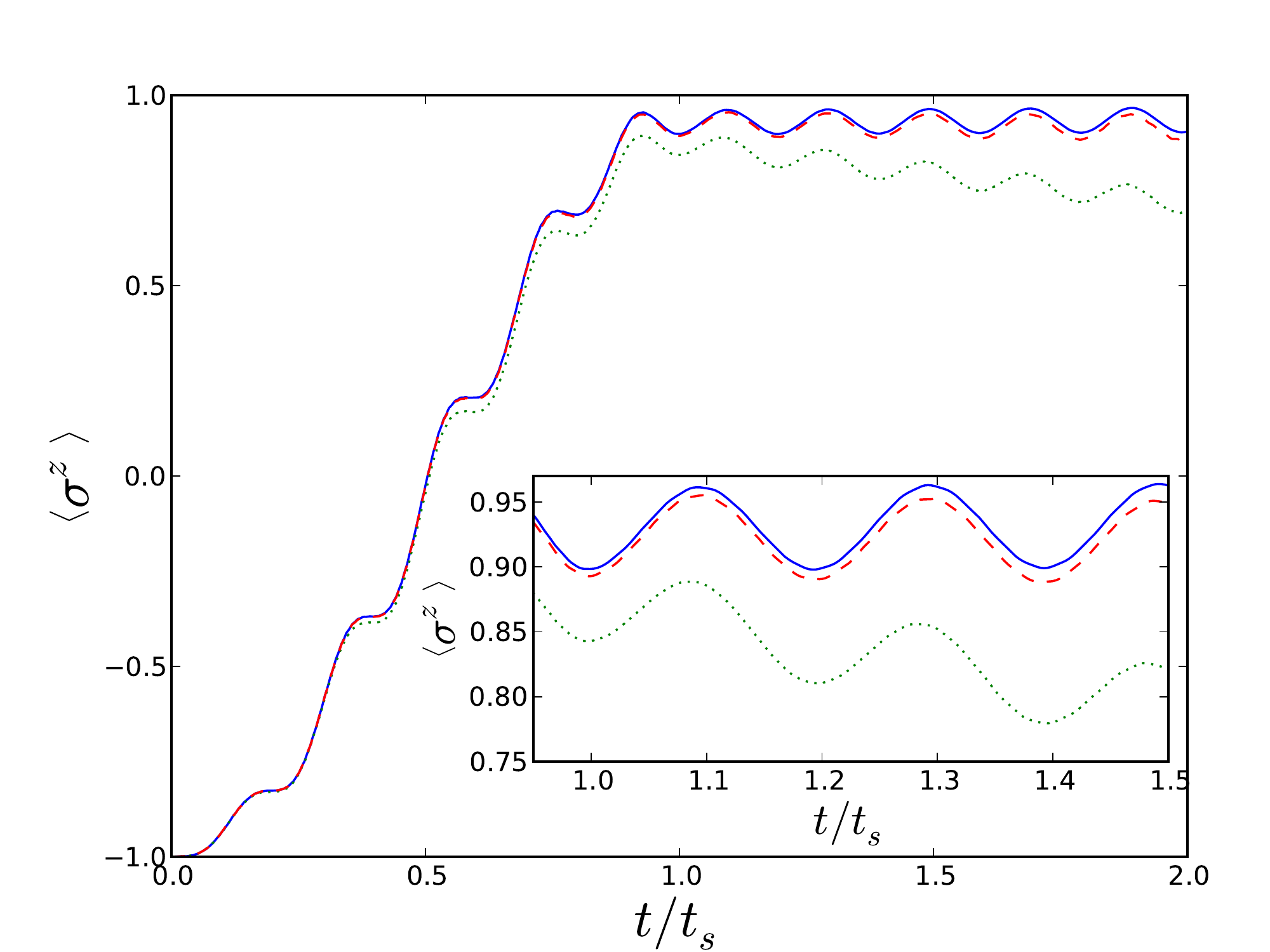}
\caption{Left panel: Dynamics of the mean number of polaritons $\langle N(t) \rangle$ in the weak-coupling $g$ limit without dissipation. Parameters are  $g/\omega_0=0.02$, $\Delta/\omega_0=0.9$ and $V_0/\omega_0=0.1$.  From the ground state, the system is brought into a non-trivial polaritonic final state by driving the cavity. The black dashed line refers to the moment when the AC coherent drive is switched off. Inset: Standard deviation with the same parameters. Right panel: Dynamics of $\langle \sigma^z \rangle$. The blue curve is the ideal dissipationless case. The dashed red curve is for  $\Gamma=10^{-5}$ and the dotted green curve for $5. 10^{-4}$.}
\label{drived_N}
\end{figure}

One could then imagine to increase the strength of the drive. The price to pay for an increase of the drive strength is the interplay of the upper levels. The anharmonicity of the JC ladder makes it possible to quantitatively reach the first polariton beyond the linear response limit. We use both the anharmonicity of the JC ladder (resulting in ``polariton blockade") and destructive interference for upper levels. The mean number of polaritons $\langle N\rangle = \langle a^{\dagger} a\rangle + (\langle \sigma^z\rangle+1)/2$ and the standard deviation associated with this observable is shown in Fig. (\ref{drived_N})-left panel. We check this result with the SSE approach and study the influence of dissipation (see right panel of Fig. (\ref{drived_N})).The order of magnitude of the time $t_s$ at which we stop the drive enables us to minimize the effect of dissipation. At weak dissipation (essentially $\Gamma\lesssim 10^{-4}$), we can see that it is possible to realize temporarily an almost pure polaritonic state on one cavity (it corresponds to an ``atom''-like polariton). The protocol that we have presented here is an analog of a standard $\pi$-pulse in terms of polaritons. The more anharmonic the JC ladder, the more efficient it is. This is not surprising as this driving scheme becomes exactly a standard atomic $\pi$ pulse in the infinitely detuned limit. \\

This protocol could be used simultaneously on all the site of a cavity array, to ``implement" the system in a non-trivial polaritonic state. If the hopping between cavities is weak enough, its effect can be disregarded during the rather short time needed to bring each driven cavity in a non-trivial polariton state \cite{Karyn:CR}. Dealing with the driven array problem at a general level remains however difficult, and no known protocols permits to explore precisely the expected out-of-equilibrium Mott physics.\\

In this section, we have tested the SSE framework on the well-known Rabi model and its driven version. We showed that numerical convergence is ensured for the free dynamics during several periods, until quite large light-matter coupling $g/\omega_0 \simeq 1$.  Above this threshold, the dynamical sign problem hampers numerical convergence, even in this simple case of one mode. The situation is even more problematic for continuous spectral functions associated with an infinite number of modes. In most cases, it is not possible to access the long-time dynamics with the SSE. There exists however situations for which we can overcome this dynamical sign problem, for example for the ohmic spinboson model in the scaling regime characterized by $\Delta/\omega_c \ll1$ and $0<\alpha<1/2$.  %  by using a na\"{i}ve sampling of the stochastic fields.

% We now turn to the more complex case of the ohmic spinboson model corresponding to an infinite number of modes.

\section{Application: dynamics in the ohmic spinboson model}

The ohmic spinboson model is actually the first model on which the SSE framework was tested, see Refs. \cite{2010stoch,Peter}. In the scaling regime, results presented in Sec. \ref{result_method} simplify and one only needs one unique \textit{purely imaginary} stochastic field $h$ to access quantitatively the dynamics, as we show below, following Refs. \cite{2010stoch,Peter}. \\

For an ohmic spectral density given by Eq. (\ref{J_ohmic}) with $s=1$, $Q_1$ and $Q_2$ functions read at zero temperature,
\begin{align}
 Q_1(t)&=\int_0^{\infty} d\omega\frac{J(\omega)}{\omega^2}\sin \omega t=2 \pi \alpha \tan^{-1} (\omega_c t), \label{q1_ohmic} \\
  Q_2(t)&=\int_0^{\infty} d\omega\frac{J(\omega)}{\omega^2}\left(1-\cos \omega t\right) = \pi \alpha \log (1+\omega_c^2 t^2).\label{q2_ohmic}
\end{align} 
We know from the analysis carried out in Chapter I that the characteristic spin-flip time $t_s$ is given by $t_s \sim 1/\Delta_r \geq 1/\Delta$ at finite $\alpha$. For times $t\sim t_s$, Eq. (\ref{q1_ohmic}) then simplifies to $Q_1(t)\simeq \pi^2  \alpha $. A more refined analysis, carried out in Appendix D of Ref. \cite{leggett:RMP}, shows that this approximation is well controlled only for $0\leq \alpha<1/2$.\\

Eq. (\ref{Q}) is then changed to
\begin{align}
 &\mathcal{F}_{n}= \exp \left[i\pi \alpha\sum_{k=0}^{2n-1}\sum_{j=k+1}^{2n} \Xi_j \Upsilon_k  \right] \exp \left[ \frac{1}{\pi} \sum_{k=1}^{2n-1}\sum_{j=k+1}^{2n} \Xi_j \Xi_k  Q_2(t_j-t_k) \right]\label{Q_ohmic}. 
 \end{align}
 We only need field $h$ to decouple the interaction, and the summation over blips and sojourns in the first term of the right-hand side of Eq. (\ref{Q_ohmic}) can be accounted for by considering for $V$ the following matrix,
\begin{equation}
V= \left( \begin{array}{cccc}
0&e^{-h}&-e^{h}&0 \\
e^{i\pi \alpha}e^{h}&0&0&-e^{-i\pi \alpha}e^{h}\\
-e^{-i\pi \alpha}e^{-h}&0&0&e^{i\pi \alpha}e^{-h}\\
0&-e^{-h}&e^{h}&0
\end{array} \right).
\label{eq:spin_hamiltonian_ohmic}
\end{equation}\\

As was stated in Ref. \cite{Peter}, the Fourier coefficients of $Q_2$ in Eq. (\ref{q2_ohmic}) are all negative, leading naturally to a \textit{purely imaginary} field $h$. Similarly, the second sampling procedure presented in Appendix \ref{appendix:sampling} also leads to a \textit{purely imaginary} field $h$. This simplification enabledthe authors of Refs. \cite{2010stoch,Peter} to study quantitatively the free dynamics of the ohmic spinboson model in the scaling regime. We can refine this scaling regime simplification by writing $Q_1(t)=\pi^2 \alpha+[Q_1(t)-\pi^2 \alpha]$. We take into account the constant part as exposed above while the remaining part $[Q_1(t)-\pi^2 \alpha]$ is decomposed into Fourier series and generates a field $k$.\\

\subsection{Free dynamics and comparison with Bethe Ansatz results}
We first study the dynamics of the Rabi-Spinboson model of Ref. \cite{Rabi} which describes the dynamics of the Rabi model with dissipation described microscopically,
\begin{align}
\mathcal{H}_{R-SB}&= \frac{\Delta}{2} \sigma^x +\sigma^z \sum_k \frac{\lambda_k}{2} (b_k +b_k^{\dagger})+ \sum_k  \omega_k \left(b_k^{\dagger}b_k+\frac{1}{2}\right)+\sigma^z \frac{g}{2} (a +a^{\dagger})+\omega_0 \left(a^{\dagger}a+\frac{1}{2}\right).
\label{Hamiltonian_spin_boson} 
\end{align}
This Hamiltonian $\mathcal{H}_{R-SB}$ would correspond to a spectral function
\begin{align}
J(\omega)=\pi g^2 \delta(\omega-\omega_0)+2\pi \alpha \omega e^{-\frac{\omega}{\omega_c}}.
\end{align}

\begin{figure}[h!]
\center
\includegraphics[scale=0.4]{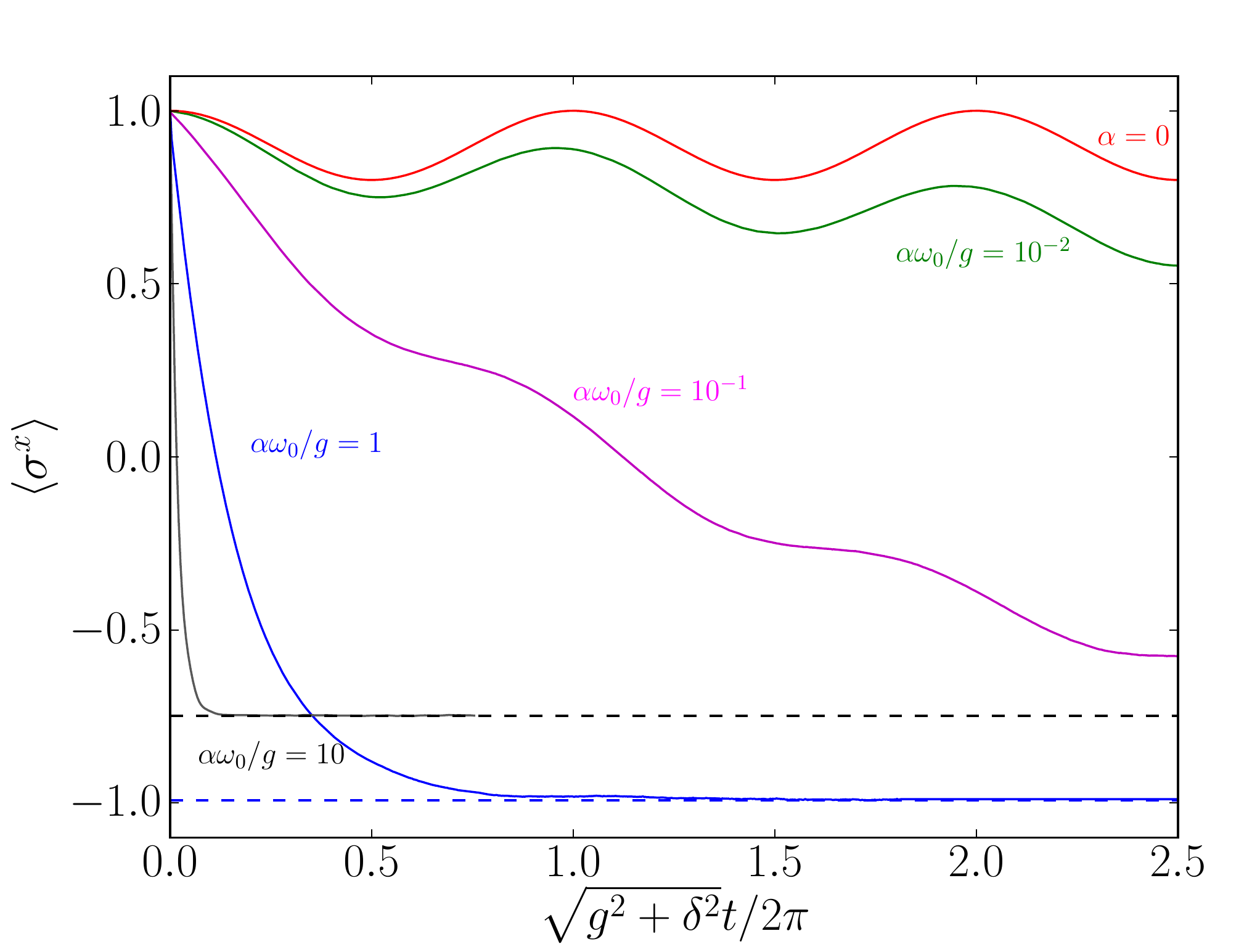}  
\caption{Dynamics of $\langle \sigma^x \rangle$ with the initial condition $|\uparrow\rangle$ which is a linear superposition of the first two polaritons (note the axis difference for the Rabi coupling). We consider $g/\omega_0=0.01$, $\Delta/ \omega_0=0.97$, $\omega_c/\omega_0=100$ and several values of $\alpha$ until $\alpha\approx 0.1$. We observe a relaxation towards a non-trivial final state by increasing $\alpha$ and the value of $\langle \sigma^z \rangle$ is in accordance with Bethe Ansatz calculations \cite{filyov,Cedraschi:PRL,Cedraschi:Annals_of_physics,kopp,Hur}.}
\label{damped_1}
\end{figure}

After some damped Rabi oscillations, the system relaxes to a final state with $\langle \sigma^x\rangle=0$, and $\langle \sigma^z\rangle$ which can be compared to Bethe Ansatz calculations \cite{filyov,Cedraschi:PRL,Cedraschi:Annals_of_physics,kopp,Hur}. Both the short time dynamics and the equilibrium properties are then accessible within the SSE framework.\\

We present now an additional analysis carried out close to the coherent-incoherent crossover zone (around $\alpha=1/2$). 
\subsection{Coherent-incoherent crossover around  $\alpha=1/2$}

We now focus on the same dynamical problem than in Chapter I Sec. \ref{NIBA_from_EOM}, i.e. the time evolution of $\langle \sigma^z(t) \rangle$ from the initial state $|\uparrow_z \rangle \otimes |0 \rangle$ for the ohmic spinboson model. It is known that the NIBA predictions concerning the incoherent behaviour of $\langle \sigma^z(t) \rangle$ are erroneous. By contrast, it is established that the time evolution of $\langle \sigma^z(t) \rangle$ changes from a coherent oscillatory regime for $\alpha<1/2$ to a incoherent monotonically decaying regime for $1/2 \leq \alpha<1$. The point $\alpha=1/2$, also known as Toulouse point, is exactly solvable and one finds $\langle \sigma^z(t) \rangle=\exp[-\pi\Delta^2/2\omega_c (t-t_0)]$ \cite{Weiss:QDS}. \\

 Conformal Field Theory calculations \cite{CFT} predicted a purely exponential decay for $\alpha<1/2$, while corrections to NIBA \cite{NIBA_corrections} predicted a incoherent part of the form $-2(1/2-\alpha)\exp[-\Delta_r t/2]/[\Delta_r (t-t_0)]^{2(1-\alpha)} $. Recent works \cite{Kennes1,Kennes2} have focused on this issue using real-time RG and functional RG techniques.  For $1/2-\alpha \ll 1$ they found in particular a result in agreement with the presence of an incoherent dynamics, as $\langle \sigma^z(t) \rangle$ shows only a finite number of zeros. \\

\begin{figure}[h!]
%\vskip -0.2cm
\center
\includegraphics[scale=0.3]{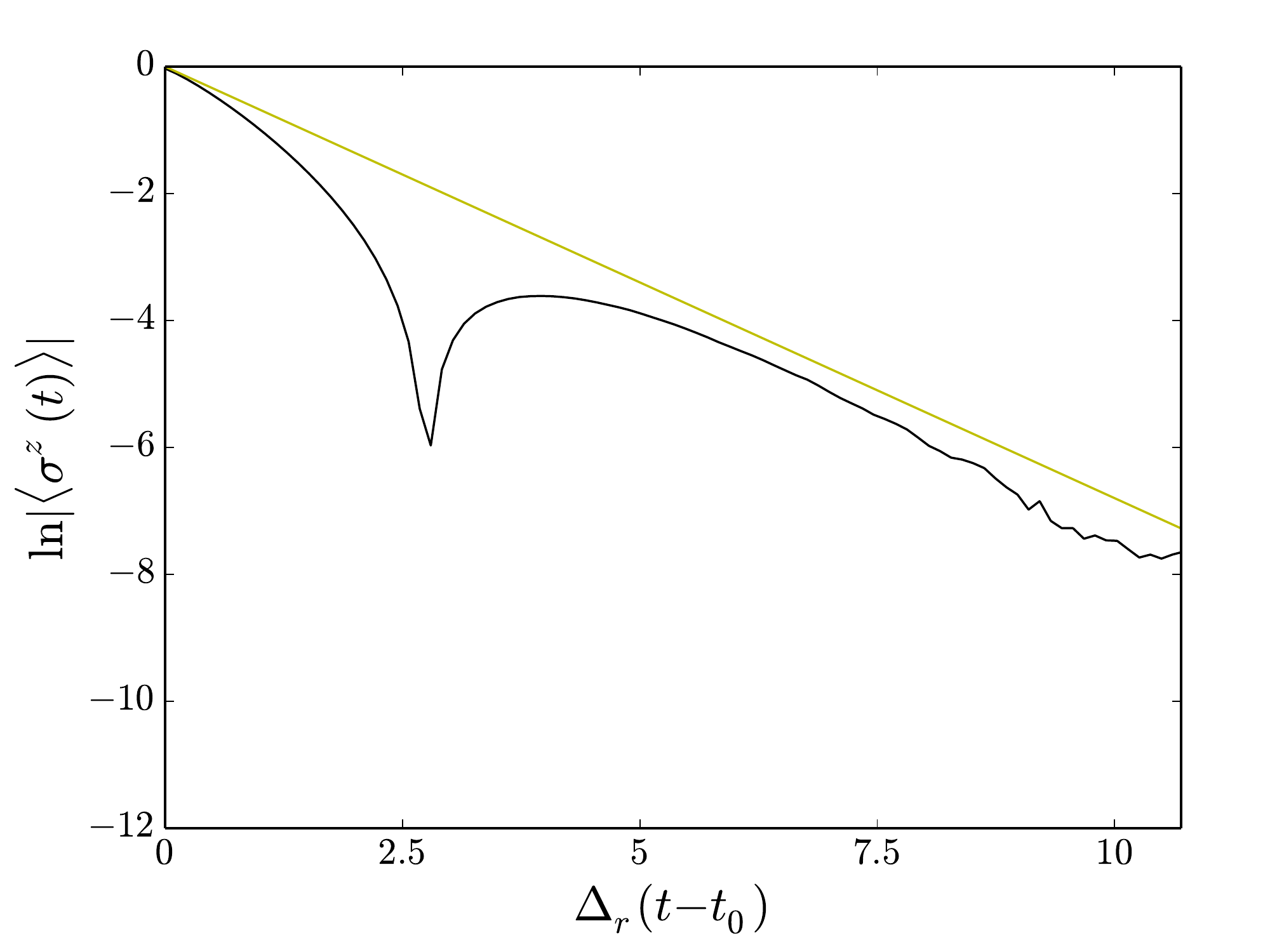} 
\caption{Coherent-to-incoherent crossover. Evolution of $\ln|\langle\sigma^z(t)\rangle|$. The black curve corresponds to $\alpha=0.45$, and the yellow curve corresponds to the Toulouse limit. }
\label{plot_coherent_to_incoherent}
\end{figure}

We show in Fig. (\ref{plot_coherent_to_incoherent}) results obtained with the SSE equation in this region.
These results would also suggest that $\langle\sigma^z(t)\rangle$ only vanishes a finite number of times when approaching the $\alpha=1/2$ point. Obtaining reliable numerical results at long times is however difficult as one needs to reach a very high precision. As stated earlier, the case of two spins will be easier to investigate as the dissipative quantum phase transition occurs for a smaller value of $\alpha$. We will study the incoherent regime in more details for this particular case in Chapter 3.

\subsection{Landau-Zener transitions}
 In Ref. \cite{Peter}, the authors investigated the non-equilibrium behavior of the system system under a linear driving term $\epsilon (t)/2 \sigma^z$, when the system is initially prepared in the ground state of the system. The linear passage, corresponding to $\epsilon (t) =\epsilon_0+v(t-t_0)$ with $(v>0)$, $-\epsilon_0/\Delta\gg 1$, is known as Landau-Zener problem. Landau \cite{Landau}, Zener \cite{Zener}, Stueckelberg \cite{Stueckelberg} and Majorana \cite{Majorana} provided an analytical description of this problem in the case of an isolated two-level system subject to a linear sweep ($\alpha=0$). The survival probability $p_{lz}$ that the spin remains in its initial state after the sweep, is fully determined by the velocity of the sweep $v$, and we have $p_{lz}=\exp [-\pi \Delta^2/2v]$. It was shown in Refs. \cite{kayanuma_1,kayanuma_2} that the presence of a gaussian dissipative bath does not affect the transition probability in the case of the Landau-Zener sweep for one single spin, as long as the coupling is along the z-direction. This result was confirmed numerically with the SSE approach in Refs. \cite{Peter} and \cite{2010stoch}. \\

\subsection{Mean field Landau Zener transitions in a dissipative spin array}
\label{section_N_spins}

We complement this Landau-Zener study and consider now an array of $M$ spins coupled to the same dissipative bath. We note recent progress concerning the dynamical study of dissipative spin arrays by coupling Matrix Product States and Operators techniques to quantum trajectories formalism \cite{MPS_QT_1,MPS_QT_2}. In our case of the SSE, the problem is not directly tractable numerically and we treat the interactions between spins at a mean-field level. We will show in that case that the presence of a gaussian dissipative bath \textit{does} affect the transition probability. This observation will be explained by the effect of the Ising-like interaction between spins, mediated by the bath. We will quantify this effect with the help of Kibble-Zurek mechanism. This study was done in our Ref. \cite{Ohmic_systems_article}.\\

%For greater values of $M$, the problem becomes rapidly untractable numerically, as the density matrix of the spin system becomes too large. We will then extend the method at a mean field level in the case of the array ($M\to \infty$) in Subsection \ref{Mean_field_approx}. In Subsection \ref{LZ_array}, we investigate Landau-Zener sweeps for the array and interpret the results with a Kibble-Zurek type argument. \\

%Recent developments linked non non-equilibrium physics in these lattice systems involve Matrix Product States \cite{Garrahan,Marco,Sanchez}; stochastic mean-field methods also allow to describe non-equilibrium light-matter systems\cite{Keeling}.\\

Let us first show how one can describe mean-field effects in the SSE formalism, to compute the spin dynamics for the multi spinboson problem described by the following Hamiltonian,
\begin{align}
\mathcal{H}_{SB}^M=&\frac{\Delta}{2}\sum_{p=1}^M  \sigma_p^x+ \sum_{p=1}^M \sum_{k} \lambda_{k} e^{ik x_p} \left(b^{\dagger}_{-k}+b_k \right) \frac{\sigma_p^z}{2} +\sum_{k}  \omega_{k} b^{\dagger}_{k} b_{k}.
\label{ising_2}
\end{align}

\subsubsection{Mean-field approximation in the SSE framework}
\label{Mean_field_approx}
For clarity, we consider the case where all the spins initially in the state $|\uparrow_z\rangle$ so that $\rho_{S} (t_{0})=\prod_{j=1}^M | \uparrow_{z,j} \rangle\langle \uparrow_{z,j} | $, and we seek to compute elements $\langle \Sigma_f | \rho_S (t) | \Sigma_f'\rangle$ at a given time $t\geq t_0$, where we define the $M$-dimensional spin vector $|\Sigma\rangle=|\sigma_1,\sigma_2,..,\sigma_M\rangle$. The time-evolution of the spin reduced density matrix can be then re-expressed as, 

%\textbf{Faire avec un general k et k'}
\begin{equation}
\langle \Sigma_f | \rho_S (t) | \Sigma_f'\rangle=  \int D\Sigma D\Sigma' \exp\left\{i\left[ S_{\Sigma}-S_{\Sigma'}\right]\right\} \mathcal{F}_{[\Sigma, \Sigma']}.
\label{eq:densitymatrixelement_N_spins}
\end{equation}

The integration runs over all $M$-dimensional constant by part paths $\Sigma$ and $\Sigma'$ such that $\Sigma_j (t_0)=\Sigma_j' (t_0)=+1$ for all $j$ and $\sigma_j^z|\sigma_{j,f}\rangle=\Sigma_j(t)|\sigma_{j,f}\rangle$, $\sigma_j'^z|\sigma_{j,f}'\rangle=\Sigma_j'(t)|\sigma_{j,f}'\rangle$. $ S_{\Sigma}$ denotes the free action to follow one given $M$-dimensional spin path without the environment. This free action contains the transverse field terms, and the Ising interaction terms. The effect of the environment is fully contained in the influence functional $\mathcal{F}_{[\Sigma, \Sigma']}$, which reads in this case:

\begin{small}
\begin{align}
\mathcal{F}_{[\Sigma, \Sigma']}=&e^{\int_{t_0}^t ds \int_{t_0}^s ds' \sum_{i,j}\left\{\frac{ i \mathcal{L}_1(s-s',x_i-x_j)}{\pi}\xi_i(s)\eta_j(s') -\frac{ i \mathcal{L}_2(s-s',x_i-x_j)}{\pi}\xi_i(s) \xi_j(s')\right\}} \mathcal{G}[\Sigma,\Sigma'],
\label{eq:influence_N_spins}
\end{align}
\end{small}
with $\xi_j(s)=[\Sigma_j (s)-\Sigma_j '(s)] /2$ and $\eta_j(s)=[\Sigma_j (s)+\Sigma_j '(s)] /2$. We have,
\begin{align} &\mathcal{L}_1(t,x)=\frac{1}{2}\left[L_1\left(t-\frac{x}{v_s}\right)+L_1\left(t+\frac{x}{v_s}\right) \right]   \notag \\
&\mathcal{L}_2(t,x)=\frac{1}{2}\left[L_2\left(t-\frac{x}{v_s}\right)+L_2\left(t+\frac{x}{v_s}\right) \right].
\label{Ls}
\end{align}

\begin{figure}[t!]
  \center
 \begin{minipage}[b]{0.95\linewidth}
 \center
%\begin{flushleft}
\includegraphics[scale=0.43]{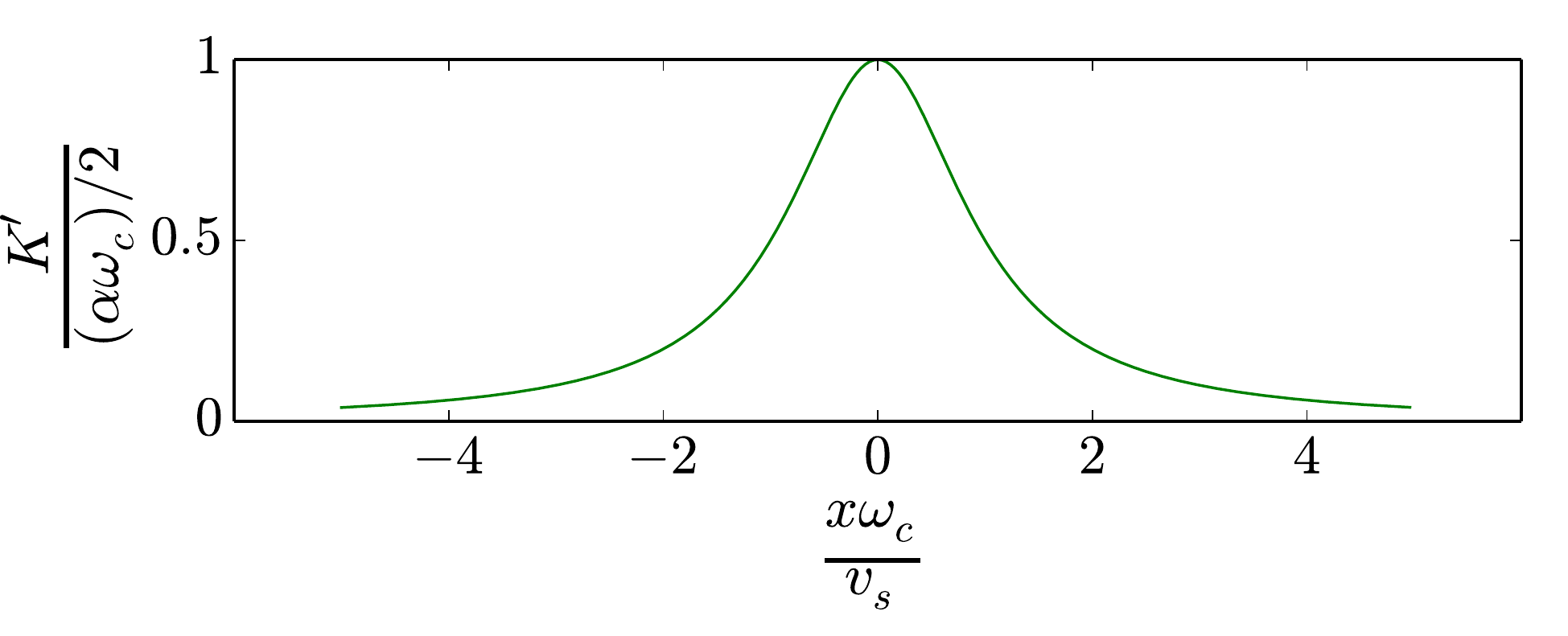} 
%\end{flushleft}
\end{minipage}\\
\begin{minipage}[b]{0.52\linewidth}
 \center
\begin{flushright}
\includegraphics[scale=0.24]{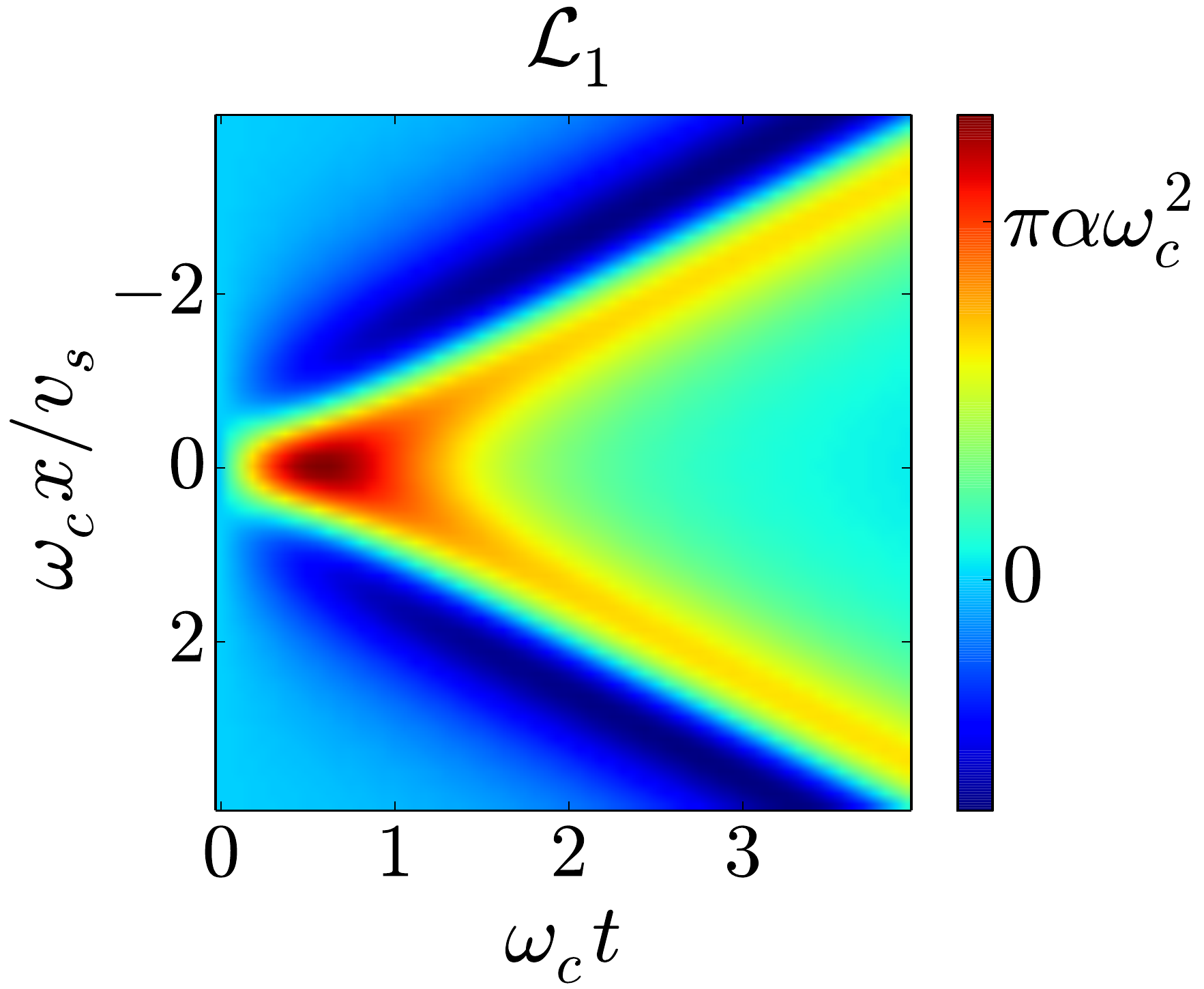} 
\end{flushright}
\end{minipage}
%\begin{minipage}[b]{0.1\linewidth}
%\end{minipage}
\begin{minipage}[b]{0.4\linewidth}
 \center
\begin{flushleft}
\includegraphics[scale=0.24]{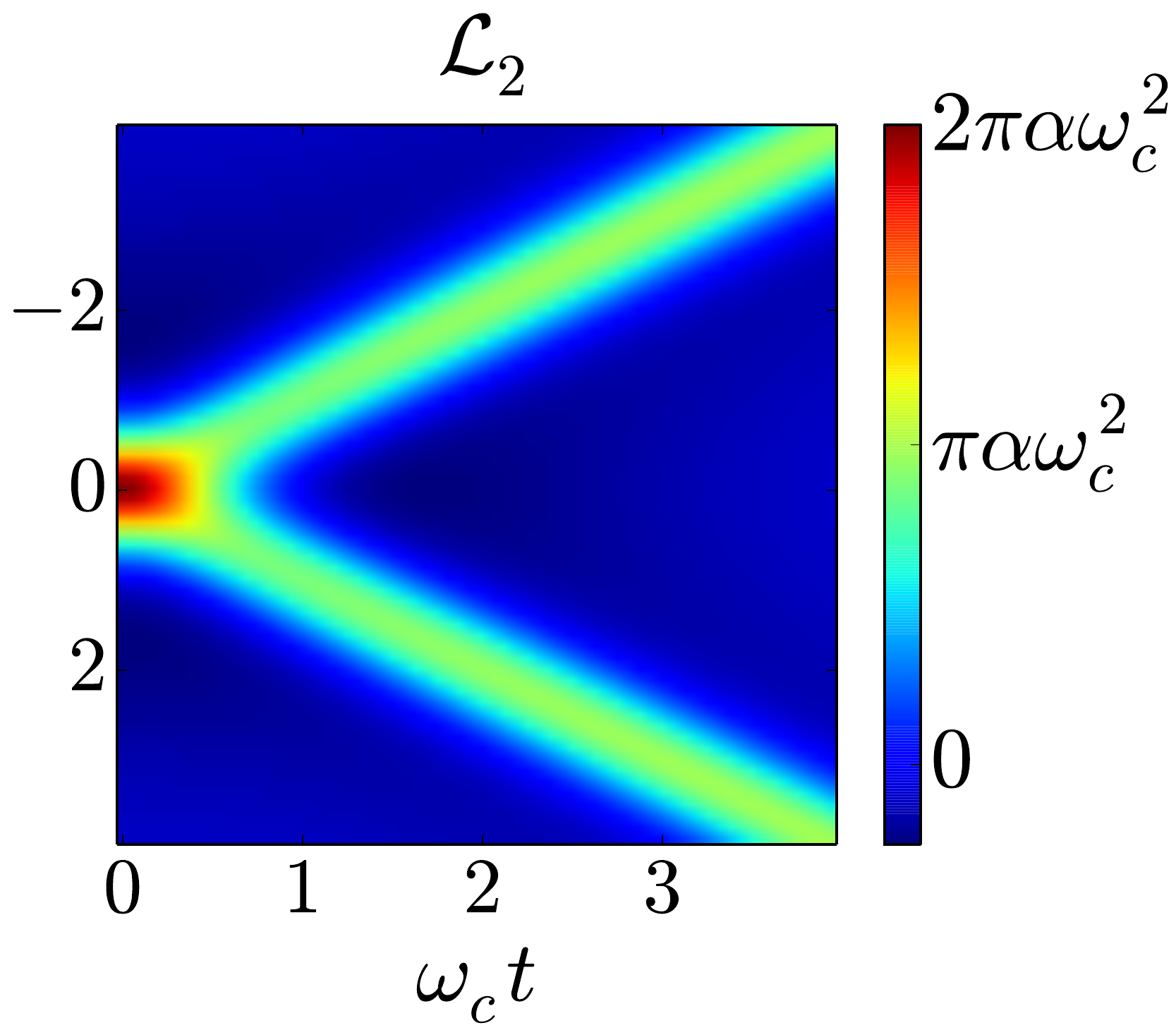}     % Logo or a photo of you, adjust its dimensions here
\end{flushleft}
\end{minipage}
\caption{Top: Evolution of the direct Ising interaction which is induced by the presence of the bath between two spins distant of $x$, as a function of  $x \omega_c/v_s$. Bottom: Space-time dependency of the coupling functions $\mathcal{L}_1$ (left) and $\mathcal{L}_2$ (right). The bath induces a long-range interaction between spins.}
\label{couplin_array}
\end{figure}

The bosonic environment couples the symmetric and anti-symmetric spin paths $\eta^p(t)=1/2[\Sigma_p(t)+\Sigma_p'(t)] $ and $ \xi^p(t)=1/2[\Sigma_p(t)-\Sigma_p'(t)]$ at different times and different lattice sites. In Fig.~\ref{couplin_array}, we plot the space and time coupling functions $\mathcal{L}_1$ (bottom left) and $\mathcal{L}_2$ (bottom right). We see that the bosons induce a long-range interaction between spins. The maximal effect between two spins separated by a distance $x$ occurs after a time $x/v_s$, due to the finite sound velocity $v_s$ of the excitations. \\

The last term of Eq. (\ref{eq:influence_N_spins}) reads
\begin{equation}
\mathcal{G}[\Sigma,\Sigma']=e^{ i \frac{\mu}{2} \int_{t_0}^t ds \left[\sum_{j} \frac{\Sigma_j (s)}{2} e^{i k x_j} \right]^2-\left[\sum_{j} \frac{\Sigma_j ' (s)}{2} e^{i k x_j} \right]^2},
\label{eq:influence_N_spins_2}
\end{equation}
with  $\mu=2/\pi \int_0^{\infty} J(\omega)/\omega$. We recover that the bath is responsible for an {\it ferromagnetic} Ising-like interaction between the spins $K'_{|j-p|}=1/(2\pi) \int_0^{\infty} J(\omega)/\omega \cos[(x_i-x_j)/v_s]$. We plot on the top panel of Fig.~\ref{couplin_array} the value of $K'_{|j-p|}$ with respect to $x\omega_c/v_s$, where $x=x_i-x_j$ is the distance between the two sites $i$ and $j$. \\

The bath is responsible for two distinct types of interactions. The first one is a retarded interaction mediated by the bosonic excitations, which travel at the speed $v_s$. The second one is an instantaneous interaction $K'$, which we already found thanks to the polaronic transformation in Eq.~(\ref{N_spins}). \\

We now proceed to the mean-field decoupling. The spins are coupled through three different terms: the instantaneous direct Ising interaction of strength $K$, the instantaneous interaction mediated by the bath in $\mathcal{G}$, and the retarded interaction mediated by the bath whose expression is given by the first term of the right hand side of Eq. (\ref{eq:influence_N_spins}). We will treat instantaneous spin-spin interactions at a mean field level, and in the limit $\omega_c a/v_s \ll 1$, where $a$ is the lattice spacing, we see that the retarded interactions have no effect between \textit{different} spins at a mean field level, since we have $\int_{-\infty}^{\infty} dx \mathcal{L}_1 (s,x)=\int_{-\infty}^{\infty} dx \mathcal{L}_2 (s,x)=0 $. In the following, we then neglect the retarded interaction between \textit{different} spins, and only conserve the retarded self-interaction. Finally the propagation integral can be factorized in a product of individual matrix elements, so that it is possible to write: 

\begin{align}
\langle \sigma_{p,f}   | \rho_{S,p} (t) | \sigma_{p,f}'\rangle= \int &D \Sigma_p D \Sigma'_p  A_p[\Sigma_p] A_p[\Sigma_p']^* \mathcal{F}_p[\Sigma_p,\Sigma_p'] e^{-i K_r \int_{t_0}^t ds  \left[\Sigma_p (s)-\Sigma_p' (s)\right]  \langle \sigma^z_p (s) \rangle },
\label{propagation_array}
\end{align}  
where $\rho_{S,p}$ denotes the density matrix of spin $p$. $A_p[\Sigma_p]$ denotes the amplitude to follow a given path for the spin $p$ in the sole presence of the transverse field. We have $K_r=2\sum_{j=1}^{\infty} K'_{j}$. The remaining term $\mathcal{F}_p[\Sigma_p,\Sigma_p']$ encapsulates the effect of the bosonic bath on the spin $p$,

\begin{align}
\mathcal{F}_p[\Sigma_p,\Sigma_p']=\exp\Big\{ \int_{t_0}^t ds \int_{t_0}^s ds' &\frac{i}{\pi} L_1(s-s') \xi^p(s)\eta^p(s') -\frac{1}{\pi} L_2(s-s')\xi^p(s)\xi^p(s')  \Big\}.
\label{eq:influence_3}
\end{align}

 We will drop the $p$ index in the following, as all the sites are equivalent in the mean-field description. Following the same steps as for the one-spin case, we focus on the computation of $\langle \uparrow_z|\rho_S(t)|\uparrow_z\rangle$ and reach the same expression than for one spin (see Eq. (\ref{Q})), with the influence functional being
 \begin{align}
 &\mathcal{F}_{n}= \exp \left[ \frac{i}{\pi} \sum_{k=0}^{2n-1}\sum_{j=k+1}^{2n} \Xi_j \Upsilon_k  Q_1(t_j-t_k) \right] \exp \left[ \frac{1}{\pi} \sum_{k=1}^{2n-1}\sum_{j=k+1}^{2n} \Xi_j \Xi_k  Q_2(t_j-t_k) \right]\mathcal{Q}_3. 
  \end{align}
with
  \begin{align}
      &\mathcal{Q}_3 =\exp \left[ -2i K_r \sum_{j=1}^{2n} \Xi_j \int_{t_0}^{t_j} ds \langle \sigma^z (s) \rangle  \right]\label{Q_3} . 
 \end{align}

We then reach for $\langle \uparrow_z|\rho_S(t)|\uparrow_z\rangle$ the same expression as the one obtained in Eq.~(\ref{scalar_prod_SSE}), with the same final vector and $|\phi\rangle$ solution of the SSE (\ref{SSE}), with the effective Hamiltonian given by (\ref{eq:spin_hamiltonian}) provided that we add to the stochastic field $h$ the field $h_I$ defined by $h_{I} (t)=-2iK_r\int_{t_0}^t ds \langle \sigma^z (s) \rangle $. We have then reached a self consistent equation, as $\langle \sigma^z(t) \rangle$ enters in the expression of $h_I (t)$. The numerical procedure requires a larger number of realizations of the field $h$ and $k$ compared to the one-spin case. For each realization, we solve the stochastic equation and $\langle \sigma^z(t) \rangle$ is obtained by averaging over the results. The effect of $\langle \sigma^z(t) \rangle$ in $h_I (t)$ is dynamically updated with the number of samplings. \\

%We can use our method to compute the free spin dynamics in the limit of $M\to \infty$. We check that the bath causes a decay towards one of the two equilibrium states in the ferromagnetic phase as well as a renormalization of both the tunneling element and the Ising coupling. However, it does not affect the dynamical properties of the quantum phase transition as long as the direct Ising term $K$ is \textit{not} zero. This behavior can be understood thanks to a thermodynamic analysis of the action at low wave-vectors $q$ and low frequency $\omega$, which is dominated by the contribution of the long range Ising interaction, as shown in Appendix F.

We use this mean-field method to study the effect of a Landau-Zener sweep simultaneously applied to all the spins. It corresponds to adding a term of the form $\sum_{j=1}^M \epsilon(t) \sigma^z_j /2 $ with $\epsilon (t) =\epsilon_0+v(t-t_0)$ with $(v>0)$, $-\epsilon_0/\Delta\gg 1$ to the Hamiltonian.  

\subsubsection{LZ transitions : Array}
\label{LZ_array}
Let us first underline that this protocol is different from the dynamical transition of the quantum Ising model in transverse field with nearest neighbours interactions studied in the literature \cite{sengupta_powell_sachdev,dzarmaga} (and references therein), where the driving parameter is the transverse field and which can be studied elegantly in $k$ space. Here, we are interested in the dynamics of local spin variables at a mean field level. A rigorous description of the dynamics should involve all the energy levels of the system, and their respective avoided crossings. Our mean-field description greatly simplifies the problem and the interplay of all the levels is in fact reduced to a single avoided crossing governed by the local self-consistent Hamiltonian,
\begin{align}
H_j=&\frac{\Delta}{2}\sigma^x_j+\left[\frac{\epsilon(t)}{2}-K_r \langle \sigma^z(t) \rangle \right] \sigma^z_j+\sum_{k} \left[ \lambda_{k} e^{ik x_j} \left(b^{\dagger}_{-k}+b_k \right) \frac{\sigma_j^z}{2}+\omega_k b^{\dagger}_k b_k \right].
\label{N_spin}
\end{align}

 \begin{figure}[t!]
\center
\includegraphics[scale=0.4]{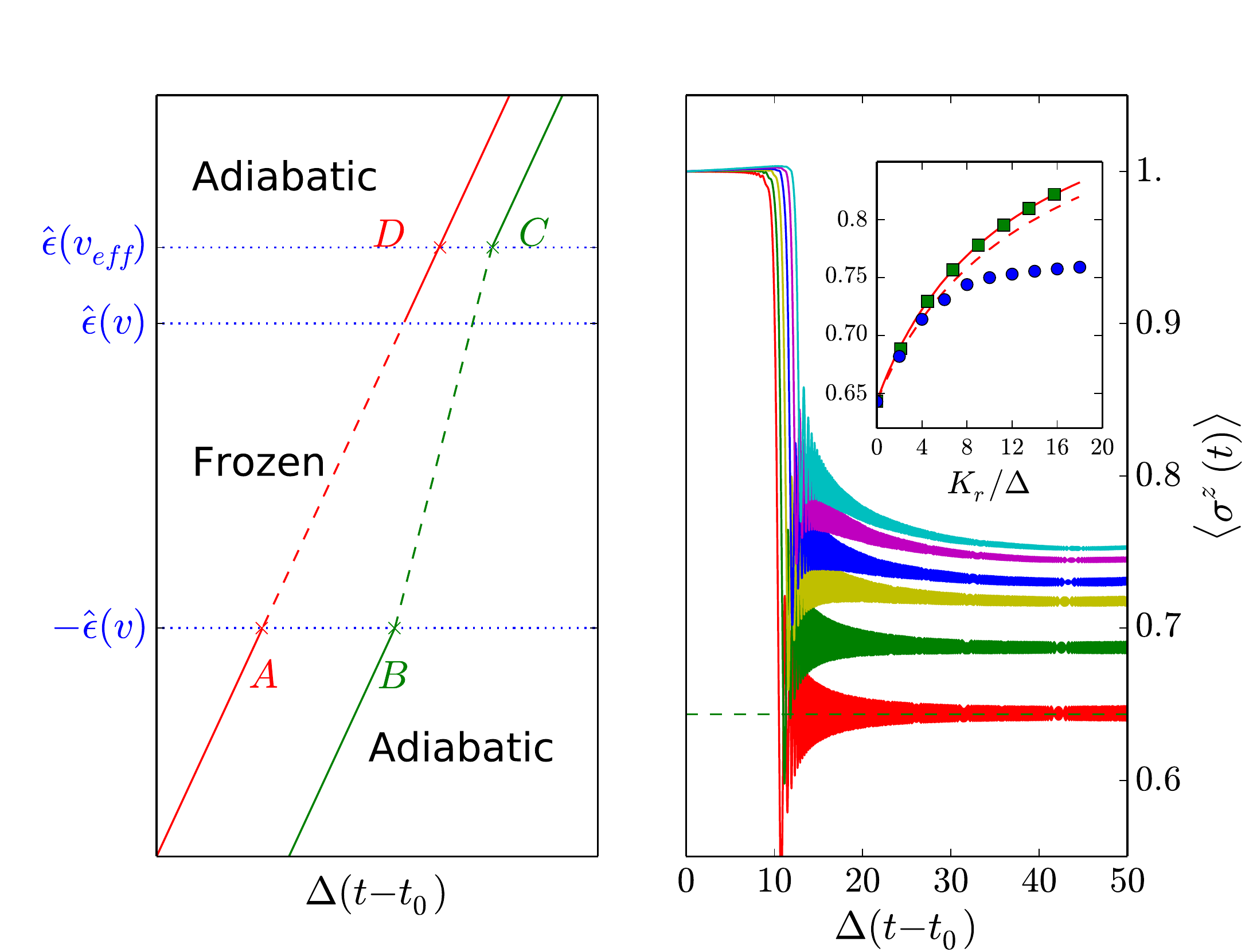}  
\caption{Left: schematic interpretation of the Landau-Zener sweep for the array in the framework of the Kibble-Zurek mechanism. The line $(AD)$ shows the evolution of the bare bias field with respect to time, while the broken line connecting points $B$ and $C$ represents the effective bias field. The lines are full during the adiabatic stages, and dashed during the frozen (non-adiabatic) period. Right: Fast sweep ($v/\Delta^2=8$) in the array, for different values of $\alpha$ corresponding to $K_r=0$ (red curve), $K_r=2$ (green curve), $K_r=4$ (yellow curve), $K_r=6$ (blue curve), $K_r=8$ (magenta curve) and $K_r=10$ (cyan curve). We have $\omega_c=100 \Delta$. Inset: the blue points show the values of $\langle \sigma^z(t\to \infty) \rangle$ with respect to $K_r/\Delta$, corresponding to the parameters of the main plot. The green squares correspond to a direct Ising interaction of strength $K_r$ between spins, without dissipative bath. Such a long-range ferromagnetic Ising interaction between spins can for example be the result of the Van der Waals interaction in Rydberg media \cite{rydberg_1,rydberg_2}. The full (dashed) red line shows the expectation value of $\langle \sigma^z(t\to \infty) \rangle$ with respect to $K_r/\Delta$ ($K_r/\Delta_r$) deduced from the Kibble-Zurek mechanism.}
\label{fig_KBZ}
\end{figure}

The presence of the Ising interaction $K_r$ lead to a change in the final value of $\langle \sigma^z (t \to \infty) \rangle$. This effect can be described by a Kibble-Zurek argument \cite{Kibble,Zurek} in the case of rapid drive. The single site fast Landau-Zener transition can indeed be described thanks to the Kibble-Zurek mechanism, which predicts the production of topological defects in nonequilibrium phase transitions\cite{dzarmaga,damski,review_KBZ}. This description splits the dynamics into three consecutive stages: it is supposed to be adiabatic in the first place, then evolves in a non-adiabatic way near the transition point, and finally becomes adiabatic again. The impossibility of the order parameter to follow the change applied on the system provokes this non-adiabatic stage, where the dynamics is said to be ``frozen". To determine the boundary between adiabatic and frozen stages, we follow the argument of Ref. \cite{damski}, inspired by the equation proposed by Zurek in Ref. \cite{Zurek}. The inverse of the energy difference between ground and excited states defines a characteristic time scale $\tau(t)$ for the system. Following Refs. \cite{Zurek,damski}, the dynamics stops being adiabatic when this time scale matches the time $t_c-t$, when $t_c$ is the crossing time defined by $ \epsilon(t_c)=0$. This argument leads to a characteristic energy scale \cite{damski} 
\begin{equation}
\hat{\epsilon}=\Delta/\sqrt{2}\left\{ \left[1+16v^2/(\pi^2\Delta^4)\right]^{1/2}-1\right\}^{1/2}, 
\end{equation}
which sets the limit between adiabatic and frozen stages (see left panel of Fig.~\ref{fig_KBZ}).\\

The effective field felt by one site is the sum of the bias field $\epsilon (t)$ and the Ising interaction, and will be denoted $\epsilon_{eff} (t)$.  The dynamics always enters in the frozen stage with $\langle \sigma^z \rangle \simeq 1$, so that we have $\epsilon_{eff} (t)=\epsilon (t)-K_r$ during the first adiabatic stage. At the end of the frozen stage, the spin expectation value has changed, and the effective field becomes $\epsilon_{eff} (t)=\epsilon (t)-K_r \langle \sigma^z (t) \rangle$. This leads to a change of the effective speed at which the frozen zone is crossed through, and ultimately of the transition probability. This can be seen on the left panel of Fig.~\ref{fig_KBZ}, where we show the evolution of both the bare and the effective bias fields with respect to time. We can estimate the renormalization of the effective speed self-consistently thanks to basic geometrical considerations in the trapezoid $(ABCD)$ of Fig.~\ref{fig_KBZ} (left panel). The effective crossing speed is given by
\begin{equation}
v_{eff}=\frac{\hat{\epsilon}(v)+\hat{\epsilon}(v_{eff})}{t_C(v_{eff})-t_B}.
\label{veff_1}
\end{equation}
The denominator can be simplified by writing that $t_C(v_{eff})-t_B=\left[t_C(v_{eff})-t_D\right]+(t_D-t_A)-(t_B-t_A)$. We know that $(t_D-t_A)=\left[\hat{\epsilon}(v)+\hat{\epsilon}(v_{eff})\right]/v$, and $\left[t_C(v_{eff})-t_D\right]-(t_B-t_A)$  can be expressed as $-K_r\left[1-\langle \sigma^z (t_C,v_{eff}) \rangle\right]/v$. Next we suppose that we can approximate $\langle \sigma^z (t_C,v_{eff}) \rangle$ by $\langle \sigma^z (t \to \infty) \rangle$. Altogether, we get 
\begin{equation}
\frac{v_{eff}}{v}=\frac{\hat{\epsilon}(v)+\hat{\epsilon}(v_{eff})}{\hat{\epsilon}(v)+\hat{\epsilon}(v_{eff})-2K_r\left[1-p_{lz}(v_{eff})\right]}.
\label{veff_2}
\end{equation}

It allows us to know the variation of the effective speed $v_{eff}$ at which the transition is crossed with respect to the Ising interaction $K_r$. The spin expectation value $\langle \sigma^z (t \to \infty) \rangle$ is then estimated thanks to the Landau-Zener formula, and its evolution with respect to $K_r$ is shown by the red curve in the inset of the right part of Fig.~\ref{fig_KBZ}.\\

We plot on the right panel of Fig.~\ref{fig_KBZ} the dynamics obtained with the SSE. We see in the inset that, at small $\alpha$, the estimation of the final value of the spin variable thanks to Eq. (\ref{veff_2}) is correct. However, it breaks down when the dissipation strength is increased because the assumption $\langle \sigma^z (t_C,v_{eff}) \rangle \simeq\langle \sigma^z (t \to \infty) \rangle$ used to derive $v_{eff}$ is no longer correct. Relaxation processes occur after the crossing of the frozen zone which lower $\langle \sigma^z(t \to \infty)\rangle$. This can be seen on the behaviour of the curves obtained at large values of $\alpha$ (the cyan curve for example), where the spin expectation value continues to go down during a rather long time after the crossing. The dotted red curve takes into account the renormalization of the tunneling frequency $\Delta_r$ due to the presence of the bath. The bath is also known to affect the dynamical critical exponent $z$ in similar situations \cite{orth_stanic_lehur}, which may also contribute to the devitations observed.\\

In this Chapter, we introduced the SSE framework and applied it to the Rabi model and the ohmic spinboson model. In this latter case, the SSE is however unable to investigate the quantum phase transition occuring at $\alpha \sim 1$. As justified in Chapter 1, the critical coupling decreases with the number of spins coupled to the environment. In the next chapter, we extend the SSE to the case of two spins, which enables us to compute the non-equilibrium dynamics in both phases.

\chapter{Dynamics of two coupled spins in an ohmic environment}
\label{dimer}
We focus in this chapter on a spin dimer coherently coupled to one common bath of harmonic oscillators:
\begin{align}
\mathcal{H}_{dimer} =&\frac{\Delta}{2}\sum_{p=1}^2  \sigma_p^x+\sum_{p=1}^M \sum_{k} \lambda_{k} e^{ik x_p} \left(b^{\dagger}_{-k}+b_k \right) \frac{\sigma_p^z}{2}-K\sigma^z_1 \sigma^z_{2}+\sum_{k} \omega_{k} b^{\dagger}_{k} b_{k}.
\label{ising_dimer}
\end{align}
We recall that this Hamiltonian can be devised in a cold atomic setup (see Chapter 1 Sec. \ref{spinboson_cold_atoms}). Coupling two CPB to a common semi-infinite transmission line could also lead to the same Hamiltonian (see Chapter 1 Sec. \ref{transmission_line}). We studied this system in our Ref. \cite{Ohmic_systems_article}.\\

In this case of two spins, it is possible to reach an exact linear stochastic differential equation describing the dynamics of the ($4\times4$) spin reduced density matrix. The case of two spins is particularly interesting as the quantum phase transition from the unpolarized phase to the polarized phase occurs for a smaller value of $\alpha$, as justified after the polaron transformation (\ref{N_spins}). While the quantum phase transition was not accessible with the SSE method in the case of one spin ($\alpha_c=1$), it will be possible to investigate this regime for two spins ($\alpha_c \simeq 0.2$), as shown in Sec. \ref{quantum_phase_transition_two_spins}. Synchronization effects are studied in Sec. \ref{synchronization}. 
\section{Extension of the SSE framework to two spins}
For the derivation we neglect the spatial separation between sites and consider that the two spins are initially in the state $|\uparrow_z\rangle$ so that $\rho_S (t_0)=|\uparrow_z,\uparrow_z\rangle \langle \uparrow_z, \uparrow_z |$. The time-evolution of a given element $x=\langle \sigma_{1,f},\sigma_{2,f} | \rho_S (t) | \sigma_{1,f'},\sigma_{2,f'}\rangle$ of the spin reduced density matrix can be then re-expressed as, 

\begin{align}
x=\int \prod_{p=1}^2 \left( D\Sigma_p D\Sigma_p'\right) \prod_{p=1}^2 \left(\mathcal{A}[\Sigma_p] \mathcal{A}^* [\Sigma_p']\right) \mathcal{F}[\Sigma_1,\Sigma_2,\Sigma_1',\Sigma_2']\notag \\
\times \exp\left\{i \int_{t_0}^{t} ds K \left[ \Sigma_1(s) \Sigma_2(s)- \Sigma_1'(s) \Sigma_2'(s)\right]\right\}.
\label{eq:densitymatrixelement_two_spins}
\end{align}

The integration runs over all paths $\Sigma_1$, $\Sigma_2$, $\Sigma_1'$ and $\Sigma_2'$ such that $\Sigma_1(t_0)=\Sigma_2(t_0)=\Sigma_1'(t_0)=\Sigma_2'(t_0)=+1$ and $\sigma^z_p|\sigma_{p,f}\rangle=\Sigma_p(t)|\sigma_{p,f}\rangle$, $\sigma^z_p|\sigma_{p,f'}\rangle=\Sigma_p'(t)|\sigma_{p,f'}\rangle$. As in the one-spin case, the terms of the form $\mathcal{A} [\Sigma_p]$ denote the amplitude to follow one given spin path $\Sigma_p$ in the sole presence of the transverse field term acting on the spin $p$. The last term of the right hand side of Eq. ~(\ref{eq:densitymatrixelement_two_spins}) comes from the Ising interaction between the two spins. The influence functional $f=\mathcal{F}[\Sigma_1,\Sigma_2,\Sigma_1',\Sigma_2']$ reads :
\begin{small}
\begin{align}
f=&e^{-\frac{1}{\pi} \int_{t_0}^t ds \int_{t_0}^s ds' \sum_{i,j=1}^2\left\{-i L_1(s-s')\frac{ \Sigma_i (s)-\Sigma_i '(s) }{2} \frac{ \Sigma_j (s')+\Sigma_j '(s') }{2} +L_2(s-s')\frac{\Sigma_i (s)-\Sigma_i '(s) }{2} \frac{ \Sigma_j (s')-\Sigma_j '(s')}{2}\right\}}\notag \\
&\times \mathcal{G}[\Sigma_1,\Sigma_2,\Sigma_1',\Sigma_2'].
\label{eq:influence_two_spins}
\end{align}
\end{small}

The additional term $\mathcal{G}$ in Eq. (\ref{eq:influence_two_spins}) reads : 

\begin{align}
\mathcal{G}[\Sigma_1,\Sigma_2,\Sigma_1',\Sigma_2'] &= \exp\left\{ i \frac{\mu}{2} \int_{t_0}^t ds \left[\sum_{j=1}^2 \frac{\Sigma_j (s)}{2} \right]^2-\left[\sum_{j=1}^2 \frac{\Sigma_j ' (s)}{2} \right]^2\right\},
\label{eq:influence_3_two_spins}
\end{align}
with  $\mu=2/\pi \int_0^{\infty} J(\omega)/\omega=4 \alpha \omega_c$. We recover in Eq.~(\ref{eq:influence_3_two_spins}) that the bath renormalizes the direct Ising interaction between the spins. The term above is indeed similar to the one coming from the direct Ising interaction $K$ (last term of the right hand side of Eq. ~(\ref{eq:densitymatrixelement_two_spins})). In the following we gather these two contributions in a functional $\tilde{\mathcal{G}}$ which reads
\begin{align}
\tilde{\mathcal{G}}[\Sigma_1,\Sigma_2,\Sigma_1',\Sigma_2'] = \exp\left\{i \int_{t_0}^{t} ds K_r \left[ \Sigma_1(s) \Sigma_2(s)- \Sigma_1'(s) \Sigma_2'(s)\right]\right\},
\end{align}
with $K_r=K+\alpha \omega_c$ the renormalized Ising interaction. \\%The ferromagnetic bath-induced interaction can also be recovered in an Hamiltonian point of view \cite{orth_stanic_lehur}.

\begin{figure}[t!]
\center
\includegraphics[scale=1.05]{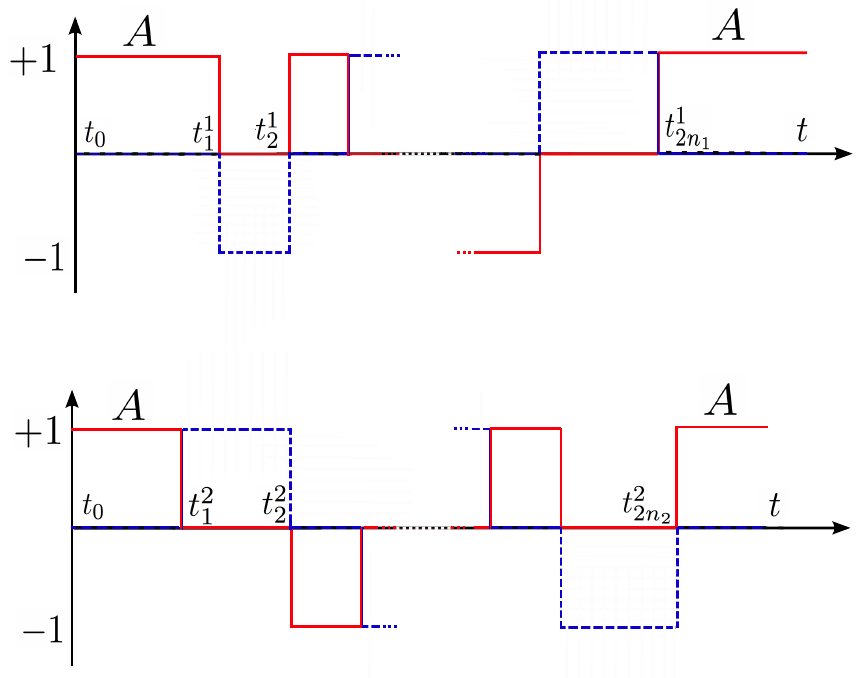}  
\caption{Spin path for the dimer problem- The upper part shows the spin path in terms of blips and sojourns for the first spin, while the lower part shows the spin path of the second spin. $\eta^p(t)=\sum_{j=0}^{2n} \Upsilon_j^p\theta(t-t_j)$ in red; $\xi^p(t)=\sum_{j=1}^{2n} \Xi^p_j\theta(t-t_j)$ in dashed blue. The system starts in the state AA, jumps to the state AB at $s_1=t_1^2$, then to the state CB at $s_2=t_1^1$. It finally ends in the state AA at $t$.}
\label{spin_path_2_main_text}
\end{figure}

The paths introduced in Eq. (\ref{eq:densitymatrixelement_two_spins}) can be viewed as one single path that visits the sixteen states corresponding to the matrix elements of the spin-reduced density-matrix. We will note $\mathcal{E}=\{$AA, AB, AC, AD, BA, BB, BC, BD, CA, CB, CC, CD, DA, DB, DC, DD$\}$ the set of these states - the states A, B, C and D have been defined for one spin (see \ref{Blips_sojourns}). The four states AA, AD, DA and DD correspond to the diagonal elements of the canonical density matrix, while other states correspond to off-diagonal elements. Let us focus on the computation of $\langle \uparrow_z,\uparrow_z | \rho_S (t) | \uparrow_z,\uparrow_z\rangle$, corresponding to the probability to come back in the state $| \uparrow_z,\uparrow_z \rangle$ at time $t$. Then, both the first and the second spin path make an even number of transitions along the way at times $t^p_{j}$, $j \in \{1,2,..,2n_p\}$ for $p \in \{1,2\}$ such that $t_{0}<t^p_1<t^p_2<...<t^p_{2n_p}<t$. We can write these spin paths as $\xi^p(t)=\sum_{j=1}^{2n_p} \Xi^p_j\theta(t-t^p_j)$ and $\eta^p(t)=\sum_{j=0}^{2n_p} \Upsilon^p_j\theta(t-t^p_j)$ where the variables $\Xi^p_j$ and $\Upsilon^p_j$ take values in $\{-1,1\}$. Such a path can be visualized in Fig.~\ref{spin_path_2_main_text} as a couple of one-spin paths. \newline

$\langle \uparrow_z,\uparrow_z | \rho_S (t) | \uparrow_z,\uparrow_z\rangle$ is given by a series in $\Delta^2$:
%\begin{equation}
%p_1(t)=\sum_{n_1,n_2=0}^{\infty}\left(\frac{i\Delta}{2} \right)^{2N}  \int_{s_0}^{t} ds_{2N} .. \int_{s_0}^{s_2} ds_{1} \sum_{\{\Xi^p_j\},\{\Upsilon^p_j%\}' } \mathcal{F}_{n_1,n_2}.
%\label{eq:p(t)_two_spins}
%\end{equation}

\begin{equation}
\langle \uparrow_z,\uparrow_z | \rho_S (t) | \uparrow_z,\uparrow_z\rangle=\sum_{\substack{n_1,n_2\\
                  \{\Xi^p_j\},\{\Upsilon^p_j\}'}}\left(\frac{i\Delta}{2} \right)^{2N}  \int_{s_0}^{t} ds_{2N} .. \int_{s_0}^{s_2} ds_{1}  \mathcal{F}_{n_1,n_2},
\label{eq:p(t)_two_spins}
\end{equation}
where $N=n_1+n_2$ and $\{s_0, s_1, ..., s_{2(n_1+n_2)}\}$ is the ordered reunion of the two sequences $\{t^1_j\}$ and $\{t^2_j\}$. The summation over $n_1$ and $n_2$ goes from $0$ to infinity. The prime in $\{\Upsilon^p_j\}'$ in Eq. (\ref{eq:p(t)_two_spins}) indicates that the initial and final states are fixed according to $\Upsilon^1_0=\Upsilon^2_0=\Upsilon^1_{2n_1}=\Upsilon^2_{2n_2}=1$. The influence functional can be written explicitly in terms of $\Xi^p_j$ and $\Upsilon^p_j$ variables:

  \begin{align}
 &\mathcal{F}_{n_1,n_2}= \left( \prod_{p=1}^2  \mathcal{Q}^p_1 \mathcal{Q}^p_2 \mathcal{M}^p_1 \mathcal{M}^p_2 \right) \tilde{\mathcal{G}}[\sigma_1,\sigma_2,\sigma_1',\sigma_2'] \label{10_two_spins},
   \end{align}
   with
   \begin{align}
  &\mathcal{Q}^p_1 =\exp \left[ \frac{i}{\pi} \sum_{k=0}^{2n_p-1}\sum_{j=k+1}^{2n_p} \Xi^p_j \Upsilon^p_k  Q_1(t^p_j-t^p_k) \right], \label{Q_1_2} \\
  &\mathcal{Q}^p_2 =\exp \left[ \frac{1}{\pi} \sum_{k=1}^{2n_p-1}\sum_{j=k+1}^{2n_p} \Xi^p_j \Xi^p_k  Q_2(t^p_j-t^p_k) \right],\label{Q_2_2}   \\
   &\mathcal{M}^p_1 =\exp \left[ \frac{i}{\pi} \sum_{k=0}^{2n_{\overline{p}}-1}\sum_{j: t^p_j>t^{\overline{p}}_k} \Xi^p_j \Upsilon^{\overline{p}}_k Q_1(t^p_j-t^{\overline{p}}_k) \right], \label{Q_1_m_2} \\
  &\mathcal{M}^p_2 =\exp \left[ \frac{1}{\pi} \sum_{k=1}^{2n_{\overline{p}}-1}\sum_{j: t^p_j>t^{\overline{p}}_k} \Xi^p_j \Xi^{\overline{p}}_k  Q_2(t^p_j-t^{\overline{p}}_k) \right].\label{Q_2_m_2}
 \end{align}

In Eqs. (\ref{Q_1_m_2}) and (\ref{Q_2_m_2}), ${\overline{p}}=2$ if $p=1$ and ${\overline{p}}=1$ if $p=2$. Functions $Q_1$ and $Q_2$ are defined by Eqs. (\ref{q1}) and (\ref{q2}). Terms $\mathcal{M}^p_1$ and $\mathcal{M}^p_2$ account for retarded interactions between the two spins, mediated by the bath. Their expression in terms of blip and sojourn variables is very similar to the ones of $\mathcal{Q}^p_1$ and $\mathcal{Q}^p_2$ and the principle of their derivation is the same as in the case of one spin done in Chapter II Sec. \ref{derivation_Q1_Q2}. The situation differs however slightly since the blip variables corresponding to one spin and the sojourn variable corresponding to the other one can be simultaneously both non-zero. A detailled derivation in this particular case is provided in Appendix \ref{appendix_M1_M2}. The (renormalized) Ising interaction (in $\tilde{\mathcal{G}}[\Sigma_1,\Sigma_2,\Sigma_1',\Sigma_2']$) can be expressed in a convenient way in this description, as we have

\begin{equation}
 \Sigma_1(s) \Sigma_2(s)- \Sigma_1'(s) \Sigma_2'(s)=2\left[ \eta^1 (s) \xi^2(s)+ \eta^2 (s) \xi^1(s) \right].
\label{eq:F1}
\end{equation}

 As for the one-spin case, we can proceed to a stochastic unravelling of the influence functional (see Chapter II Sec. \ref{unravelling}), and we have

  \begin{align}
\mathcal{F}_{n_1, n_2}=& \overline{  \prod_{i=1}^{2n_1} \exp\left[ h(t^1_i) \Xi_j^1+k(t^1_{i-1}) \Upsilon^1_{i-1}    \right]} \overline{  \prod_{j=1}^{2n_2} \exp\left[ h(t^2_j) \Xi_j^2+k(t^2_{j-1}) \Upsilon^2_{j-1}    \right]}\tilde{\mathcal{G}}[\Sigma_1,\Sigma_2,\Sigma_1',\Sigma_2'].
\label{functionnal_2}
\end{align}
The fields $h$ and $k$ verify the correlations of Eqs. (\ref{height_1}), (\ref{height_2}), and (\ref{height_3}). Eq.~(\ref{eq:p(t)_two_spins}) together with Eq.~(\ref{functionnal_2}) has now the form of a time ordered product, averaged over the noise variables.\\

 The summation over the variables $\{\Xi^p_j\}$ and $\{\Upsilon^p_j\}'$ for $p\in \{1,2\}$ can be incorporated by considering an effective Hamiltonian $H_{1}(t)$ for the spin density matrix, acting on the space $\mathcal{E}$. It can be written as a sum of two terms $H_{1}(t)=U_{1}+V_{1}(t)$. The (renormalized) Ising interaction is contained in the first term $U_{1}$, while the second term $V_{1}(t)$ accounts for tunneling events. 
 
 $U_{1}$ is a diagonal matrix, whose elements are $\left(U_{1}\right)_{i,i}=2K_r (\eta^1_i \xi^2_i+ \eta^2_i  \xi^1_i )$, where $\eta^p_i$ and $\xi^p_i$ are the value of $\eta^p$ and $\xi^p$ for the state in the position $i$ in the set $\mathcal{E}=\{$AA, AB, AC, AD, BA, BB, BC, BD, CA, CB, CC, CD, DA, DB, DC, DD$\}$. The sequence $\left(U_{1}\right)_{i,i}$ gives explicitely $(0,k,-k,0,k,0,0,-k,-k,0,0,k,0,-k,k,0)$ with $k=2 K_r$.

The 16 by 16 matrix $V_{1}(t)$ accounts for tunneling elements and has the following form,
 
 \begin{equation}
V_{1}(t)=\frac{\Delta}{2} \left( \begin{array}{cccc}
W&D_{\textrm{B}\to \textrm{A}}&D_{\textrm{C}\to \textrm{A}}&(0 )\\
D_{\textrm{A}\to \textrm{B}}&W&(0)&D_{\textrm{D}\to \textrm{B}}\\
D_{\textrm{A}\to \textrm{C}}&(0)&W&D_{\textrm{D}\to \textrm{C}}\\
(0)&D_{\textrm{B}\to \textrm{D}}&D_{\textrm{C}\to \textrm{D}}&W
\end{array} \right).
\label{eq:system_two_spins}
\end{equation}
 Each term of this matrix corresponds to a transition from one state in $\mathcal{E}$ to another, induced by one spin-flip. It is written in Eq. (\ref{eq:system_two_spins}) in a block structure. Each block is a 4 by 4 matrix that can be given a physical interpretation. The diagonal matrices correspond to flips of the second spin, the first one left unchanged. As a result the matrix $W(t)$ has the same structure as in the one-spin case,

\begin{equation}
W(t)= \left( \begin{array}{cccc}
0&e^{- h+k }&-e^{ h+k }&0 \\
e^{h-k }&0&0&-e^{h+k }\\
-e^{- h-k}&0&0&e^{- h+k }\\
0&-e^{- h-k }&e^{h-k }&0
\end{array} \right).
\end{equation}
All the elements of the 4 by 4 matrices on the diagonal running from the lower left to the upper right are zero, because the corresponding states are not coupled by one single spin-flip. The eight matrices $D_{\textrm{B}\to \textrm{A}}$, $D_{\textrm{C}\to \textrm{A}}$, $D_{\textrm{A}\to \textrm{B}}$, $D_{\textrm{D}\to \textrm{B}}$, $D_{\textrm{A}\to \textrm{C}}$, $D_{\textrm{D}\to \textrm{C}}$, $D_{\textrm{B}\to \textrm{D}}$ and $D_{\textrm{C}\to \textrm{D}}$ describe spin flips of the first spin (the precise transition corresponds to the subscript), the second one left unchanged. They read respectively $e^{- h+k }\times I_4$, $-e^{ h+k } \times I_4$, $e^{h-k }\times I_4$, $-e^{h+k}\times I_4$, $-e^{- h-k}\times I_4$, $e^{- h+k}\times I_4$, $-e^{- h-k}\times I_4$ and $e^{h-k}\times I_4$ ($I_4$ is the identity). The first transition at $s_1=t_1^2$ corresponds to the transition AA$\to$AB. Its amplitude is given by the term of the first column and the second row of the top left matrix $W$. The next transition at $s_2=t_1^1$ corresponds to the transition AB$\to$CB. Its amplitude is given by the term of the second column and the second row of the matrix $D_{\textrm{A} \to \textrm{C}}$.\\

 Finally, the dynamics of the 16 dimensional spin reduced density matrix is governed by an effective SSE with Hamiltonian $H_1$:
\begin{equation} 
\langle \uparrow_z,\uparrow_z | \rho_S (t) | \uparrow_z,\uparrow_z\rangle=\overline{\langle \Phi_f | \Phi (t) \rangle},
\label{eq:p1:two_spins}
\end{equation}
where $\langle \Phi_f |=(e^{-2k(s_{2N})},0,0,0,0,0,0,0,0,0,0,0,0,0,0,0)$ and  $| \Phi \rangle$ is the solution of the stochastic Schr\"{o}dinger equation 

\begin{equation} 
i \partial_t | \Phi \rangle = H_1 (t) | \Phi \rangle
\label{eq:SSE:two_spins}
\end{equation}
with initial condition 
\begin{equation}
|\Phi_i \rangle=(e^{2k(t_0)},0,0,0,0,0,0,0,0,0,0,0,0,0,0,0)^T.
\end{equation}

Other initial and final conditions lead to a different choice for the vectors $|\Phi_i \rangle$ and $|\Phi_f \rangle$, in analogy with the case of one spin. Simplifications also occur in the scaling regime, as shown in Appendix \ref{appendix_M1_M2}. 
\section{Nonequilibrium dynamics and quantum phase transition in the dimer model} 
\label{quantum_phase_transition_two_spins}

Next, we use the SSE approach to investigate the dissipative quantum phase transition for this dimer problem and address the non-equilibrium dynamics of the system both in the unpolarized and in the polarized phase. This problem is indeed well-known to exhibit a dissipative quantum phase transition \cite{Garst_Vojta,sougato,Peter_two_spins,Winter_Rieger} where the bath entirely polarizes the two spins by analogy to a ferromagnetic phase. The transition line can be located thanks to the evolution of the entanglement entropy with respect to $\alpha$ (see Fig.~5 of Ref. \cite{Peter_two_spins}) or to the evolution of the connected correlation function $C=\langle \sigma^z_1 \sigma^z_2 \rangle-\langle \sigma^z_1 \rangle \langle \sigma^z_2 \rangle$ (see Fig.~10 of Ref. \cite{Winter_Rieger}).\\
%Here, we apply the SSE methodology in order to tackle the non-equilibrium spin dynamics in the presence of strong dissipative interactions in the case of two spins. 

We define the triplet subspace spanned by the three states $\{|T_-\rangle=|\downarrow_z, \downarrow_z\rangle,|T_0\rangle=1/\sqrt{2}\left[|\uparrow_z, \downarrow_z\rangle+|\downarrow_z,\uparrow_z\rangle\right],|T_+\rangle=|\uparrow_z,\uparrow_z\rangle\}$, while the singlet state is\linebreak $|S\rangle=1/\sqrt{2}\left[|\uparrow_z,\downarrow_z\rangle-|\downarrow_z,\uparrow_z\rangle\right]$ and remains isolated in the dynamics if $\Delta_1=\Delta_2$.\\

\begin{figure}[h!]
\center
\includegraphics[scale=0.45]{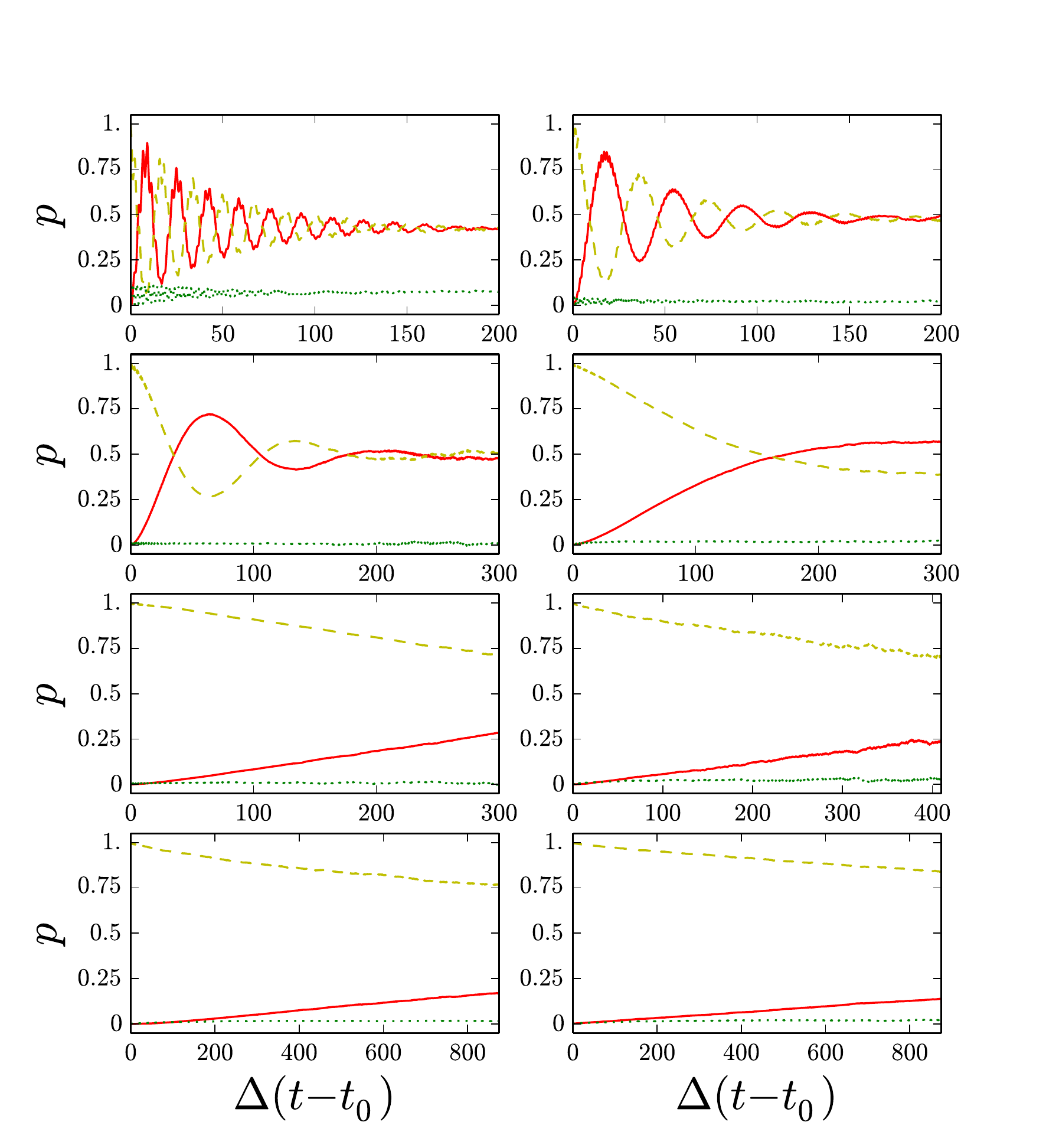}  
\caption{Dynamics of the dimer model in the unpolarized phase: the dashed yellow line represents $p_{|T_+\rangle}$, the full red line represents  $p_{|T_-\rangle}$ and the dotted green line represents $p_{|T_0\rangle}$. From the top left to the bottom right, we have $\alpha=0.01$, $\alpha=0.02$, $\alpha=0.04$, $\alpha=0.06$, $\alpha=0.08$, $\alpha=0.1$, $\alpha=0.12$ and , $\alpha=0.14$. The system starts in the state $|T_+\rangle$ for all the plots. We have taken $\omega_c/\Delta=100$ and $K=0$ for all plots.}
\label{free_dynamics_21_10_2015}
\end{figure}

 We consider that the system initially starts from the state $|T_+\rangle$ at time $t_0$, when spin and bath are brought into contact. We show in Fig.~\ref{free_dynamics_21_10_2015} the time evolution of $p_{|T_0\rangle}$, $p_{|T_+\rangle}$ and $p_{|T_-\rangle}$, which are the occupancies of the states $|T_0\rangle$ , $|T_+\rangle$ and $|T_-\rangle$.

The different panels correspond to different values of $\alpha$ from $\alpha=0.01$ (top left) to $\alpha=0.14$ (bottom right). All these values corrrespond to the unpolarized phase in the range of parameters used ($K=0$ and $\omega_c=100$). 

% At small $\alpha$ we remark that the equilibrium value of $\langle \sigma^z_1\sigma^z_2 \rangle_{eq}$ is smaller than expected, and it becomes larger than the predictions when increasing $\alpha$. We see for example that for $\Delta/\omega_c=0.2$, $\langle \sigma^z_1\sigma^z_2 \rangle_{eq}$ reaches $1$ at around $\alpha=0.25$. As we will see later (see Fig.~\ref{caracteristic_time_phase_diagram}), this point correspond to the location of the quantum phase transition for $\Delta/\omega_c=0.2$. We do not observe a jump of $\langle \sigma^z_1\sigma^z_2 \rangle_{eq}$ at the phase transition but rather a continous evolution.\\

 \begin{figure}[h!]
%\center
\includegraphics[scale=0.32]{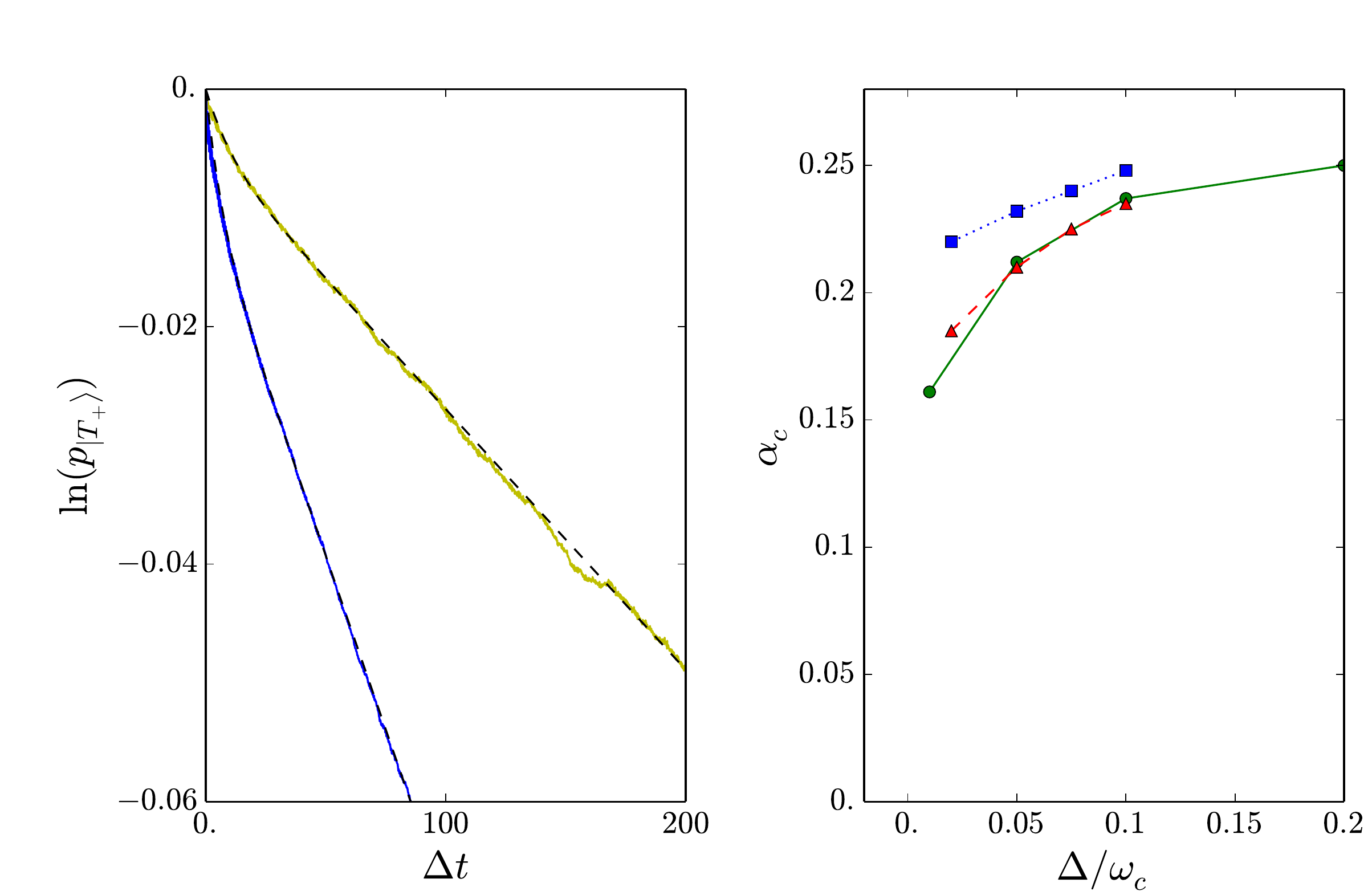}  \includegraphics[scale=0.28]{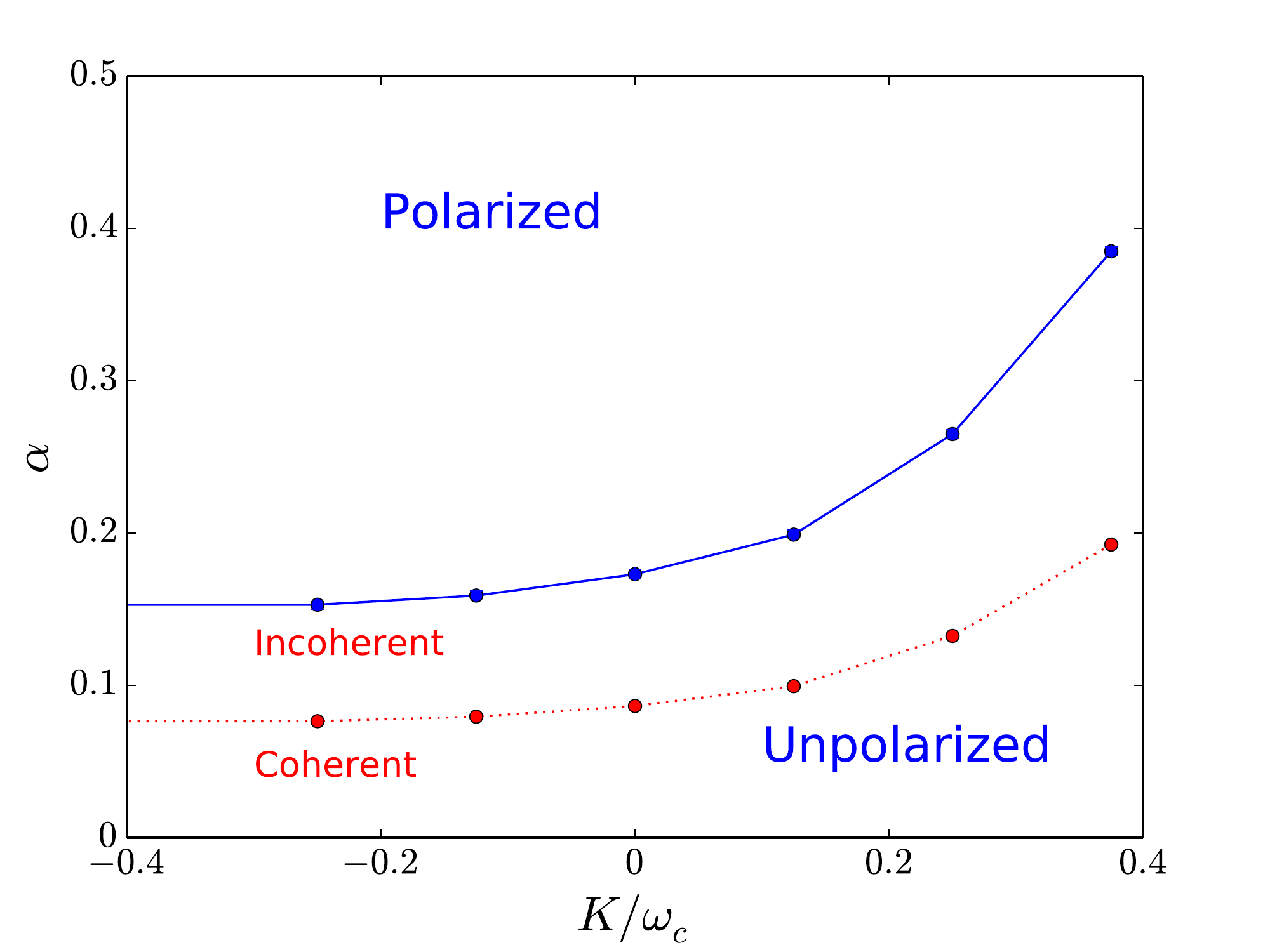}  
\caption{Left panel: evolution of $\ln(p_{|T_+\rangle})$ at $\alpha=0.14$, for $\Delta/\omega_c=0.01$ (yellow line-top) and $\Delta/\omega_c=0.05$ (blue line-bottom), and bi-exponential fit (dashed black line). Middle panel: Critical line with respect to $\Delta/\omega_c$ at $K=0$ (green dots and full green line) and comparison with the results obtained in Ref. \cite{Peter_two_spins} (TDNRG) (red triangles and red dashed line) and Ref. \cite{Winter_Rieger} (QMC) (blue squares and dotted blue line). Right panel: Critical line with respect to $K$ for $\Delta/\omega_c=0.01$ (blue points and full blue line). Above the line, the system relaxes to a polarized steady-state. The red dots and the dotted red line show the location of the crossover line from coherent to incoherent behaviour for the spin oscillations.}
\label{decay_phase_diagram}
\end{figure}

%\begin{figure}[h!]
%\center
%\includegraphics[scale=0.34]{figure6.pdf}  
%\caption{(Color online) Critical line with respect to $K$ for $\Delta/\omega_c=0.01$ (blue points and full blue line). Above the line, the system relaxes to a polarized steady-state. The red dots and the dotted red line show the location of the crossover line from coherent to incoherent behaviour for the spin oscillations.}
%\label{phase_diagram}
%\end{figure}

% Evolution of the characteristic time scale of the relaxation with respect to $\alpha$ for $K=0$, $\Delta/\omega_c=0.01$. 

We first note in Fig.~\ref{free_dynamics_21_10_2015} a progressive suppression of the Rabi oscillations between the two states $|T_+\rangle$ and $|T_-\rangle$ when increasing the parameter $\alpha$. This behavior is similar to the one observed in the case of the single spin-boson model, where the crossover from coherent oscillations to an incoherent dynamics occurs at $\alpha_c/2$. At high values of $\alpha$, the relaxation from the initial state $|T_+\rangle$  becomes slower due to the strong ferromagnetic interaction, and it is numerically harder to investigate the dynamics in the zone $\alpha \geq 0.1$, due to the time scales involved (other initial states lead to an easier numerical investigation, allowing to determine accurately the equilibrium density matrix at long times). In the zone $\alpha_c/2<\alpha<\alpha_c$, we find a monotonic relaxation towards the equilibrium. In this zone, for the case of one spin, conformal field theory has predicted that several timescales are involved in the dynamics, leading to a multi-exponential decay \cite{Lesage_Saleur} (which has not been seen in NRG \cite{KLHQPT}). A bi-exponential decay was found in this case thanks to a multilayer multiconfiguration time-dependent Hartree method \cite{Wang_Thoss}. Other studies have predicted more complicated forms for the relaxation, without any pure exponential decay (see for example the results of Ref. \cite{Kashuba_Schoeller} obtained with renormalization group methods). \\

We are then able to locate the phase transition from the divergence of the associated time scale. The transition line is shown on the center panel of Fig.~\ref{decay_phase_diagram}, together with the previous results obtained with a time dependent Numerical Renormalization Group (TDNRG) method \cite{Peter_two_spins}, or with a Quantum Monte-Carlo (QMC) method \cite{Winter_Rieger}. This plot corresponds to a vanishing direct Ising interaction $K=0$, and different values of $\omega_c$. Here, for two spins and at small to intermediate times, we obtain results which are also consistent with a bi-exponential relaxation, as shown on the left panel of Fig.~\ref{decay_phase_diagram}. The phase diagram of the system with respect to the parameter $K$ is shown on the right panel. The full blue line shows the phase transition line between the polarized and the unpolarized phase, while the dotted red line shows the crossover line from coherent to incoherent Rabi oscillations in the dynamics \cite{Peter_two_spins}.\\

%For $\alpha > \alpha_c$, the system stays localized in the state $|T_+\rangle$ (i.e. there is no evolution of the probabilities). It is interesting to study the evolution of the equilibrium correlation function $C^{z_1 z-2}=\langle \sigma^z_1 \sigma^z_2  \rangle_{eq}-\langle \sigma^z_1 \rangle \langle \sigma^z_2  \rangle_{eq}$. 

%We then confirm that, at small dissipation, the main effect of the bath is to induce a ferromagnetic Ising-like interaction between the spins of strength $K_r=\alpha \omega_c$. 

%For larger dissipation strength, this is no longer true. As shown in Fig.~6,  the value of $\langle \sigma^z_1 \sigma^z_2  \rangle_{eq}$ rapidly drops to zero when increasing $\alpha$. From $\alpha_c /2 \lesssim \alpha \leq \alpha_c $, we have $\langle \sigma^z_1 \sigma^z_2  \rangle_{eq} \simeq 0$. 

%We notice that the parameter zone $[\alpha_c /2, \alpha_c]$ corresponds to the suppression of the Rabi oscillations between $|T_+\rangle$ and $|T_-\rangle$ (for  $\alpha \gtrsim \alpha_c /2$, we always have $p_{|T_+\rangle} > p_{|T_-\rangle}$ during the evolution). As shown in Ref.~\onlinecite{Peter_two_spins}, this zone corresponds to a maximum plateau in the entanglement entropy. \textbf{developper un peu} . At $\alpha=\alpha_c$, there is a jump of the order parameter from $\langle \sigma^z  \rangle_{eq} = 0$ to $|\langle \sigma^z \rangle_{eq}| = 1$, as in the one-spin case. We show in Fig.~7 the phase diagram of our system, and compare to the values found in the litterature.\\

%\textbf{TEMPORARY}

Next, we show results concerning the dynamics in the polarized phase ($\alpha>\alpha_c$), corresponding to a quantum quench across the critical line, from $\alpha=0$ to $\alpha>\alpha_c$. Some theoretical studies have focused on this question in spins \cite{essler,gambasi,delcampo} or bosonic systems \cite{roux_kollath,sciolla_biroli,rancon}.  For example, at $K=0$ and $\alpha=0$, the initial state of the system is given by $|\psi\rangle=|-_x\rangle \otimes |-_x\rangle=1/2(|T_+\rangle+|T_-\rangle)-1/\sqrt{2} |T_0\rangle$. The associated spin density matrix is

\begin{equation}
\rho_{S} (t_{0})=\frac{1}{4}\left(\begin{array}{cccc}
1 & -1 &-1 & 1   \\
-1 & 1 &1 & -1   \\
-1 & 1 &1 & -1   \\
1 & -1 &-1 & 1  \end{array} \right).
\label{eq:density_matrix_two_spins}
\end{equation}

\begin{figure}[t!]
\center
\includegraphics[scale=0.36]{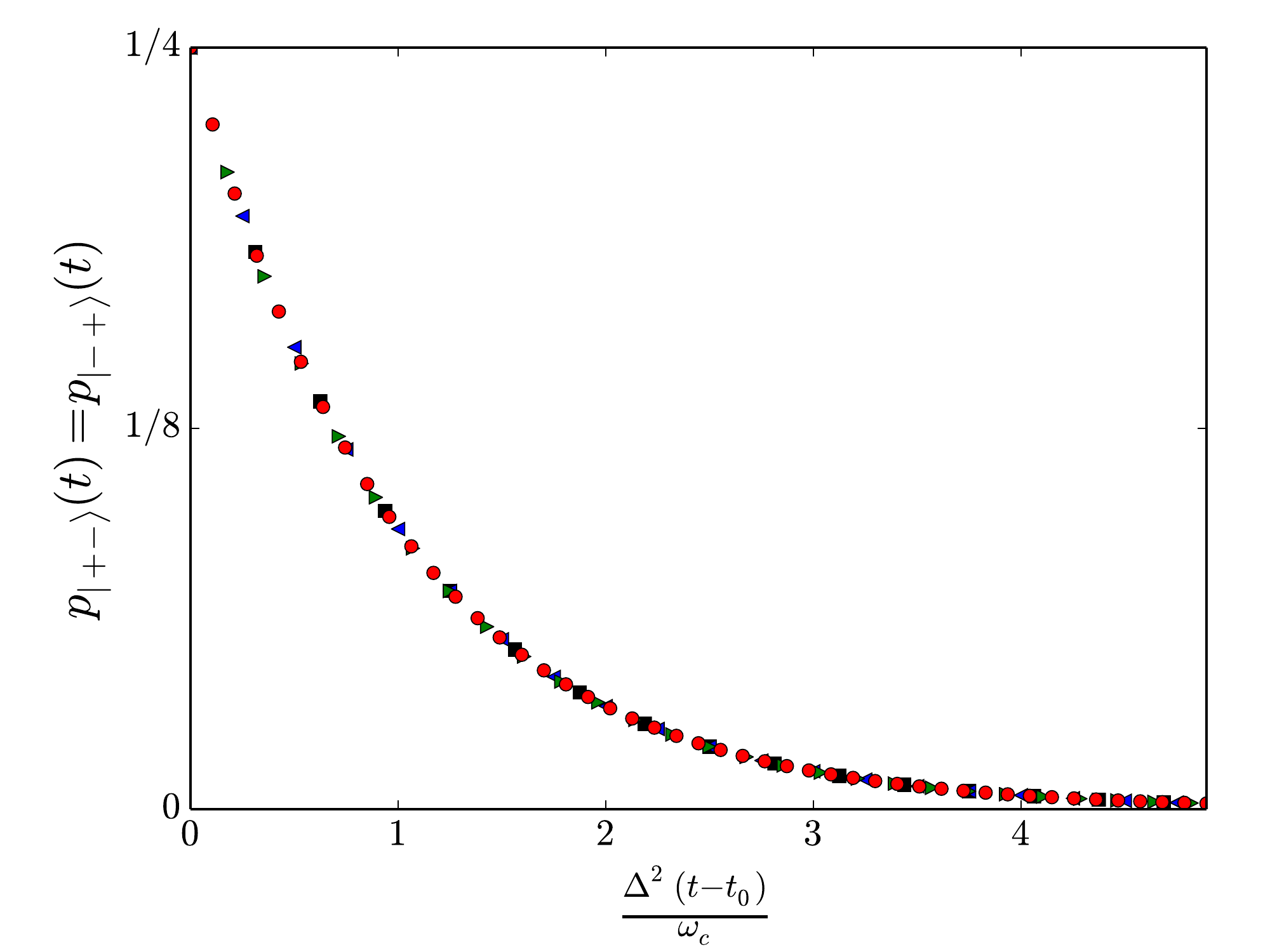}  
\caption{Universal dynamics of the dimer model in the polarized phase. The system starts in the nonequilibrium state described by the density matrix of Eq. (\ref{eq:density_matrix_two_spins}), and relax towards a statistic superposition of $|T_+\rangle$ and $|T_-\rangle$. The parameters are $\alpha=0.2$, $\omega_c/\Delta=100$ (red points); $\alpha=0.25$, $\omega_c/\Delta=50$ (right pointing green triangles); $\alpha=0.22$, $\omega_c/\Delta=80$ (left pointing blue triangles); $\alpha=0.3$, $\omega_c/\Delta=20$ (black squares). Taking $K \neq 0$ gives the same exponential relaxation.}
\label{fig_localized}
\end{figure}

 After a sudden change of the parameter $\alpha$, the system is in a nonequilibrium state. We compute the spin dynamics for different values of $\alpha>\alpha_c$ and for different values of $\Delta/\omega_c$. We find numerically that the system evolves towards the final density matrix 
  
 \begin{equation}
\lim_{t \rightarrow \infty} \rho_s(t) = \frac{1}{2}\left(\begin{array}{cccc}
1 & 0 &0 & 0   \\
0 & 0 &0 & 0   \\
0 & 0 &0 & 0   \\
0 & 0 &0 & 1  \end{array} \right),
\label{eq:density_matrix_two_spins_final}
\end{equation}
corresponding to a statistical superposition of the states $|T_+\rangle$ and $|T_-\rangle$ (up to an error of around $10^{-4}$). We find moreover that the spin dynamics is universal in the polarized phase, in the sense that it does not depend on $\alpha$ and $K$. More precisely, we find that 
 \begin{equation}
p_{|+-\rangle}(t)=p_{|-+\rangle}(t)=p_0 \exp\left[-\frac{\Delta^2 (t-t_0)}{\omega_c} \right],
\label{Eq:relaxation}
 \end{equation}
 as shown in Fig.~\ref{fig_localized}, for a quench from $\alpha=0$ to $\alpha>\alpha_c$. $p_{|+-\rangle}(t)$ ($p_{|-+\rangle}(t)$) is the probability to find the system in the state $|\uparrow_z,\downarrow_z\rangle$ ($|\downarrow_z,\uparrow_z\rangle$) at time $t$, given by the diagonal term of the density matrix $[\rho_S]_{22}$ ($[\rho_S]_{33}$). The strong bath-induced Ising interaction and the orthogonality between the polarized state seem to provoke a rapid evolution independent of the other external parameters.\\
 
\begin{figure}[h!]
\center
\includegraphics[scale=0.42]{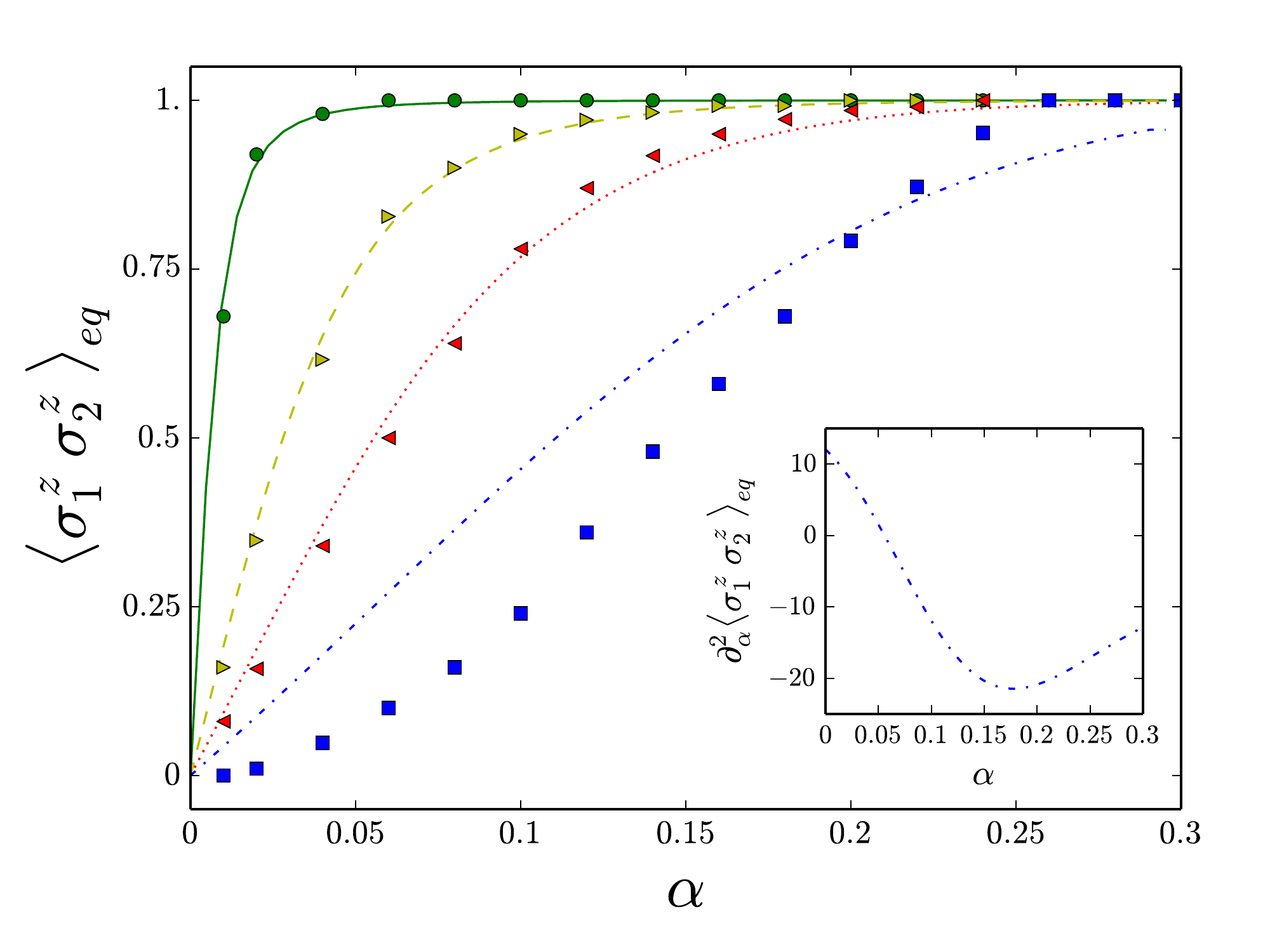}  
\caption{Equilibrium value of $\langle \sigma^z_1 \sigma^z_2  \rangle$ as a function of $\alpha$ for $\Delta/\omega_c=0.01$ (green circles), $\Delta/\omega_c=0.05$ (yellow right-pointing triangles), $\Delta/\omega_c=0.1$ (red left-pointing triangles) and $\Delta/\omega_c=0.2$ (blue squares). We have $K=0$. The lines correspond to the value predicted by a toy-model of two interacting spins with tunneling element $\tilde{\Delta}_r$ and Ising interaction $\tilde{K}_r$ obtained thanks to a variational procedure. Parameters are $\Delta/\omega_c=0.01$ (full green line), $\Delta/\omega_c=0.05$ (yellow dashed line), $\Delta/\omega_c=0.1$ (red dotted line) and $\Delta/\omega_c=0.2$ (blue mixed line). The inset shows the evolution of $\partial^2_{\alpha} \langle \sigma^z_1 \sigma^z_2  \rangle $ with $\alpha$ for $\Delta/\omega_c=0.2$. The sign of this quantity changes when increasing $\alpha$.}
\label{Correlation_function_two_spins}
\end{figure}

We also remark that, in the unpolarized phase, the value of $\langle \sigma^z_1\sigma^z_2 \rangle_{eq}$ is non-zero due to the strong ferromagnetic interaction mediated by the bath. We compute this quantity as the limit of $tr_B\left[\rho_S (t) \sigma^z_1 \sigma^z_2\right]$ at long times, and plot its evolution with respect to $\alpha$ for different values of $\omega_c$ in Fig.~\ref{Correlation_function_two_spins}. At very small $\Delta/\omega_c$ we have roughly $\langle \sigma^z_1\sigma^z_2 \rangle_{eq}=\alpha \omega_c /\sqrt{(\alpha \omega_c)^2+\Delta_r^2}$, which would be the equilibrium value of this quantity in a two-spins Ising model governed by the Hamiltonian 

\begin{align}
H_I=\frac{\Delta_r}{2}\left( \sigma_1^x+\sigma_2^x \right)-K_r \sigma_1^z \sigma_2^z,
\label{toy_model_ising_two_spins}
\end{align}
where $\Delta_r=\Delta (\Delta/\omega_c)^{\alpha/(1-\alpha)}$ is the renormalized tunneling element obtained by an adiabatic renormalization procedure.

There are notable deviations with respect to this toy-model, especially when $\Delta/\omega_c$ becomes larger ($\Delta/\omega_c \geq 0.02$). In this case, the adiabatic renormalization procedure is no longer valid, as the bath and spin degrees of freedom evolution time scales are not well separated. The assumption of fully polarized bath states associated with one given spin polarization no longer holds and we need to refine the analysis, for example by using a variational technique on the ground state wavefunction following the ideas of Refs. \cite{Silbey_Harris,sougato}. We write the Hamiltonian of the system in a displaced oscillator basis defined by the four states $\{|B_{\uparrow \uparrow}\rangle \otimes |\uparrow_z,\uparrow_z\rangle,|B_{0}\rangle \otimes|\uparrow_z,\downarrow_z\rangle,|B_{0}\rangle \otimes|\downarrow_z,\uparrow_z\rangle,|B_{\downarrow\downarrow}\rangle \otimes|\downarrow_z,\downarrow_z\rangle \}$, with

\begin{align}
|B_{\uparrow \uparrow}\rangle&=\prod_{k} \exp\left[-\frac{f_k}{\omega_k}\left( b^{\dagger}_k- b_k\right) \right]|B_{0}\rangle   \\
|B_{\downarrow \downarrow}\rangle&=\prod_{k} \exp\left[\frac{f_k}{\omega_k}\left(b^{\dagger}_k- b_k\right) \right] |B_{0}\rangle,
\end{align}
where $|B_{0}\rangle$ is the ground state of the bosonic bath taken in isolation at zero temperature. $f_k$ are variational parameters with $f_k \neq \lambda_k$ at a general level. With this ansatz we do not specify the amplitude with which a given mode is displaced \textit{ab initio}, but these coefficients are found by minimizing the free energy of the total system. The displacement from the equilibrium position of a given oscillator may then depend on other parameters. Following Ref. \cite{sougato}, we find self-consistent equations for the bath-induced Ising interaction $\tilde{K}_r$ and the renormalized tunneling element $\tilde{\Delta}_r$,

\begin{align}
\tilde{\Delta}_r&=\Delta \exp\left[-\alpha\int_{0}^{\infty} d\omega \frac{G(\omega)^2}{\omega} e^{-\omega/\omega_c} \right],\\
\tilde{K}_r&=\alpha \int_{0}^{\infty} d\omega G(\omega)[2-G(\omega)]  e^{-\omega/\omega_c},\\
G(\omega)&=\frac{\sqrt{\tilde{K}_r^2+\tilde{\Delta}_r^2}+\tilde{K}_r}{\sqrt{\tilde{K}_r^2+\tilde{\Delta}_r^2}+\tilde{K}_r+\frac{\tilde{\Delta}_r^2}{\omega}}.
\end{align}

We plot the corresponding evolution of $\langle \sigma^z_1  \sigma^z_2 \rangle_{eq}=\tilde{K}_r /\sqrt{(\tilde{K}_r)^2+\tilde{\Delta}_r^2}$ with respect to $\alpha$ for different values of $\omega_c$ in Fig.~\ref{Correlation_function_two_spins}. We find a good agreement with the exact results given by the SSE method as long as $\Delta/\omega_c$ remains small ($\Delta/\omega_c \leq 1$). We notably recover a change of the concavity of $\langle \sigma^z_1  \sigma^z_2 \rangle_{eq}$ with respect to $\alpha$, as shown in the inset of Fig. \ref{Correlation_function_two_spins} where we plot the evolution of the second derivative of $\langle \sigma^z_1  \sigma^z_2 \rangle_{eq}$ for $\Delta/\omega_c=0.2$. This feature cannot be recovered by the adiabatic renormalization procedure, but we see that this effect is far more pronounced in the results of the SSE than in the variational treatment. The dynamical adjustment of both the bath and spin degrees of freedom can thus explain some features of the results obtained numerically, especially at small $\Delta/\omega_c \leq 0.1$ but this variational approach fail at quantitatively describing the regime of strong coupling and the dissipative quantum phase transition. %{\color{red} As seen in Fig.~\ref{Correlation_function_two_spins}, the main effect at large $\omega_c/\Delta$  is to induce a large ferromagnetic interaction. We will use this effect concerning synchronization and LZ interferometry. }

\section{Synchronization}
\label{synchronization}

Synchronization phenomena occur spontaneously in a wide range of physical systems \cite{synchronization}. Here we quantitatively describe driveless synchronization mechanisms between two spins 1/2 with different bare frequencies, which are coupled to a common bath of harmonic oscillators. A comparison between classical and quantum regimes for this kind of problems without dissipation was recently done in Ref. \cite{synchronization_fuchs}.\\

\subsection{First mechanism}

\begin{figure}[t!]
\center
\includegraphics[scale=0.4]{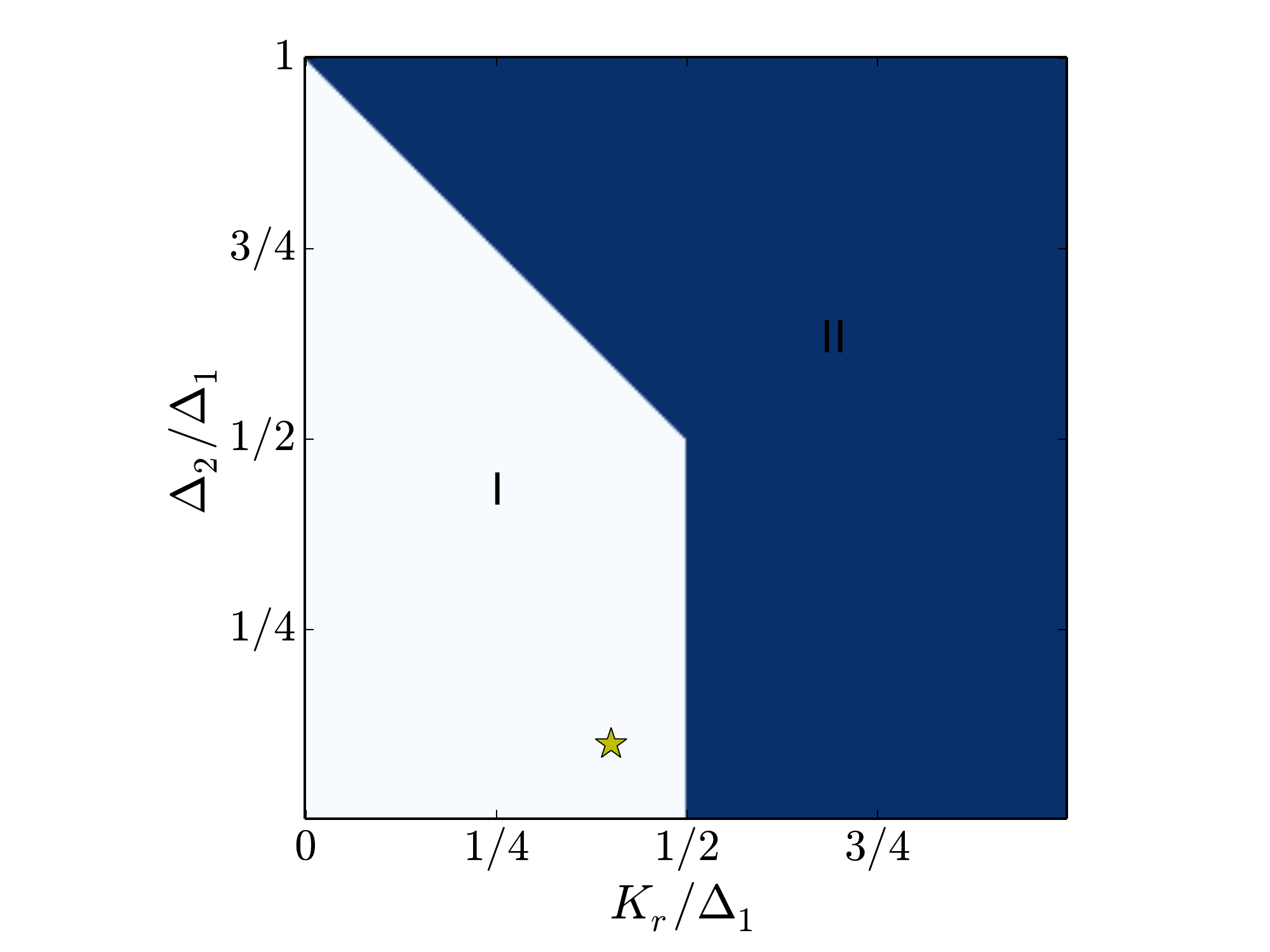}
\caption{Synchronization phase diagram in the case of direct Ising interaction $K_r=K$. Region I (in white) corresponds to the unsynchronized regime : $\langle \sigma^z_1 \sigma^z_2 \rangle$ vanishes periodically. The yellow star shows the point for which we compare direct and bath-induced interaction (see text). Region II (in blue) corresponds to the synchronized regime : $\langle \sigma^z_1 \sigma^z_2 \rangle>0$ at all times.
}
\label{Synchronization_phase_diagram_1}
\end{figure}

We first consider the dynamics of two interacting spins whose dynamics is governed by $\mathcal{H}_{dimer}$ in Eq. (\ref{ising_dimer}), with different bare oscillation frequencies $\Delta_1$ and $\Delta_2$ (with $\Delta_1>\Delta_2>0$). We focus on the free dynamics of the dimer starting from the same initial state $|\uparrow_z,\uparrow_z \rangle$. \\

The bath-induced Ising interaction will tend to synchronize the oscillations of the two spins. We quantify this effect, thanks to spin-spin correlations in time and compare the case of direct versus bath-induced interaction. We denote by $K_r$ the effective strength of the interaction between the spins. In the case of a coupling through the bath we identify $K_r=\alpha \omega_c$ while we have $K_r=K$ in the case of a direct Ising interaction. Some efforts were done to study this effect in Ref.~\cite{Peter_two_spins}. \\

Let us first consider the case of direct Ising interaction $K$. A quantitative description of this type of synchronization can be done by studying the time-evolution of $\langle \sigma^z_1 \sigma^z_2 \rangle$. The system starts in the state $|\uparrow_z,\uparrow_z\rangle$, so that  $\langle \sigma^z_1 \sigma^z_2 \rangle (t_0)=1$ at the initial time. We define the synchronized regime as the region in the parameters space for which $\langle \sigma^z_1 \sigma^z_2 \rangle$ stays positive at all times. We show in Fig~\ref{Synchronization_phase_diagram_1} the synchronization phase diagram with respect to $\Delta_2/\Delta_1$ and $K_r/\Delta_1=K/\Delta_1$. In the region I (in white), the two spins are not synchronized and the correlation function $\langle \sigma^z_1 \sigma^z_2 \rangle$ changes sign periodically. In the other region (region II in blue in Fig.~\ref{Synchronization_phase_diagram_1}) $\langle \sigma^z_1 \sigma^z_2 \rangle$ always stays positive. For $K_r/\Delta_1>1/2$ the Ising interaction dominates and the dynamics is synchronized for all values of $\Delta_2$. When $\Delta_2$ approaches $\Delta_1$, the two spins have comparable oscillating frequencies and the synchronization is then easier. It necessitates a smaller value of $K_r/\Delta_1$ for the two spins to synchronize.\\

\begin{figure}[h!]
\center
\includegraphics[scale=0.2]{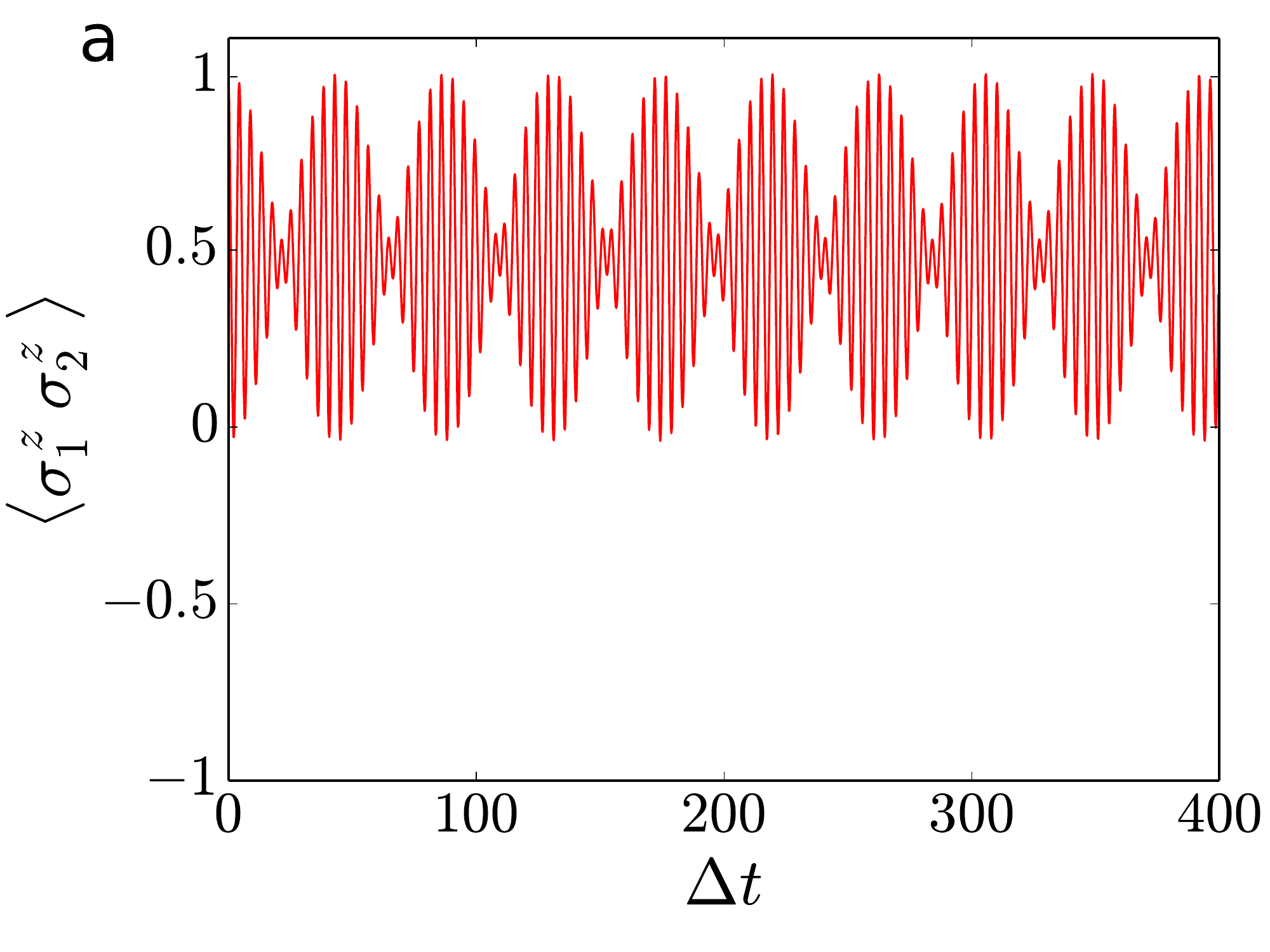}  \includegraphics[scale=0.2]{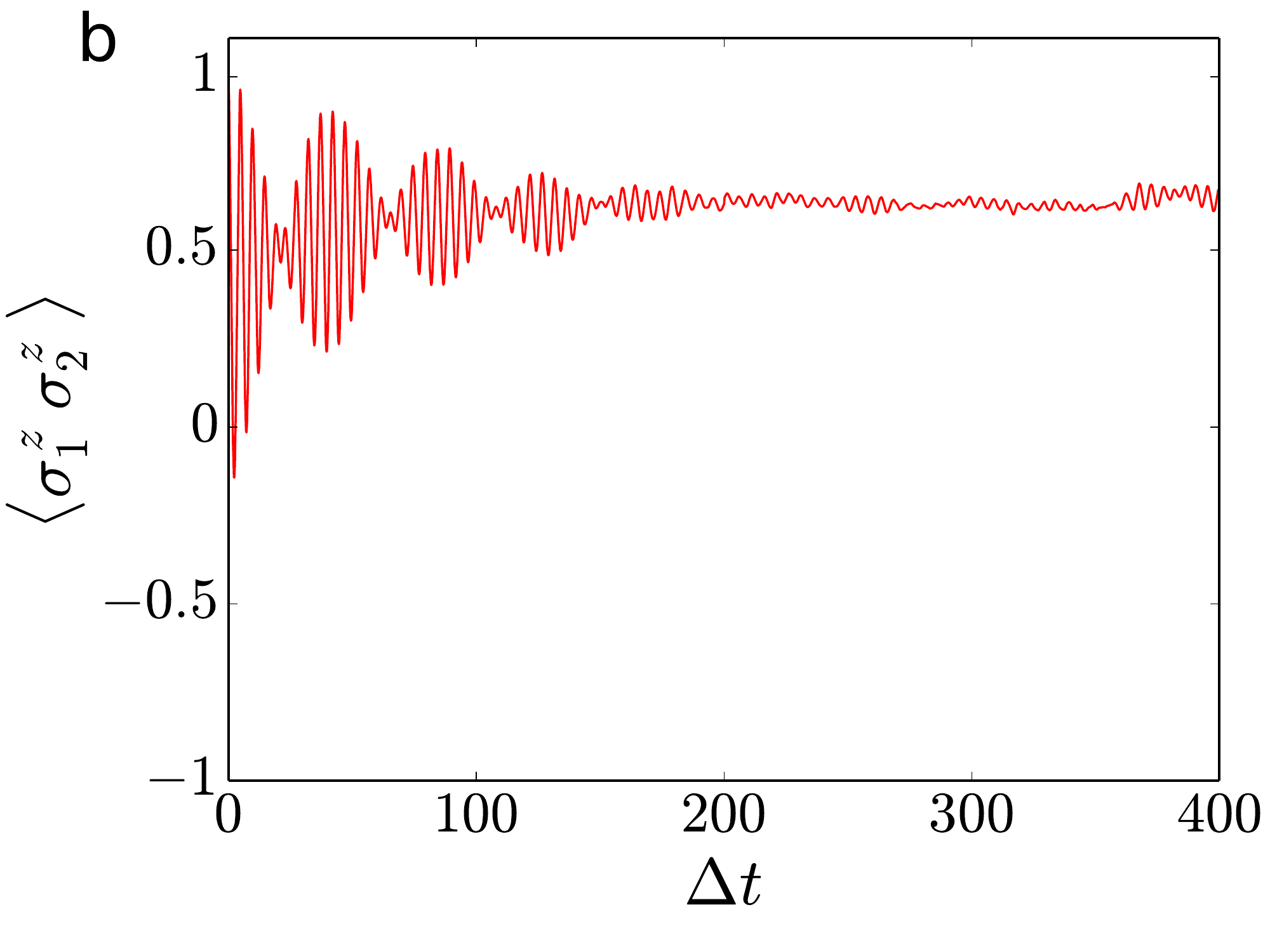}  \\
\includegraphics[scale=0.2]{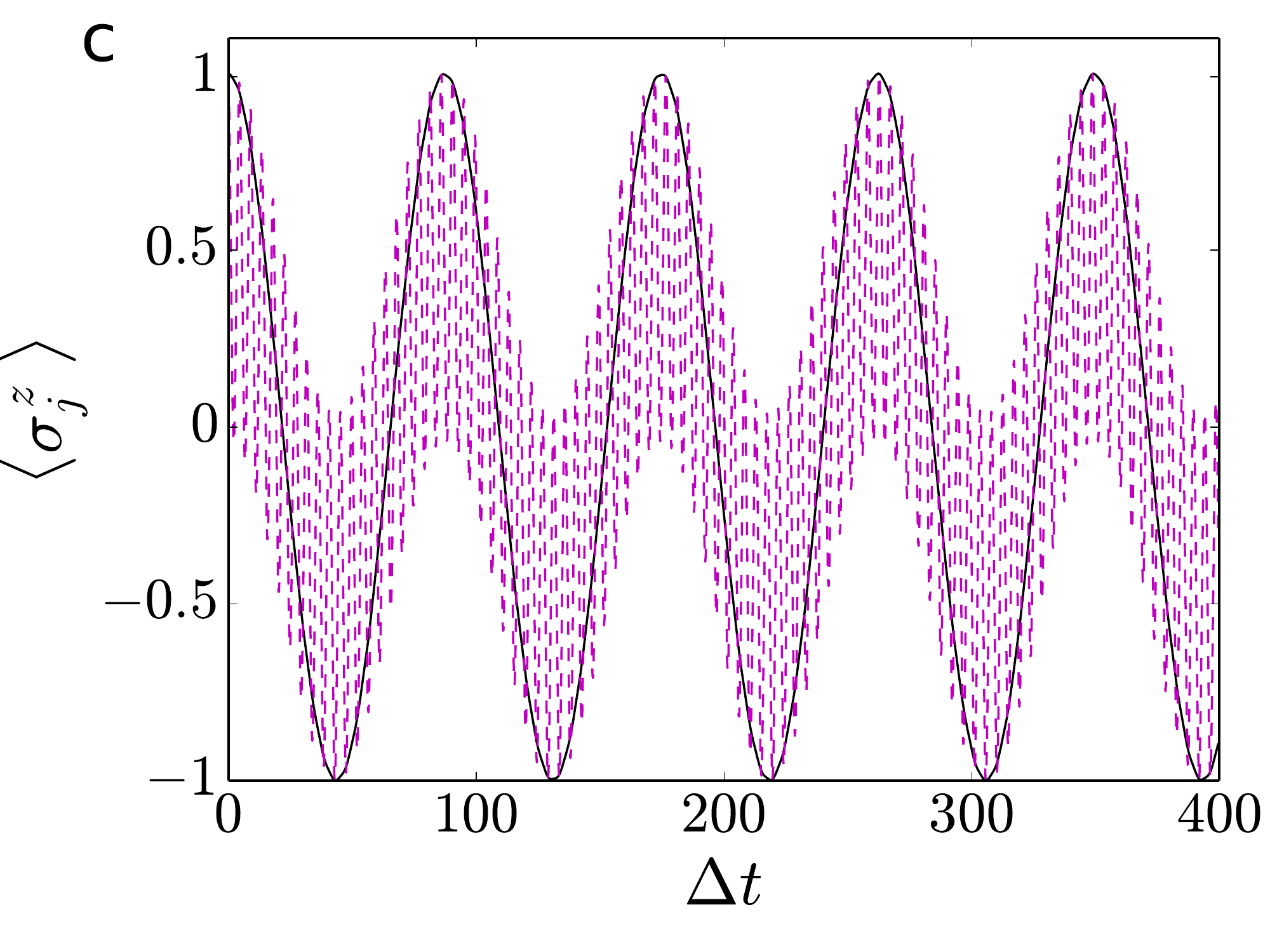}  \includegraphics[scale=0.2]{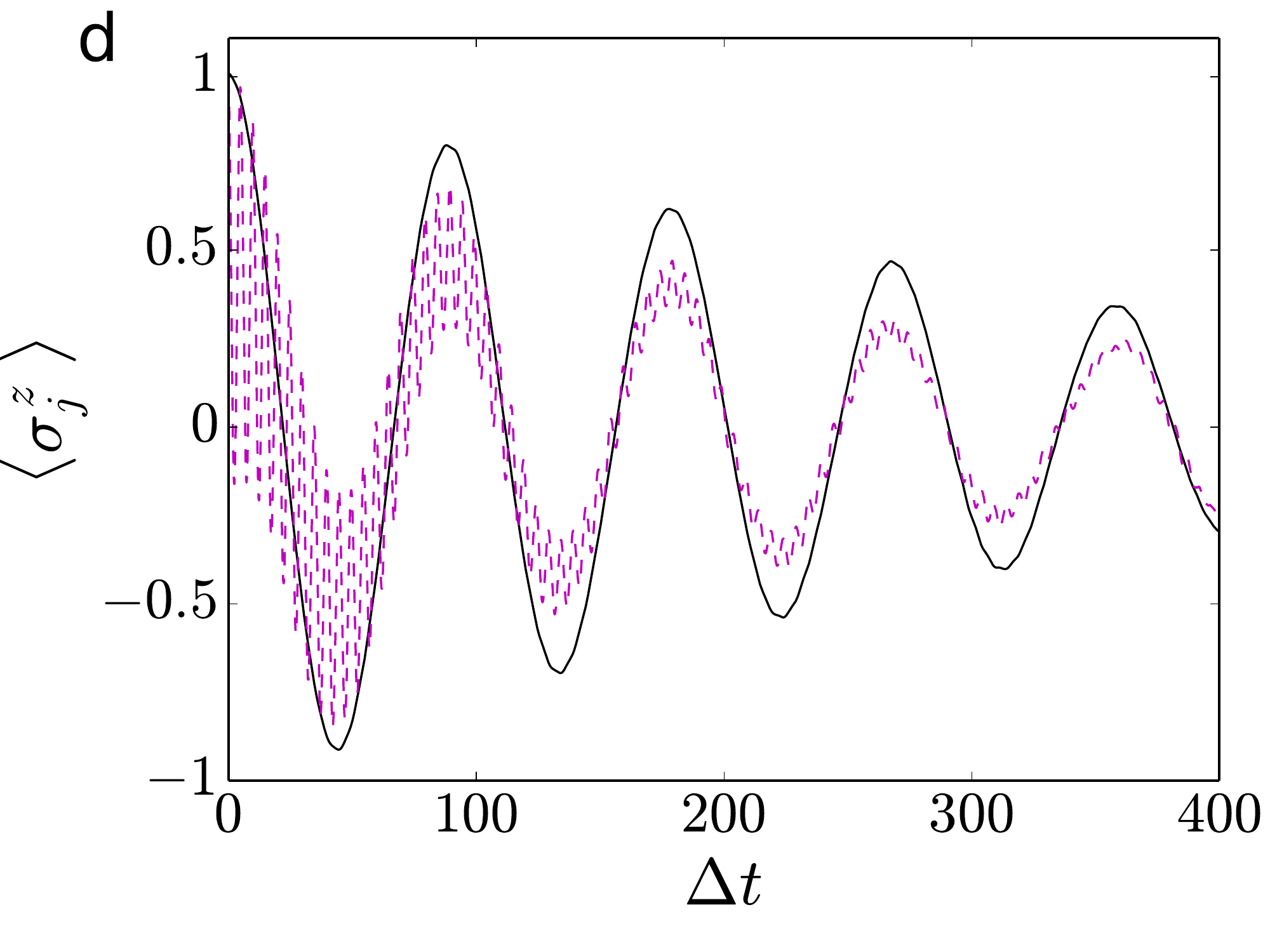}  
\caption{Panels a and b: time evolution of $\langle \sigma^z_1 \sigma^z_2 \rangle$ for a direct Ising interaction interaction $K_r=K$ (panel a) and for a bath-induced interaction  $K_r=\alpha \omega_c$ (panel b). Panels c and d: time evolution of $\langle \sigma^z_1\rangle $ and $\langle \sigma^z_2 \rangle$ for a direct Ising-like interaction $K_r=K$ (panel c) and a bath-induced interaction $K_r=\alpha \omega_c$ (panel d). We have $K_r/\Delta_1=0.4$, $\Delta_2/\Delta_1=0.1$ and $\omega_c=20 \Delta_1$. 
}
\label{Synchronization_two_spins_1}
\end{figure}

The dissipative case, for which the interaction originates from the interaction with the bath, shows a similar phase diagram. There are however notable differences in the unsynchronized regime close to the transition line. In this region, the interaction with the bath leads to an effective synchronization after a short time unsynchronized dynamics. To exemplify this effect, we focus on the spin dynamics at $K_r/\Delta_1=0.4$ and $\Delta_2/\Delta_1=0.1$ in both cases. These parameters correspond to the yellow star in Fig.~\ref{Synchronization_phase_diagram_1}. The evolution of $\langle \sigma^z_j\rangle $ and $\langle \sigma^z_1 \sigma^z_2\rangle $ is shown in Fig.~\ref{Synchronization_two_spins_1} in both cases. We remark that in the case of direct Ising coupling (panel a), there is no synchronization transition as $\langle \sigma^z_1 \sigma^z_2\rangle $ changes sign periodically. By contrast, we remark that $\langle \sigma^z_1  \sigma^z_2 \rangle$ only vanishes a finite number of times (see panel b). After this short time behaviour, the system enters a synchronized regime for which $\langle \sigma^z_1  \sigma^z_2 \rangle$ no longer vanishes and tends to a non-zero equilibrium value corresponding to a polarized equilibrium state. \\

This synchronization effect is the sole consequence of the Ising-like interaction between spins mediated by the bosonic bath. In this respect, it could be understood qualitatively by a comparison of the dynamics to the one corresponding to two free spins coupled via an Ising interaction. Next, we present another synchronization mechanism induced by the presence of the bath, which cannot be interpreted as the effect of an effective interaction between the two spins.

\subsection{Second mechanism}
 We consider a system of two spin 1/2 with different frequencies, whose relative motion is coupled to a common environment. We find that, under certain initial conditions, the joint dynamics of the two spins enters a \textit{dissipationless} synchronized regime when the coupling to the environment is increased. \\

A synchronization regime was recently observed in Ref.~\cite{synchronization_salomon} in the oscillatory dynamics of a mixture of bosonic and fermionic species. In this paper, the authors suggested that the appearance of synchronization was due to the coupling of the relative motion of the two clouds to a dissipative environment. Here, we build a toy model related to this problem by restricting the dynamics of each species to only two motional states. The resulting Hamiltonian is given by
\begin{align}
H=\frac{\Delta_1}{2}\sigma^x_1+\frac{\Delta_2}{2}\sigma^x_2-K\sigma^z_1\sigma^z_2+ \sum_k\left[ \left(\sigma^z_1-\sigma^z_2\right)\frac{\lambda_k}{2} (b_k +b_k^{\dagger})+\omega_k b_k^{\dagger} b_k\right],
\label{hamiltonien}
\end{align}
$\Delta_1 \neq \Delta_2$ are the two bare frequencies of the two species. $K$ denotes the interaction strength between the two spins. We assume that the relative motion of the two species is coupled to an external bath. As before, we consider the case of ohmic dissipation.\\

We should take into account a larger number of motional states to fully describe the system of  Ref.~\cite{synchronization_salomon}. But this simplified toy-model is sufficient to exhibit a synchronization effect, as shown below. This toy-model would rather a double well problem, which could also be implemented in a cold-atomic setups \cite{recati_fedichev,orth_stanic_lehur}.\\
 
% The synchronization phenomenon under consideration could also shed light on long-lived coherence observed in quantum biology (see Ref.~\onlinecite{huelga_plenio} and references therein), where the coupling of several excitation states to the same Non-Markovian environment is believed to play an important role.

\begin{figure}[h!]
\center
\includegraphics[scale=0.24]{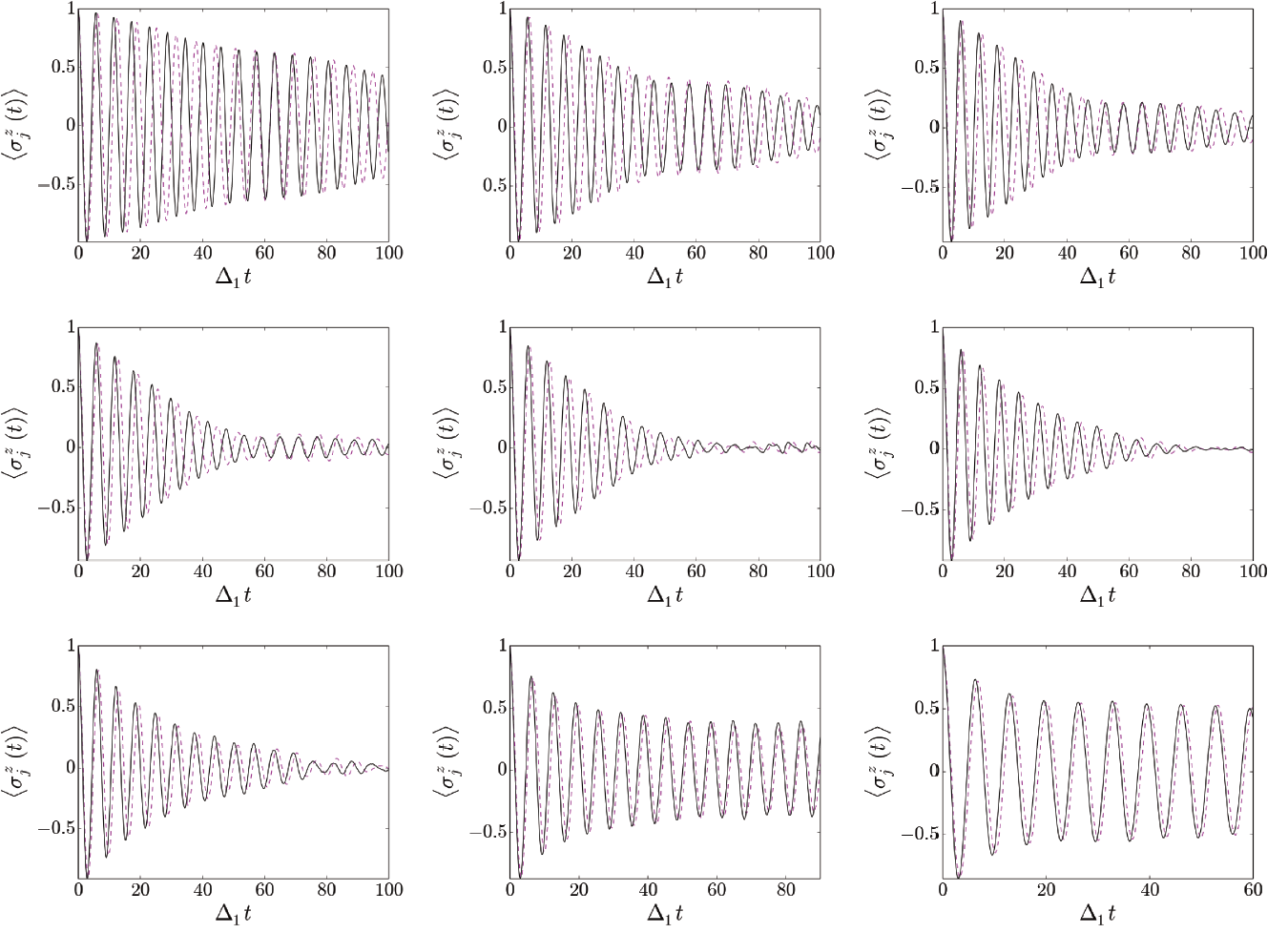}   
\caption{Dynamics of $\langle \sigma^z_1 \rangle$ and $\langle \sigma^z_2 \rangle$ for different values of $\alpha$. From left to right and top to bottom, we have $\alpha=0.005$, $\alpha=0.01$, $\alpha=0.015$, $\alpha=0.02$, $\alpha=0.025$, $\alpha=0.03$, $\alpha=0.035$, $\alpha=0.05$ and $\alpha=0.06$. We have $\Delta_2/\Delta_1=1.1$, $\omega_c=20 \Delta_1$, $T=0$ and $K=0$.
}
\label{Synchronization_two_spins}
\end{figure}

\begin{figure}[h!]
\center
\includegraphics[scale=0.4]{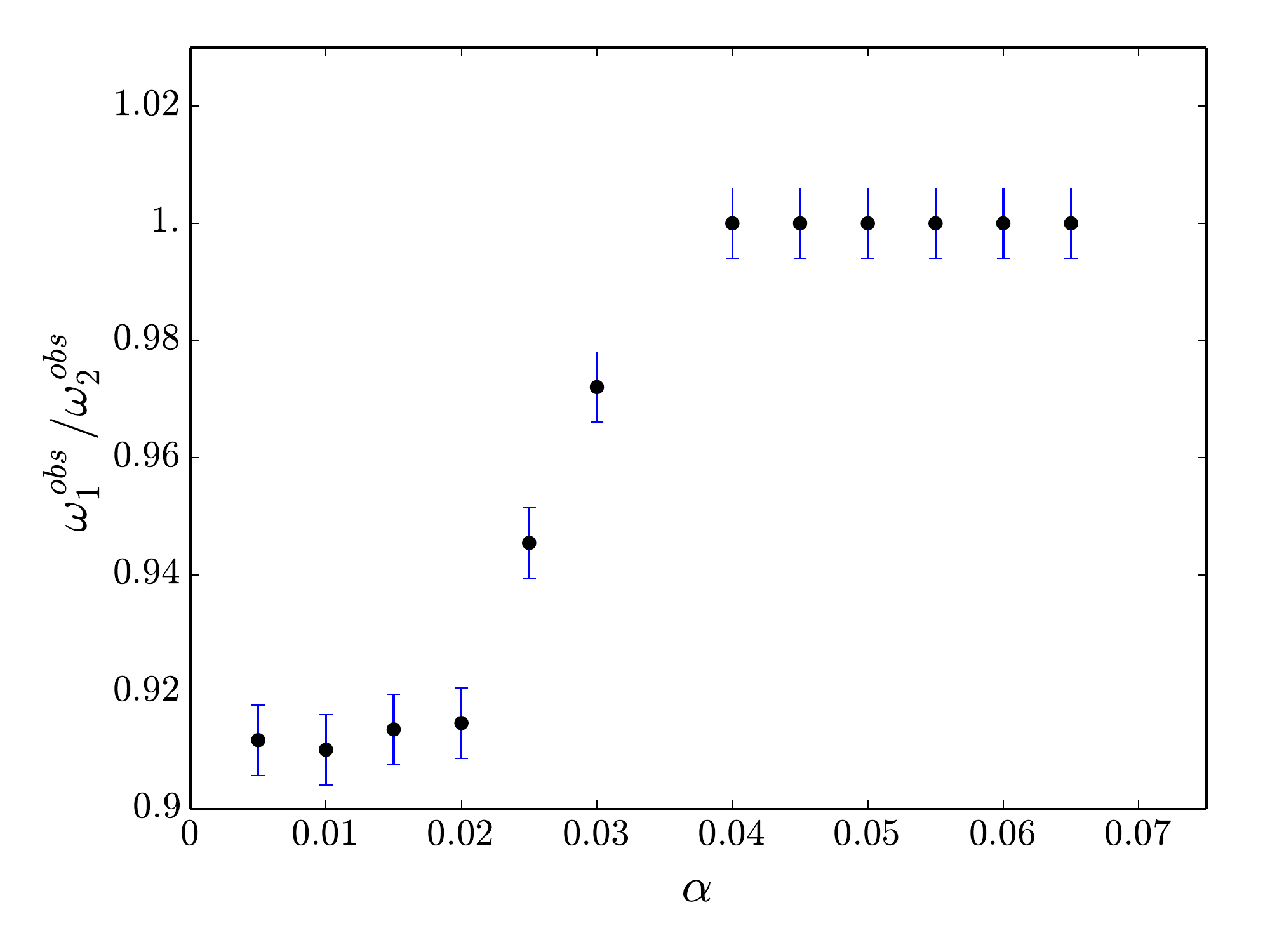}  
\caption{Ratio $r=\omega^{\textrm{obs}}_1/\omega^{\textrm{obs}}_2$ of the two observed oscillating frequencies with respect to $\alpha$. Same parameters as in Fig.~\ref{Synchronization_two_spins}. It seems that the frequency of the synchronized oscillations corresponds to the smallest frequency.
}
\label{Relative_frequency}
\end{figure}

Fig. \ref{Synchronization_two_spins} shows the dynamics of the two spins, with initial state $|\uparrow_z,\uparrow_z\rangle$ for increasing strengths of the system-environment coupling (from left to right and top to bottom). At very weak coupling ($\alpha<0.02$), we observe an asynchronous decay of the two spin oscillations towards an equilibrium state with $\langle \sigma^z_1 \rangle=\langle \sigma^z_2 \rangle=0$. Interestingly, we remark that the damping of the oscillations is temporarily smaller when the oscillators are in phase (see top left panel between $\Delta t=50$ and $\Delta t=80$), signaling the onset of synchronization. In this regime of very weak coupling, the life-time of the oscillations diminishes with $\alpha$. Then, above a certain coupling strength, we observe the appearance of long-lived synchronized oscillations of the two spins. In this regime, the two spins oscillate at the same frequency and these oscillations acquire an infinite life-time. We plot in Fig. \ref{Relative_frequency} the evolution of the observed frequencies of the oscillations, obtained by Fourier analysis, with respect to the coupling strength $\alpha$.\\

Other initial conditions do not lead to the same synchronized regime, signaling the presence of an attractor in the phase space.\\

\begin{figure}[h!]
\center
 \includegraphics[scale=0.22]{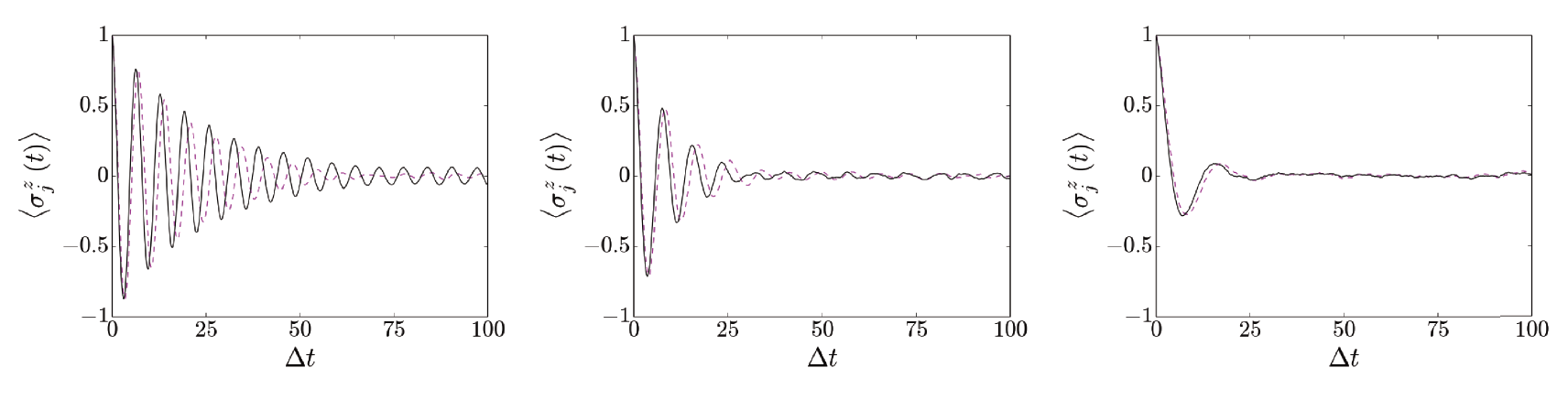} 
\caption{Dynamics of $\langle \sigma^z_1 \rangle$ and $\langle \sigma^z_2 \rangle$ for different values of $\beta$. From left to right, we have $\beta=5$, $\beta=2.5$ and $\beta=1$. We have $\Delta_1/\Delta_2=1.1$, $\omega_c=20 \Delta_1$, $\alpha=0.02$ and $K=0$. We do not observe a synchronization of the oscillations.
}
\label{Synchronization_two_spins_3}
\end{figure}

One could wonder if Markovian dissipation could be at the origin of a similar effect. We provide now evidence that it is not the case for a thermal bath. We show in Fig.~\ref{Synchronization_two_spins_3} the joint dynamics of the two clouds for the same protocol, at $\alpha=0.02$, and with different values of the temperature. \\

We see that increasing dissipation (meaning lowering $\beta$) only leads to a faster decay rate, without synchronization between the two spins. It seems then that the quantumness and coherence of the bath is important for such an effect to arise. Interestingly, the Ising interaction does not seem to be a key ingredient in this synchronization effect. A change in $K$ modifies the critical coupling at which synchronization occurs, but does not modify qualitatively the collective behavior.  \\

A more precise analytical investigation of bath effects would be necessary to understand the mechanism at the origin of this synchronization phenomenon. This preliminary numerical result constitutes however a strong hint that coupling two spins to the same quantum environment may have drastic effects on their joint dynamics. In particular, synchronization seems to appear at a rather low value of the coupling $\alpha$.\\

We studied in this chapter the dynamics of two spins coupled to a common ohmic environment. We investigated the dissipative quantum phase transition induced by the bath, as well as many-body synchronization effects. The dissipative transition is associated with a suppression of spin coherence, quantified by the vanishing of off-diagonal elements of the spin-reduced density matrix. One could then wonder how the bath modifies the intrinsic properties of the phase-space accessible for the spin, and in particular its topology.

\chapter{Dissipative topological transition}

In this Chapter, we will study the effect of an ohmic environment on the topology of a spin 1/2. After a brief introduction on the topological and geometrical properties associated with a spin 1/2 and their importance for condensed matter systems, we will study the deformation of the Berry curvature induced by the presence of the environment by using a variational ansatz for the ground state. Then, we investigate the effect of the environment with a non-equilibrium protocol where the spin is subject to a time-dependent drive. This dynamical protocol was introduced in Ref. \cite{polkovnikov:PNAS} and allows to have access to the Berry curvature through a measure of the spin observables, because the non-adiabatic response of a slowly driven quantum system depends on the geometry of its instantaneous ground state.  We describe quantitatively the features of the bath-induced transition with the SSE method. The so-called Toulouse point at $\alpha=1/2$ being exactly solvable, we study the dynamical scaling of the dynamical Chern number at the transition within Keldysh formalism. We also study the evolution of the final Entanglement entropy of the system with $\alpha$ and introduce an effective thermodynamical description of the transition, which appears then as a temperature inversion point. Results have been published in Refs. \cite{topo_loic}.

\section{Topology of a spin 1/2}
\begin{figure}[h!]
\center
\includegraphics[scale=0.2]{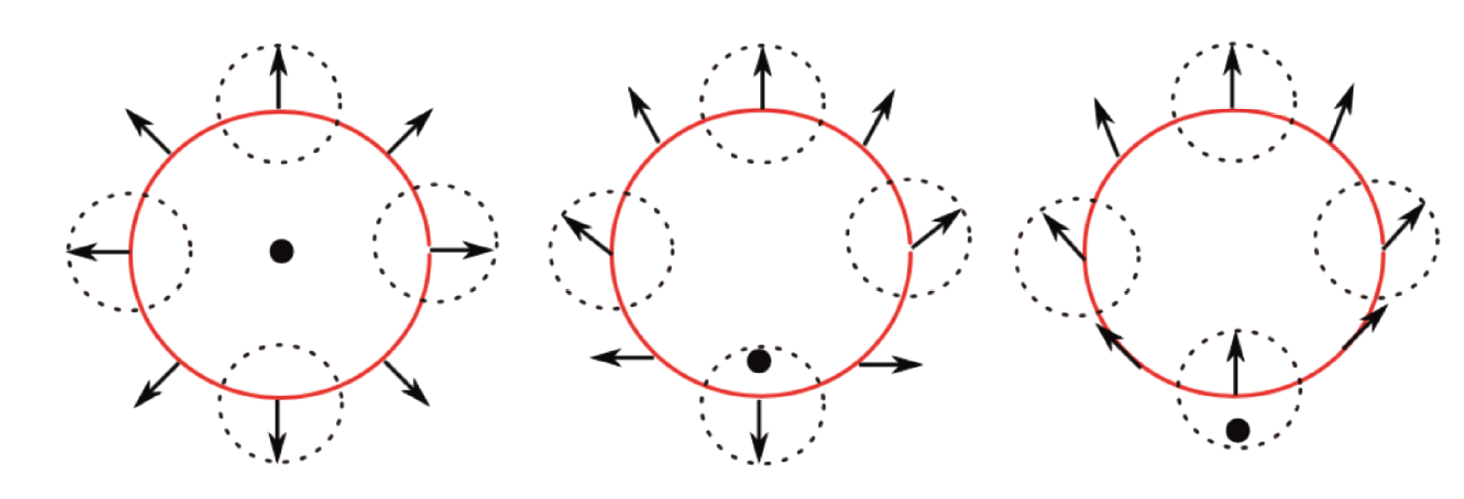}
\caption{The red circles are parametrized by $(H \sin \theta,H_0+H \cos \theta)$ with $H\neq 0$. We have $H_0/H=0$ (left), $0<H_0/H<1$ (middle) and $H_0/H>1$ (right) and the black dot shows the position of the origin in each case. The arrows show the orientation of the Bloch vector for each value of $\theta$. The vertical component is given by $\langle \sigma^z \rangle_{eq}$ while the horizontal component is given by $\langle \sigma^x \rangle_{eq}$.}
\label{topo_TLS_isole}
\end{figure}

Let us consider a spin-1/2 in a magnetic field, corresponding to the Hamiltonian
\begin{align}
\mathcal{H}_{TLS}=-\frac{1}{2}\vec{d}.\vec{\sigma},
\label{Htls_topo}
\end{align}
where $\vec{d}=(H \sin \theta \cos \phi , H \sin \theta \sin \phi, H_0+H \cos \theta )^T$. We show in Fig. \ref{topo_TLS_isole} the orientation of the equilibrium Bloch vector for all the values of $\theta \in [0,2\pi[$ when $\phi=0$. The behaviour of this vector changes depending on the sign of $H_0-H$. When $H_0<H$, we see that the angle between this vector and the vertical axis goes from $0$ to $2 \pi$ when $\theta$ goes from $0$ to $2 \pi$. By contrast, this angle goes from $0$ to $0$ when $\theta$ goes from $0$ to $2 \pi$ when $H_0>H$.\\

For any $2\pi$ periodic Hamiltonian in terms of a given variable $\theta$, the Bloch vector is the same for $\theta=0$ and $\theta=2 \pi$ up to a phase factor. The way it winds when $\theta$ varies from $0$ to $2\pi$ is characterized by a relative integer $n \in \mathcal{Z}$ (which counts the number of turns around the way). In our case characterized by Hamiltonian (\ref{Htls_topo}), one has $n=1$ for $H_0<H$ and $n=0$ for $H_0>H$.\\

\subsection{Chern number and Berry curvature}

A more rigorous characterization can be done with homotopy theory, as exposed in Ref. \cite{Mermin_topo}, by considering a group $G$ of transformations between possible equilibrium Bloch vectors. In our case, a suitable group $G$ would correspond to the two-dimensional rotation group $SO(2)$, which is a continuous group with a well-defined topology. One can show in particular that to each loop in $SO(2)$, is associated a relative integer essentially equivalent to the relative integer $n$ that we introduced above\footnote{Formally, the fundamental group of $SO(2)$ is the additive group of integers $\mathcal{Z}$}. The continuous path parametrized by $\theta$ from $\theta=0$ to $\theta=2 \pi$ defines a loop in $SO(2)$, and the cases $H_0/H<1$ or $H_0/H>1$ correspond to a different value of $n$. Let us now make a link with the widespread notions of Berry curvature and Chern number.\\

The Berry curvature is a local quantity (in terms of the variables $\theta$ and $\phi$ introduced in Hamiltonian (\ref{Htls_topo})) which characterizes the local geometry of a state upon infinitesimal variation of $\theta$ and $\phi$. The Berry curvature $\mathcal{F}_{\phi\theta }$ associated with a state $|g\rangle$ is defined by 
\begin{align}
\mathcal{F}_{\phi\theta}=\partial_{\phi} \mathcal{A}_{\theta}-\partial_{\theta} \mathcal{A}_{\phi},
\label{Berry_curvature}
\end{align}
where $\mathcal{A}_{\phi}$ and $\mathcal{A}_{\theta}$ are called Berry connections and defined by 
\begin{align}
\mathcal{A}_{\phi}=\langle g|i\partial_{\phi}|g \rangle,\\
\mathcal{A}_{\theta}=\langle g| i\partial_{\theta}|g \rangle.
\label{Berry_connection}
\end{align}
The Berry curvature is a gauge-independent quantity, which equals to $\mathcal{F}_{\phi \theta}=1/2 \sin \theta$ when we chose $|g\rangle$ to be the ground state of Hamiltonian (\ref{Htls_topo}) for $H_0=0$. The Chern number $C$ is an integer, which characterizes the global topology of the system, and it is related to the Berry curvature as
\begin{align}
C=\int_{0}^{2\pi} d \phi \int_{0}^{\pi} d \theta \mathcal{F}_{\phi \theta}.
\end{align}
In our precise case, $\mathcal{F}_{\phi \theta}$ is independent of $\phi$. For $H_0=0$, we recover $C=1$. By contrast for $H_0>H_1$, we find $C=0$. The Chern number matches exactly with the relative integer $n$ introduced above to describe the evolution of the Bloch vector when $\theta$ goes from $0$ to $2 \pi$ at $\phi=0$. We show explicitly in Appendix F.\ref{appendix_index} that we have for Hamiltonian (\ref{Htls_topo})\\
\begin{align}
C=\frac{\langle \sigma^z(\theta=0) \rangle-\langle \sigma^z(\theta=\pi) \rangle}{2}.
\label{C_sigma_z}
\end{align}\\

\subsection{Relation with topological properties of condensed matter systems}

We illustrated the notions of Berry curvature/connections and Chern number over this simple example of a single spin in a magnetic field. These definitions are however very general and non-trivial topological effects arise for any Hamiltonian with periodic properties in terms of a given variable. These notions play in particular a very important role in the investigation of electronic properties in crystalline solids, such as the integer quantum Hall effect \cite{PRL_quantum_hall,theory_quantum_hall}, as the periodicity of the cristalline potential leads to a periodic invariance of the Hamiltonian in terms of the reciprocal wave-vector.  Let us for example consider a general simple model of non-interacting electrons on a lattice, where there is a gap between energy bands and the Fermi energy lies in one gap. Each energy band is characterized by a given Chern number upon variation of the reciprocal wave-vector. These topological invariants have a striking physical manifestation, as the Hall conductivity $\sigma_{xy}$ of this so-called Chern insulator model is given by
\begin{align}
\sigma_{xy}=\frac{e^2}{2\pi \hbar}\sum_{\alpha} C_{\alpha},
\end{align}
where $C_{\alpha}$ denotes the Chern number of the band $\alpha$, and $\alpha$ runs over the filled bands. This is the famous Thouless-Kohomoto-Nightingale-den Nijs (TKNN) formula\cite{TKNN} which can be shown by the application of Kubo formula. In this context, one may see Hamiltonian (\ref{Htls_topo}) as a single-particule Hamiltonian of the simplest class of Chern insulators with only two bands.   \\

Recent experiments have focused on the realizations of topological phases with photons in artificial systems \cite{Karyn:CR} together with developments in ultra-cold atoms. While in nature this is achieved by a magnetic field coupled to charged particles, artificial gauge fields have been realized with ultracold atoms in optical lattices \cite{cold_atoms_topo_1,cold_atoms_topo_2} and photonic systems \cite{photonic_topo}. We may think of a topological phase of noninteracting particles as characterized by a bulk topological invariant which necessarily implies the existence of edge states protected against backscattering at the system boundary. The lure of optical lattices and photonic systems is that band topology, which underlies the characterization of quantum Hall-like phases, as well as edge state transport, can be probed. Moreover, a strong magnetic field at a suitably chosen filling in the presence of interactions leads to the fractional quantum Hall effect.\\

\subsection{Coupling to an environment}
An interesting question that emerges naturally concerns the effect of a quantum dissipative environment on the topology of the spin. To describe dissipation, we will work in this Chapter with Hamiltonian $\mathcal{H}=\mathcal{H}_{TLS}+\mathcal{H}_{diss}$, where \\
\begin{align}
\mathcal{H}_{diss}&= \sigma^z \sum_k \frac{\lambda_k}{2} (b_k +b_k^{\dagger})+ \sum_k \omega_k \left(b_k^{\dagger}b_k+\frac{1}{2}\right).
\label{Hamiltonian_spin_boson} 
\end{align}

 One can give a first answer by considering the two phases of the ohmic spinboson model. Above a critical coupling strength $\alpha_c \sim 1$, the spin is known to be in a localized phase associated with a loss of coherence \cite{leggett:RMP,Weiss:QDS}, with vanishing off-diagonal elements of the spin-reduced density matrix. In such a phase, spin tunneling is forbidden and the spin is trapped in a polarized state along the z-axis. The possible equilibrium Bloch vectors in such a localized phase cannot be mapped one onto the other by continuous transformations of the Hamiltonian, thus corresponding to a trivial topology. We expect then a change of the system topology induced by the presence of the bath at large coupling. To understand better the effect of the bath, let us develop a variational approach in relation with the polaron ansatz introduced in Sec. \ref{polaron}.\\
 
\section{Shifted oscillators approach}

In this Section, we explore the effect of the bath on the geometrical properties of the spin by using a variational ansatz for the ground state. At weak dissipation, one may indeed safely approximate the ground state $|g \rangle$ of Hamiltonian $\mathcal{H}$ for a given $\theta$ in the ``shifted oscillators" picture \cite{Silbey_Harris,leggett:RMP,Weiss:QDS,Hur,bera:PRB} (or polaron picture), by
\begin{align}
|g \rangle=\frac{1}{\sqrt{p^2+q^2}}\left[p e^{-i \phi} |\uparrow_z\rangle \otimes |\chi_{\uparrow}\rangle+q |\downarrow_z\rangle \otimes |\chi_{\downarrow}\rangle\right],
\label{ansatz}
\end{align}
where states $|\chi_{\uparrow}\rangle$ correspond to multi-mode coherent states. These states, as well as the real numbers $p$ and $q$, are determined \textit{a posteriori} by minimizing the energy of the system. By contrast to the ansatz introduced in Sec. \ref{polaron}, here we do not fix $p$ and $q$ but determine them variationally. Due to the symmetry of the Hamiltonian, $p$ and $q$ do not depend on $\phi$.\\ 

Coefficients $p$, $q$ and states $|\chi_{\uparrow \downarrow}\rangle=\exp\left[\sum_k f_k^{\uparrow \downarrow} (b_k -b_k^{\dagger}) \right]|0\rangle$ are determined variationally by minimizing $E=\langle g | \mathcal{H} |g\rangle$ with respect to $p$, $q$ and the sets of real numbers $\{f_k^{\uparrow }\}$ and $\{f_k^{\downarrow }\}$. It is also important to note that the displacements associated with different spin polarizations may differ in absolute value ($f_k^{\uparrow } \neq f_k^{\downarrow }$ in general). Here $|0\rangle$ denotes the vacuum with all oscillators in equilibrium. As the $\phi$ dependency is solely contained in the phase factor of Eq. (\ref{ansatz}), we work at $\phi=0$. We have, 
\begin{align}
E=\frac{1}{p^2+q^2}\Bigg[&\frac{H}{2} \cos \theta (p^2-q^2)+H p q\sin \theta  e^{-\sum_k \frac{\left(f_k^{\uparrow}-f_k^{\downarrow}\right)^2}{2}}+ \sum_k \lambda_k \left(p^2 f_k^{\uparrow}-q^2 f_k^{\downarrow}\right)\\
&+\sum_k \omega_k \left(p^2 |f_k^{\uparrow}|^2+ q^2 |f_k^{\downarrow}|^2 \right)  \Bigg].
\label{Energy}
\end{align}
Minimizing $E$ with respect to $f_k^{\uparrow}$ and $f_k^{\downarrow}$ gives for all $k$,
\begin{align}
p^2 \lambda_k  +2p^2 f_k^{\uparrow}\omega_k -p q H\delta \sin \theta(f_k^{\uparrow}-f_k^{\downarrow})=0 \label{minimization_f_k_g_k_0}\\
-q^2 \lambda_k  +2q^2 f_k^{\uparrow}\omega_k +p q H \delta \sin \theta(f_k^{\uparrow}-f_k^{\downarrow})=0 
\label{minimization_f_k_g_k}
\end{align}
where $\delta=e^{-\sum_k \frac{\left(f_k^{\uparrow}-f_k^{\downarrow}\right)^2}{2}}$.  Minimizing $E$ with respect to $p$ or $q$ gives the same equation, namely
\begin{align}
H \delta \sin \theta \left[ q^2 -p^2 \right]+2 p q \left[H \cos \theta +\sum_k \lambda_k (f_k^{\uparrow}+f_k^{\downarrow})+\sum_k \omega_k (|f_k^{\uparrow}|^2-|f_k^{\downarrow}|^2) \right]=0.
\label{minimization_alpha}
\end{align}

\begin{figure}[h!]
\center
\includegraphics[scale=0.35]{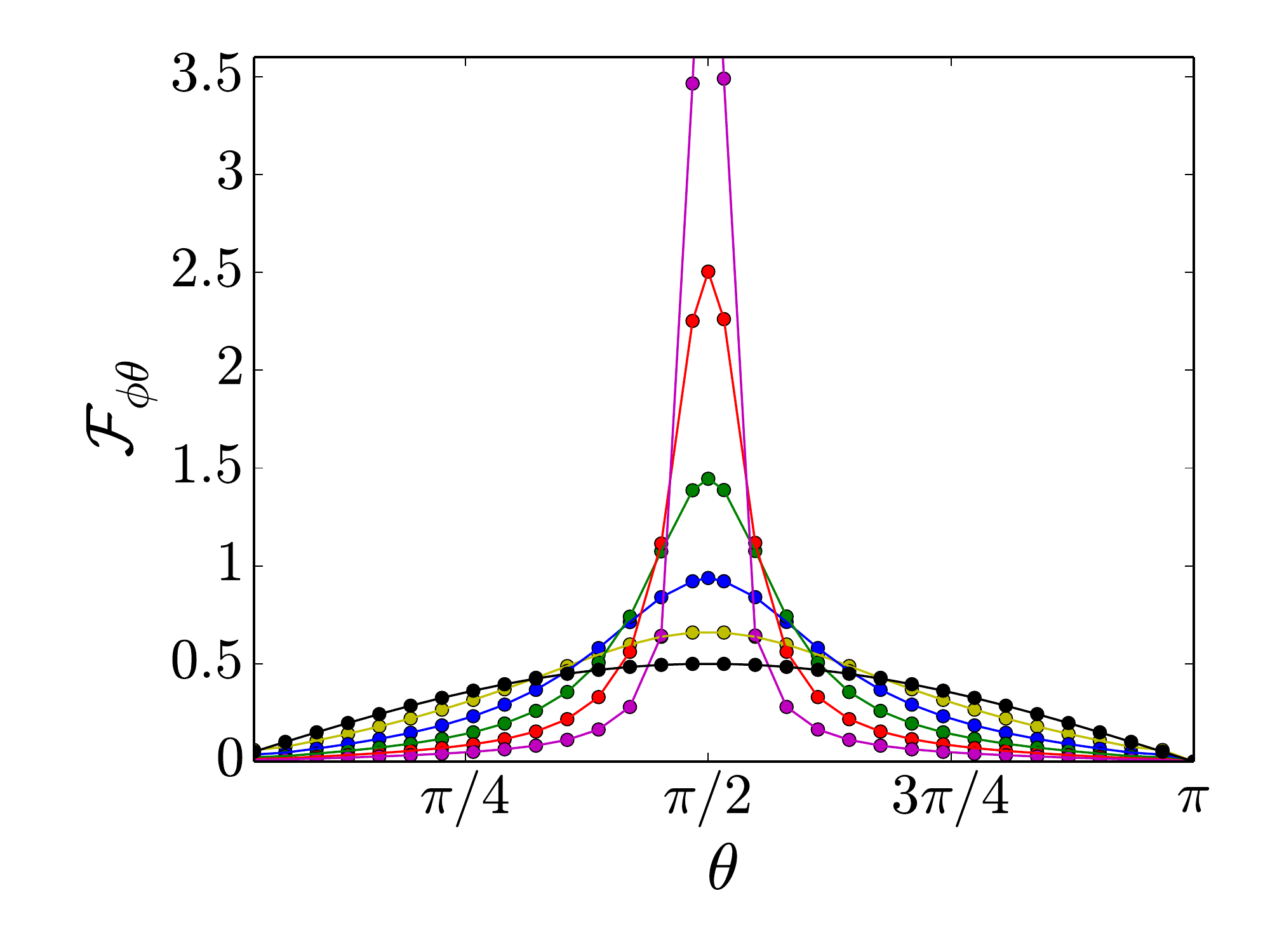}
\caption{Evolution of the Berry curvature $\mathcal{F}_{\phi \theta}$ with respect to $\theta$ for $\alpha=0$ (black), $\alpha=0.1$ (yellow), $\alpha=0.2$ (blue), $\alpha=0.3$ (green), $\alpha=0.4$ (red) and $\alpha=0.5$ (magenta). We have $H/\omega_c=0.02$.}
\label{Berry_curvature_shifted_oscillators}
\end{figure}

Solving self-consistently the set of equations determined by Eqs. (\ref{minimization_f_k_g_k_0}--\ref{minimization_alpha}) allows to compute $p$ and $q$ and their evolution with respect to $\theta$ for different values of $\alpha$. This ansatz allows to reach convenient expressions for the Berry connections, as we have
\begin{align}
&\mathcal{A}_{\phi}=\langle g|i\partial_{\phi}|g \rangle=\frac{p^2}{p^2+q^2},\\
&\mathcal{A}_{\theta}=\langle g| i\partial_{\theta}|g \rangle=0.
\label{Berry_connection_ansatz}
\end{align}
One then easily compute the Berry curvature $\mathcal{F}_{\phi \theta}$ according to Eq. (\ref{Berry_curvature}), and we show its evolution with respect to $\theta$ for different values of $\alpha$ in Fig. \ref{Berry_curvature_shifted_oscillators}. We see that the environment gradually deforms the manifold spanned by the ground state upon a variation of $\theta$. This corresponds to a Berry curvature which becomes more and more peaked around $\theta=\pi/2$ as $\alpha$ increases. \\

Above $\alpha=0.5$, this variational method does not give physical results. It is shown in Ref. \cite{topo_loic} that the Berry curvature diverges exactly at the quantum phase transition $\alpha=1$, and can be related to the divergence of the magnetization susceptibility.

\section{Experimental protocol}
Recent circuit-QED experiments \cite{Schroer:PRL,Roushan:Nature} have shown that it was possible to probe geometrical and topological properties of artificial spin 1/2 systems with a dynamical protocol. This protocol relies on the Adiabatic Rapid Passage (ARP) technique, widely used in the magnetic resonance community to invert the population of two-level systems \cite{Grynberg_Aspect_Fabre}.\\

 It consists in devising Hamiltonian (\ref{Htls_topo}), where $\theta(t)=v(t-t_0)$ grows linearly from $\theta(t_0)=0$ to $\theta(t_f)=\pi$ and $\phi$ is stationary, for a spin starting from the initial ground state $|\uparrow_z\rangle$. When $v/H$ is small enough, the dynamics becomes nearly adiabatic. At $H_0=0$ for example (case of the standard ARP protocol), the Bloch vector spirals around the field, following a characteristic cyclo\"{i}d curve from north to south pole. For simplicity we take $\phi=0$ in the following. The dynamics is conveniently described in the rotating frame, using a rotation $U(t)$ corresponding to a rotation of axis (0y) of angle $vt$. Defining $|\psi'(t)\rangle=U(t)|\psi(t)\rangle$, the Schr\"{o}dinger equation gives,
 \begin{align}
&i\partial_t \left[U(t)|\psi(t)\rangle \right]=\mathcal{H}_{TLS}(t) |\psi(t)\rangle\notag\\
&i \partial_t |\psi'(t)\rangle=\underbrace{\left[i \dot{U}U^{\dagger}+U \mathcal{H}_{TLS}U^{\dagger}   \right]}_{H_{eff}}|\psi'(t)\rangle
\label{rotated_basis}
\end{align}
The dynamics of $|\psi'(t)\rangle$ is governed in the rotating frame by $H_{eff} =-H \sigma^z/2 -v\sigma^y/2$.  Starting from $|\psi'(t_0)\rangle=|\psi(t_0)\rangle=|\uparrow_z \rangle$, we find that $|\psi'(t)\rangle$ rotates around the rotation vector $\vec{\Omega}=(H,0,v)$. We recover that the dynamics is static in the rotating frame when $v/H\to 0$ (adiabatic limit). The non-adiabatic response is characterized by the angle of $\vec{\Omega}$ with the z-axis, and we find in particular that it leads to a non-zero expectation value of $\langle \sigma^y (t) \rangle$ whose amplitude is proportional to $v$.\\
 
 The key point is that the non-adiabatic response of a slowly driven quantum system can be related to the geometry of its instantaneous groundstate, as shown in Ref. \cite{polkovnikov:PNAS}. Measuring the non-adiabatic response provides then a way to probe the geometrical and topological properties the system. As shown in Refs. \cite{polkovnikov:PNAS,polkovnikov:course}, we have
\begin{align}
H/2 \sin \theta (t)  \langle \sigma^y (t) \rangle=v \mathcal{F}_{ \phi=0 \theta(t)}+o(v/G),
\label{curvature_sigma_y}
\end{align}
where $\mathcal{F}_{\theta \phi }$ stands for the Berry curvature of the isolated system and $G$ denotes the energy difference between ground and excited states. This result can be shown in time-dependent perturbation theory, as shown in Refs. \cite{polkovnikov:PNAS,polkovnikov:course}. We explicitely provide this derivation in Appendix \ref{appendix:TDperturbation_theory_topo}.II. This can also be seen as an application of the time-dependent version of the Hellmann-Feynman theorem \cite{Hayes:chem_phys}.
\begin{align}
\langle \partial_{\phi} \mathcal{H}_0 \rangle=i\partial_t \langle \psi |\partial_{\phi}|\psi \rangle
\label{Hellmann_feynman_time_dependent}
\end{align}\\
As $\partial_t=v \partial_{\theta}$ and $|\psi \rangle=|g_t \rangle+o(v)$, where $|g_t \rangle$ is the instantaneous ground state of Hamiltonian $\mathcal{H}_0$, Eq. (\ref{Hellmann_feynman_time_dependent}) estimated at $\phi=0$ gives immediately the result.\\

\begin{figure}[h!]
\center
\includegraphics[scale=0.24]{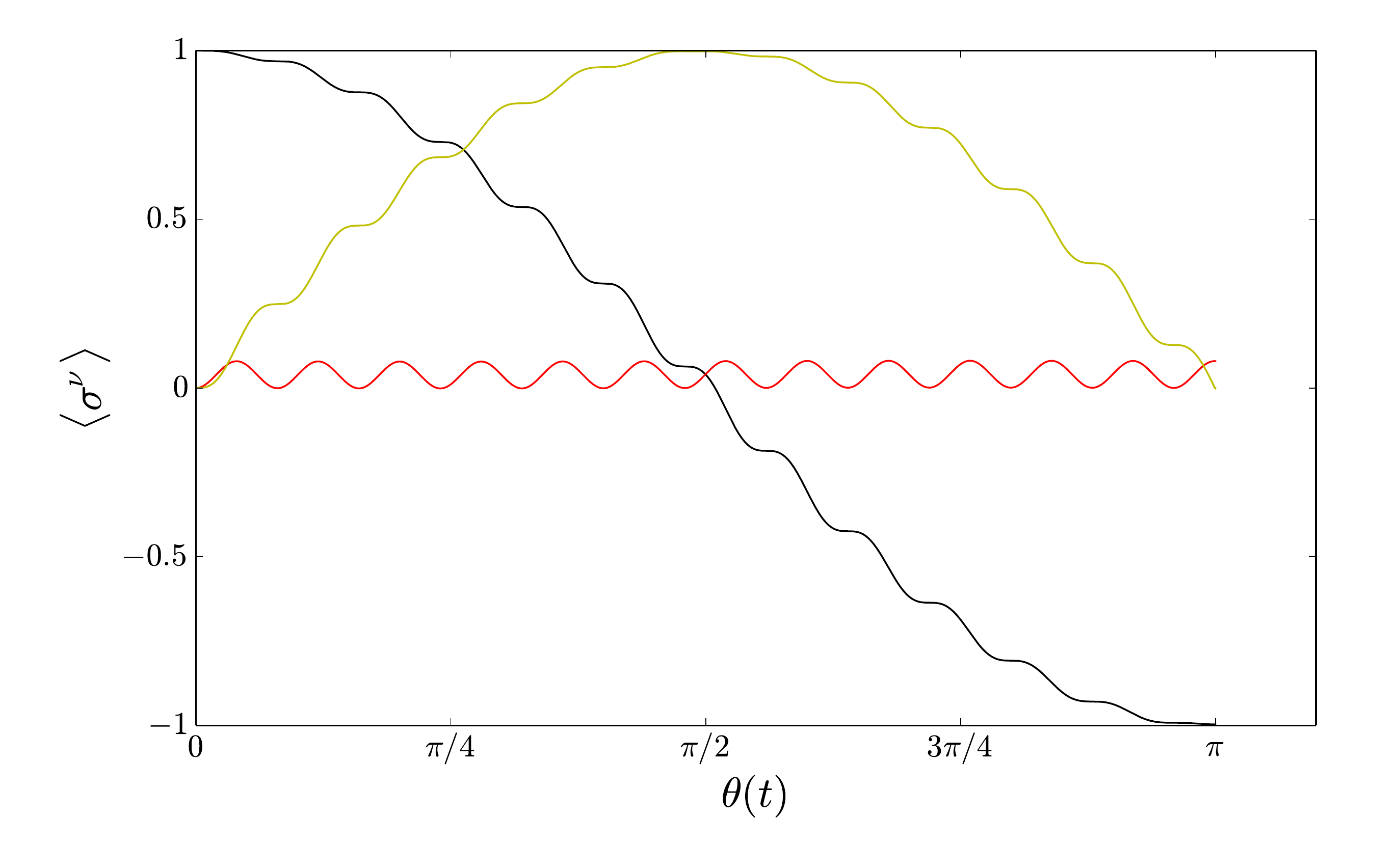}  \includegraphics[scale=0.28]{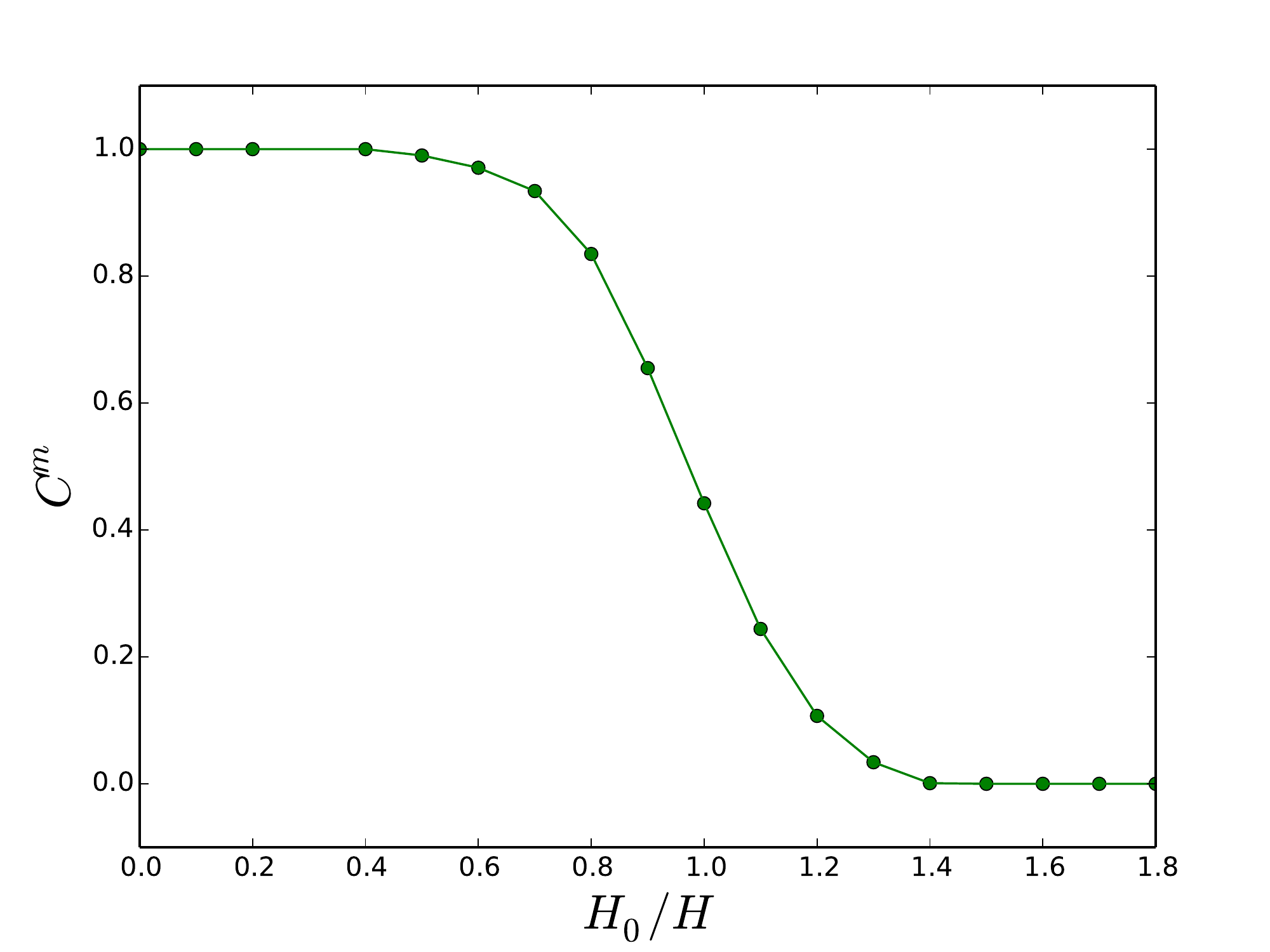} 
\caption{Left: Spin variables $\langle \sigma^z \rangle$ (black), $\langle \sigma^x \rangle$ (yellow) and $\langle \sigma^y \rangle$ (red) and their time-evolution during the sweep. We have $H_0=0$, $H=1$ and $v/H=0.04$. Right: Measured Chern number $C^m$ with respect to $H_0/H$ for $v/H=0.04$.
}
\label{spin_dynamics_topo}
\end{figure}

As an example, we show in Fig. \ref{spin_dynamics_topo} (left panel) the dynamics of $\langle \sigma^{\nu} (t) \rangle$ for $\nu \in \{x,y,z\}$ during a sweep from the north pole to the south pole with $H_0=0$.
We see that $\langle \sigma^y  \rangle \neq 0$ during the protocol. From its measurement, we have access to
\begin{align}
\mathcal{F}_{\phi \theta(t)}^m=\frac{H}{2v} \sin \theta (t)  \langle \sigma^y (t) \rangle,
\label{curvature_sigma_y_measured}
\end{align}
where $\mathcal{F}_{ \phi \theta(t)}^m$ is the measured Berry curvature. From the Berry curvature of the ground state at parameters $(\theta(t),  \phi)$, we can access the measured Chern number (note that $\mathcal{F}_{\phi \theta }$ is independent of $\phi$),
\begin{align}
C_m=\int_{t_0}^{t_f} dt \frac{H}{2} \sin vt   \langle \sigma^y (t) \rangle.
\end{align}

This technique was recently used in circuit QED experiments \cite{Schroer:PRL,Roushan:Nature}, where measurements of the transverse spin components were achieved using $\pi/2$ tomographic pulses. From the time integration of $\langle  \sigma^y (t)  \rangle$, the authors accessed an estimation for the Chern number $C^m$ for each value of $H_0$ and addressed quantitatively the Haldane topological transition occuring at $H_0=H$. We show in Fig. \ref{spin_dynamics_topo} (right panel) the corresponding evolution of the measured Chern number $C^m$ with respect to $H_0/H$ for $v/H=0.04$. We do not observe a clear jump of the measured Chern number $C^m$ at the Haldane transition, but rather a continuous evolution. This feature is due to the fact that we can no longer use pertubation theory in the vicinity of $H_0/H=1$ (used to derive Eq. (\ref{curvature_sigma_y})). In this region, when $\theta$ is close to $\pi$, the energy difference between ground and excited state is very small. \\

 It may be noticed that one can reach an even more convenient expression for $C^m$ using the Heisenberg equation of motion for $\sigma^z$, $\langle \dot{\sigma}^z (t) \rangle=-H \sin \theta (t) \langle \sigma^y (t) \rangle $, to derive the dynamical generalization of Eq. (\ref{C_sigma_z}),
\begin{align}
C^m=\frac{\langle \sigma^z(t_0) \rangle-\langle \sigma^z(t_f=t_0+\pi/v) \rangle}{2}+o(v/G).
\label{C_m_sigma_z}
\end{align}
We recover easily that $C^m=1$ at first order in $v/H$, as we find from Eq. (\ref{rotated_basis})\\
 \begin{align}
\langle \sigma^z (t_f=t_0+\pi/v) \rangle=- \frac{1+\left(\frac{v}{H}\right)^2 \cos \left(\frac{\sqrt{H^2+v^2}}{v} \pi\right) }{1+\left(\frac{v}{H}\right)^2}.
\label{sigmaz_final_analytical}
\end{align} \\

Equation (\ref{C_sigma_z}) and (\ref{C_m_sigma_z}) allow to understand that topology is robust to the environment at weak coupling and in the adiabatic regime $v/H\to 0$. In this regime, one may safely approximate the ground state $|g \rangle$ of Hamiltonian $\mathcal{H}$ by the ``shifted-oscillators" ansatz, Eq. (\ref{ansatz}). For $\theta=0$, one has $p=1$ while $q=1$ for $\theta=\pi$. From Eq. (\ref{C_m_sigma_z}), we then deduce that $C^m=1$ at weak non-zero dissipation: the environment do not affect the topology of the spin, and the Chern number stays equal to one. This argument is quite general and could be generalized for other kinds of environments, such as non-ohmic spectral densities or fermionic baths. As long as the system-environment coupling is weak enough to write a wavefunction of the form of Eq. (\ref{ansatz}), the global topology of the spin is protected. It should be noted that an additional random dephasing noise on the variable $\phi$ would not change the result as well, as long as $\langle \phi \rangle$ remains a constant. A study of this exact problem was realized in Ref. \cite{vavilov}, where the authors used a Bloch-Redfield pertubative approach to compute the spin dynamics. They notably confirmed that the ohmic environment does not affect the measurement of the topological properties at low coupling. This result must be contrasted with the one obtained in Ref. \cite{whitney_gefen}, where the authors studied the effect of an ohmic bosonic bath on the Berry phase acquired by the spin for a path characterized by a variation of $\phi$ at fixed $\theta$. They showed in this precise case that the environment affected the Berry phase, which is a local observable. Here, we will confirm that, despite having effects localy on the Berry curvature, the environment does not globally affect the topology of the spin at low coupling.   \\ 

When the coupling strength gets larger, entanglement grows between spin and bath and a quantitative study is necessary. Below, we investigate numerically this high coupling region with a Stochastic Schr\"{o}dinger Equation approach.

\section{Dissipative Spin dynamics and topology}

\subsection{Results for spin dynamics}

\begin{figure}[h!]
\center
\includegraphics[scale=0.35]{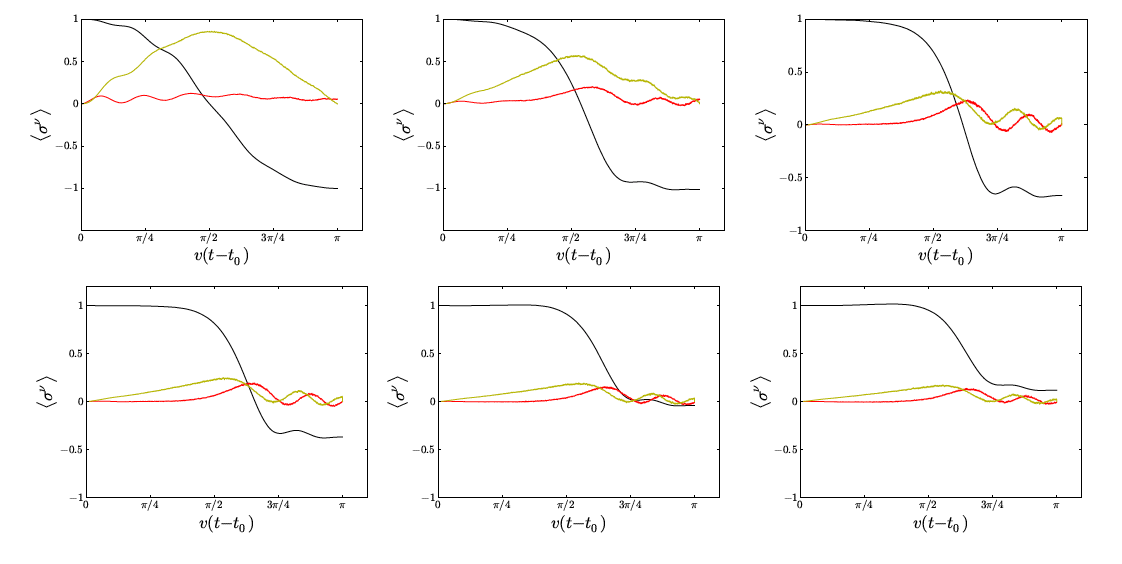}
\caption{Spin variables $\langle \sigma^z \rangle$ (black), $\langle \sigma^x \rangle$ (yellow) and $\langle \sigma^y \rangle$ (red) and their time-evolution during the sweep, with $v/H=0.08$. From left to right and top to bottom, we take $\alpha=0.05$, $\alpha=0.15$, $\alpha=0.26$, $\alpha=0.3$, $\alpha=0.34$ and $\alpha=0.36$. In the low coupling regime $\alpha \ll 1$ we recover the two different frequencies for the dynamics $v$ and $\sqrt{H^2+v^2}$, as shown by the description in the rotating frame (Eq. (\ref{rotated_basis}) and discussion below).
}
\label{dynamics_topo}
\end{figure}

We show in Fig. \ref{dynamics_topo} the evolution of the spin variables $\langle \sigma^{\nu}(t) \rangle$ at $H_0=0$ for $v/H=0.08$ when one progressively increases the coupling to the environment. 

At greater values of $\alpha$, the bath strongly affects the spin dynamics.  Here are the main conclusions that we draw from Fig. \ref{dynamics_topo}:
\begin{itemize}
\item Below $\alpha \simeq 0.15$, $\langle \sigma^z(t) \rangle$ shows a complete oscilation from $+1$ to $-1$, leading to a measured Chern number $C^m=1$. The continuous change from $+1$ to $-1$ becomes however sharper when one increases the coupling to the environment from $\alpha=0$ to $\alpha \simeq 0.15$, as can be seen from the first two top panels of Fig. (\ref{dynamics_topo}).
\item Above $\alpha \simeq 0.2$ the bath progressively leads to decoherence: we remark that the length of the spin at the end of the protocol is no longer equal to $1$. This is associated with a final value of $\langle \sigma^z \rangle$ greater than $-1$, and a Chern number $0<C^m<1$. Around $\alpha \simeq 0.35$, the bath completely destroys the spin coherence.
\item  Above this value, the bath ultimately leads to a final spin expectation value $\langle \sigma^z(t_f=t_0+\pi/v) \rangle>0$ associated with a restoring of the coherence.
\end{itemize}

\begin{figure}[h!]
\center
\includegraphics[scale=0.3]{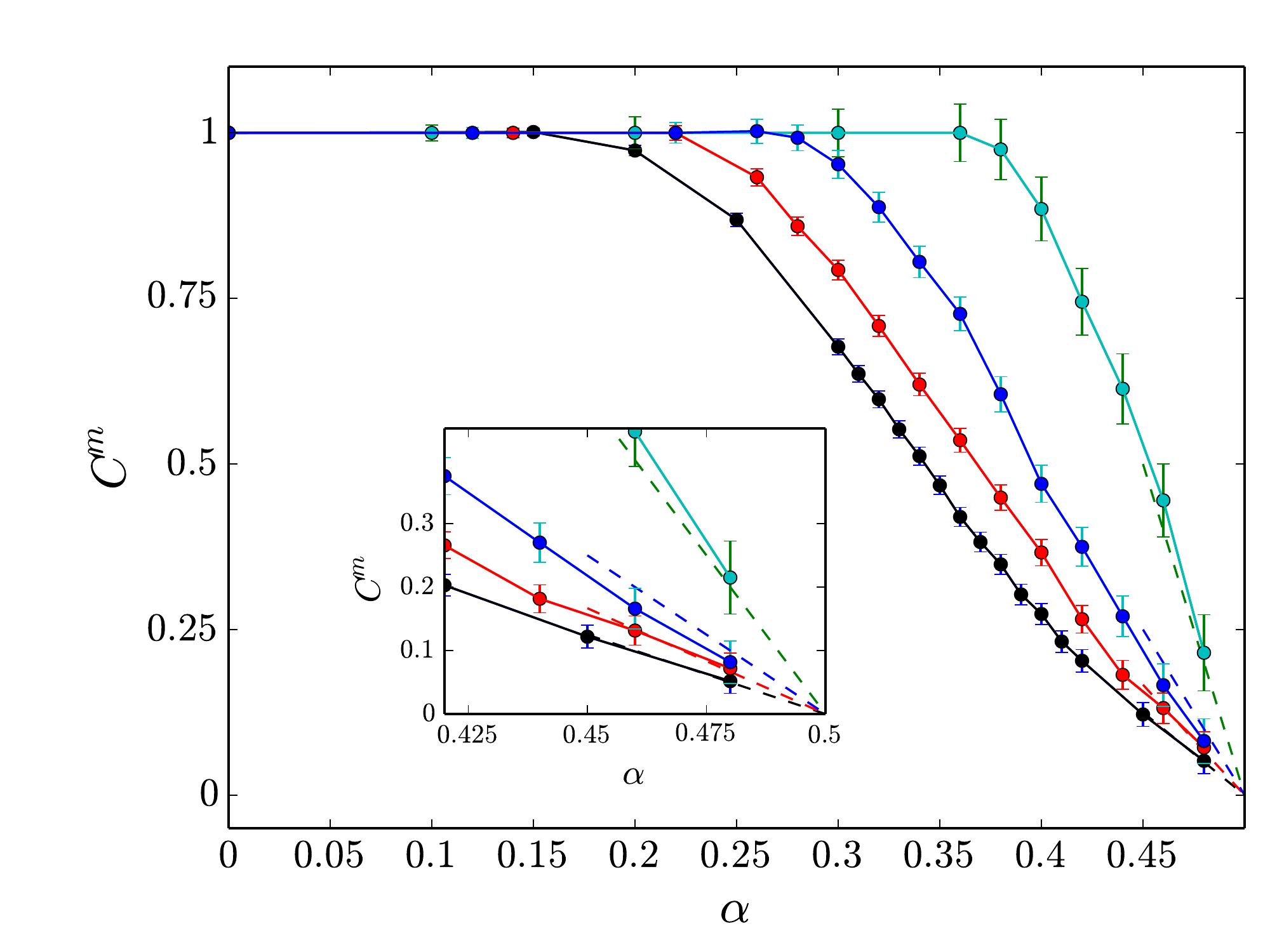}\includegraphics[scale=0.32]{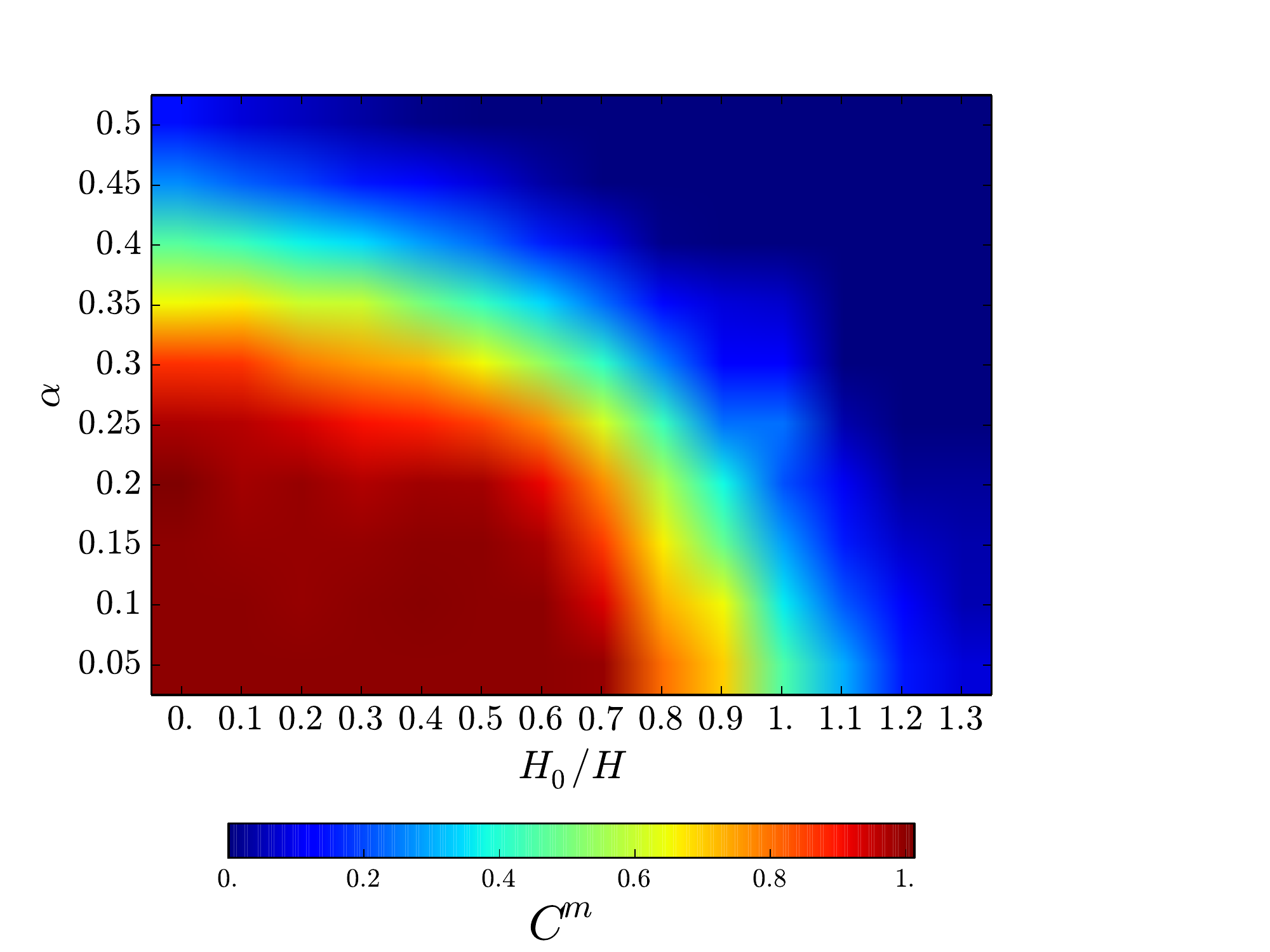}
\caption{Left: The main panel shows the evolution of the measured Chern number $C^m$ with respect to $\alpha$ for $v/H=0.08$ (black), $v/H=0.06$ (red), $v/H=0.04$ (blue), and $v/H=0.02$ (green). The inset zooms on the region around $\alpha=1/2$, and the dashed lines represent predictions based on the mapping with a non-interacting resonant level valid at $\alpha=1/2$ (see Sec. \ref{Toulouse}). Right : Evolution of $C^m$ with respect to $\alpha$ and $H_0/H$ for $v/H=0.08$. 
}
\label{chern_transition_1D}
\end{figure}

The measured Chern number $C^m$ can be computed by looking at the non-adiabatic response of the spin $\langle \sigma^y(t) \rangle$ (or equivalently from the final value of $\langle \sigma^z \rangle $), and we show its evolution with respect to $\alpha$ for different velocities in Fig. \ref{chern_transition_1D}, left panel. Increasing the coupling to the bath triggers a continuous transition from $C^m=1$ to $C^m=0$. 

Interestingly, this transition becomes sharper at low velocities. More precisely, it seems that $C^m\to 0$ when $\alpha \to 0.5$ in the regime considered here. We also show in Fig. \ref{chern_transition_1D} the complete phase diagram with the evolution of $C^m$ when both $\alpha$ and $H_0/H$ varies. An extended discussion is given in Ref. \cite{topo_loic}.\\

\subsection{Radiative cascade of photons and effective magnetic field}

We give now a simple physical picture to account for the evolution of the spin observables in Fig. \ref{dynamics_topo} at large spin-bath coupling. Driving the spin at velocity $v$ leads to the emission of bosons of frequencies close to $v$, which in return affect the spin dynamics. From Eq. (\ref{Hamiltonian_spin_boson}), we indeed see that the boson coherence $h=\sum_k \lambda_k (b_k+b_k^{\dagger})$ acts on the spin as an effective field along the $z$-direction. At sufficiently high coupling, the quantitative number of bosons emitted induce a large negative value of this quantity $h$, which compensate the field along the $z$-direction and force the spin to point upwards at the end of the dynamical protocol.\\

\begin{figure}[h!]
\center
\includegraphics[scale=0.3]{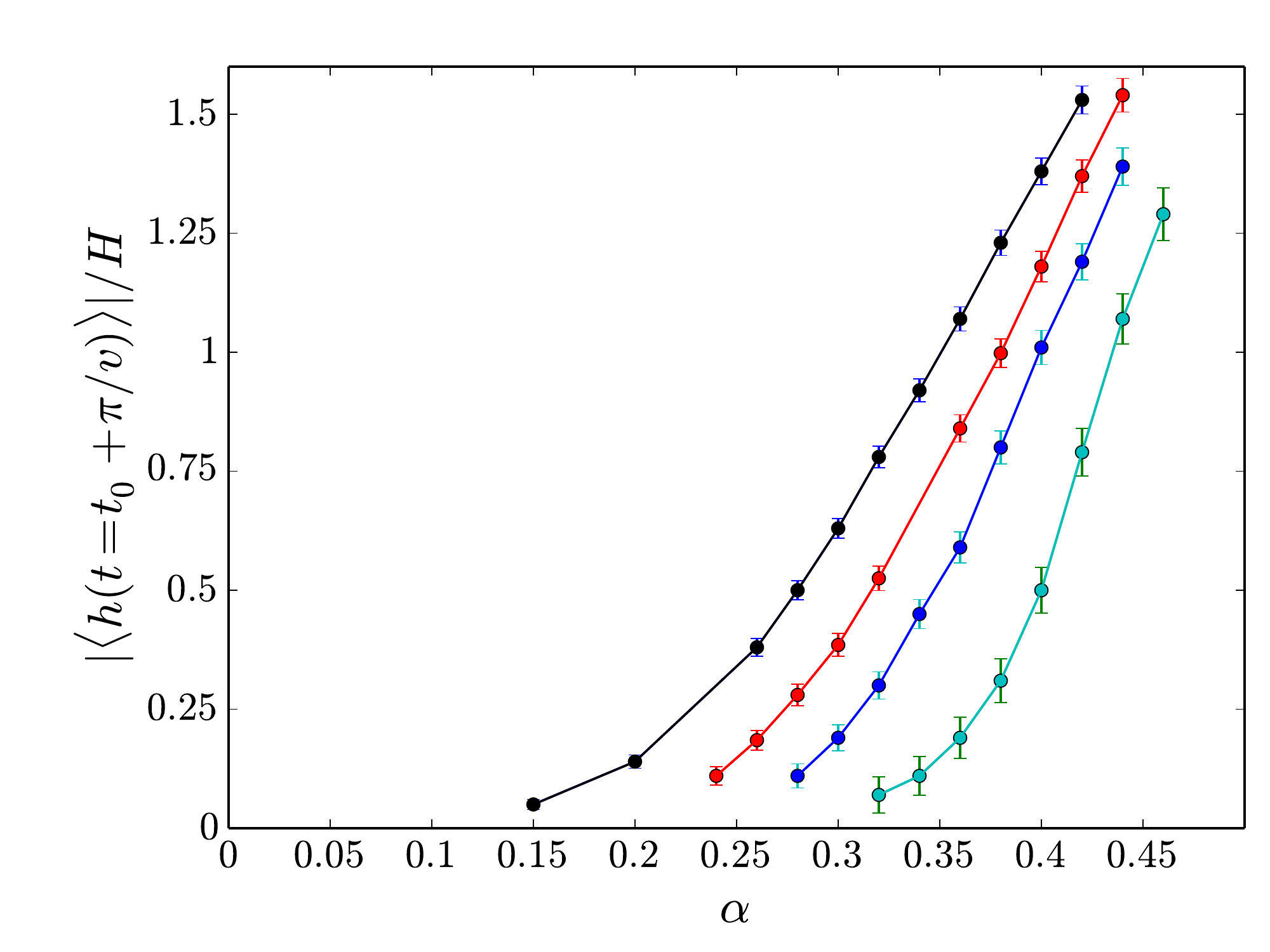}    
\caption{Absolute value of the boson-induced field $|\langle h \rangle|/H$ felt by the spin at the end of the dynamical protocol as a function of $\alpha$, for $v/H=0.08$ (black), $v/H=0.06$ (red), $v/H=0.04$ (blue), $v/H=0.02$ (green).}
\label{Results_toy_model}
\end{figure}

To make the discussion more quantitative we study numerically the toy model
\begin{align}
\mathcal{H}_{toy}=v b^{\dagger} b +\frac{\lambda}{2} \langle \sigma^z \rangle \left(b+b^{\dagger}\right),
\end{align}\\
where we extract $\langle \sigma^z (t) \rangle$ from the SSE results, and we take $\lambda=\sqrt{2 \alpha v}$. We plot in Fig. \ref{Results_toy_model} the absolute value of the boson-induced field felt by the spin at the end of the dynamical protocol $\langle h \rangle=\lambda/2 \langle b+b^{\dagger}\rangle(t_f=t_0+\pi/v)$ as a function of $\alpha$, for different values of the velocity. 
For a given velocity, the final value of the boson-induced field grows in modulus with $\alpha$. Above a certain coupling strength this field compensates $H$, which is responsible for the positive value of $\langle \sigma^z(t_f=t_0+\pi/v) \rangle>0$. From Fig. \ref{Results_toy_model}, we confirm that this simple toy-model gives a correct estimate for the coupling at which $\langle \sigma^z(t_f=t_0+\pi/v) \rangle$ changes sign (or equivalently, the coupling for which one gets $C^m=0.5$) for the velocities considered.  \\

In this section, we showed that the contact of the spin with a many-body quantum environment destroys the spin topology at high coupling. We studied the dynamical transition within a dynamical framework, which could be readily implemented in modern quantum simulation platforms. We provided a basic dynamical interpretation of the transition in terms of a radiative cascade of bosons, whose feedback strongly affects the spin dynamics. By analogy to the dissipative transition, one can consider that the bath makes the system more ``classical" at high coupling and destroys the topology of the quantum state. In particular, this finding does not seem restricted to our dynamical protocol but seems rather to have a link with equilibrium properties, as the effect subsists and becomes sharper when $v/H\to 0$.  Next, we use the exact mapping between the ohmic spinboson model at $\alpha=1/2$ and Toulouse Hamiltonian \cite{Toulouse,Anderson_Yuval_Hamann,Weiss:QDS,kondo_spinboson}, describing an isolated electronic level uniformly coupled to a bath of spinless electrons. We will also study the scaling properties of $C^m$ around $\alpha=1/2$ at non-zero velocities within Keldysh formalism. 

\section{Study around the Toulouse point}
\label{Toulouse}

%The ohmic spinboson model is known to be equivalent to the Toulouse Hamiltonian \cite{Toulouse,Anderson_Yuval_Hamann}, describing an isolated electronic level uniformly coupled to a bath of spinless electrons with a bandwidth $2\omega_c$ around the Fermi level. 

Toulouse Hamiltonian reads
\begin{align}
\mathcal{H}_T=\sum_k \epsilon_k c_k^{\dagger} c_k +\epsilon_d d^{\dagger}d +V \sum_k \left( c_k^{\dagger}d+d^{\dagger} c_k \right),
\label{H_toulouse}
\end{align}\\
where $c_k$ denotes the annihilation operator of an electron with energy $\epsilon_k$. We assume a constant density of states over a bandwidth $D$, and $d$ corresponds to the annihilation operator of the isolated electron of energy $\epsilon_d$. $V$ quantifies the hybridization between the isolated level and the surrounding electrons. The equivalence between the two models in the limit $H/\omega_c \ll 1$ can be shown by an explicit computation of the partition function \cite{leggett:RMP,weiss}, or with bosonization \cite{kondo_spinboson}. One has the following correspondence,
\begin{align}
& (H \sin \theta) \equiv V \sqrt{\frac{D}{4\omega_c}}\notag \\
&(-H \cos \theta) \equiv \epsilon_d.
\label{mapping}
\end{align}
The two cutoffs can be related by $D=4\omega_c/\pi$ \cite{Hur}. This model is exactly solvable, and we have at equilibrium \cite{Hur,kopp_hur},
\begin{align}
& \langle \sigma^z \rangle=\frac{2}{\pi} \tan^{-1}  \frac{ 4 \omega_c \cos \theta}{\pi H \sin^2 \theta} \label{spin_observables_1} \\
&\langle \sigma^x \rangle=-2 \left\lvert \frac{H\sin \theta}{4\omega_c}\right\rvert \left\{ 2+ \ln \left[  \left(\frac{\pi H\sin \theta}{4 \omega_c}\right)^4   +  \left(\frac{\pi H\cos \theta}{4 \omega_c}\right)^2    \right]\right\}.
\label{spin_observables_2}
\end{align}\\
When $\theta$ varies from $0$ to $2 \pi$, the equilibrium Bloch vector describes a curve with trivial topology due to the symmetry $\pi-\theta \leftrightarrow \pi+\theta$ of equations (\ref{spin_observables_1}) and (\ref{spin_observables_2}). This statement is true for all $H/\omega_c$ and remains valid in the limit $H/\omega_c\to 0$ where Toulouse Hamiltonian is equivalent to the ohmic spinboson model. This confirms that the point $\alpha=1/2$ of the ohmic spinboson model corresponds to $C=0$. In this limit $H/\omega_c\to 0$, one gets $\langle \sigma^z \rangle=+ 1$ when $\theta<\pi/2$ and $\langle \sigma^z \rangle=- 1$ when $\theta>\pi/2$.  \\

For the dynamical protocol $\theta(t)=v(t-t_0)$, the quantity $\langle \sigma^z(t) \rangle$ can be related to the occupation of the central level $\langle d^{\dagger}d (t) \rangle$. Such a quantity can be computed with Keldysh non-equilibrium techniques in an exact manner, following Ref. \cite{antti_pekka_wingreen_meir} where the authors investigated the time-dependent transport for Hamiltonian $\mathcal{H}_T$ in Eq. (\ref{H_toulouse}). From the equations of motion, we have
\begin{align}
\partial_t \langle d^{\dagger}d \rangle= iV \sum_k \left[ \langle c_k^{\dagger} d \rangle-\langle d^{\dagger} c_k  \rangle \right].
\label{EOM_toulouse}
\end{align}
The computation of the right hand side of Eq. (\ref{EOM_toulouse}) can be performed exactly within Keldysh formalism, and we reach an expression similar to the one obtained by the authors of Ref. \cite{antti_pekka_wingreen_meir} for the lead currents (see Eq. (42) of this reference),
\begin{align}
\partial_t \langle d^{\dagger}d \rangle= -\Gamma(t) \langle d^{\dagger}d \rangle -\int d\epsilon f(\epsilon)/\pi \int_{t_0}^t dt_1 \Gamma(t_1,t) \Im m \left\{ e^{-i \epsilon (t_1-t)} G^r (t,t_1) \right\}.
\label{EOM_toulouse_2}
\end{align}
In Eq. (\ref{EOM_toulouse_2}), $f$ denotes the Fermi distribution and one has 
\begin{align}
&\Gamma(t,s)=\frac{2\pi}{D} V(t) V(s),\\
& G^r(t,s)=-i \theta(t-s) \exp\left[-i \int_{s}^t du \epsilon_d (u) \right] \exp\left[-\frac{1}{2} \int_{s}^t du \Gamma (u) \right],
\label{coeffs_EOM_toulouse}
\end{align}
and we identified $\Gamma(t)=\Gamma(t,t)$. One can solve Eq. (\ref{EOM_toulouse_2}) and we recover that $\langle d^{\dagger}d \rangle (t_f=t_0+\pi/v) \to 1$ corresponding to $C^m=0$, in the validity limit of the correspondance $H/\omega_c \ll 1$ . Next we explore the region where $u=(1/2-\alpha) \ll 1$. Deviations from the exact mapping point result in a additional term $\mathcal{H}_t$ in $\mathcal{H}_T$, of the form \cite{kondo_spinboson}
\begin{align}
\mathcal{H}_t=U\sum_{k,k'} \left( c^{\dagger}_k c_{k'}-\frac{1}{2}\right) \left( d^{\dagger} d-\frac{1}{2}\right).
\label{additional_term}
\end{align}
In Eq. (\ref{additional_term}), one has $U=\pi \left(1-\sqrt{2\alpha} \right)=\pi u +o(u)$. To go further, we assume that the main contribution of Eq. (\ref{additional_term}) comes from the terms with $k=k'$ and we decouple the interaction in a mean-field manner. This results in a time-dependent shift of the chemical potential for the electrons depending on the level occupation $\langle d^{\dagger}d \rangle (t)$, which can be accounted for in the resolution of Eq. (\ref{EOM_toulouse_2}). We find numerically a linear behaviour in terms of $u$, $C^m(u)=\beta(v) u$, where $\beta(v)$ is proportional to $H/v$. We show in fig. \ref{chern_transition_1D} the corresponding linear predictions, obtained with a one-parameter fit.
This linear dependency can be interpreted in the light of the Fermi-liquid behaviour of the ohmic spinboson model, or its Kondo analogue \cite{Nozieres}. In this description, the local susceptibility $\chi=-\partial_{d_z} \langle \sigma^z \rangle $ is known to be constant with respect to $d_z$. As shown above, a deviation $u>0$ from the point $\alpha=1/2$ may be seen as a shift of the electronic level energy by a factor proportional to $u$, or equivalently a shift of $d_z$ in the spinboson description. This argument would confirm then a linear dependency of $\langle \sigma^z (t_f=t_0+\pi/v) \rangle$ with respect to $u$, and thus the scaling $C^m(u)=\beta u$. The form of $\beta(v)\propto H/v$ bears similarities with the dependency of $\chi \propto 1/T_K$ in the anisotropic Kondo model, which diverges at the antiferromagnetic-ferromagnetic transition $\alpha=1$.\\

In this Section, we studied the scaling of $C^m$ close to the transition point. It is instructive to study in more details the evolution of the entanglement between spin and bath when increasing $\alpha$.

\section{Entanglement entropy and effective thermodynamics}

\begin{figure}[h!]
\center
\includegraphics[scale=0.3]{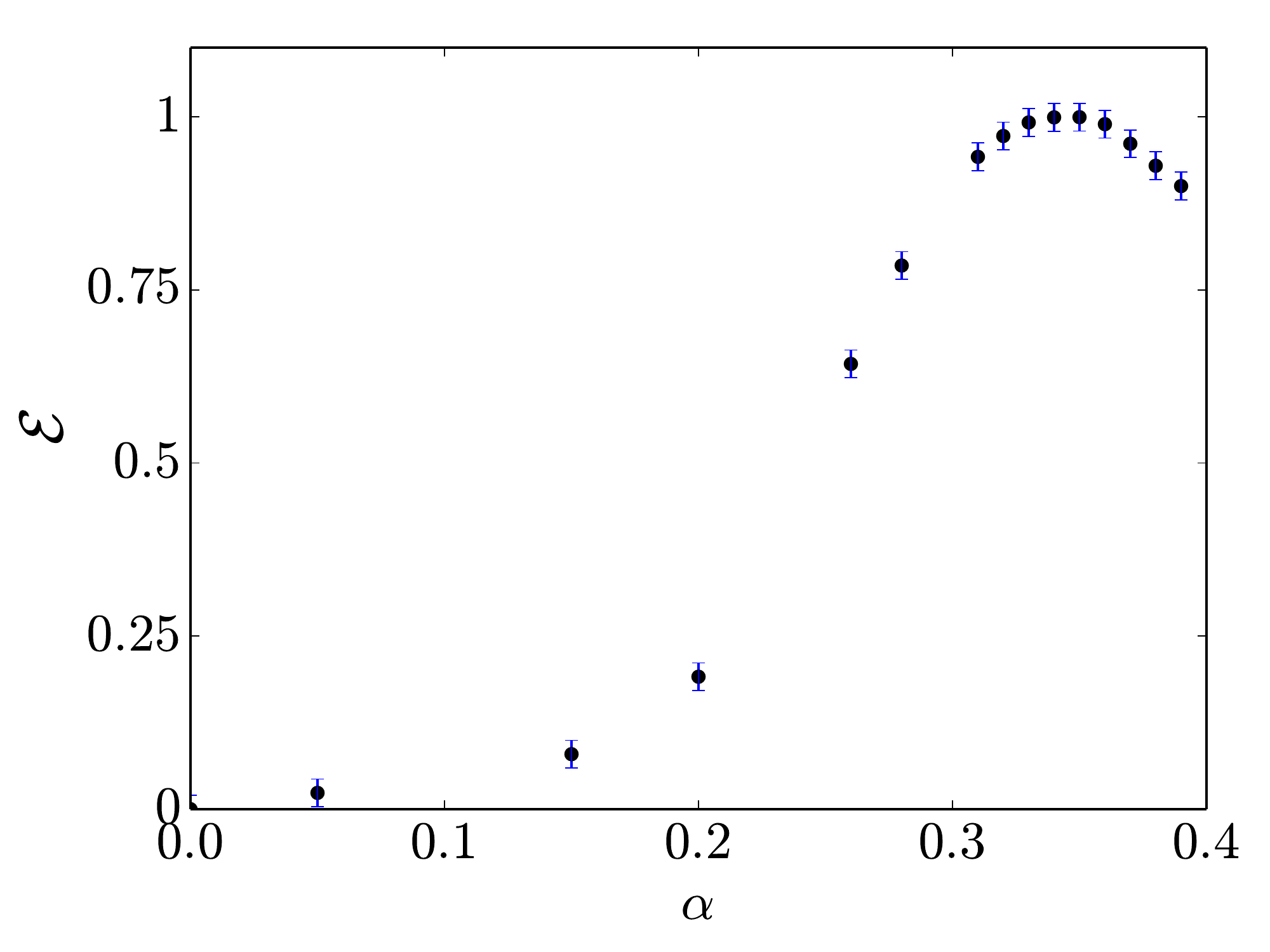}
\includegraphics[scale=0.3]{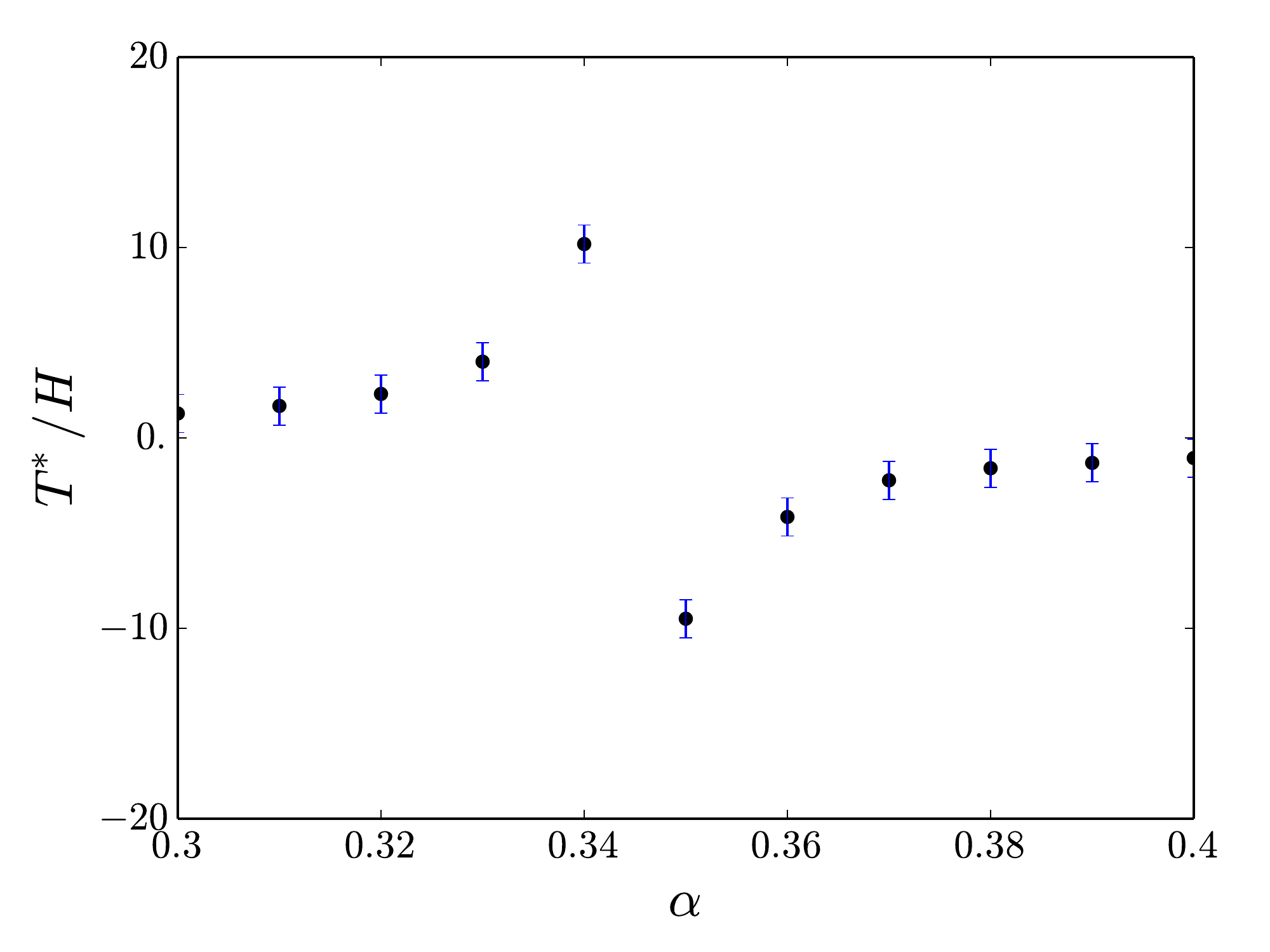}
\caption{Final entanglement entropy (Left panel), and final spin length (right panel) for $v/H=0.08$. Inset: Effective temperature with respect to $\alpha$. We have $H/\omega_c=0.01$.}
\label{criticality}
\end{figure}

For completeness, we study the entanglement between spin and bath at the end of the dynamical protocol. To this end, we introduce the entanglement entropy \cite{entanglement_entropy} $\mathcal{E}=-Tr\left[ \rho_S \log_2 \rho_S\right]$, where $\rho_S$ is the spin-reduced density matrix. $0\leq \mathcal{E}\leq 1$ quantizes the degree of entanglement between the spin and its environment. For a pure state, spin and bath can be factorized and one has $\mathcal{E}=0$. On the contrary, the case $\mathcal{E}=1$ corresponds to a maximally entangled state. We plot on the left panel of Fig. (\ref{criticality}) the evolution of the final entanglement entropy (i.e. the entanglement entropy at the end of the dynamical protocol) with respect to $\alpha$ for $v/H=0.08$. We remark that the entanglement progressively grows between spin and bath to reach a maximal value of $\mathcal{E}=1$ at the coupling for which $\langle \sigma^z (t_f=t_0+\pi/v) \rangle=0$. This correspond to the point for which $C^m=0.5$ in Fig. \ref{chern_transition_1D}, and when the boson-induced field exactly compensates the field $H$ in the radiative cascade picture. As shown in Fig. \ref{chern_transition_1D}, this coupling tends to $1/2$ when one decreases the value of $v/H$.\\

Another interesting related quantity is the spin length  $S=(\langle \sigma^x \rangle^2+\langle \sigma^y \rangle^2+\langle \sigma^z \rangle^2)^{1/2}$ at the final time, which goes to zero when $\mathcal{E}$=1. One can also indeed interpret the spin reduced density matrix at the end of the protocol as the one of an isolated spin in thermal equilibrium, with effective Hamiltonian and temperature determined by $\langle \sigma^x \rangle$, $\langle \sigma^y \rangle$ and $\langle \sigma^z \rangle$ \cite{williams_hur_jordan}. More precisely, we interpret the spin reduced density matrix as the one of a factorizable state $\rho=\rho_S \otimes \rho_B$ where the spin is in thermal equilibrium with a bath at temperature $T^*$.\\

Treating the system within this analogy allows to define an effective temperature $T^*$, whose evolution with respect to $\alpha$ is shown in Fig. \ref{criticality}. For $\alpha>\alpha_c$, we find negative temperatures, which is not surprinsing as the protocol leads to population invertion. Negative temperatures have notably been realized in localized spin systems \cite{negative_spin_1,negative_spin_2}, and were recently measured experimentally for motional degrees of freedom in a cold atomic setup \cite{negative_temp}.\\

 Interestingly, this effective temperature diverges at the transition and one has the leading behaviour $T^*/H \propto (\alpha-\alpha_c)^{-1}$, independently of $v/H$. This scaling behaviour is intimately related to the scaling of $\langle \sigma^z \rangle$ (or $C^m$) at the transition, see Section \ref{Toulouse}. This critical behaviour may be of interest for future experimental studies. \\

In this Chapter, we first provided a discussion of the effect of an environement on the topology of a spin $1/2$ in the ``shifted oscillators" picture, and we found that the bath gradually deforms the spin ground state manifold. At $\alpha=1$, the Berry curvature becomes infinite and this singularity signals the dissipative topological transition \cite{topo_loic}. Beyond the characteristic vanishing of the off-diagonal elements of the density matrix, decoherence is thus also responsible for global changes in the spin space topology.\\

Then, we investigated this transition in a dynamical protocol, by focusing on the non-adiabatic response of the slowly driven spin, which is known to be related to the geometric properties of the ground state \cite{polkovnikov:PNAS}. The use of the SSE method enabled us to study quantitatively the progressive deformation of the system topology for various values of the drive. We complemented this findings by analytical and numerical argument close to the point $\alpha=1/2$, where the ohmic spinboson model can be mapped onto the exactly solvable Toulouse Hamiltonian. Beyond its theoretical interest, an experimental evidence of this transition seems accessible in state-of-the-art experimental platforms.\\

As exposed in Chapter I, this model can be engineered in cold-atomic or circuit QED setups, which are great platforms for the exploration of many-body effects. In the next Chapter, we focus on recently developped hybrid platforms interfacing mesoscopic electronic physics with circuit QED elements.  

\chapter{Hybrid electron-photon systems}

Recent years have seen the developments of experimental setups and theory proposals interfacing superconducting circuits with the quantized electronic levels of quantum dots \cite{Dousse,Xiang:RMP,Delbecq:PRL,Petersson:Nature,Frey:PRL,Frey:PRB,Toida:PRL,Delbecq:NC,Deng:2013,Liu:PRL,Deng:2014,Childress:PRA,Lin:PRL,Bergenfeldt:PRL,Viennot:PRB,Basset:PRB,Hu:PRB,Bergenfeldt:PRB,Lambert:EPL,Pulido:NJP,Basset:APL,Liu:Science,Stockklauser:arxiv,LoicNano,Karyn:CR}. These hybrid platforms offer a new way to address interaction effects between photons and electrons on-chip, and typically involve an electronic nano-circuit, such as a single or double quantum dot (DQD) coupled to source/drain leads and to an electromagnetic resonator. \\

In a non-extended system like a quantum dot coupled to metallic leads, interactions play an important role. This can be understood physically as the charge tends to be localized on the nanostructure. The Coulomb interaction hinders the flow of charges through the nanostructure for a small quantum dot, as incoming electrons on the quantum dot do not have enough energy to change the charge on the dot. This results in the suppression of the conductance through the device, with the exception of the charge degeneracy points in which two different charge occupations on the dot become degenerate in energy, a phenomenon named Coulomb blockade. Taking into account the many-body interaction of the quantized dot levels with its surroundings is necessary to describe accurately the strong correlations induced by these charging effects \cite{bruus}.\\

Coupling these many-body electronic systems to microwave light provides then a new manner to probe many-body phenomena, or devise new nanotechnology devices. Experimental groundings have notably triggered substantial work linked to thermoelectrics \cite{Rafa,Sothmann2012,Jordan2013,Bjoern,Casati,Dutt_LeHur,Lim,experi1,experi2,experi3,experi4}, with possible applications for energy harvesting and cooling. The subject of quantum thermodynamics has then been rapidly growing in interest \cite{Jukka}. In Sec. \ref{nanoengine}, we explore the possibility of rectification of electrical current from quantum vacuum fluctuations in the specific case of electrical transport through a quantum dot, and its use as a nano-heat engine, following our Ref. \cite{LoicNano}. \\

These hybrid systems are also of fundamental interest as platforms to explore exotic quantum impurity physics with light and matter \cite{Marco:PRB}. In Sec. \ref{su4_kondo} we present evidence for exotic Kondo correlations in the microwave response of a DQD-cavity coupled to leads, a unique signature of many-body quantum impurity physics with light, following our Ref. \cite{Deng_Henriet}.\\

\section{Nanoengine from quantum vacuum fluctuations}
\label{nanoengine}

\begin{figure}[h!]
\center
\includegraphics[width=9.8cm]{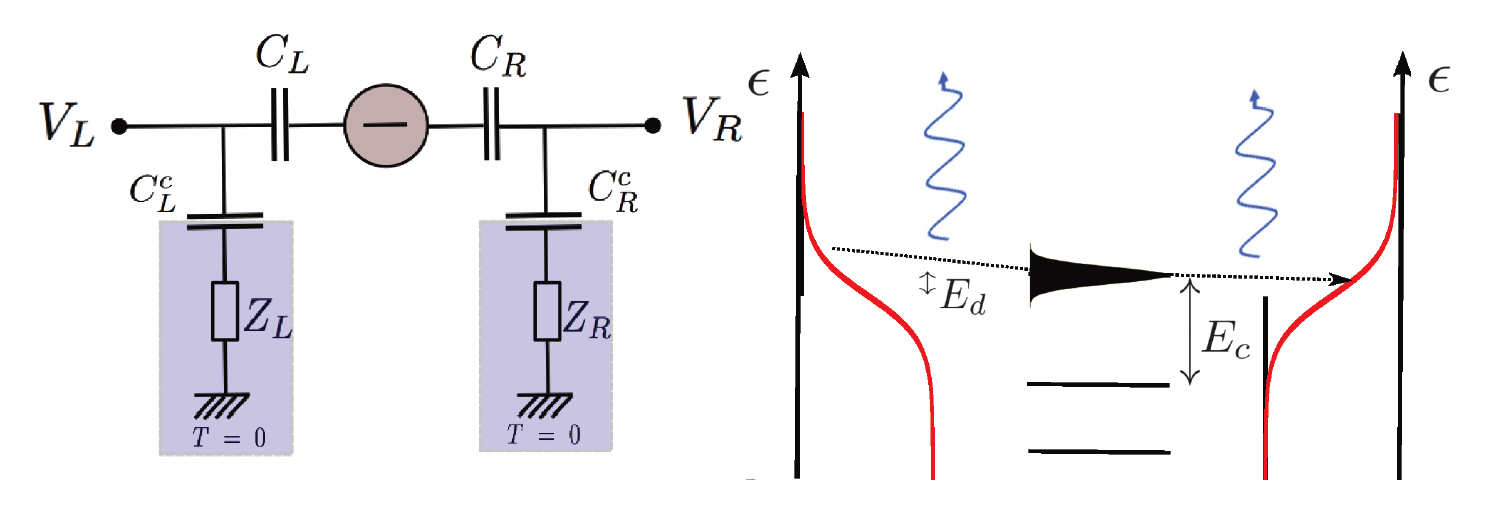}%\includegraphics[width=4.1cm]{hot_electrons_24_04_2.pdf}
\label{fig:2figsB}
\caption{Quantum vacuum nano-engines with a quantum dot in the Coulomb blockade regime. Left: circuit. Right: pictorial representations of the tunneling processes at stake - the lateral parts correspond to the energy diagram of the electrons in the corresponding lead and the central part represents the energy levels of the quantum dot. The straight dotted arrows represent a tunneling event from left to right, associated with curvy arrows which correspond to photon emission.}
\label{figsetup}
\end{figure}

We consider the system displayed on the left of Fig.~\ref{figsetup}. The left and right electronic leads contain electrons at the same temperature $T$, tunnel coupled to one another via a quantum dot, which is described with coupling capacitances $C_L, C_R$. The system is in the Coulomb blockade regime, with the lowest unoccupied quantum dot level at an energy $E_d$ above the electrons chemical potential. Additionally, this system is part of an electrical circuit with finite impedances $Z_L$ and $Z_R$ connected to the system via the capacitances $C_L^c$ and $C_R^c$ respectively. \\

Naturally, no current is rectified when the system is put in state of overall zero temperature $T=0$ and without electric bias: it is necessary to give some energy to the system for a current to flow. We consider the case where the electrons in the two leads are in thermal equilibrium, both at the same small temperature $T$. Some electrons are then thermally populated into the energy of the quantum dot level.\\

 Giving some energy to the system is however not sufficient to induce a current. We also need to break the spatial symmetry in order to favor one direction of tunneling events with respect to the other. This can be done by coupling the leads to zero temperature electrical circuits with different impedances $Z_L$ and $Z_R$ (or the same impedance but with different coupling capacitances $C_L$ and $C_R$). As seen in Sec. \ref{transmission_line} of Chapter I, the impedances can be modeled by a chain of LC oscillators, which when quantized, behave as a system of quantum harmonic oscillators, whose frequency $\omega_i$ depends on the impedance and inductance of the chain. The vacuum energy of the chain is $E_0 = (1/2) \sum_i \hbar \omega_i$\footnote{We restore the $\hbar$ in this section.}, and leads to fluctuations of the charge (or voltage) on the plate of the coupling capacitor to the quantum dot \cite{Milonni}. If the impedances coupled to the two leads are different, the fluctuations of the vacuum will break Left/Right symmetry, giving a preferred direction for the particles to flow - thus giving current rectification, without the application of a bias voltage across the system. \\

%The paper is organized as follows. In Sec. II, we introduce the quantum many-body physics corresponding to Fig.~1, and discuss the physics in terms of $P(E)$ theory\cite{IN}. In Sec. III, we derive the quantum vacuum-induced rectified current for the different geometries considered, and show its dependence on the physical parameters of the system. In Sec. IV, we add a load voltage to the system, and consider its operation as a heat engine, finding the power produced and thermodynamic efficiency versus temperature and asymmetry. We give our conclusions in Sec. V. In Appendix A, we quantize the voltage fluctuations through a transmission line description.

\subsection{Analysis}
We let $c_k$ be a fermionic annihilation operator on the left lead of momentum $k$, similarly, $d$ and $c_q$ describe electrons on the dot, and the right lead. The system Hamiltonian is, up to a constant term and assuming spin-polarized electrons,
\begin{align}
H = \sum_k (\epsilon_k + e v_L) c_k^\dagger c_k + \sum_q (\epsilon_q + e v_R) c_q^\dagger c_q + E_c d^\dagger d+ H_{ph} + H_C,
\label{Hamiltonian_nano}
\end{align}
$H_{ph}$ is the Hamiltonian of the environment in isolation and $v_{L,R}$ are the quantum voltage induced by the presence of the Left/Right environments. We define the zero of energy as the Fermi level of the leads. $E_c>0$ is the charging energy of this level, which corresponds to the difference of energy when the dot is charged, versus uncharged \cite{charging_1,charging_2}. The external impedances are coupled to the electrical circuit through coupling capacitances, allowing then an efficient thermal isolation. \\

The tunneling terms are
$H_C  = \sum_k t_{kd}( d^\dagger c_k + h.c.) +   \sum_q (t_{dq} c_q^\dagger d + h.c.)$.
Here, $t_{kd}$ is the tunneling matrix element onto the dot from the left lead with momentum $k$, $t_{dq}$ is the tunneling matrix element from the dot to the right lead with momentum $q$. $H_{ph}=\sum_{l \in \{L, R\}} \sum_n \hbar \omega_{n,l} \left( b_{n,l}^{\dagger}b_{n,l}+1/2 \right)$ describes the environmental degrees of freedom. The voltages on the tunnel junctions are related to the bosonic degrees of freedom by $v_l=\sum_{n} \lambda_{n,l} (b_{n,l}^{\dagger}+b_{n,l})$. The effect of these quantum voltages on the tunnel junctions and coupling capacitors can be conveniently taken into account by a unitary transformation $U = \exp ( i \phi_{L} \sum_k c_k^\dagger c_k + i \phi_{R} \sum_q c_q^\dagger c_q )$.
 This transformation $H \to U^{-1} H U$ has the effect of eliminating the voltages on the bare system Hamiltonians and putting a phase on the tunneling elements,
\begin{eqnarray}
t_{kd} \rightarrow t_{kd} \exp(i \phi_{L}),\\
t_{dq} \rightarrow t_{dq} \exp(- i\phi_{R}).
%t_{kd} \rightarrow t_{kd} e^{i \phi_{L,0} - i\phi_{d,0}},
%t_{dq} \rightarrow t_{kd} e^{i \phi_{d,0} - i\phi_{R,0}}.
\end{eqnarray}
Here, we have introduced the phase $\phi_{l} = (e/\hbar) \sum_n \lambda_{n,l}/\omega_{n,l}(b_{n,l}-b_{n,l}^{\dagger})/i $ for  $l \in \{L, R\}$.\\

We treat the problem within the so-called $P(E)$ theory \cite{IN}, assuming that tunneling on and off of the dot can be described with tunneling rates described within a golden rule picture in perturbation theory. This assumes that environments relax faster than the tunneling time scale rates. In this picture, we do not take into account the backaction of the electrons on the excitations of the impedances. The tunneling rates are thus controlled by the probability of the electrical circuit to either absorb or emit a photon. We recall that within a golden rule picture, the rate of transitions $\Gamma_{i \to f}$ between an initial state $|i\rangle$ and a final state $|f\rangle$ is given by $\Gamma_{i \to f}=2\pi/h\times |\langle f |H|i\rangle|^2 \delta(E_f-E_i)$, where $H$ is the Hamiltonian describing the system and $E_i$ and $E_f$ are initial and final state energies. \\

We seek first to compute the tunneling rate from the left lead to the dot in the golden rule framework, and summing over all the possible final bath states. This tunneling rate $T_{+,L}$ (the subscript $L$ denotes the Left lead, and the index $+$ refers to a positive direction chosen to be from left to right) is given by
\begin{align}
T_{+, L}=\frac{2 \pi}{h} \nu_L V_L \int_{-\infty}^{\infty} d\epsilon f(\epsilon) |t_L|^2 \sum_{|B\rangle} |\langle B| e^{i\phi_L}|0 \rangle|^2 \delta(\epsilon+\mu/2-E_d -E_{|B\rangle}+E_{|0\rangle}),
\end{align}
where $\nu_L$ and $V_L$ denotes density of state and volume of the lead $L$; state $|B\rangle$ is an arbitrary one-excitation bath-state with energy $E_{|B\rangle}$ and $|0 \rangle$ is the vacuum. We have included the possibility of a bias $V$ $(\mu=eV)$ on the system. We use then the identity $\delta(E)=1/h \int_{-\infty}^{\infty} \exp[iEt/\hbar]$ and reach the following expression
\begin{align}
T_{+, L}=&\frac{2\pi}{h} \nu_L V_L |t_L|^2 \int_{-\infty}^{\infty} d\epsilon f(\epsilon) P_L(\epsilon+\mu/2-E_d)\notag\\
=&\frac{2\pi}{h} \nu_L V_L |t_L|^2 \int_{-\infty}^{\infty} d\epsilon f(\epsilon + E_d - \mu/2) P_L(\epsilon),
\end{align}
where the function $P_L$ is defined by
\begin{align}
P_L(E) &= (1/h) \int_{-\infty}^\infty dt e^{i E t/\hbar}  \underbrace{\sum_{|B\rangle} \langle B| e^{i\phi_L}|0 \rangle e^{-i/\hbar \left[E_{|B\rangle}-E_{|0\rangle}\right]t} \langle 0| e^{-i\phi_L}|B \rangle}_{\langle e^{i\phi_L(t)}  e^{-i\phi_L(0)} \rangle}       ,\notag \\
&= (1/h) \int_{-\infty}^\infty dt e^{i E t/\hbar} e^{k_L(t)}.
\label{pE}
\end{align}
In Eq. (\ref{pE}), $k_L(t) = \langle (\phi_L(t) - \phi_L(0)) \phi_L(0)\rangle$ is the lead correlation functions taken in isolation\footnote{We could go from the first line of Eq. (\ref{pE}) to the second line because the bath is gaussian.}. A similar type of calculation holds for the right lead, and for the negative direction (from right to left), and one generally gets $T_{\pm, L}=T_{\pm, L} (E_d)$ with the $T_{\pm, L}$ function defined as
\begin{eqnarray}
T_{\pm, L} (\Omega) = T_{0,L} \int d\epsilon f(\epsilon \pm \Omega \mp \mu/2) P_L(\epsilon), \label{Gamma_L}\\
T_{\pm, R} (\Omega) = T_{0,R} \int d\epsilon f(\epsilon \mp \Omega \mp \mu/2)  P_R(\epsilon)\label{Gamma_R},
\end{eqnarray}
where $T_{0,l} = 2\pi/h |t_l|^2 \nu_l V_l$. From Eqs. (\ref{Gamma_L}) and (\ref{Gamma_R}), one remarks that the $P$ functions can be interpreted as the probability of the electrical circuit to either absorb or emit a photon (see Ref. \cite{IN} for a detailed description). \\

Now that we have the expressions of the tunneling rates, we can compute the current flowing through the device for different kinds of environments.

 \subsection{Rectified current}
 
 \subsubsection{Coupling to an ohmic environment}

We first consider the case when the external impedances are ideal Ohmic resistor described by the frequency-independent
impedance $Z_l=R_l$. The lead correlation function reads \cite{IN}
\begin{align}
k_l(t)&=\int_{0}^{\infty} \frac{d \omega}{\omega} 2 Re  \left[ \frac{R_l}{ R_q}\right]  (e^{-i\omega t}-1), \notag \\
\label{k}
\end{align}
where $R_l(=\sqrt{L_l/C^c_l})$ is the impedance of the transmission line coupled to the lead $l$ (see also Chapter I Sec. \ref{transmission_line}) and $R_q = h/e^2$ is the resistance quantum. Let us define $\alpha_l = R_q/R_l$ (we will mainly consider the regime  $\alpha_l \gg1$). In this ohmic case, we have at long times \cite{IN} $k_{l}(t) = -(2/\alpha_l)[ \ln(\alpha_l E_c t/\pi \hbar) + i\pi/2 + \gamma_e]$, where $\gamma_e$ is the Euler constant.\\

This form of $k_{l}$ gives a $P_l(E)$ function which vanishes for negative energies and has a power-law divergence for small positive energies \cite{IN},
\begin{align}
P_l(E) = \frac{e^{-\frac{2\gamma_e}{\alpha}}}{\Gamma(\frac{2}{\alpha})}\frac{1}{E}\left(\frac{\pi}{\alpha} \frac{E}{E_c} \right)^{\frac{2}{\alpha}}.
\label{Pe_ohmic}
\end{align}\\

This from of $P_l(E)$ enables us to calculate the tunneling rates $T_{\pm,l}$,
\begin{eqnarray}
T_{\pm,L}(\Omega)  = -T_{0,L} e^{- \frac{2\gamma_e}{\alpha_L}} \left(\frac{\pi k_B T}{\alpha_L E_c}\right)^{\frac{2}{\alpha_L}}  {\rm Li}_{\frac{2}{\alpha_L}} \left(-e^{\frac{\mp \Omega \pm \mu/2}{k_B T}}\right)\notag \\
T_{\pm,R}(\Omega)  = -T_{0,R} e^{-\frac{2\gamma_e}{\alpha_R}} \left(\frac{\pi k_B T}{\alpha_R E_c}\right)^{\frac{2}{\alpha_R}}  {\rm Li}_{{\frac{2}{\alpha_R}}} \left(-e^{\frac{\pm \Omega \pm \mu/2}{k_B T}}\right)\notag,
\end{eqnarray}
where ${\rm Li}_x(z)$ is the Polylogarithm function. \\

\paragraph{Non-resonant sequential tunneling} Let us first assume non-resonant sequential tunneling of electrons, as described in \cite{RO}. Physically, a whole process participating to the current corresponds in this case to two distinct and successive photo-assisted tunneling events. An incoming electron first hops from one lead onto the dot while emitting a photon. Later, this electron exits the dot to the other lead while emitting another photon. These two tunneling events are considered independently. The total current results then from the net inbalance of electrons going from left to right and from right to left. The steady state current is then given by\cite{RO},
\begin{equation}
I =e ~\frac{T_{+,L}(E_d) T_{+,R}(E_d) - T_{-,L}(E_d) T_{-,R}(E_d)}{ T_{+,L}(E_d) + T_{+,R}(E_d) +  T_{-,L}(E_d) + T_{-,R}(E_d) }.
\label{rectified_current}
\end{equation}\\
We recover that $I=0$ when the bare tunneling rates and the impedances are equal. In this case, $T_{+,L}(E_d)= T_{-,R}(E_d)$ and $T_{-,L}(E_d)= T_{+,R}(E_d)$.\\

\begin{figure}[h!]
\center
\includegraphics[scale=0.36]{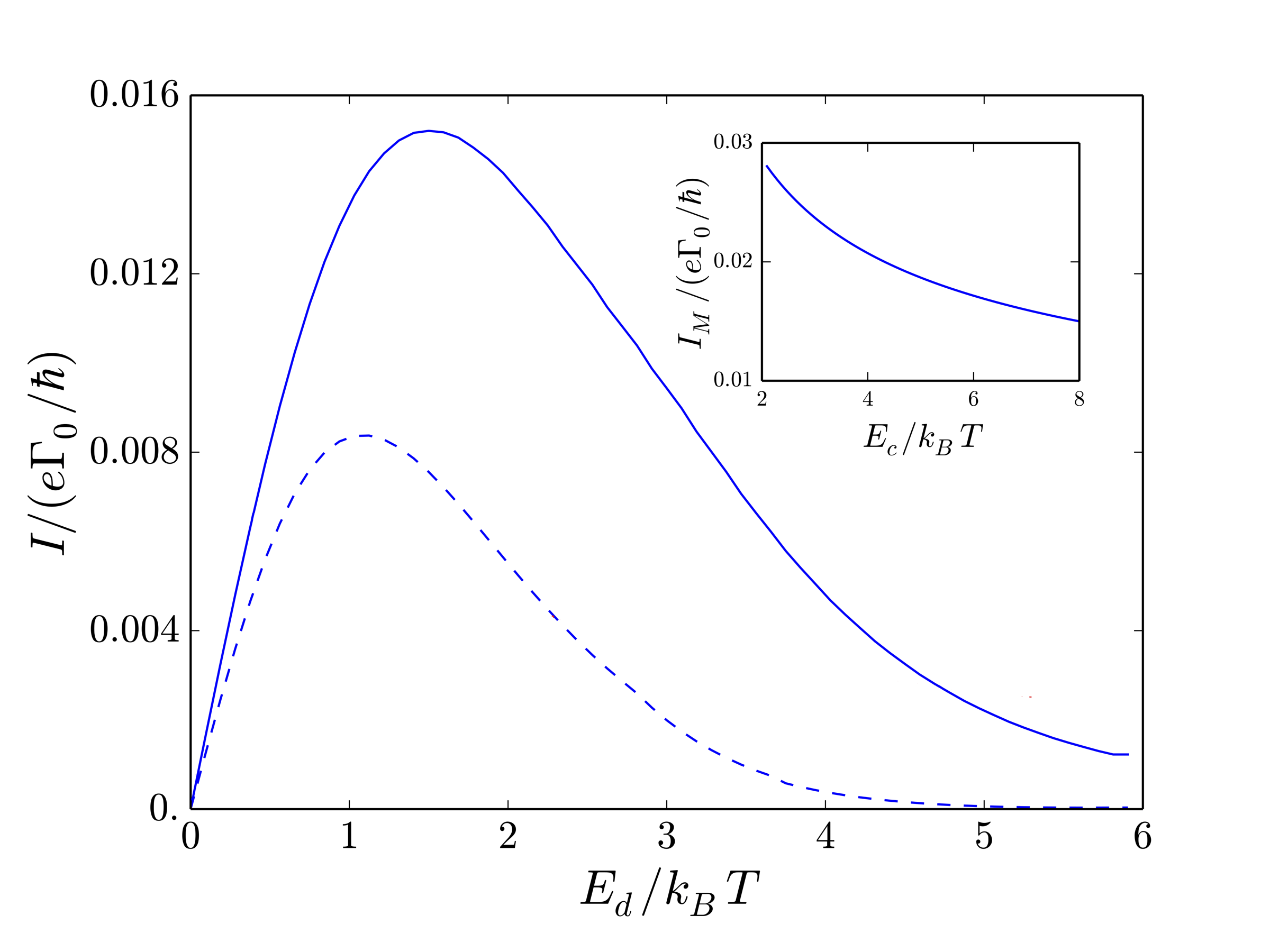}  
\caption{Rectified current with respect to $E_d/k_B T$ at zero bias. The dashed blue and the full blue curves correspond respectively to the non-resonant and the resonant case of the setup displayed in Fig.~\ref{figsetup}(b). We have $k_B T=0.125 E_c$, $\alpha_L=5$, $\alpha_R=30$, $\Gamma_0=0.125 E_c$. We do not calculate the current at temperatures of the order or greater than the charging energy. Above this threshold, the description we made is no longer accurate as the other levels of the quantum dot must be taken into account. Inset: Evolution of the maximum value of the current $I_M$ at the resonance with respect to $E_c/k_B T$.}
\label{rectified_current} 
\end{figure}

\begin{figure}[h!]
\center
\includegraphics[scale=0.44]{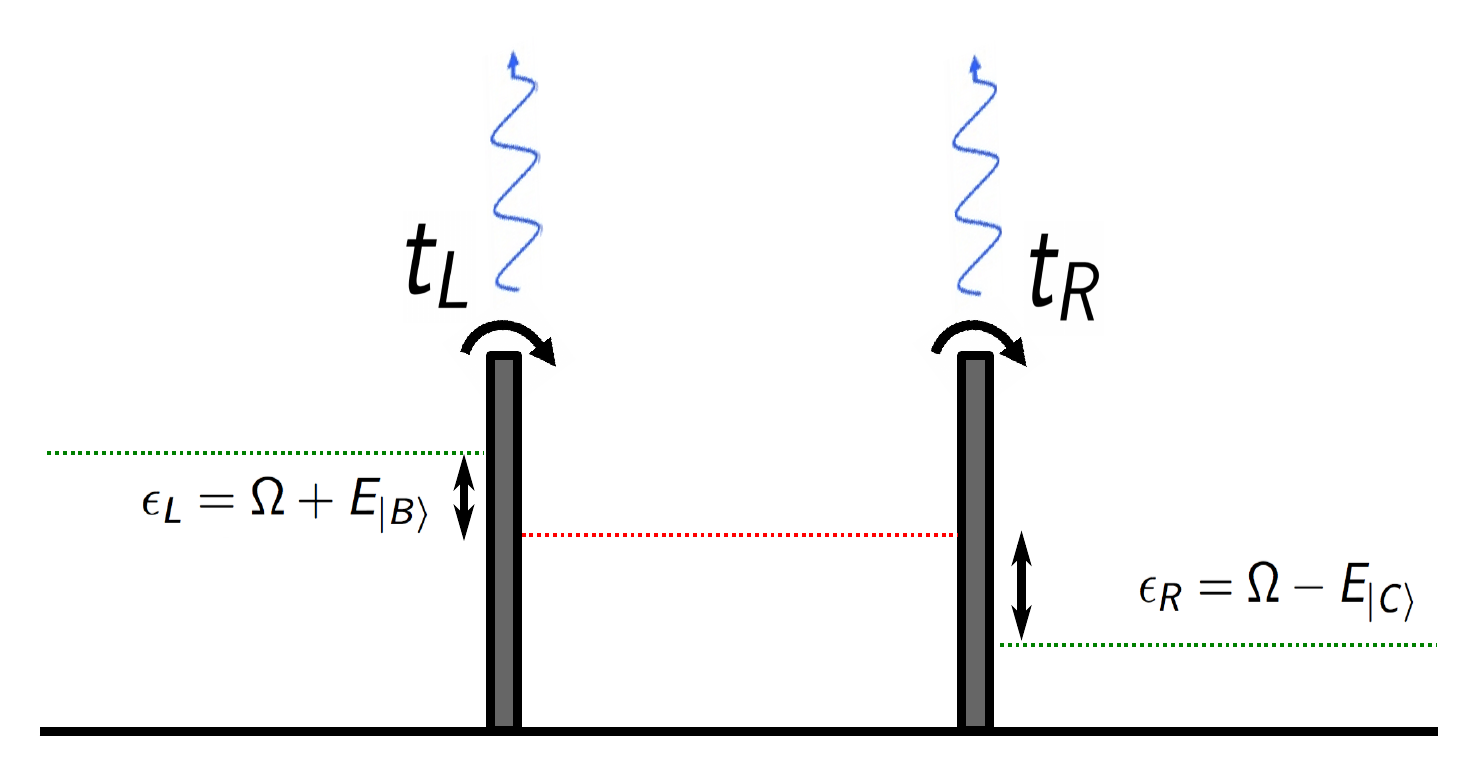}  
\caption{Resonant tunneling process. An incoming electron tunnels onto the dot while emitting a photon. If the resulting energy is close to a resonant energy of the quantum dot, the electron can coherently bounce back and forth between the two barriers before exiting the dot.}
\label{rectified_current_process} 
\end{figure}

\paragraph{Resonant sequential tunneling} As the quantum dot is well isolated from the external environment, it might be relevant to address resonant transport processes through the device. The transmitted electron can indeed coherently bounce back and forth between the two barriers and interfere constructively before it exits, as illustrated in Fig. \ref{rectified_current_process}. This case permits Fabry-Perot type resonances between the two junctions forming the quantum dot \cite{datta}. \\

In the case of a simple isolated double barrier, the transmission amplitude $t^+$ from the left to the right side results from all the possible paths with a given number of reflections between the two barriers,
\begin{align}
t^+&=t_L t_R+t_L r_R r_L  e^{i\theta}t_R+...+t_L (r_R r_L e^{i\theta} )^n t_R+... \notag \\
&= \frac{t_L t_R}{1-r_R r_L e^{i\theta}},
\label{resonant}
\end{align}
where $\theta$ is the phase acquired during one round trip between the two barriers (note that we have $t^+=t^-$). In our case, each tunneling event can be associated with the emission of one photon. Let $\epsilon_L$ be the initial energy of the electron, and $\epsilon_R$ its energy at the end of the process. During the tunneling process, emissions of photons are possible in the left and in the right lead. Let $\Omega$ be the energy of the electron after the first tunneling process and the eventual emission of photons in the left transmission line. We can write

 \begin{align}
&t^+_L (\epsilon_L,\Omega)= t_L \times \sum_{|B\rangle} \langle B| e^{i\phi_L} |0_L\rangle \delta(\epsilon_L-\Omega-E_{|B\rangle})\\
&t^+_R (\epsilon_R,\Omega)= t_R \times \sum_{|C\rangle} \langle C| e^{-i\phi_R} |0_R\rangle \delta (\epsilon_R+E_{|C\rangle}-\Omega).
\end{align}

 Making use of Eq. (\ref{resonant}), we get:
%\begin{footnotesize}
\begin{align}
t^+(\epsilon_L, \epsilon_R, \Omega)&=t^+_L(\epsilon_L,\Omega) t^+_R (\epsilon_R,\Omega)+t^+_L(\epsilon_L,\Omega) r_R r_L  e^{i\theta(\Omega)}t^+_R(\epsilon_R,\Omega)+... \notag \\
&= \frac{t^+_L(\epsilon_L,\Omega) t^+_R(\epsilon_R,\Omega)}{1-r_R r_L e^{i\theta(\Omega)}}.
\end{align}
%\end{footnotesize}
\newline

The transmission probability $\gamma^+(\epsilon_L, \epsilon_R)$ associated with this process is given by
\begin{align}
\gamma^+(\epsilon_L, \epsilon_R)=&\int d\Omega ~ t^+(\epsilon_L, \epsilon_R, \Omega) \left[t^+(\epsilon_L, \epsilon_R, \Omega)\right]^* \notag \\
=&\int d\Omega ~ \frac{|t_L|^2 |t_R|^2}{\left|1-r_R r_L e^{i\theta(\Omega)}\right|^2}\times \left\{ \sum_{|B\rangle} |\langle B| e^{i\phi_L} |0_L\rangle|^2 \delta(\epsilon_L-\Omega-E_{|B\rangle})\right\} \notag \\
&~~~~~~~~~~~~~~  ~~~~~~~~~~~ \times \left\{  \sum_{|C\rangle} |\langle C| e^{-i\phi_R} |0_R\rangle|^2 \delta (\epsilon_R+E_{|C\rangle}-\Omega)\right\}.
\label{transmission_probability}
\end{align}

The last two terms of Eq. (\ref{transmission_probability}) can be reexpressed thanks to P(E) theory (see above), and the denominator can be simplified by using the Lorentzian approximation \cite{datta}. A similar expression holds for the negative direction. We integrate now over the lead electron energies, we get the following expressions for the total transmission probabilities $\Gamma^{\pm}$
\begin{align}
\Gamma^{+}=\int d\Omega ~  \frac{T_{+,L}(\Omega) T_{+,R}(\Omega)}{\left(\frac{T_{0,L}+T_{0,R}}{2}\right)^2+\left(\frac{\Omega-E_d}{\hbar}\right)^2}, \\
\Gamma^{-}= \int d\Omega ~  \frac{T_{-,L}(\Omega) T_{-,R}(\Omega)}{\left(\frac{T_{0,L}+T_{0,R}}{2}\right)^2+\left(\frac{\Omega-E_d}{\hbar}\right)^2} .
\label{Gammas_resonant}
\end{align}
 where $T_{\pm,l}(\Omega)$ are given by Eqs. (\ref{Gamma_L}) and (\ref{Gamma_R}). In the following, we consider that the Left and Right junctions are identical. The total rectified current is $I=(e/\hbar)\left(\Gamma^{+}-\Gamma^{-}\right)$. It corresponds to a resonant sequential tunneling associated with the emission of two photons.\\

%\begin{figure}[t!]
%\center
%\includegraphics[scale=0.4]{Fig_cavity_13_04.pdf}  
%\caption{(color online) Left panel: Rectified current with respect to the left and right cavity frequencies $\omega_L$ and $\omega_R$. We have $k_B T/\Gamma_0=1.2$, $E_d/ \Gamma_0=2$, $Z_L/R_q=Z_R/R_q=0.5$ and $\mu=0$. Right panel: Rectified current with respect to $\omega_R$, for a setup with only one cavity coupled to one of the leads.}
%\label{fig_rectified_cavity_1}
%\end{figure}

Results are shown in Fig.~\ref{rectified_current}. We notice that the rectified current is maximal at a dot energy of the order of the temperature. The value of this maximal current also depends on the charging energy of the dot. The inset shows then the evolution of the maximum value of the current with respect to $E_c/k_B T$ (for $E_c/k_B T>2$). We notice that the current decreases with the charging energy, as expected from the power law evolution of $P(E)$. \\
 
 \subsubsection{Coupling to a resonator}
Let us consider the case where the lead $l$ is now coupled to a zero-temperature resonator of frequency $\omega_l=1/\sqrt{L_l C^c_l}$ (composed of an inductance $L_l$ and a capacitance $C^c_l$). We then have

\begin{align}
k_l(t)= \frac{Z_l}{R_q}\left(e^{-i\omega_l t}-1\right).
\end{align}
where $Z_l= \sqrt{L_l/C^c_l}$. This allows us to reach the following expression for $P_l(E)$,

\begin{eqnarray}
%P_l(E)=&\sum_{n=0}^{\infty}\sum_{k=0}^{n} \frac{1}{n!} \binom {n} {k}(-1)^{n-k} \left(\frac{z_l}{R_q}\right)^n \delta(E-k\hbar\omega_l) \notag\\
%=&\sum_{k=0}^{\infty} \sum_{n=k}^{\infty} \frac{1}{n!} \binom {n} {k}(-1)^{n-k} \left(\frac{z_l}{R_q}\right)^n \delta(E-k\hbar\omega_l) \notag\\
P_l(E)=&\sum_{k=0}^{\infty} \frac{1}{k!}\left(\frac{Z_l}{R_q}\right)^k e^{-\frac{Z_l}{R_q}} \delta(E-k\hbar\omega_l).
\end{eqnarray}

\begin{figure}[t!]
\center
\includegraphics[scale=0.36]{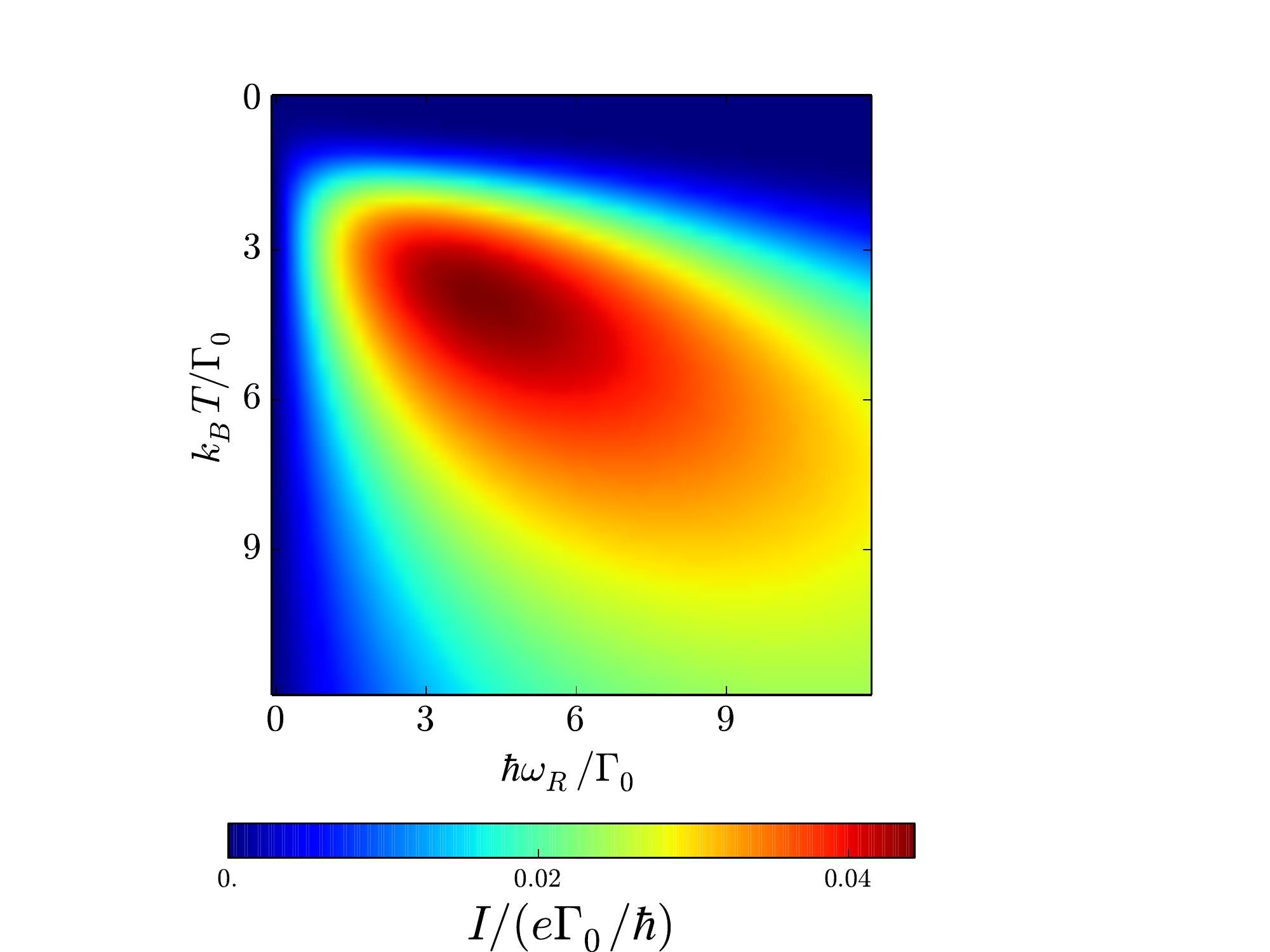}  
\caption{Rectified current with respect to the temperature, and the cavity frequency. We have $\Gamma_{0,L}=\Gamma_{0,R}=\Gamma_{0}$, $E_d/\Gamma_0=6$, $Z_R/R_q=0.5$. }
\label{fig_rectified_cavity_2}
\end{figure}

The left and right tunneling rates given by Eqs. ( \ref{Gamma_L}) and ( \ref{Gamma_R}) read

\begin{eqnarray}
T_{\pm, L} (\Omega) = T_{0,L} \sum_k  \frac{\left(\frac{Z_L}{R_q}\right)^k}{k!} e^{-\frac{Z_L}{R_q}} f(k\hbar\omega_L \pm \Omega) \\
T_{\pm, R} (\Omega) = T_{0,R} \sum_k  \frac{\left(\frac{Z_R}{R_q}\right)^k}{k!} e^{-\frac{Z_R}{R_q}} f(k\hbar\omega_R \mp \Omega).
\end{eqnarray}
These rates take into account tunneling events with the emission of several photons, and the ratio $Z_l/R_q$ determines the dominant processes occuring in the device. We will mainly consider small values of $Z_l/R_q$ where one-photon processes are the most relevant. % and we only couple the right lead to a cavity. In this case, the current is simply given by

%\be
%I=\frac{e z_R}{\hbar R_q}\int d\Omega \frac{\Gamma_0^2}{\Gamma_0^2+(\Omega-E_d)^2} F(\Omega),
%\label{current_one_cavity_one_photon}
%\ee
%where
%\be
%F(\Omega)=\left[ f\left(\Omega\right)f\left(\hbar \omega_R-\Omega\right)-f\left(-\Omega\right)f\left(\hbar \omega_R+\Omega\right)\right].
%\ee

%Here we first fix the temperature $T$ and $\Gamma_0$ such that $k_B T/\Gamma_0=1.2$. In Fig.~\ref{fig_rectified_cavity_1} we plot the rectified current for different values of $\omega_L$ and $\omega_R$, but with constant $Z_L=Z_R$. 

 The best situation corresponds to very asymmetric configurations, where one of the two cavities is suppressed. In this case, the coupling to a frequency-selective cavity leads to a rectified current substantially greater than in the case of two standard resistances in a highly asymmetric configuration. We have then photon-assisted tunneling and current rectification without any drive on the cavity, in contrast to Refs. \cite{tien_gordon,tucker,pedersen_buttiker}. We study the dependence of the rectified current with respect to the temperature and the cavity frequency in this highly asymmetric case, where only the right lead is coupled to a cavity. In this case, the current is simply given by

\begin{align}
I=\frac{e Z_R}{\hbar R_q}\int d\Omega \frac{\Gamma_0^2}{\Gamma_0^2+(\Omega-E_d)^2} \left[ f\left(\Omega\right)f\left(\hbar \omega_R-\Omega\right)-f\left(-\Omega\right)f\left(\hbar \omega_R+\Omega\right)\right].
\label{current_one_cavity_one_photon}
\end{align}

 The rectified current is written in Eq. (\ref{current_one_cavity_one_photon}) as a difference of two terms, which are products of two Fermi functions. At zero temperature both terms are zero (because one of the Fermi function is zero in both terms), and no tunneling events can occur. In the limit of high temperature, the flattening of the Fermi functions erase the asymmetry of the setup introduced by $\omega_R$, which also leads to a zero rectified current. There exists an optimal lead temperature between these two limits, for which current is maximal.\newline
 
In Fig.~\ref{fig_rectified_cavity_2} we plot the rectified current with respect to the temperature and to the cavity frequency. In this case of one-photon processes, we naturally find that the resonance occurs at a temperature $T$ such that $k_B T\simeq \hbar \omega_R$. For greater values of $Z_R/R_q$, this resonance is shifted to a greater value of the temperature $T$, which can be estimated by $k_B T\simeq k \hbar \omega_R$, where $k$ is the number of photons of the dominant process at this value of $Z_R/R_q$\footnote{The photon number follow more precisely a discrete Poisson distribution. We would however need to take into account higher energy levels of the resonator for a more complete description.}. \\

The heat engines can be characterized thanks to two physical quantities which are the maximum power that can be generated, and the efficiency of the engine. Next, we consider briefly the different characteristics of the engine.
 \subsection{Heat engine characteristics}

\label{characteristics}

\begin{figure}[h!]
\center
\includegraphics[scale=0.4]{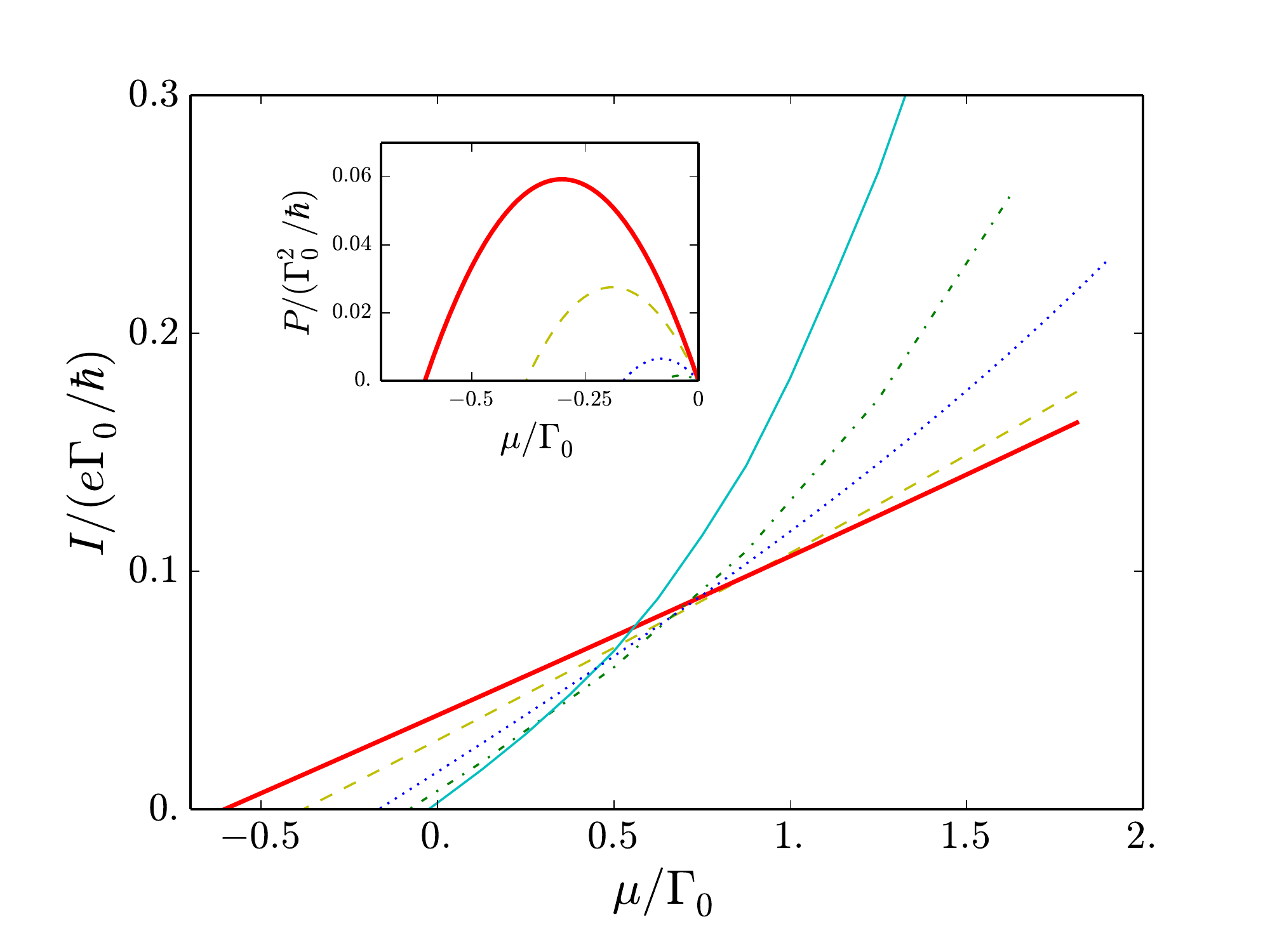}  
\caption{Rectified current with respect to an external bias at different lead temperatures, ranging linearly from the limit of zero temperature (full cyan curve) to $k_B T/\Gamma_0=3$ (full red curve). We have taken the resonant case of the setup displayed in the Fig.~\ref{figsetup} (and adjusted the value of $E_d/k_B T$ at small bias). Inset: generated power with respect to the external bias. We have $\alpha_L=5$, $\alpha_R=30$ and $\Gamma_0=0.125 E_c$.}
\label{fig_power}
\end{figure}

The generated power is given by the product $I \times V$. When $I>0$ and $V<0$, the device produces power: the generated current flows against a load potential. In Fig.~\ref{fig_power} we plot the evolution of the resonant current with respect to an external bias field at different lead temperatures. We see that the increase of the temperature has two effects: it leads to a drop in the conductance, and to the appearance of the zero-voltage rectification current. We show the maximum power that can be generated in the inset of Fig.~\ref{fig_power}. \\

\begin{figure}[h!]
\center
\includegraphics[scale=0.4]{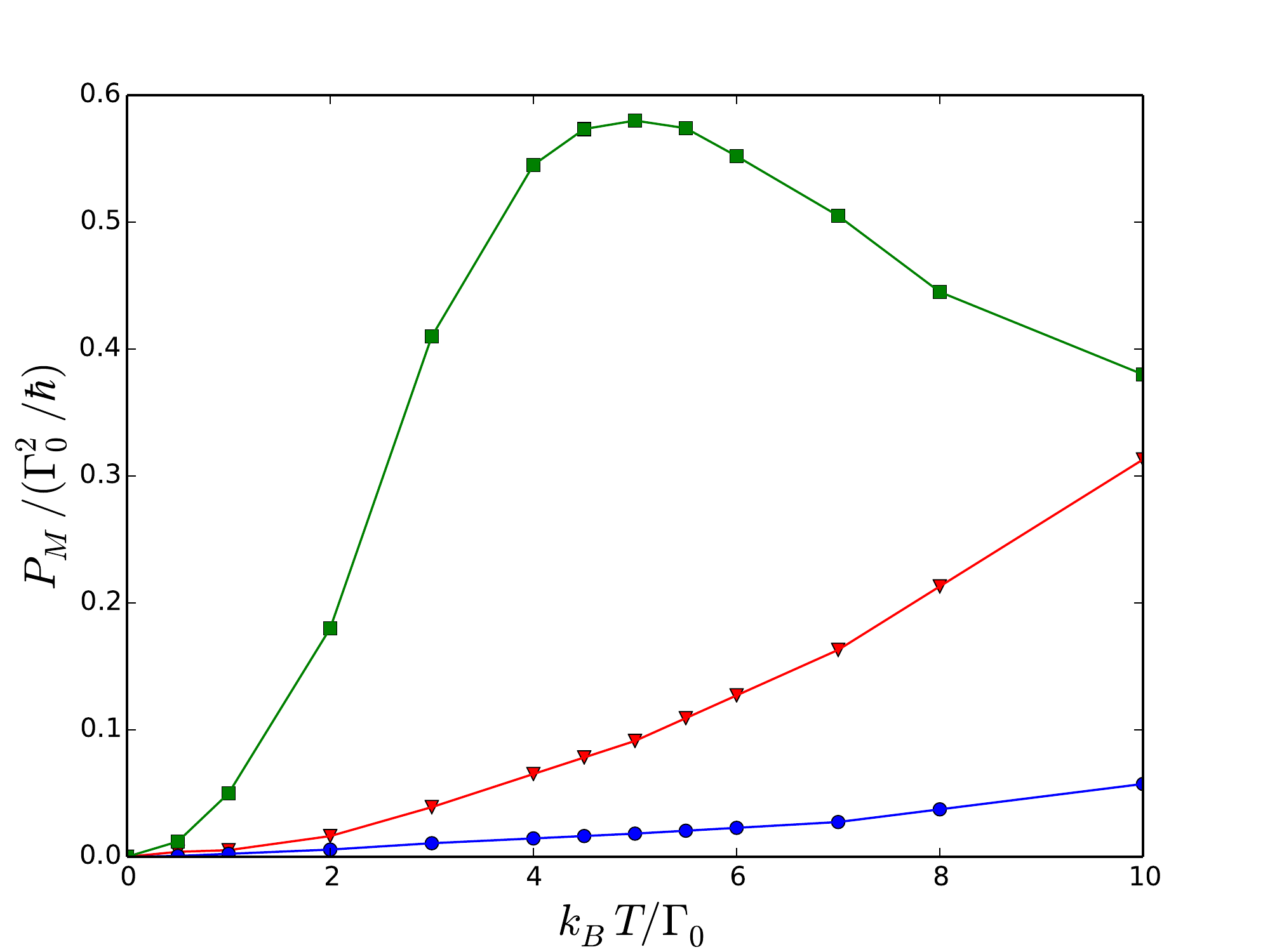}  
\caption{Maximal power $P_M$ with respect to $k_B T/\Gamma_0$. The red curve with down-pointing triangles and the blue curve with circles correspond respectively to the resonant and the non-resonant case of the setup displayed in Fig.~\ref{figsetup}. For these setups, we have $E_c/\Gamma_0=20$, $\alpha_L=5$ and $\alpha_R =30$, and $E_d$ is adjusted to maximize the power. The green curve with squares corresponds to the case of one cavity, and $\omega_R$ and  $E_d$ are adjusted to maximize the power.}
\label{maximal_power}
\end{figure}

We plot in Fig.~\ref{maximal_power} the maximum power as a function of temperature. In the single cavity case it exhibits a peak at a given temperature, while on the other hand the power generated by the circuit coupled to ohmic impedances is increasing with temperature. In all cases we must consider thermal energies which are smaller than the level spacing of the dot. Above this limit, our calculations no longer hold.\\

\begin{figure}[h!]
\center
\includegraphics[scale=0.4]{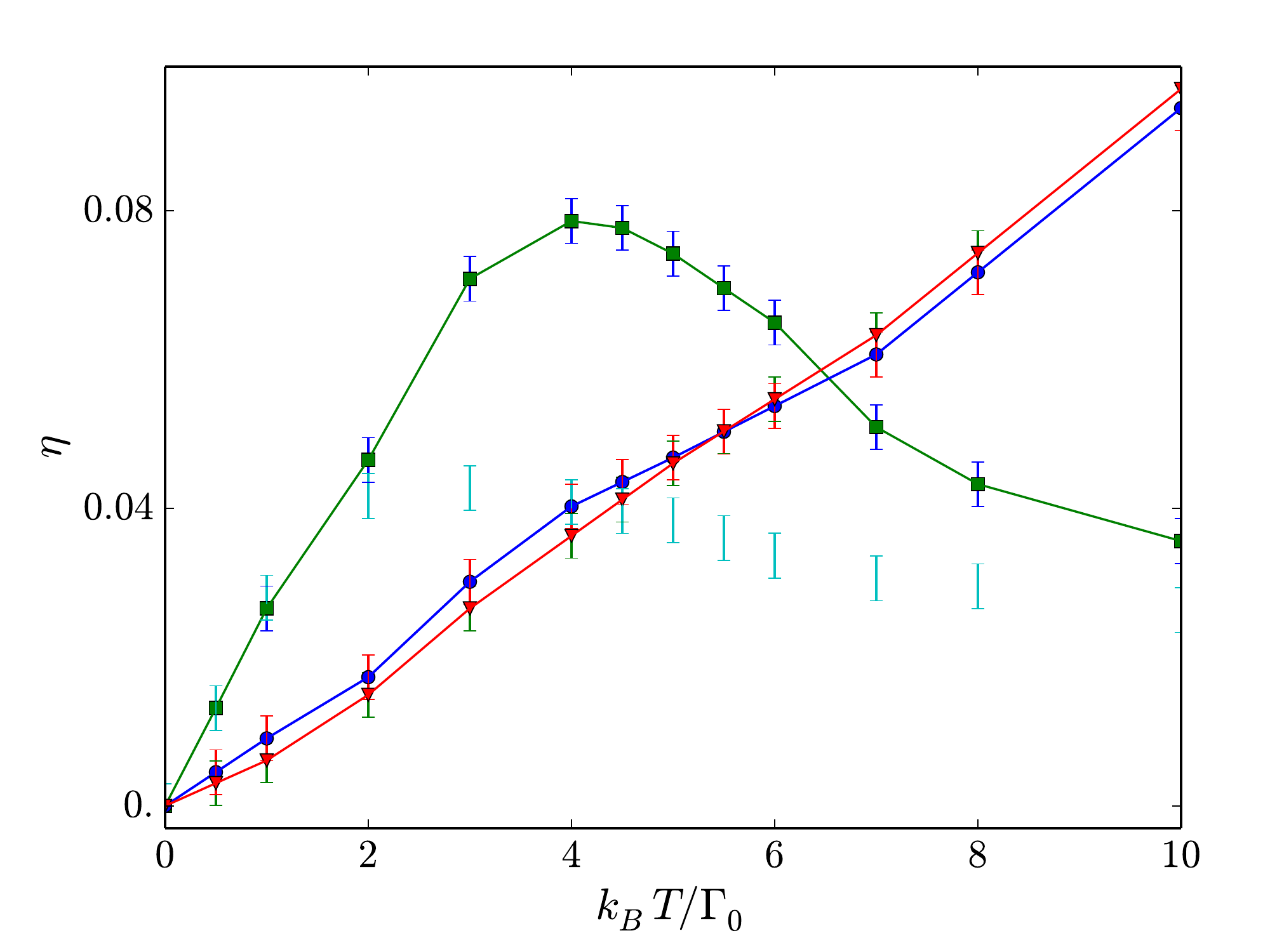}  
\caption{Efficiency of the engines with respect to $k_B T/\Gamma_0$. We use the same color and symbol legend as in Fig.~\ref{maximal_power}. Parameters are unchanged.}
\label{efficiency}
\end{figure}

The efficiency is given by $\eta=IV/(IV-J)$, where $-J$ is the heat expelled to the cold environment, which is then given by the sum of all the energy carried by the emitted photons. The heat currents associated with the tunneling events are

\begin{eqnarray}
J_{\pm, L} (\Omega) = \Gamma_{0,L} \int d\epsilon f(\epsilon \pm \Omega \mp \mu/2) P_L(\epsilon) \epsilon , \label{J_L_appendix}\\
J_{\pm, R} (\Omega) = \Gamma_{0,R} \int d\epsilon f(\epsilon \mp \Omega \mp \mu/2)  P_R(\epsilon) \epsilon \label{J_R_appendix}.
\end{eqnarray}

Defining  $J(\Omega)=J_{+, L}(\Omega)+J_{-, L}(\Omega)+J_{+, L}(\Omega)+J_{+, R}(\Omega)$, the emitted heat current is simply given by $J(E_d)$ in the case of non-resonant tunneling, while it is
\begin{align}	
J=\int d\Omega ~  J(\Omega)\frac{\left(\frac{\Gamma_{0,L}+\Gamma_{0,R}}{2}\right)}{\left(\frac{\Gamma_{0,L}+\Gamma_{0,R}}{2}\right)^2+(\Omega-E_d)^2}\notag \\
\end{align}
in the case of resonant tunneling. We plot in Fig.~\ref{efficiency} the efficiency of the Nano-engines with respect to $k_B T/\Gamma_0$. Interestingly, we remark that the efficiency of the setup displayed in Fig.~\ref{figsetup} is similar for the non-resonant and the resonant case.\\

Another heat-engine setup is introduced in Appendix \ref{appendix_characteristics}. \\

\subsection{Conclusion and discussion about higher order terms}

We explored in this section coherent transport phenomena in a quantum dot device. We notably discussed resonant tunneling processes leading to current leakage out of the device.\\

\begin{figure}[h!]
\includegraphics[scale=0.3]{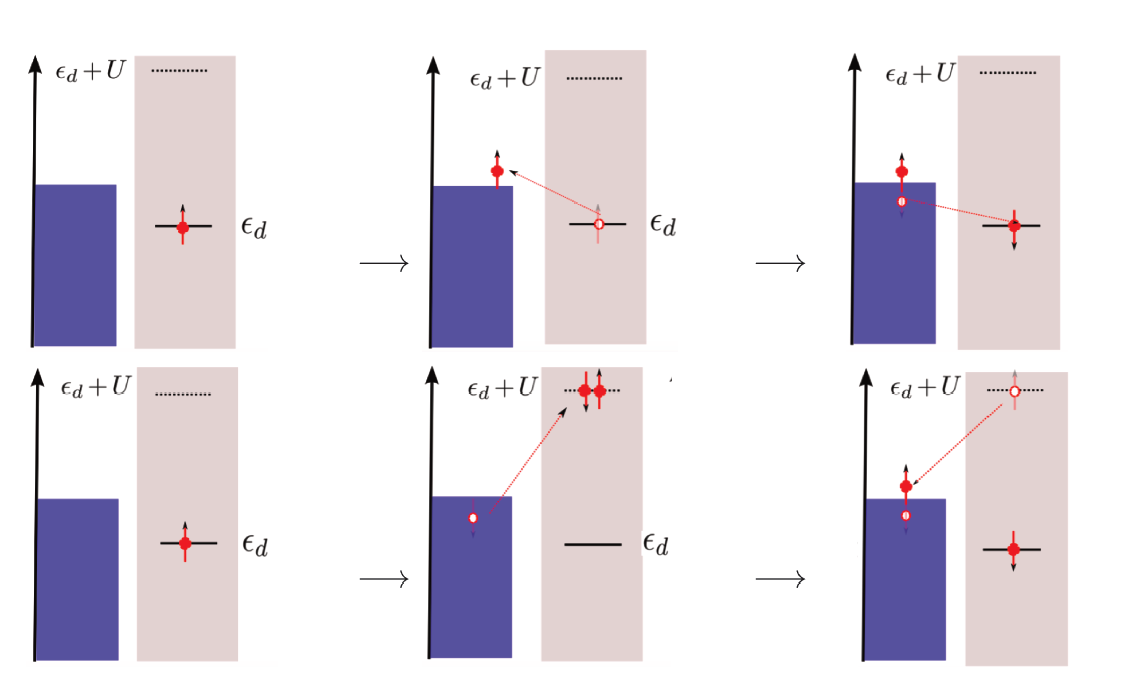}
\caption{Spin flip scattering involving a quantum dot in light grey and one reservoir lead in blue. The top panels show processes with zero intermediate occupancy, while the bottom panels show processes with double intermediate occupancy.
}
\label{Kondo_processes}
\end{figure}

 When the dot levels are strongly coupled to the surrounding environment, high-order tunneling processes become relevant. One very interesting example of the importance of high-order tunneling terms comes from the realization of the Kondo model in such quantum dot devices, occuring when the quantum dot has an average occupation close to an odd number of electrons. In a regime of low temperatures, the effect of spin-flip processes occuring via intermediate states with double or zero occupancies, which are illustrated in Fig. \ref{Kondo_processes}, becomes important. 
One may formally derive the Kondo Hamiltonian from the standard Anderson Hamiltonian, and restricting the dynamics to the single occupied states. This link was achieved in Ref. \cite{Schrieffer_wolff} by J. R. Schrieffer and P. A. Wolff, with the help of a unitary transformation. The derivation starts from the standard Anderson Hamiltonian,

\begin{align}
\mathcal{H}_{A}=\sum_{k \sigma} \epsilon_k c_{k \sigma }^{\dagger}c_{k \sigma } +\sum_{\sigma} \epsilon_d d_{\sigma }^{\dagger}d_{\sigma }+U n_{d\uparrow} n_{d\downarrow}+ \sum_{k \sigma} t \left[ c_{k \sigma}^{\dagger}  d_{\sigma} +d_{\sigma} ^{\dagger}  c_{k \sigma}  \right].
\label{Anderson}
\end{align}
The operator $c_{k \sigma}$ annihilates an electron with spin $\sigma$ in mode $k$, and the operator $d_{\sigma}$ annihilates an electron on the dot with spin $\sigma$. The last term in the right-hand side of Eq. (\ref{Anderson}) describes the hybridization between the dot and the lead. We take a tunneling term $t$ independent of $k$ for simplicity. At sufficiently low temperatures, processes described in Fig. \ref{Kondo_processes} become highly relevant. The idea developped in Ref. \cite{Schrieffer_wolff} to describe this effect is to seek an effective low-energy Hamiltonian derived from $\mathcal{H}_{A}$ in the subspace of $n_d=1$. This can be done with the help of a unitary transformation $\tilde{\mathcal{H}}_{A}=e^{U}\mathcal{H}_{A}e^{-U}$, for which the Hermitian operator $A$ has a perturbative expansion in $t$. As processes in Fig. \ref{Kondo_processes} involve two tunneling events, we build $U$ so that $\tilde{\mathcal{H}}_{A}$ do not contain terms in $\mathcal{O}(t)$. The leading processes would then be of order $\mathcal{O}(t^2)$, and describe the relevant processes. This criterion is satisfied by choosing \cite{Schrieffer_wolff},
\begin{align}
U=\sum_{k \sigma} t \left[ \frac{1}{\epsilon_k - \epsilon_d} c_{k \sigma}^{\dagger}d_{\sigma}+\frac{U}{(\epsilon_d - \epsilon_k)(\epsilon_d+U-\epsilon_k)} d_{\overline{\sigma}}^{\dagger}d_{\overline{\sigma}} c_{k\sigma}^{\dagger}d_{\sigma} \right]-h.c.
\label{Anderson_S_W}
\end{align}
The effective Hamiltonian in the subspace $n_d=1$ at order $\mathcal{O}(t^2)$ yields Kondo model,
\begin{align}
\mathcal{H}_{A}|_{n_d =1}=\sum_{k \sigma} \epsilon_k c_{k \sigma }^{\dagger}c_{k \sigma }+J \sum_{k k'} \vec{s}_{k,k'}.\vec{S},
\label{Kondo_anderson}
\end{align}
where $\vec{S}$ is an effective spin operator for the impurity, $S^z=d^{\dagger}_{\uparrow} d_{\uparrow}-d^{\dagger}_{\downarrow} d_{\downarrow} $, $S_+=d^{\dagger}_{\uparrow} d_{\downarrow}$ and $S_-=d^{\dagger}_{\downarrow} d_{\uparrow}$. The operators $\vec{\sigma}_{k,k'}$ are conduction band spin operators $\vec{s}_{k,k'}=\sum_{\sigma \sigma'} c_{k \sigma}^{\dagger} \vec{\sigma}_{\sigma \sigma'} c_{k' \sigma'}^{\dagger}$ where $\vec{\sigma}=(\sigma^x,\sigma^y,\sigma^z)^T$. The coupling $J$ is independent of $k,k'$ if only the electrons close to the Fermi level participate to the effect and we have $|\epsilon_k|,|\epsilon_{k'}| \ll |\epsilon_d|, |\epsilon_d+U|$. The interaction term in Hamiltonian (\ref{Kondo_anderson}) is now SU(2) symmetric.\\

The manifestation of the formation of the SU(2) Kondo resonance in quantum dots differs greatly from the effects in metals. At very low temperatures, the Kondo resonance develops in this case at the Fermi level, and the corresponding states in the resonance allow electrons to pass through freely when the dot is coupled to two leads. By contrast to the Kondo effect in dilute magnetic alloys, it leads in this precise case to an increase of the zero-bias conductance between the leads, which was observed experimentally in Refs. \cite{Goldhaber-Gordon:Nature,Wiel:science}. \\

Coupling several dots together may lead to more involved Kondo physics, as we will show below in the results of a recent collaboration.

\section{Experimental evidence of SU(4) Kondo physics in a DQD device}
\label{su4_kondo}
\begin{figure}[h!]
\center
\includegraphics[scale=0.5]{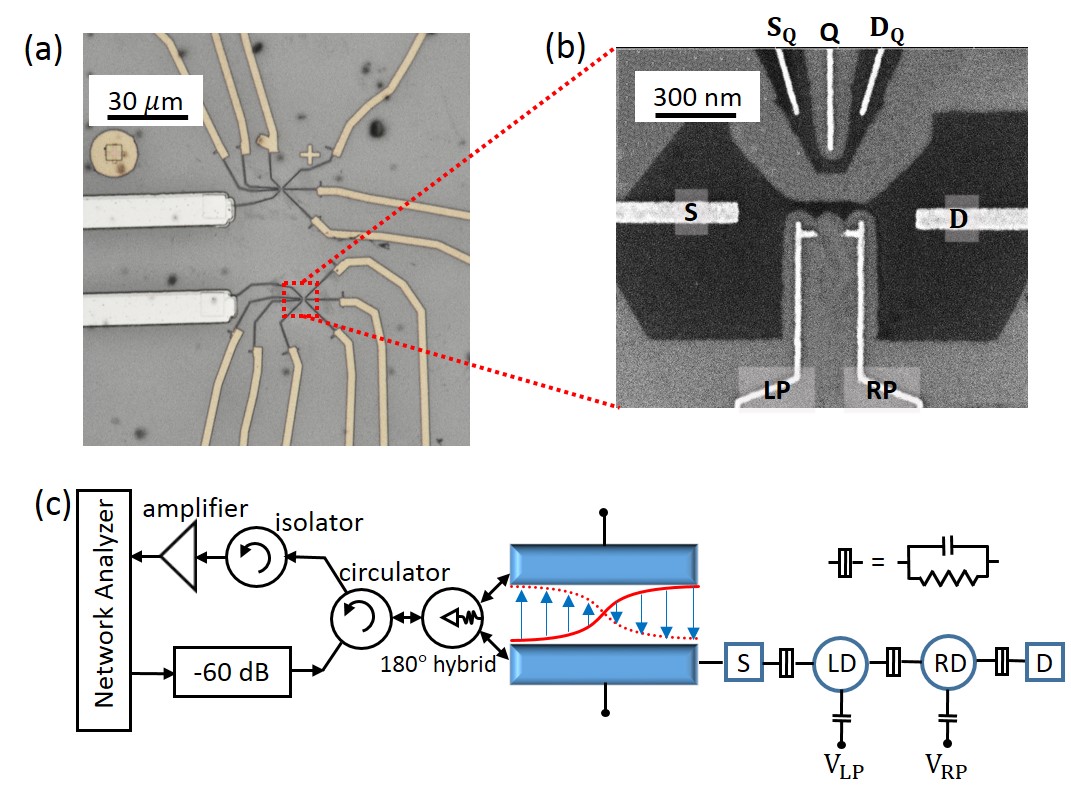}
\caption{(a) Micrograph of the DQD gate structure. (b) Sample structure of a typical etched graphene DQD. The dc voltages used to control the charge numbers in the DQD are applied via left and right plunger (LP and RP) gates. A quantum point contact with a source ($S_Q$) and drain ($D_Q$) channel and a tuning gate (Q) is integrated near the DQD. (c) Circuit schematic of the hybrid device. The half-wavelength reflection line resonator is connected to DQD's left dot (LD) at one end of its two stripelines. The right dot (RD) is connected to the drain. A microwave signal is applied to the other end of the resonator, and the reflected signal is detected using a network analyzer.
}
\label{figure1}
\end{figure}

We present here the result of a collaboration studying a Double Quantum Dot (DQD) coupled to a superconducting resonator. The device (shown in Fig.~\ref{figure1}) is mounted in a dry dilution refrigerator, its base temperature is about 30 mK. Two DQDs, made of few-layer etched graphene, are coupled to the resonator through their sources \cite{Deng:2014}, however, only one of them is used while the other is grounded all over the experiment. The DQD geometry is achieved by adjusting the gate voltages $V_{LP}$ and $V_{RP}$. A vector network analyzer (VNA) is used to apply coherent microwave driving tone and measure the reflection signal $S_{11}$ (see Fig.~\ref{figure1}(c)). The reflection signal can be measured by its amplitude ($A= |S_{11}|$) and phase ($\phi=arg(S_{11})$) components through the VNA. \\ 

\subsection{SU(4) Kondo physics in a DQD device}
\label{Kondo_DQD}

\begin{figure}
\center 
\includegraphics[scale=0.2]{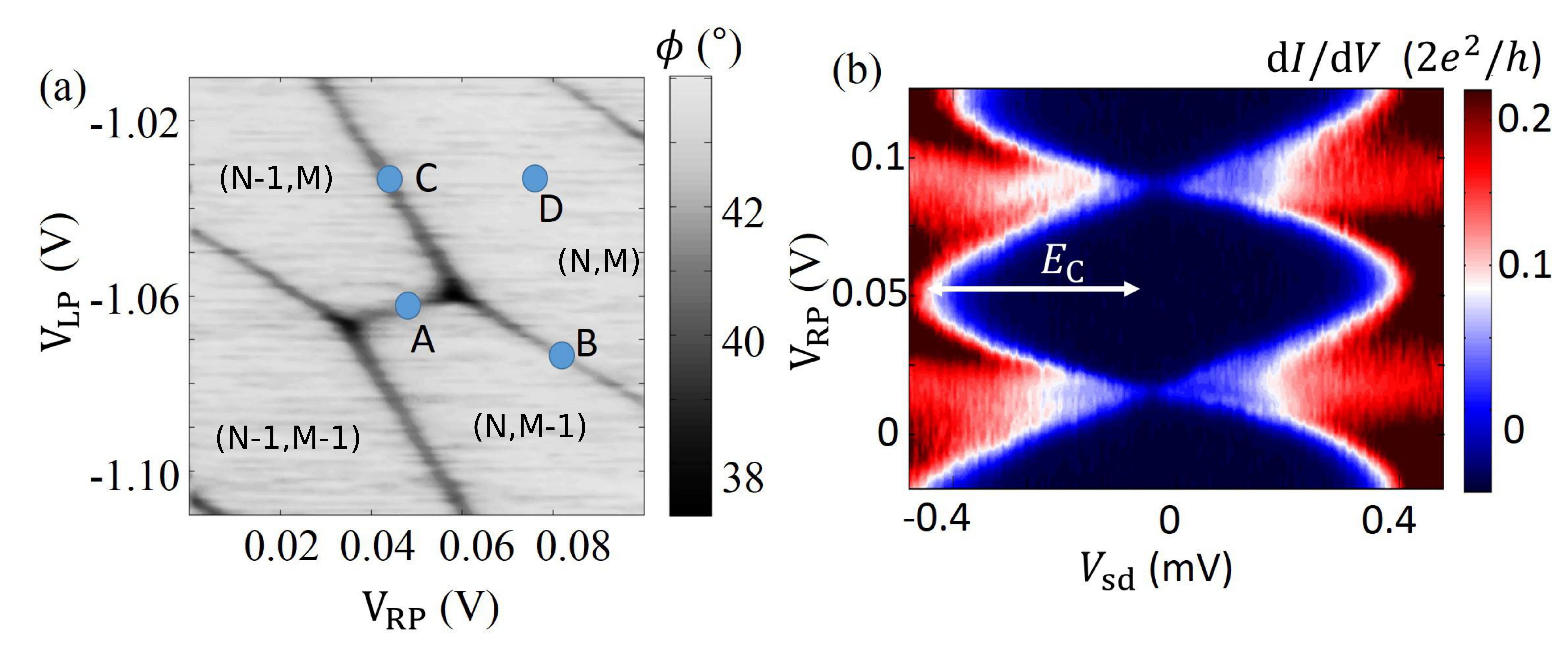}
\caption{Charge stability diagram of the DQD measured by dispersive readout using the resonator. The gate lever arms are $\sim 10\%$.}
\label{charge_stability_diagram}
\end{figure}

The charge stability diagram of a DQD shows the equilibrium charge state of the two dots as a function of the gate voltages $V_{LP}$ and $V_{RP}$ which control the Left and Right dot energy levels. Such a diagram exhibits a typical hexagonal structure \cite{Wiel:RMP}, as shown in Fig. \ref{charge_stability_diagram}. Inside each hexagon, the number of electrons in the dots is fixed and well-defined. At point D for example, the electronic configuration $(N,M)$ corresponds to $N$ electrons in the left dot and $M$ electrons in the right dot. Along the grey lines (points A, B and C), two charge configurations are degenerate. Points B and C are equivalent and lie on a degeneracy line between states with the same number of electrons in one dot and a difference of one electron in the other dot.\\

 In contrast, the point A corresponds to a charge degeneracy between two states $(N-1,M)$ and $(N,M-1)$ with the same total charge. It costs the same energy to have this extra electron on the left or on the right dot \cite{Wiel:RMP}. Other configurations on the DQD cost higher energy. The low-energy theory describing this situation can be built by introducing, in addition to the spin 1/2 $S^z$ of the electron delocalized on the DQD, the orbital pseudospin 1/2 $T^z$, projecting on these two allowed states (see Fig. \ref{Kondo_hybrid}), which are further coupled by a finite tunneling term $t' T^x$. These four quantum states generate exotic Kondo physics when coupled to the two channels conduction electrons \cite{Borda:PRL}. \\

To describe the low-energy physics, we introduce a set of operators. The operators in $\vec{S}$ correspond to the spin operators of the electron delocalized on the DQD; and the operators in $\vec{\sigma}$ and $\vec{\tau}$ act respectively on the spin space and orbital (lead) space of the electron operators. Operators $P_{\pm}=(1\pm 2T^z)/2$  are projectors on the DQD states  $(N,M-1)$ and $(N-1,M)$, while operators $p_{\pm}=(1\pm 2\tau^z)/2$ are projectors related to right and left leads. \\

 As a result of second-order tunneling processes (as described above in the SU(2) case with one lead), we reach the low energy Hamiltonian $\mathcal{H}_1=H_{kin}+H_{kondo}+ H_{orbital}+H_{assist} $ \cite{Borda:PRL}. Here, $H_{kin}$ represents the kinetic energy in the two leads. Let us present the other terms in more detail. Second-order tunneling processes are classified through

 \begin{itemize}
 \item pure Kondo terms involving spin flips contained in $H_{kondo}$, with
 \begin{align}
 &H_{Kondo}=\frac{J}{2} P_+ \vec{S} (\psi^{\dagger}\vec{\sigma}p_+ \psi )+\frac{J}{2} P_- \vec{S} (\psi^{\dagger}\vec{\sigma}p_- \psi ).
\end{align}
 Here, $\psi_{\sigma l}=\int_{-D}^D d\epsilon c_{\epsilon \sigma l}$, where $D>0$ is a cutoff in energy and $c_{\epsilon_k \sigma l}$ is the annihilation operator of an electron with enery $\epsilon$ with spin $\sigma$ in lead $l$. Formally, the operator $c$ refers to a symmetric superposition of $A$ and $B$ sub-lattice electron operators in graphene. 
 \item orbital contributions changing the lead index from say $l=L$ (Left) to $l=R$ (Right) and flipping the charge state on the DQD contained in $H_{orbital}$, with
 \begin{align}
 H_{orbital} = \frac{1}{2} \left\{ V_z T^z ( \psi^{\dagger}\tau^z \psi) +V_{\perp} \left[ T^+  (\psi^{\dagger}\tau^- \psi) +h.c.\right]\right\}.
\end{align}
 \item assisted tunneling processes entangling the charge and spin $H_{assist}$, with
 \begin{align}
 H_{assist} = Q_z T^z \vec{S} \psi^{\dagger}\tau^z\vec{\sigma}+ Q_{\perp} \left( T^+  \vec{S} \psi^{\dagger}\tau^-\vec{\sigma}\psi +h.c.\right).
 \end{align}
 \end{itemize}
 The couplings above can be written in terms of the tunneling rate to the right/left lead $\Gamma_{\pm}$ and the charging energy $E_c$ of the DQD respectively as~\cite{Borda:PRL} $J=Q_z=(\Gamma_++\Gamma_-)/4E_c$, $Q_{\perp}=V_{\perp}=\sqrt{\Gamma_+\Gamma_-}/E_c$.\\

The addition of the orbital pseudo-spin degree of freedom changes the symmetry of the problem compared to Hamiltonian (\ref{Kondo_anderson}). We have here a SU(4) symmetry for the low-energy interaction term. This new symmetry will impact the properties of the Kondo resonance.\\

As studied in Refs. \cite{Borda:PRL,Simon:PRB,Le_Hur:PRB,Rosa,Nozieres_ILTP,Ian,Mora}, the system flows to a Kondo Fermi-liquid fixed point where the spectral function on the DQD can be modeled by an effective resonant level model, quite robust to (charge) noise effects \cite{Meirong}. In this description the spectral function of the DQD can be modeled by an effective resonant level model \cite{Le_Hur:PRB}, associated with spinless fermionic operators $d$ and $d^{\dagger}$. These operators $d$ and $d^{\dagger}$ are defined in correspondance with the processes involving the orbital operators $T^-$ and $T^+$ respectively. We stress that even though we describe the effective resonant level with spinless operators, the spin of the electrons is a highly relevant quantity: the joint effect of the orbital and spin degrees of freedom lead to a SU(4) Kondo resonance at $\epsilon_0$ of the order of $T_K$, above the Fermi surface. The position of the resonance must be contrasted with the case of SU(2) Kondo effect, for which it is located at the Fermi level. This Fermi-liquid picture constitutes a phenomenological description which greatly simplifies the problem and proved to describe well the low-temperature physics \cite{Nozieres_ILTP}.  \\
 
 Experiments have reported the occurrence of this emergent $SU(4)$ symmetry at the low-energy fixed point \cite{Keller:NatPhys,Finkelstein,Tarucha,TakisNoise,Pablo}. \\

%The remarkable fact is that the bare parameters of the model are replaced by a single parameter describing the low-energy fixed point, namely the Kondo energy scale $T_K$.

In this strong coupling regime, the system can be described by a the spectral function of the form :
\begin{equation}
\rho(\omega) = \frac{1}{\pi} \frac{\Gamma(V_{sd})}{(\hbar\omega-\epsilon_0)^2 + \Gamma(V_{sd})^2},
\label{spectral_function}
\end{equation}
where the position of the resonance $\epsilon_0$ is of the order of $T_K$ \cite{Le_Hur:PRB,Mora}. In this phenomenological description, the width $\Gamma(V_{sd})$ for the pseudo-fermion $d$ is known to be voltage dependent \cite{Rosch}, due to decoherence effects.\\

\subsection{Light-matter coupling and Input-Output theory}

\begin{figure}
\center
\includegraphics[scale=0.2]{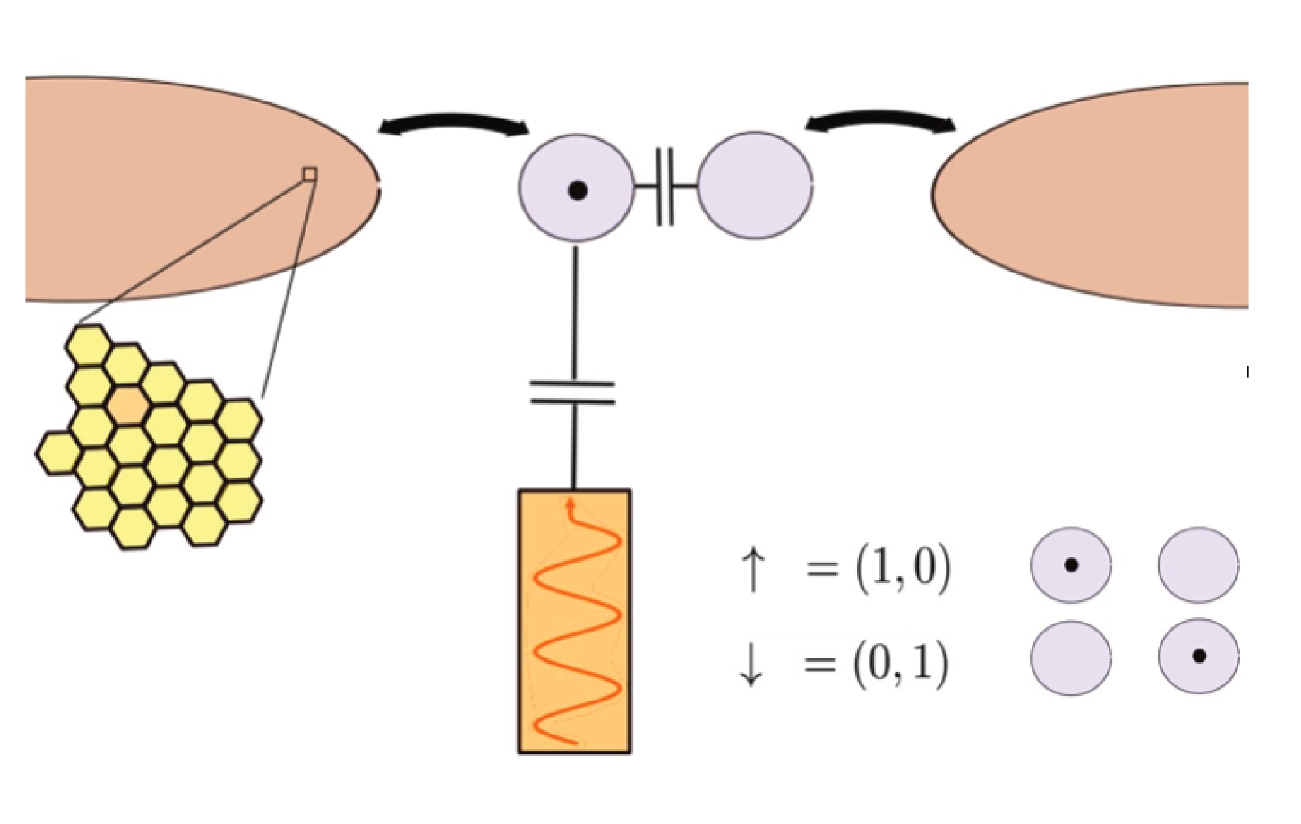}  \\
\caption{Schematic representation of the Kondo DQD device}
\label{Kondo_hybrid}  
\end{figure}

The DQD device is coupled on its left side to a resonator (see Fig. \ref{figure1}). The description of the coupling between the electronic degrees of freedom and the bosonic modes of the resonator requires some care, as described in Ref. \cite{Cottet:PRB_2}. A full analogy with the capacitive coupling described in circuit QED is not possible due to the presence of the orbital degree of freedom in this precise case. Assuming that the dots are much smaller than the wavelength of the resonator excitation, the interaction between microwave photons and the device can be approximated by a dipolar coupling \cite{Childress:PRA,Petersson_Peta,Cottet:PRB_2}, and the hybrid Hamiltonian reads $\mathcal{H}_1+\mathcal{H}_{2}$ with
\begin{align}
\mathcal{H}_{2}=\lambda (a+a^{\dagger}) T^z+\omega_0 \left(a^{\dagger}a+\frac{1}{2}\right),
\label{light_coupling}
\end{align}
where $\lambda$ quantifies the strength of the light-matter coupling. Fig. (\ref{Kondo_hybrid}) shows a pictorial representation of the setup at point A.   \\

In order to probe this hybrid system, we can consider that the bosonic mode $a$ is capacitively coupled to a very long transmission line, and the Hamiltonian reads

\begin{align}
\mathcal{H}_{total}=\sum_k \omega_k b_k^{\dagger} b_k +(a+a^{\dagger})\sum_k \lambda_k (b_k +b_k^{\dagger})  +  \mathcal{H}_1+\mathcal{H}_{2},
\label{light_coupling_2}
\end{align}
where $b_k$ describes photonic modes of the transmission lines. We follow Refs. \cite{Clerk:RMP,Le_Hur:PRBR,Marco:PRB} and derive Input-Output relations by writing the equation of motion of the operators $b_k$.
\begin{align}
\dot{b}_k=-i \omega_k b_k -i g_k (a+a^{\dagger}).
\label{EOM_bk_input_output}
\end{align}
We integrate this equation both forwards and backwards in time and reach 
\begin{align}
b_k (t)=e^{-i \omega_k (t-t_0)}b_k (t_0)-i \lambda_k\int_{t_0}^t d\tau e^{-i\omega_k (t-\tau)} (a+a^{\dagger})(\tau),\label{EOM_1}\\
b_k (t)=e^{-i \omega_k (t-t_1)}b_k (t_1)-i \lambda_k \int_{t}^{t_1} d\tau e^{-i\omega_k (t-\tau)} (a+a^{\dagger})(\tau),\label{EOM_2}
\end{align}
with $t_0<t<t_1$. We now introduce the input and output voltages as,
\begin{align}
&V_{in} (t)=\sum_k \lambda_k \left[ e^{-i \omega_k (t-t_0)} b_k (t_0)+e^{i \omega_k (t-t_0)} b_k^{\dagger} (t_0) \right]\\
&V_{out} (t)=\sum_k \lambda_k \left[ e^{-i \omega_k (t-t_1)} b_k (t_1)+e^{i \omega_k (t-t_1)} b_k^{\dagger} (t_0) \right].
\end{align}
Using Equations (\ref{EOM_1}) and (\ref{EOM_2}), one can relate the input and output fields, and we find that
\begin{align}
V_{out} (t)= V_{in} (t)-2 \sum_k g_k^2 \int_{t_0}^{t_1} d\tau \sin \omega_k (t-\tau)(a+a^{\dagger})(\tau).
\end{align}
Introducing a susceptibility defined by $\langle( a+a^{\dagger}) (\tau) \rangle=\int_{t_0}^{\infty} \chi_{xx} (\tau-t') \langle V_{in} (t') \rangle$ we reach,
\begin{align}
\langle V_{out} (t) \rangle=\int_{t_0}^{\infty} dt' \underbrace{\left[\delta(t-t')-2 \int_{t_0}^{t_1} d\tau L_1(t-\tau)  \chi_{xx} (\tau-t') \right]}_{r(t-t')} \langle V_{in} (t') \rangle.
\end{align}
The expression above defines a reflection coefficient $r(t-t')$. Taking the particular case of $t_0 \to -\infty$ and $t_1 \to +\infty$ and after a change of variables $\tau \to \tau -t'$, we get
\begin{align}
r(t-t')=\delta(t-t')-2(L_1 * \chi_{xx}) (t-t').
\end{align}
From the convolution theorem we recover \cite{Le_Hur:PRBR,Marco:PRB}
\begin{align}
r(\omega)= 1 - 2iJ(\omega)\chi_{xx}(\omega).
\label{reflection_coeff}
\end{align}
The susceptibility $\chi_{xx}$ defined above is the response of the photon coherence to an input signal $V_{in}$, which is also the quantity studied in the linear response regime by the standard Kubo formalism. Next, we present light measurements at point A and show how they relate to this SU(4) Kondo model. From Eq. (\ref{reflection_coeff}), we indeed see that the reflection coefficient permits to probe light-matter interaction through $\chi_{xx}(\omega)$. Note that in the case of an open line, a factor $-1$ multiplies result (\ref{reflection_coeff}).

\subsection{Experimental results at the charge degeneracy point A: Phase and Amplitude of Reflected Microwave Signal}

\begin{figure}[h!]
\center
\includegraphics[width=0.35\columnwidth]{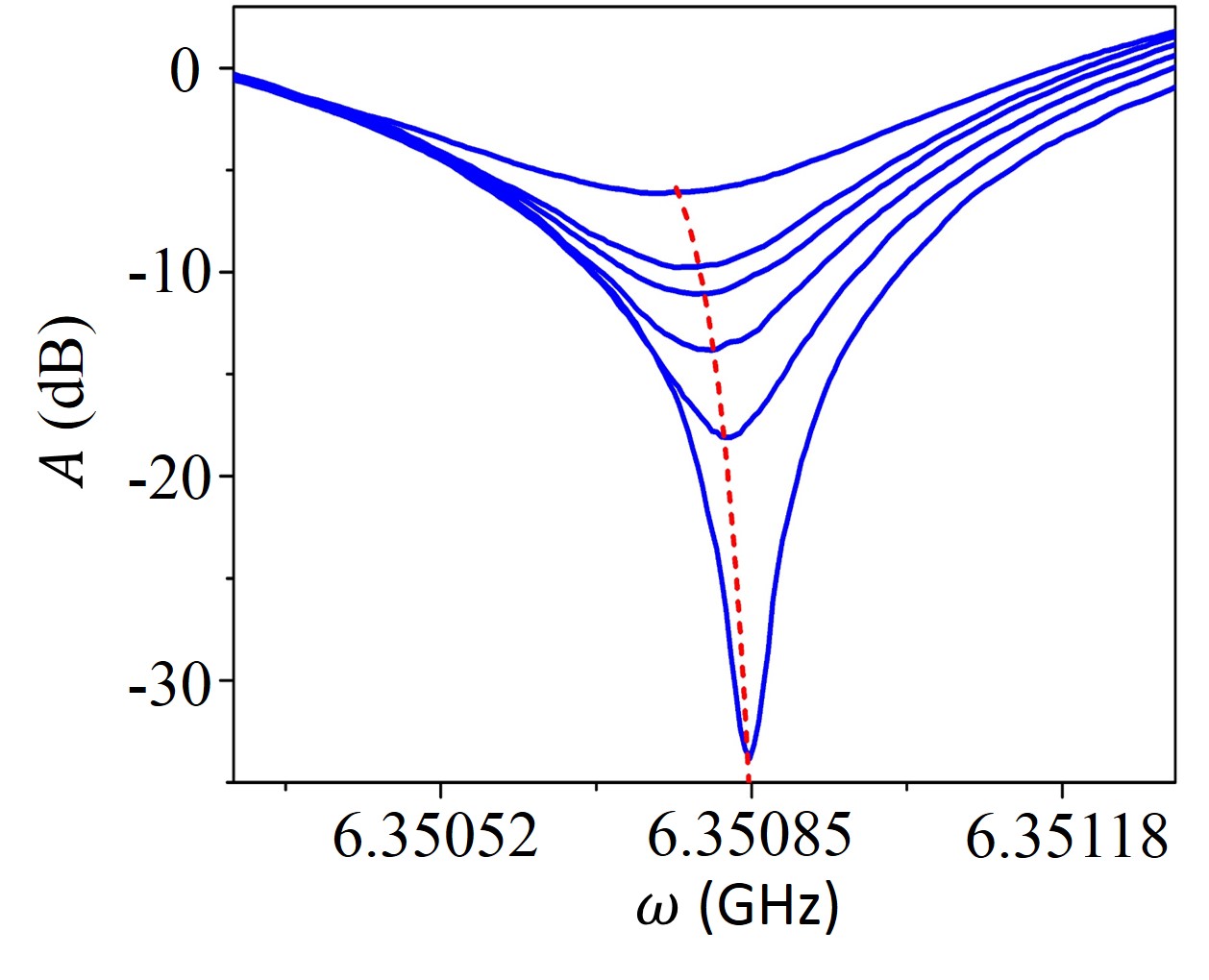}
\caption{Amplitude response $A=|r|$ as a function of the driving frequency $\omega$, for various bias voltages (0, 0.5, 0.6, 0.7, 0.8, 1 mV) at point A. The red dashed line shows the fitted resonance frequency, shifted by the electron transport.}
\label{figure3}
\end{figure}

We can first characterize the effect of the DQD on the resonator at point A by measuring the amplitude response as a function of the driving frequency, for various bias voltages, as shown in Fig.~\ref{figure3}. The coupling to the electronic system leads to a renormalization of the bare resonator frequency $\omega_0$ to a voltage-dependent value $\omega_0^*(V_{sd})$, which can be seen in the amplitude response of the cavity (and associated with the red dashed line in Fig. \ref{figure3}) \cite{Marco:PRB}. All light measurements will be carried out at this precise value of the driving frequency $\omega=\omega_0^*(V_{sd})$. \\

\begin{figure}[h!]
\center
\includegraphics[scale=0.3]{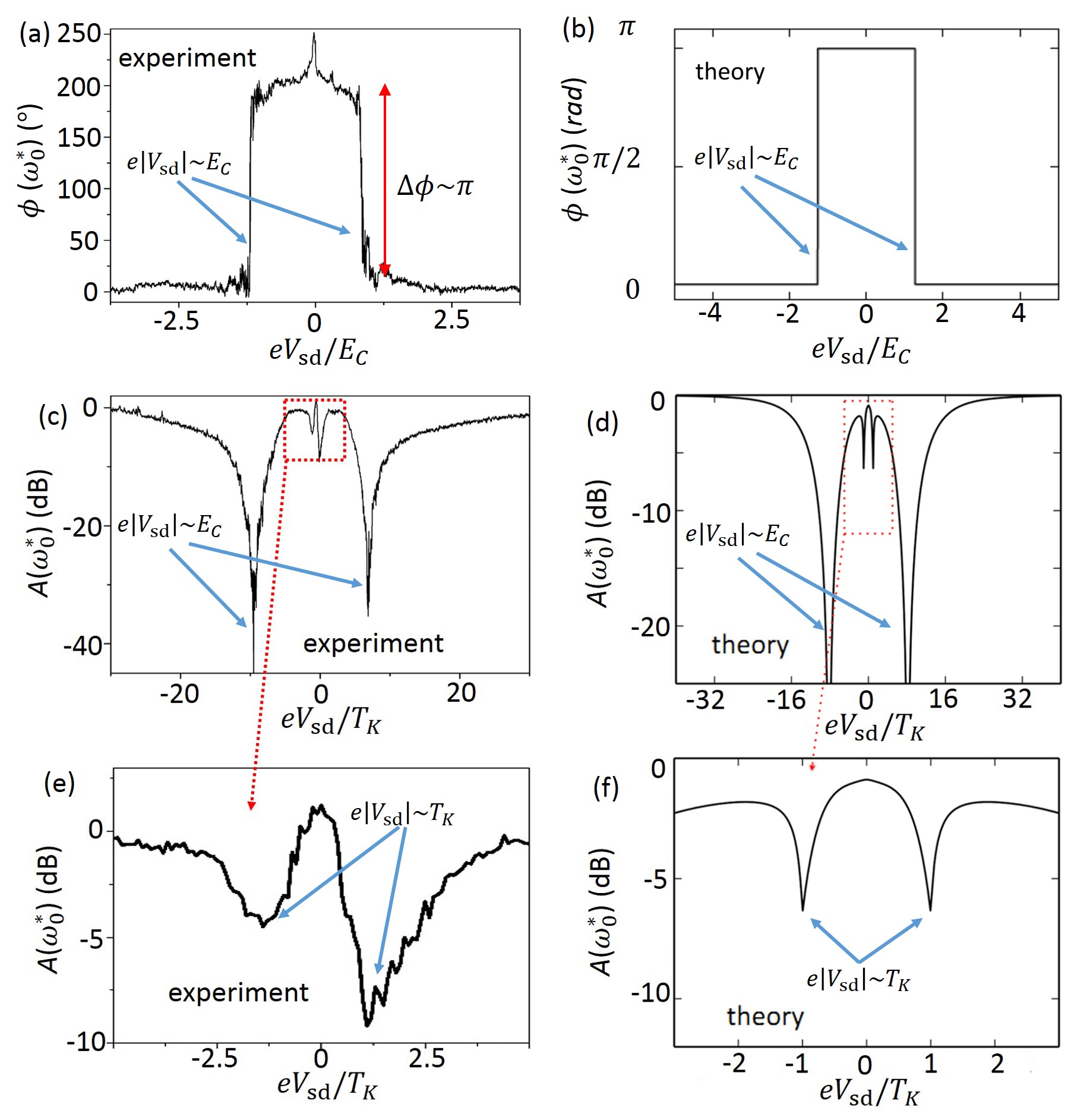} %\includegraphics[scale=0.32]{figure4_29_01.jpg}
\caption{Experimental result for the the reflection coefficient $r(\omega_0^*)=A(\omega_0^*) e^{i\phi(\omega_0^*}$ at point the charge degeneracy point A. Phase $\phi(\omega_0^*)$ (a) and Amplitude $A(\omega_0^*)$ (b) and (c), as a function of the bias voltage; the zoom focuses on low-energy features. (c)-(d)-(e) Theory results based on the effective quantum impurity model (see text). }
\label{figure5}
\end{figure}

Measurements of the reflection coefficient are first carried out at a temperature of 30mK. We show in Fig. \ref{figure5} the evolution of the phase $\phi$ and amplitude $A$ responses at the resonance frequency $\omega_0^*(V_{sd})$ with respect to bias voltage. A very robust phase of $\pi$ is observed from bias voltage $V_{\rm sd}\simeq 0$ to $V_{\rm sd}\simeq \pm$0.4 mV, where the phase drops to zero. The amplitude (in dB) shows also two pronounced dips around $V_{\rm sd}\simeq \pm$0.4 mV. We notice that these features in the experimental results can be associated with the charging energy $E_c$. Interestingly, we also notice that the low bias regime shows a (slightly asymmetric) dip structure in the amplitude response. \\

In the following, we interpret the measurements presented above in the light of the Fermi-liquid picture of the SU(4) Kondo effect. We use input-output theory  \cite{Gardiner:PRA,Clerk:RMP,Zhang:APL} and show that this picture explains well the robust $\pi$ phase as well as the shape of the amplitude measurements .

\subsection{Interpretation of the experimentals measurements}
The input-output theory (see Refs. \cite{Gardiner:PRA,Clerk:RMP,Zhang:APL,Le_Hur:PRBR,Marco:PRB} and above) enables us to compute the reflection coefficient of the microwave signal. We follow Ref. \cite{Marco:PRB} where the authors considered a similar microwave resonator coupled to an electronic system described by an Anderson-Holstein Hamiltonian with one central quantum dot. The reflection coefficient $r$ at the driving frequency 
$\omega$ and bias voltage $V_{sd}$ takes the form :
\begin{equation}\label{eqn:S11}
r(\omega,V_{sd}) = -1 + 2iJ(\omega)\chi_{xx}(\omega,V_{sd}),
\end{equation}
The susceptibility $\chi_{xx}^R$ for the photon is defined as \cite{Marco:PRB}
\begin{equation}\label{eqn:chi}
\chi_{xx}(\omega) = \frac{\omega_0}{\omega^2 - \omega_0^2 -  \omega_0 \Pi^R(\omega) + i J(\omega)\omega_0},
\end{equation}
where the photon self-energy $\Pi^R = \Re e \Pi^R - i \Im m \Pi^R$ absorbs the light-matter coupling. In Eqs.~(\ref{eqn:S11}) and (\ref{eqn:chi}), $J(\omega)$ characterizes the photonic dissipation due to the coupling of the resonator to an external long transmission line. From Eq. (\ref{eqn:chi}), the renormalized cavity frequency is found to verify the self-consistent equation
\begin{equation}
\omega_0^*(V_{sd})^2=\omega_0^2+\omega_0 \Re e\Pi^{R}\left[\omega_0^*(V_{sd}) \right].
\label{renormalized_frequency}
\end{equation}\\

At resonance $\omega=\omega_0^*$, we find for $S_{11}$ at bias voltage $V_{sd}$, 
\begin{eqnarray}\label{eqn:Eq_fit}
S_{11}[\omega_0^*(V_{sd})]&=& \frac{\kappa_{\rm i}-\kappa_{\rm e}}{\kappa_{\rm i}+\kappa_{\rm e}},
\end{eqnarray}
where $\kappa_i = J(\omega_0^*(V_{sd}))$ and $\kappa_e = \Im m\Pi^R(\omega_0^*(V_{sd}))$. $\kappa_i$ represents the dissipation from the photon system only, while $\kappa_e$ characterizes the light-matter coupling. It is important to notice that in general $\kappa_i$ and $\kappa_e$ at $\omega=\omega_0^*(V_{sd})$ both depend on the applied bias voltage $V_{sd}$.\\

In the absence of light-matter coupling ($\kappa_e=0$), we would have $r(\omega_0^*(V_{sd}))=1$, in agreement with a phase of zero for an open line. A phase of $\pi$ on the other hand characterizes a rather strong light-matter interaction ($\kappa_e>\kappa_i$). The phase of  $\pi$ at low bias fields at point A is then in accordance with a creation of a bound state \cite{Karyn:CR}. \\

On the other hand, at large bias voltages, the current through the DQD is large and results in strong decoherence effects for the electronic excitations. As a consequence we expect the photonic dissipation through the electronic system, $\kappa_e$, to be shorter than the typical dissipation rate through the transmission line, resulting in $\kappa_e< \kappa_i$ and a phase which is zero (or $2\pi$). The main question that arises from this reasoning is the following: for which value of the bias voltage does this transition from $\pi$ to $0$ occur? \\

%We now argue that the experimental results in Figs. \ref{figure4} and \ref{figure5} are not compatible with a simple analysis of the electronic system (biased leads plus double quantum dots) based on a resonant level model as was done in Ref.~\cite{Marco:PRB}.
%Indeed in this picture the electronic lifetime would be essentially bias independent and set by the width of the resonance $\Gamma$.  Results of Figs.~\ref{figure4} and \ref{figure5} would then suggest $\Gamma\simeq E_c=0.4$meV. Such a large value of $\Gamma$ seems unphysical (in the context of a DQD weakly coupled to graphene leads). Moreover, the electronic levels in the Coulomb diamond would not be well defined. This is also in contradiction with the presence of two distinct energy scales in the amplitude of the reflected microwave signal. 
%In contrast, we attribute the robustness of the $\pi$ phase and the low bias features of the amplitude response to a formation of a bound state between light and matter in the Kondo regime. 

Following the description provided in the previous section, the system can be treated by analogy with the case of one single dot studied in Ref. \cite{Marco:PRB}, the only difference being that the position $\epsilon_0$ and the width $\Gamma(V_{sd})$ of the electronic level are determined by the SU(4) fixed point. We can compute the electron-induced photon lifetime, $\Im m\Pi^R (\omega_0^*(V_{sd})) $, and its evolution with bias voltage to answer the question about the phase of transmitted photons. Using a resonant level model for the electronic system at the charge degeneracy point A, then gives  \cite{Marco:PRB}:
\begin{align}
\label{imp_pi}
\Im m\Pi^R (\omega_0^*) &= \lambda^2 f_{\Gamma}(\omega_0^*) \sum_{\alpha ,a=\pm} \alpha \arctan \left(\frac{\mu_{a} -\epsilon_0 + \alpha\hbar\omega_0^*}{\Gamma}\right) \notag \\
+& \lambda^2 f_{\Gamma}(\omega_0^*)
\sum_{\alpha ,a=\pm} \frac{\Gamma}{\omega_0^*}\ln \left(\frac{(\mu_{a} -\epsilon_0+\alpha\hbar\omega_0^*)^2 + \Gamma^{2}}{(\mu_{a}-\epsilon_0)^2+\Gamma^{2}} \right),
\end{align}
where $f_{\Gamma}(\omega_0^*)=\Gamma/(4\pi^2\Gamma^2+\pi^2\hbar^2\omega_0^{*2})$. In addition,  $\mu_{a}=a e V_{sd}/2$ are the chemical potentials associated with each lead, where formally $a=+1$ for left (L) lead and $a=-1$ for right (R) lead. From the (weak) light shift in Fig. 3 (given by the red dashed curve), we estimate a relatively small light-matter coupling $\lambda/\omega_0\sim 10^{-2}$ which justifies a perturbative calculation in $\lambda$.  Formally, this form of $\Im m\Pi^R (\omega_0^*)$ is obtained by considering the limit $|t'| \lambda/\omega_0^* \ll \lambda$.\\

We take a bias dependent width $\Gamma(V_{sd})$ for the pseudo-fermion $d$ \cite{Rosch} computed with perturbative approaches, 

%Decoherence effects on the Kondo resonance will lead to a bias dependent Kondo width $\Gamma(V_{sd})$. In particular, following the Non-Crossing Approximation calculation in Ref.~\cite{Rosch} and taking $\Gamma(V_{sd})$ of the form
\begin{align}
\Gamma&=T_K \textrm{ for } eV_{sd} \ll T_K,\label{gamma_Vsd_1}   \\
\Gamma &\sim e V_{sd}/\ln^2 (eV_{sd}/T_K) \textrm{ for } e V_{sd} \gg T_K \label{gamma_Vsd_2}.
\end{align}
At large bias voltages, the current produces dissipation and decoherence effects on the Kondo resonance. A polynomial interpolation is performed between small and large biases. 
\begin{figure}[h!]
\center
\includegraphics[width=0.5\columnwidth]{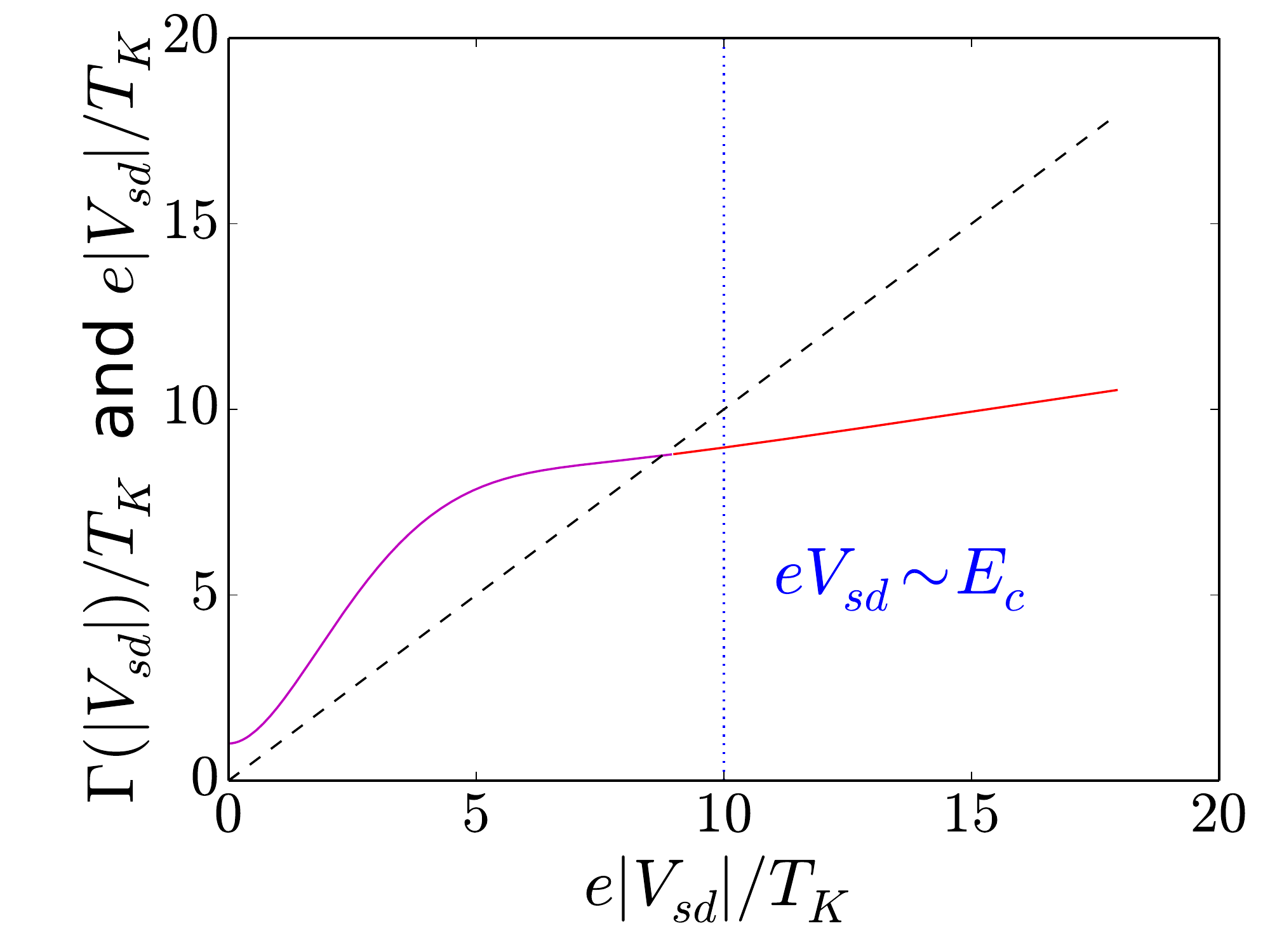}
\caption{Evolution of $\Gamma/T_K$ with respect to $e V_{sd}/T_K$. At low bias voltages, we have $\Gamma/T_K \sim 1$, while $\Gamma \sim V_{sd}/\left[\ln(V_{sd}/T_K)\right]^2$ for large biases $eV_{sd}/T_K\gg 1$ (full red line). This expression is valid asymptotically and for $\Gamma<V_{sd}$ \cite{Rosch}. We use a polynomial (here degree 6) interpolation between these two limits (full magenta line). The vertical dotted blue line shows the value of the charging energy $E_c$. }
\label{figure6}
\end{figure}

Using this form of bias-dependent $\Gamma$, with a Kondo temperature $T_K\simeq 550$mK, seems to explain (almost quantitatively) the experimental observations in Figs. \ref{figure5}, by evaluating the photon self-energy with this effective (Kondo) resonant level model. The validity of such a form of Kondo resonance (as a function of bias voltage) has also been justified in the context of other exotic Kondo fixed points \cite{Chung}.  \\

As shown in Fig.~\ref{figure5}, this analysis successfully corroborates the experimental results at point A if we take a Kondo temperature of the order of 550mK. For small bias voltages, the photon field is sensitive to the Kondo electronic level producing $\kappa_e > \kappa_i$ and a phase of $\pi$ in the microwave reflected signal. Interestingly, the decoherence effects on the Kondo resonance induced by the bias voltage enlarge the value of $\Gamma$ for $eV_{sd}\geq T_K$. This is sufficient to maintain the $\pi$ phase shift until large values of $e V_{sd}$. The low bias features in the amplitude signal can also be accounted for by our model. As can be seen in Fig.~\ref{figure6}, ($\Gamma(V_{sd})-eV_{sd}$) (given by the space between the magenta line and the dashed black line) initially decreases until $e|V_{sd}|\sim T_K$. For  $e|V_{sd}| \geq T_K$, $\kappa_e$ stays greater than $\kappa_i$, leading to a robust $\pi$ phase in the reflected signal. At $e|V_{sd}|/T_K \sim 10$, the quantity $\Gamma(V_{sd})$ becomes smaller than $V_{sd}$, leading to a drop of $\kappa_e$ and a phase shift from $\pi$ to $0$. Above $e|V_{sd}| \sim E_c$ (vertical dotted blue line) the model ceases to be valid as we should take into account other energy levels. \\

We argue that the experimental results in Figs. \ref{figure5} are not compatible with a simple analysis of the electronic system (biased leads plus double quantum dots) based on a resonant level model as was done in Ref.~\cite{Marco:PRB}.
Indeed in this picture the electronic lifetime would be essentially bias independent and set by the width of the resonance $\Gamma$.  Experimental results would then suggest $\Gamma\simeq E_c=0.4$meV. Such a large value of $\Gamma$ seems unphysical (in the context of a DQD weakly coupled to graphene leads). Moreover, the electronic levels in the Coulomb diamond would not be well defined. This is also in contradiction with the presence of two distinct energy scales in the amplitude of the reflected microwave signal. \\

We remark that there exists a low bias anomaly concerning the phase response in Fig.~\ref{figure5}, which is not present theoretically. From the input-output approach we expect that the phase remains fixed and equal to $\pi$ at low bias. Near this point, the reflection coefficient is indeed found to be real and negative. The Fermi liquid corrections \cite{Nozieres_ILTP}, which are not taken into account in the input-output approach, may affect the behavior in this region. \\

In the following we confirm our hypothesis with DC transport measurements.\\

\subsection{DC electron transport}

\begin{figure}[h!]
\center
\includegraphics[scale=0.31]{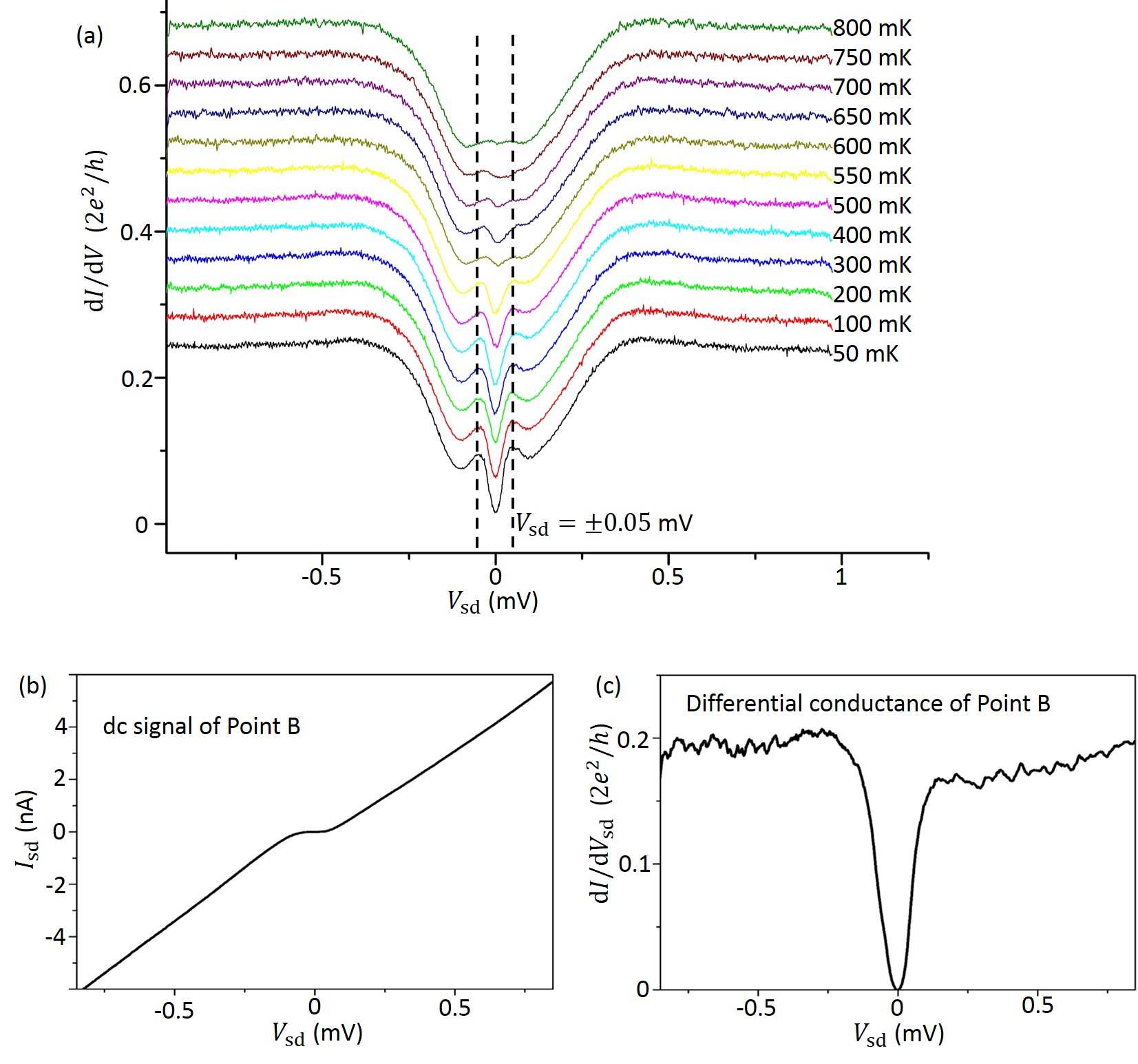}
\caption{(a) Temperature evolution of the differential conductance at point A, showing two peaks at finite voltage $V_{sd}\sim \pm 0.05mV$ which smoothly disappear above a temperature of about $500$mK. Curves are displaced from the base reference by a constant offset. For an $SU(4)$ Kondo model, the zero-bias anomaly in the conductance does not occur at $V_{sd}\rightarrow 0$, but rather at voltages in relation with Kondo physics \cite{Le_Hur:PRB}. In addition, when $V_{sd}\rightarrow 0$, at low temperatures, one predicts $G_0=dI/dV (V_{sd}=0)\sim (|t' |/T_K)^2 2e^2/h$ \cite{Borda:PRL} which confirms that $|t'| \sim 0.1-0.3 T_K \ll T_K$. Current (b) and differential conductance (c) at point B at $T$=30mK, showing no zero-bias anomaly.}
\label{figure7}
\end{figure}

We present in Fig.~\ref{figure7} the transport data at points A and B, and the evolution of conductance with respect to temperature for point A. In the cooling procedure for point A, the effective electrical sample temperature is around $100mK$ whereas the sample environment temperature of the dilution refrigerator can reach lower temperatures. Therefore, the data do not evolve much for temperatures of the refrigerator below $100mK$.\\

At point A, these measurements show at low temperatures a very rich small-bias structure, well inside the charging energy bands, with two peaks at finite voltage $V_{sd}\sim \pm 0.05$mV and a zero bias dip corresponding to a finite small residual conductance. Upon heating these features smoothly disappear above a crossover temperature of about $500$mK. The conductance across the DQD at small bias voltages essentially follows $2 e^2/h  T_K \rho(\omega=e|V_{sd}|) 
(| t' |/T_K)^2$ \cite{Le_Hur:PRB}. The spectral function gives the density of states accessible for an incoming electron, the prefactor $(|t'|/T_K)^2$ corresponds to the transition probability between the two dots. The presence of anomalies at $V_{sd}\sim \pm 0.05mV$ are thus consistent with the form of the spectral function in Eq. (\ref{spectral_function}), with $\epsilon_0\simeq T_K$ above the Fermi energy of the reservoir electron leads. By contrast, a purely spin Kondo effect would yield $\epsilon_0=0$, and transport properties would be very distinct. This also leads to a finite conductance at $V_{sd}\rightarrow 0$, as observed in Fig.~\ref{figure7}. In the Kondo limit, one predicts a differential conductance $G_0=dI/dV (V_{sd}=0)\sim (|t'|/T_K)^2 2e^2/h$ \cite{Borda:PRL}, which leads to the prediction $t'\sim 0.1-0.3 T_K$. In addition, the $SU(4)$ Kondo theory predicts that the conductance increases linearly with the bias voltage for $V_{sd}\rightarrow 0$ \cite{Le_Hur:PRB}, as found in the experiment. From the temperature analysis of the data, we confirm that the Kondo energy scale $T_K$, at which the zero-bias peaks in the differential conductance disappear is roughly around  $550$mK.\\

In contrast, the differential conductance at point B does not show such anomalies, as can be seen in the two lower panels of Fig.~\ref{figure7}. At point B, the levels of the two dots are not degenerate : this energy splitting is the analog of an orbital magnetic field along the $z$ direction, which suppresses the orbital Kondo effect. Fixing the number of electrons on one dot naturally diminishes the current at low bias voltages from left to right. \\

Next, we perform additional measurements at higher temperatures and analyze this effect of the Kondo features.

\subsection{Evolution of the microwave response with temperature}

\begin{figure}[h!]
\center
\includegraphics[scale=0.35]{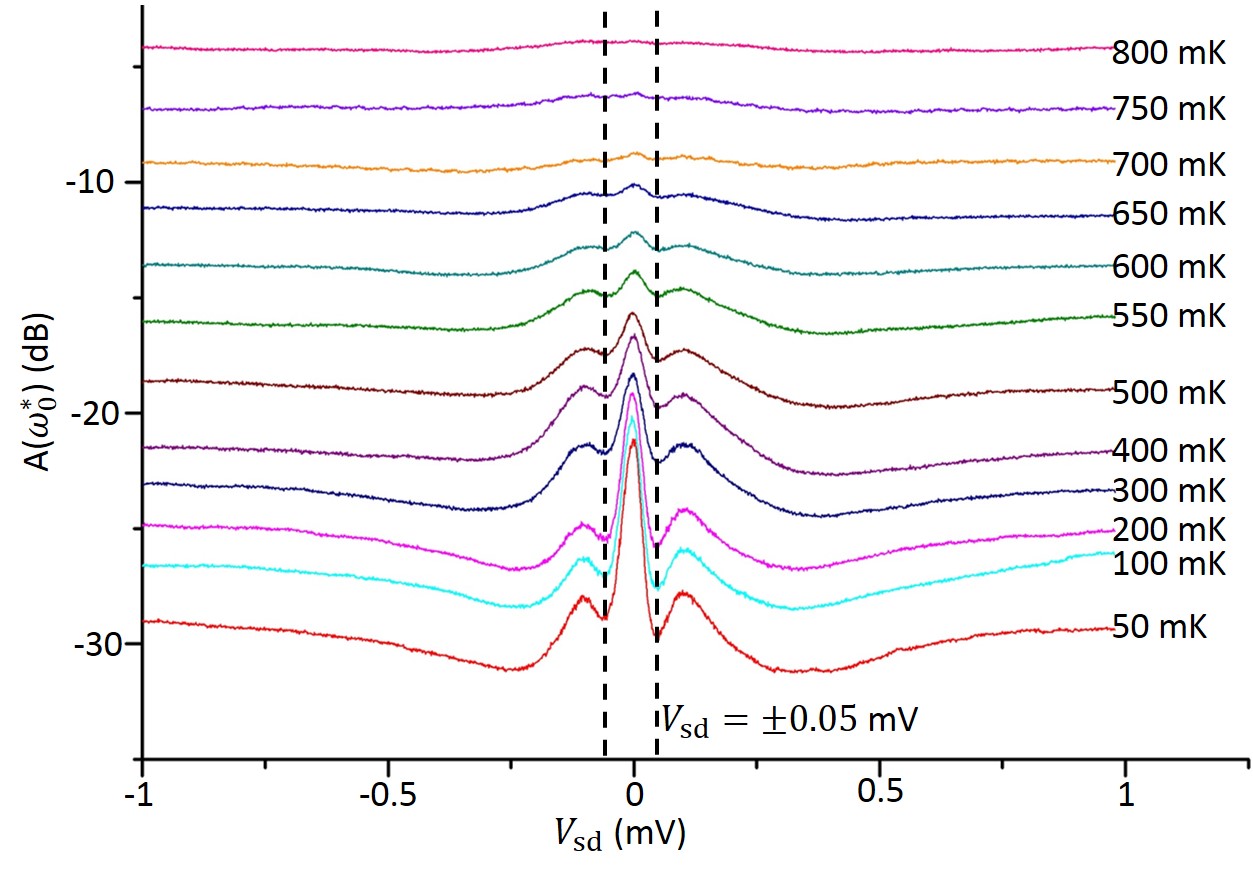}
\caption{Temperature dependence of the microwave reflected signal amplitude at point A. Curves are displaced from the base reference by a constant offset. (We emphasize that in this cooling procedure, the effective electrical temperature of the sample is around $100mK$ whereas the dilution refrigerator temperature $T$ can reach lower temperatures. The data do not change much on the figure for temperature $T$ smaller than $100mK$.)}
\label{figure8}
\end{figure}

\begin{figure}[h!]
\center
\includegraphics[width=7cm]{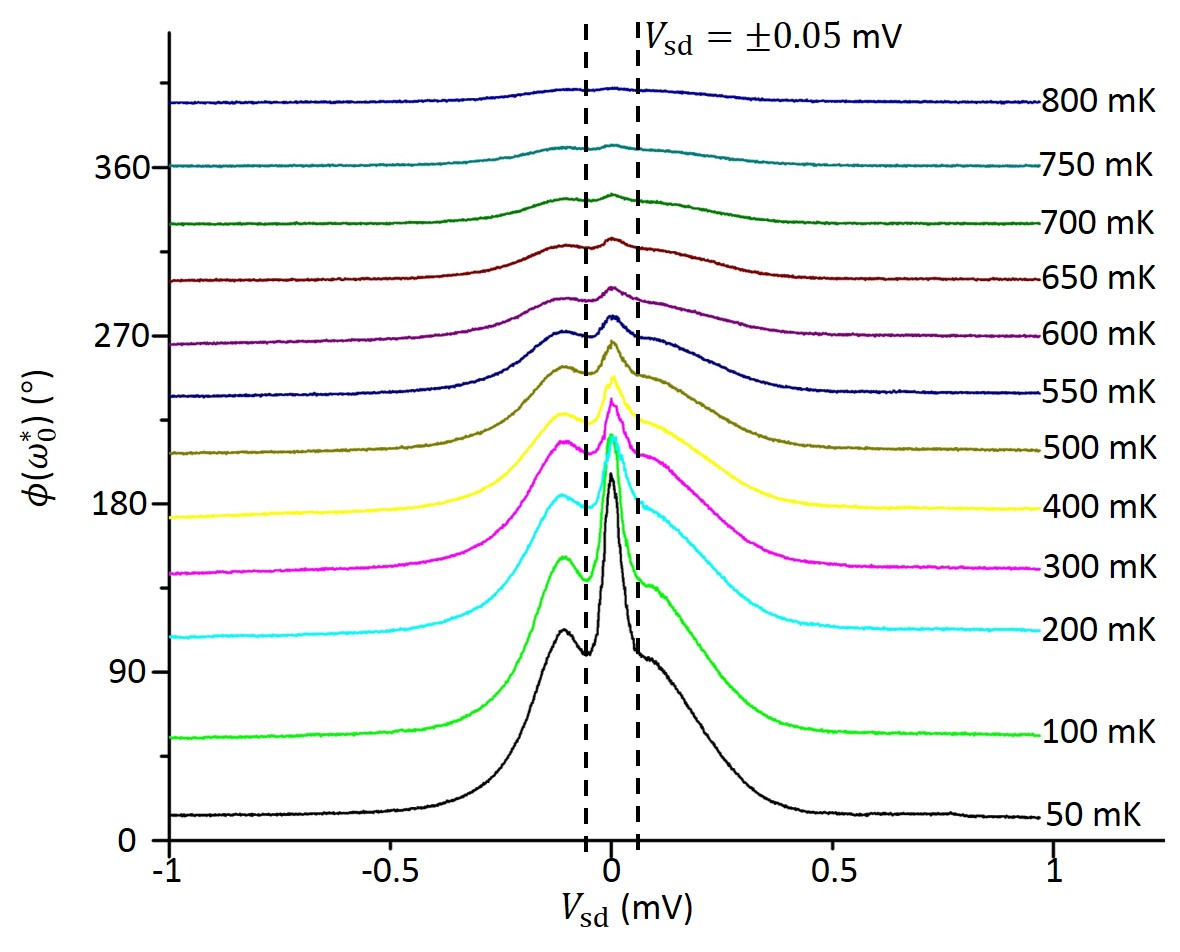}
\caption{Evolution of the phase of the microwave reflected signal at point A. Curves are displaced from the base reference by a constant offset. The cooling procedure is similar to that used in Fig.~\ref{figure8} for the amplitude. Interestingly, one observes that the phase converges to $\pi$ for $T<T_K=550mK$ for $V_{sd}\rightarrow 0$ and that the phase evolution is sensitive to the Kondo temperature. In addition, experimentally, for intermediate temperatures (we remind that the effective electrical temperature of the sample is around $100mK$ in this cooling procedure), the phase smoothly drops to zero for $e|V_{sd}|>T_K$.
}
\label{figure9}
\end{figure}

For completeness, we show in Fig.~\ref{figure8} and \ref{figure9} the temperature evolution of the amplitude and phase of the reflected signal. The cooling procedure here is the same as for Fig.~\ref{figure9}(a).\\

From Fig.~\ref{figure9}, one concludes that phase measurement allows to probe the Kondo temperature and approaches $\pi$ for $V_{sd}\rightarrow 0$ when $T<T_K$. A more precise modelling with both the bias voltage and the temperature remains an open question. The phase behavior in the regime $eV \sim T_K$ is very sensitive to the precise modelisation of the bias and temperature dependence of $\Gamma$, as also observed in
the experiment of Ref. \cite{Delbecq:PRL}. \\

In this Section, we have provided evidence for the formation of a robust state of interacting light and matter at the charge degeneracy point in relation with Kondo physics. Note that we assumed above that the density of states in graphene
is finite, meaning that the chemical potential does not lie at the charge neutrality point where the density of states vanishes. In this case the Kondo temperature would vanish, which does not seem in agreement with experimental results.\\

In this Chapter, we studied how hybrid systems coupling nanoscale electronic components with light could be useful both for the creation of thermoelectric devices and the investigation of many-body electronic effects.

\chapter*{Conclusion and perspectives}

We introduced Spin-boson models and highlighted their relevance to modern experimental platforms. Motivated by this perspective, we introduced a Stochastic Schr\"{o}dinger Equation describing the dynamics of the spin-reduced the density matrix. This approach proved to give reliable results for the Rabi model and the ohmic spinboson model for one and two spins. We applied it in Chapters 3 and 4 to study non-trivial dynamical processes for an ohmic bath, that may be measurable soon in current state-of-the-art experiments. The dissipative quantum phase transition for two spins, and the dissipative topological transition correspond to moderate values of $\alpha \simeq 0.2-0.5$. Bath-induced synchronization could be observed more easily as a dissipative parameter of the order of $\alpha \simeq 10^{-2}$ is sufficient.\\

Several interesting directions for the development of the SSE method are identified:
\begin{itemize}
\item Extension to the lattice. Coupling this approach to modern techniques able to deal with a large number of sites such as Matrix Product States/Operators techniques could be interesting. The exact treatment of the two spins may constitute the starting point of a ``clustering" point of view in this direction (\cite{DMFT} and cluster extensions).
\item  Extension to problems with disordered interactions.
\item  Computation of two-points correlation functions with the SSE. 
\item  Extension of the method to fermionic environments.
\end{itemize}
Being able to deal with the dynamical sign problem at a general level remains an open question, as the SSE method requires some care and optimization for each problem. \\

We also illustrated on a precise example how hybrid systems coupling electrons to microwave photons on-chip could be interesting to explore many-body phenomena in electronic systems and devise exotic transport devices. The development of these ``mesoscopic" QED setups motivates theoretical work to describe interacting light and matter out-of-equilibrium. 

\addcontentsline{toc}{chapter}{\protect\numberline{}Conclusion and perspectives}%

\begin{appendices}
\appendixpage
\noappendicestocpagenum
\addappheadtotoc

\chapter{Derivation of the Feynman Vernon influence functional}
\label{appendeix_FV_derivation}
For the sake of clarity, we consider throughout this proof Hamiltonian $\mathcal{H}$ with only one bosonic mode $b$ of frequency $\omega$, coupled to the spin with strength $\lambda$. Let $\{|u\rangle\}$ be the basis of coherent states of $\epsilon_B$, and $\{|\sigma_{k}\rangle\}=\{ |\uparrow_{z}\rangle,|\downarrow_{z}\rangle\}$ the canonical basis associated with the $z$-axis of $\epsilon_S$. We have
\begin{align}
\langle \sigma_f | \rho_S (t) | \sigma_{f'}\rangle&=\int d \mu (u)   \langle  u, \sigma_f | U(t,t_0) \rho(t_0) U^{\dagger}(t,t_0) |  u,\sigma_{f'}\rangle,
\label{eq_appendix:densitymatrix}
\end{align}
where 
 \begin{align} 
d \mu (u) = \frac{1}{\pi} d u_x du_y e^{-|u|^2},
\end{align}\\
with $u_x$ and $u_y$ respectively the real and imaginary part of $u$. In Eq. (\ref{eq_appendix:densitymatrix}), $U$ is the unitary time-evolution operator of the whole system. Next, we insert a closure relation (rewriting of the identity operator in terms of basis projectors)
 \begin{align} 
\mathbb{I}=\sum_{k} \int d \mu (v)   |v,\sigma_k \rangle \langle v, \sigma_k |,
\label{eq_appendix:idnetity}
\end{align}
both on the left and on the right of $\rho(t_0)$. This leads to
%\begin{footnotesize}
\begin{align}
\langle \sigma_f |\rho_S (t)|\sigma_{f'} \rangle= \int d \mu (u) d \mu (v) d \mu (w)   \sum_{k,k'} \Big\{& \langle u,\sigma_f |U(t,t_0)|v,\sigma_k \rangle  \langle v,\sigma_k |\rho(t_0)|w,\sigma_{k'} \rangle \notag \\
& \times \langle w,\sigma_{k'} |U(t,t_0)^{\dagger}|u,\sigma_{f'} \rangle \Big\}.
\end{align}
%\end{footnotesize}
We use the factorising initial condition $\rho(t_0)=\rho_B(t_0) \otimes \rho_S(t_0)$ and reach
%\begin{footnotesize}
\begin{align}
\langle \sigma_f |\rho_S (t)|\sigma_{f'} \rangle=\sum_{k,k'}  [\rho_S (t_0)]_{k,k'} J_{k,k',f,f'},
\label{eq_appendix:densitymatrix_with_identity}
\end{align}
with
\begin{align}
J_{k,k',f,f'}=\int d \mu (u) d \mu (v) d \mu (w)  & \Big\{ \underbrace{\langle v |\rho_B (t_0)|w \rangle}_{O} \underbrace{\langle u,\sigma_f |U(t,t_0)|v,\sigma_k \rangle}_{P}\underbrace{\langle w,\sigma_{k'}  |U^{\dagger}(t,t_0)|u,\sigma_{f'} \rangle}_{Q} \Big\}.
\label{eq_appendix:densitymatrix_with_identity_bis}
\end{align}

The remaining of the proof can be decomposed into three steps. First, we adopt a standard path-integral approach to write the propagators $P$ and $Q$ which appear in Eq. (\ref{eq_appendix:densitymatrix_with_identity_bis}). Then we access the bosonic trajectories thanks to the stationary phase method, which is exact as long as the Hamiltonian is quadratic. We finally integrate over the endpoints $u$, $u^*$, $v$, $v^*$ $w$, and $w^*$.\\

\section{Path integral representation of the propagators $P$ and $Q$}~\\

This consists in splitting the time interval $t-t_0$ into $N$ equal steps of size $\epsilon=(t-t_0)/N$, and inserting closure relations at each intermediate time. We shall eventually take the limit $N\to \infty$. Doing this trick of time-splitting allows to estimate the infinitesimal propagator at first order in $\epsilon$, as $e^{\epsilon(A+B)}=e^{\epsilon A} e^{\epsilon B}+\mathcal{O}(\epsilon^2)$ for $A$ and $B$ two arbitrary (non-commuting) operators. Taking then the limit $N\to \infty$ provides an exact expression for the propagator in terms of path integrals. We use the canonical basis associated with the $z$ axis for the spin and the basis of coherent states for the bosonic mode. We express $P= \langle u,\sigma_f |U(t,t_0)|v,\sigma_k \rangle$ as
 \begin{small}
 \begin{align} 
P= \int \left(\prod_{p=1}^{N-1} d \mu (\psi_p)\right) \sum_{\{k_p,0<p<N\}} & \langle \psi_{N}, \sigma_{k_{N}} | e^{-i\epsilon H} |     \psi_{N-1}, \sigma_{k_{N-1}} \rangle \langle \psi_{N-1}, \sigma_{k_{N-1}} | e^{-i\epsilon H} |     \psi_{N-2}, \sigma_{k_{N-2}} \rangle \notag\\
 &\times ...  \times \langle \psi_{1}, \sigma_{k_{1}} | e^{-i\epsilon H} |     \psi_{0}, \sigma_{k_{0}} \rangle,
\end{align}
 \end{small}
where $ \langle u,\sigma_f |= \langle \psi_{N}, \sigma_{k_{N}} |$ and $|v,\sigma_k \rangle= |     \psi_{0}, \sigma_{k_{0}} \rangle$. One can reach a more convenient expression for the infinitesimal propagator,
 \begin{align} 
\langle \psi_{p+1}, \sigma_{k_{p+1}} | e^{-i\epsilon H} |     \psi_{p}, \sigma_{k_{p}} \rangle=& \langle\psi_{p+1} | \psi_{p} \rangle X_{p+1,p},
\end{align}
with 
%=&\delta \left(\Sigma_{k_{p+1}}-\Sigma_{k_{p}}\right) \exp \left[-i\epsilon \left( \omega \psi_{p}^*  \psi_{p-1}+ \frac{\lambda}{2} \Sigma_{k_{p}} ( \psi_{p}^*+ \psi_{p-1})  \right) \right]-i \delta \left(\Sigma_{k_{p+1}}-\overline{\Sigma_{k_{p}}}\right) \frac{\Delta}{2}  \epsilon \exp \left[-i\epsilon \left( \omega \psi_{p}^*  \psi_{p-1}\right) \right]+\mathcal{O}(\epsilon^2)\notag \\
\begin{small}
 \begin{align} 
X_{p+1,p}=& \left[\delta \left(\Sigma_{k_{p+1}}-\Sigma_{k_{p}}\right) -i\epsilon \delta \left(\Sigma_{k_{p+1}}-\overline{\Sigma_{k_{p}}}\right)  \frac{\Delta}{2} \right]   e^{-i\epsilon \left( \omega \psi_{p}^*  \psi_{p-1}+ \frac{\lambda}{2} \Sigma_{k_{p}} ( \psi_{p}^*+ \psi_{p-1})  \right) }+\mathcal{O}(\epsilon^2).
\label{xp_appendix}
\end{align}
\end{small} 
In Eq. (\ref{xp_appendix}), $\delta$ is the Dirac Delta function, $\Sigma_{k_{p}}$ is the eigenvalue of $\sigma^z$ associated with the eigenvector $|\sigma_{k_{p}}\rangle$, and $\overline{\Sigma_{k_{p}}}$ is the complementary of $\Sigma_{k_{p}}$ in $\{-1,1\}$. $P$ then reads,
\begin{small}
 \begin{align} 
P= \int \left(\prod_{p=1}^{N-1}  \frac{d \psi_{p,x} d\psi_{p,y}}{\pi}\right) \sum_{\{k_p,0<p<N\}} & e^{ -\left[ \frac{|\psi_0|^2}{2}+\frac{|\psi_N|^2}{2}+\sum_{p=1}^{N-1}|\psi_p|^2-\sum_{p=1}^{N}\psi_p^*\psi_{p-1}\right] } \prod_{p=0}^{N-1} X_{p+1,p}.
\label{propagator_appendix_expanded}
\end{align}
\end{small}
We use the standard rewriting of the first term of the right-hand side of Eq. (\ref{propagator_appendix_expanded}) to make the derivative of the field appear in the continuum limit $N \to \infty$, i.e. $ 
\sum_{p=1}^{N-1}|\psi_p|^2-\sum_{p=1}^{N}\psi_p^*\psi_{p-1}=-\frac{1}{2}(\psi_N^* \psi_{N-1}+\psi_1^* \psi_{0})-\frac{\epsilon}{2} \sum_{p=1}^{N-1}\left(\psi_p \frac{\psi_{p+1}^*-\psi_{p}^*}{\epsilon}- \psi_p^* \frac{\psi_{p}-\psi_{p-1}}{\epsilon} \right)
$, and reach
\begin{align} 
P= \int \mathcal{D}[\Sigma] \mathcal{D} [\psi^*, \psi] A[\Sigma] \exp\left\{-\frac{1}{2}(|u|^2+|v|^2)+S[\psi^*, \psi,\Sigma]  \right\},
\label{P_appendix}
\end{align}
with
\begin{small}
\begin{align} 
S[\psi^*, \psi,\Sigma]=&\frac{1}{2}\left[\psi^*(t)\psi(t)+\psi^*(t_0)\psi(t_0)\right]\notag\\
&-\left[\int_{t_0}^t d\tau \frac{1}{2}\left[\dot{\psi}(\tau)\psi^*(\tau)-\dot{\psi}^*(\tau)\psi(\tau)\right]+i\left( \omega \psi^*(\tau) \psi(\tau)+ \frac{\lambda}{2}\Sigma(\tau) (\psi^*(\tau)+\psi(\tau)) \right) \right].
\label{action_appendix_forwards}
\end{align}
\end{small}
$ \mathcal{D} [\psi^*, \psi]=\lim_{N \to \infty} \left(\prod_{p=1}^{N-1}  \frac{d \psi_{p,x} d\psi_{p,y}}{\pi}\right)$ denotes the integration measure over the path $\psi^* (\tau)$ and  $\psi (\tau)$ with endpoints $\psi (t_0)=v$ and $\psi^* (t)=u^*$. $ \mathcal{D}[\Sigma] $ is the integration measure over all constant-by-parts spin path taking value in $\{-1,+1\}$ with $n$ spin flips. The $j$-th spin flip happens at the $p_j$-th discrete time. Taking the continuum limit allows to reach a convenient integral expression
\begin{align} 
 \int \mathcal{D} [\Sigma] A[\Sigma] =&\lim_{N \to \infty}\sum_{n=0}^{N} \sum_{p_n=0}^N \sum_{p_{n-1}=0}^{p_n-1}...\sum_{p_{1}=0}^{p_2-1}   \left(i\frac{\Delta}{2}\right)^n \left(\frac{t-t_0}{N}\right)^n \notag \\
 =&\underbrace{\sum_{n=0}^{\infty} \int_{t_0}^t dt_n  \int_{t_0}^{t_{n}} dt_{n-1}...  \int_{t_0}^{t_2} dt_1}_{\mathcal{D}[\Sigma]} \underbrace{\left(i\frac{\Delta}{2}\right)^n}_{A[\Sigma]}.
 \label{spin_appendix_measure}
\end{align}
where $t_j$ denotes the time of the $j$-th spin flip. The path $\Sigma$ is constrained by its initial and final values which verify $\sigma^z|\sigma_k\rangle=\Sigma(t_0)|\sigma_k\rangle$ and $\sigma^z|\sigma_f\rangle=\Sigma(t)|\sigma_f\rangle$. An integration by parts in Eq. (\ref{action_appendix_forwards}) leads to 
\begin{align} 
&S\left[\psi^*,\psi,\Sigma\right] =\psi^*(t)\psi(t)-\int_{t_0}^t d\tau \mathcal{L}[\psi^*,\psi,\Sigma] \label{action_P_appendix}\\
&\mathcal{L}[\psi^*,\psi,\Sigma] =\dot{\psi}(\tau)\psi^*(\tau)+i\omega \psi^*(\tau)\psi(\tau)+i\frac{\lambda}{2}\Sigma(\tau) [\psi(\tau)+\psi^*(\tau)].\label{lagrangien_appendix}\\
\label{actions_appendix_expressions}
\end{align}

Equation (\ref{P_appendix}) together with Eqs. (\ref{action_P_appendix}) and (\ref{lagrangien_appendix}) constitute our path integral expression for $P$. We use the same procedure for the determination of $Q$ and reach,
\begin{align} 
&Q= \int \mathcal{D}[\Sigma'] \mathcal{D} [\psi'^*, \psi'] A'[\Sigma'] \exp\left\{-\frac{1}{2}(|u|^2+|w|^2)+S'[\psi'^*, \psi',\Sigma']  \right\},\label{Q_appendix}\\
&S'\left[\psi'^*,\psi',\Sigma'\right] =\psi'^*(t_0)\psi'(t_0)+\int_{t_0}^t d\tau \mathcal{L}[\psi'^*,\psi',\Sigma'] \label{action_Q_appendix},
\end{align}
with the same functional for the Lagrangian $\mathcal{L}$. We moreover have the standard expression for the equilibrium bath density matrix element at inverse temperature $\beta$,
\begin{align} 
 O=e^{-\frac{1}{2}(|v|^2+|w|^2)+v^* w e^{-\beta \omega}}.
 \label{O} 
\end{align}\\

Altogether, inserting the derived expressions into Eq. (\ref{eq_appendix:densitymatrix_with_identity_bis}) and inverting the order of integrals, we recover Eq. (\ref{eq:densitymatrixelement}) with
%\begin{small}
\begin{align} 
F[\Sigma,\Sigma']=\int \frac{d u_x du_y}{\pi}  \frac{d v_x dv_y}{\pi}  \frac{d w_x dw_y}{\pi}\mathcal{D}[\psi^*,\psi,\psi^{*'},\psi']    &e^{-|u|^2-|v|^2-|w|^2}\notag\\
\times& e^{S\left[\psi^*,\psi,\Sigma\right]+S'\left[\psi^{*'},\psi',\Sigma'\right]+ w v^* e^{-\beta \omega} }.
\label{FV_appendix_not_integrated}
\end{align}
%\end{small}

\section{Stationary trajectories with Lagrange equations}~\\

As the Hamiltonian is quadratic in terms of bath operators, we can integrate out exactly the bosonic degrees of freedom in Eq. (\ref{FV_appendix_not_integrated}) by solving the equations of motions. We first determine the stationary fields $\psi$ and $\psi^*$, which verify: \\
\begin{align} 
&\frac{d}{d\tau} \frac{\partial \mathcal{L}}{\partial \dot{\psi}}=\frac{\partial \mathcal{L}}{\partial\psi},\label{Lagrange_equation_1_appendix}\\
&\frac{d}{d\tau} \frac{\partial \mathcal{L}}{\partial \dot{\psi^*}}=\frac{\partial \mathcal{L}}{\partial\psi^*}\label{Lagrange_equation_2_appendix},
\end{align}
with boundary conditions $\psi (t_0)=v$ and $\psi^* (t)=u^*$. We find 
\begin{align} 
&\psi^* (\tau)=u^*e^{i\omega (\tau-t)}-i\frac{\lambda}{2} \int_{\tau}^t d\tau' \Sigma(\tau') e^{i\omega (\tau-\tau')},\label{Lagrange_solution_1_appendix}\\
&\psi (\tau)=v e^{-i\omega (\tau-t_0)}-i\frac{\lambda}{2} \int_{t_0}^{\tau} d\tau'  \Sigma(\tau') e^{-i\omega (\tau-\tau')}.\label{Lagrange_solution_2_appendix}
\end{align}
$\psi'$ and $\psi'^*$ can be determined in an analog manner, by solving Eq. (\ref{Lagrange_equation_1_appendix}) and (\ref{Lagrange_equation_2_appendix}) with boundary conditions $\psi' (t)=u$ and $\psi'^* (t_0)=w^*$:
\begin{align} 
&\psi'^* (\tau)=w^*e^{i\omega (\tau-t_0)}+i\frac{\lambda}{2} \int_{t_0}^{\tau} d\tau'  \Sigma'(\tau') e^{i\omega (\tau-\tau')},\label{Lagrange_solution_1_appendix}\\
&\psi' (\tau)=u e^{-i\omega (\tau-t)}+i\frac{\lambda}{2} \int_{\tau}^{t} d\tau'  \Sigma(\tau') e^{-i\omega (\tau-\tau')}.\label{Lagrange_solution_2_appendix}
\end{align}
Inserting these stationary paths into Eq. (\ref{FV_appendix_not_integrated}) leads to

\begin{small}
\begin{align} 
F[\Sigma,\Sigma']=\int \frac{d u_x du_y}{\pi}  \frac{d v_x dv_y}{\pi}  \frac{d w_x dw_y}{\pi} &   e^{-|u|^2-|v|^2-|w|^2+w v^* e^{-\beta \omega}+u^* v e^{-i \omega (t-t_0)}+w^* u e^{i\omega (t-t_0)} }\\
&\times \exp\left[-i u^* A+iv B-iu C +iw^* D+E\right]\label{FV_integrated_psi_appendix}
\end{align}
\end{small}
with
\begin{small}
\begin{align} 
&A=\frac{\lambda}{2}\int_{t_0}^t d\tau \Sigma(\tau) e^{-i\omega (t-\tau)} ,~~~~~~~~~~~~~~~~~~ B=\frac{\lambda}{2}\int_{t_0}^t d\tau \Sigma(\tau) e^{-i\omega (\tau-t_0)},\\
&C=\frac{\lambda}{2}\int_{t_0}^t d\tau \Sigma'(\tau) e^{i\omega (t-\tau)},~~~~~~~~~~~~~~~~~~ D=\frac{\lambda}{2}\int_{t_0}^t d\tau \Sigma'(\tau) e^{i\omega (\tau-t_0)},\\
&E=\left(\frac{\lambda}{2}\right)^2\int_{t_0}^t d\tau \int_{\tau}^{t} d\tau' \Sigma'(\tau) \Sigma'(\tau') e^{i\omega (\tau-\tau')}-\left(\frac{\lambda}{2}\right)^2\int_{t_0}^t d\tau \int_{t_0}^{\tau} d\tau' \Sigma(\tau) \Sigma(\tau') e^{-i\omega (\tau-\tau')}.
\end{align}
\end{small}
\paragraph{Gaussian integration over the endpoints}~\\

We are left with standard Gaussian integrals over the endpoints, that we perform successively. This procedure generates double integrals quadratic in terms of spin path variables. After some manipulations on integrals boundaries and symmetry considerations of the integrand, we reach 
\begin{align} 
F[\Sigma,\Sigma']=&e^{ \int_{t_0}^t ds \int_{t_0}^s ds'\left[i \lambda^2  \sin \omega(s-s')\frac{ \Sigma (s)-\Sigma '(s) }{2} \frac{ \Sigma (s')+\Sigma '(s') }{2}\right]}\notag\\
&\times e^{\int_{t_0}^t ds \int_{t_0}^s ds'\left[-\lambda^2  \cos \omega(s-s') \left(1+2\frac{e^{-\beta \omega}}{1-e^{-\beta \omega}}\right)\frac{\Sigma (s)-\Sigma'(s) }{2} \frac{ \Sigma (s')-\Sigma'(s')}{2}\right]}\label{FV_integrated_total_appendix}
\end{align}
Equation (\ref{FV_integrated_total_appendix}) corresponds to the desired result in the case of one mode. The generalization to an arbitrary number of environmental mode is straightforward as we are left with a product of independent integrals, and we recover Eq. (\ref{eq:influence}) with Eq. (\ref{Ls_2}). An alternate derivation can be done with algebraic considerations in the interaction picture \cite{Brandes_course}, as shown in appendix \ref{appendix_brandes}.\\\\

\chapter{Alternate derivation of Feynman-Vernon influence functional}    
\label{appendix_brandes}
Here we derive the expression of the Feynman-Vernon influence functional (\ref{eq:influence}), using the method of Ref. \cite{Brandes_course}. In order to simplify the derivation, we first consider that one single bosonic mode is coupled to the spin. Let us call $H_S=\Delta/2 \sigma^x$ the spin-part and $H_B=\hbar \omega b^{\dagger}b+\frac{\lambda}{2} \sigma^z (b+b^{\dagger})$ the interaction part. Eq. (\ref{FV_appendix_not_integrated}) can be recast as
\begin{equation}
F[\Sigma, \Sigma']=tr_B \left\{\rho_{B}(t_0) U_{B}[\Sigma] (t) U_{B}^{\dagger}[\Sigma'](t) \right\} ,
\end{equation}
where $U_B[\Sigma]$ being the time evolution operator related to $H_B$. In order to evaluate this functional we derive the expression of the bath evolution operator. To do so, we switch to the interaction picture (where $V=\lambda/2(a+a^{\dagger})\sigma$ is the interaction term) and define $\tilde{U}_{B}[\Sigma](t)$ the corresponding time evolution operator. We have :
\begin{equation}
i \hbar \partial_t \tilde{U}_{B}[\Sigma](t)=\tilde{V}(t) \tilde{U}_{B}[\Sigma](t) 
\end{equation}
Defining $\hat{X}=\frac{b+b^{\dagger}}{\sqrt2}$, and $\hat{P}=\frac{b-b^{\dagger}}{i \sqrt2}$, the commutation relations gives:\newline

$\left\{
\begin{array}{l}
  e^{-i\omega b^{\dagger}b t} \hat{X} e^{i\omega b^{\dagger}b t}=\hat{X}+\omega t ~e^{-i\omega b^{\dagger}b t}  \hat{P} e^{i\omega b^{\dagger}b t} \\
  e^{-i\omega b^{\dagger}b t} \hat{P} e^{i\omega b^{\dagger}b t}=\hat{P}-\omega t ~ e^{-i\omega b^{\dagger}b t}  \hat{X} e^{i\omega b^{\dagger}b t}
\end{array}
\right.$\newline

which results in:
\begin{equation}
\tilde{V}(t)=\frac{\lambda}{2} \sigma^x (t) [(b+b^{\dagger}) \cos \omega t+\frac{b-b^{\dagger}}{i} \sin \omega t ].
\end{equation}

As the evolution operator $\tilde{U}_{B}[\sigma](t)$ is unitary, we suppose that we can write it as ${e^{-i \alpha (t)} ~e^{-i \beta(t)(b+b^{\dagger}) } ~e^{-i \gamma (t) \frac{(b-b^{\dagger})}{i}}}$. The Schr\"{o}dinger equation gives us the expression of $\alpha$, $\beta$, $\gamma$:\newline

$\left\{
\begin{array}{l}
\beta(t)= \int_{t_0}^t ds \frac{\lambda}{2} \Sigma (s) \cos \omega s   \\
 \gamma(t)=\int_{t_0}^t ds  \frac{\lambda}{2}  \Sigma (s) \sin \omega s    \\
 \alpha(t)=-\int_{t_0}^t ds \int_0^s ds' \left(\frac{\lambda}{2}\right)^2 \Sigma (s) \Sigma (s') \cos \omega s'  \sin \omega s 
\end{array}
\right.$\\

Then, we have :

\begin{align}
 F[\Sigma, \Sigma']=e^{i \left[\alpha'\left(t\right)-\alpha(t) \right]}\int dX \langle X|\rho_B&(0) ~e^{i\gamma'(t)\hat{P}} ~e^{i\beta'(t)\hat{X}} e^{-i\beta(t)\hat{X}} ~e^{-i\gamma(t)\hat{P}}  | X \rangle,
\end{align}
where the states $|X\rangle$ represent a complete set of position eigenstates. It simplifies into
\begin{align}
F[\Sigma, \Sigma']=e^{i\left[\alpha'(t)-\alpha(t)\right]}\int dX &\langle X|\rho_B(0) | X+\gamma(t)-\gamma'(t) \rangle e^{i\left[\beta'(t)-\beta(t)\right]\left[X+\gamma(t)\right]}.
\end{align}

In order to evaluate the element ${\langle X|\rho_B(t_0) | X+\gamma(t)-\gamma'(t) \rangle}$, we assume a thermal equilibrium at inverse temperature $\beta$ for the operator $\rho_B(t_0)$:

\begin{align}
\langle X_1|\rho_B(0) | X_2\rangle=\frac{1}{Z} &\left(\frac{1}{2\pi \sinh \beta \omega }\right)^{\frac{1}{2}} e^{-\frac{1}{2 \sinh \beta \omega }\left[\left(X_1^2+X_2^2\right) \cosh \beta \omega -2X_1X_2\right] }. 
\end{align}

Using the properties of Gaussian integrals, as well as the identity $\frac{\cosh \beta \omega -1}{ \sinh \beta \omega }=\tanh \beta \omega /2$, we get:

\begin{align}
F[\Sigma, \Sigma']=&e^{i\big[\alpha'(t)-\alpha(t)\big]+i\big[\beta'(t)-\beta(t)\big]\big[\gamma(t)+\gamma'(t)\big]  } e^{-\frac{1}{4} \coth \beta \omega / 2\big[(\beta'(t)-\beta(t))^2+(\gamma(t)-\gamma'(t))^2\big] }.
\end{align}
Hence re-inserting the expressions of $\alpha$, $\beta$ and $\gamma$ and after trigonometric calculations and using the symmetry of the integrand we finally  recover Eq.~(\ref{eq:influence}) of the main text, with $L_1(t)=\pi \lambda^2 \sin \omega t$ and $L_2(t)=\pi \lambda^2 \cos \omega t \coth \beta \omega/2$. The generalization to an infinite number of modes is straightforward.

\chapter{Blips and Sojourns decomposition}
\label{appendix_blips_sojourns}
As Eq. (\ref{eq:densitymatrix_with_identity}) is linear with respect to the elements of the initial density matrix, we can compute the different contributions separately. We begin by considering the diagonal elements of the density matrix.
\section{Diagonal elements of the spin density matrix}
We first focus on the contributions stemming from the diagonal elements of the initial spin density matrix, encapsulated in $J_{1,1,f,f}$ and $J_{2,2,f,f}$.

\subsection{Contributions $J_{k,k,f,f}$}
\label{J_{kk}}
 In this case the double spin path starts and ends in the sojourn states A or D. One path of this type makes $2n$ transitions along the way at times $t_i$, $i \in \{1,2,..,2n\}$ such that $t_0<t_1<t_2<...<t_{2n}$. We can write this spin path as $\xi(t)=\sum_{j=1}^{2n} \Xi_j\theta(t-t_j)$ and $\eta(t)=\sum_{j=0}^{2n} \Upsilon_j\theta(t-t_j)$ where the variables $\Xi_i$ and $\Upsilon_i$ take values in $\{-1,1\}$. \\

Using the explicit representation of path summation introduced in Eq. (\ref{spin_appendix_measure}), $J_{k,k,f,f}$ is given by a series in the tunneling coupling $\Delta^2$ \cite{leggett:RMP,Weiss:QDS,Peter} :
\begin{equation}
J_{k,k,f,f}=\sum_{n=0}^{\infty} \left(\frac{i\Delta}{2} \right)^{2n} \int_{t_0}^{t} dt_{2n} ... \int_{t_0}^{t_2} dt_{1} \sum_{\{\Xi_j\},\{\Upsilon_j\}' } \mathcal{F}_{n}.
\label{eq:p(t)_appendix}
\end{equation}
 The prime in $\{\Upsilon_j\}'$ in Eq. (\ref{eq:p(t)_appendix}) indicates that the initial and final sojourn states are fixed according to the initial and final conditions. The initial sojourn is A for $J_{1,1,f,f}$ ($\Upsilon_0=1$) and D for $J_{2,2,f,f}$ ($\Upsilon_0=-1$). The final sojourn is A if $|\sigma_f\rangle=|\sigma_{f'}\rangle=|\uparrow_z\rangle$ ($\Upsilon_{2n}=1$) and D if $|\sigma_f\rangle=|\sigma_f'\rangle=|\downarrow_z\rangle$ ($\Upsilon_{2n}=-1$). $\mathcal{F}_{n}$ corresponds to the evaluation of $\mathcal{F}[\Sigma,\Sigma']$ for a given path with $2n$ spin flips, introduced above. Given this path, we can evaluate Eq. (\ref{eq:influence}). This leads to:
 \begin{align}
 &\mathcal{F}_{n}= \exp \left[ \frac{i}{\pi} \sum_{k=0}^{2n-1}\sum_{j=k+1}^{2n} \Xi_j \Upsilon_k  Q_1(t_j-t_k) \right] \exp \left[ \frac{1}{\pi} \sum_{k=1}^{2n-1}\sum_{j=k+1}^{2n} \Xi_j \Xi_k  Q_2(t_j-t_k) \right]\label{Q_appendix} . 
 \end{align}
 The functions $Q_1$ and $Q_2$ verify Eqs. (\ref{q1}) and (\ref{q2})
\label{derivation_Q1_Q2}
\subsection{Contributions $J_{k,\overline{k},f,f}$}
Contributions $J_{1,2,f,f}$ and $J_{2,1,f,f}$ coming from the off-diagonal elements of the initial spin density matrix can be computed in a similar manner after the rewriting of the spin path,
\begin{equation}
J_{k,\overline{k},f,f}=\sum_{n=1}^{\infty} \left(\frac{i\Delta}{2} \right)^{2n-1} \int_{t_0}^{t} dt_{2n-1} ... \int_{t_0}^{t_2} dt_{1}\sum_{\{\Xi_j\}'} \sum_{\{\Upsilon_j\}' }   \mathcal{F}_{n}',
\end{equation} 
with
\begin{align} \mathcal{F}_{n}'=&\exp\left[ \frac{i}{\pi} \sum_{k=1}^{2n-2}\sum_{j=k+1}^{2n-1} \Xi_j \Upsilon_k  Q_1(t_j-t_k)\right] \exp \left[ \frac{1}{\pi} \sum_{k=0}^{2n-2}\sum_{j=k+1}^{2n-1} \Xi_j \Xi_k  Q_2(t_j-t_k) \right]. \label{Fbis_appendix}
  \end{align}
Here the initial blip is B for $J_{1,2,f,f}$ ($\Xi_0=1$) and C for $J_{2,1,f,f}$ ($\Xi_0=-1$), and the final sojourn value $\Upsilon_{2n-1}$ depends on $|\sigma_f\rangle=|\sigma_{f'}\rangle$.

\section{Off-diagonal elements of the spin density matrix}  
  
We compute an off-diagonal term of the density matrix in terms of a series expansion in $\Delta$, considering spin paths that end in a blip state with $|\sigma_f\rangle \neq |\sigma_{f'}\rangle$. The paths contributing to  $J_{k,k,f,f'}$ make now $2n-1$ transitions. Here the initial sojourn state is fixed, as well as the final blip state. \\%An example of such a path can be seen in Fig. \ref{spin_path_2}.
 % \begin{figure}[h!]
%\center
%\includegraphics[scale=0.42]{test_removal_A.pdf}  
%\caption{Spin path- $\eta(t)=\sum_{j=0}^{2n-1} \Upsilon_j\theta(t-t_j)$ in red; $\xi(t)=\sum_{j=1}^{2n-1} \Xi_j\theta(t-t_j)$ in dashed blue. Here the spin path ends in the blip state B. The initial state is A.}
%\label{spin_path_2}
%\end{figure}  
All blips are coupled to all previous sojourns and blips as can be seen in Eq. (\ref{Q_appendix}). Paths considered in Sec. \ref{J_{kk}} ended in a sojourn state, and the latest coupling period lasted from $t_{2n-1}$ to $t_{2n}$. The situation is different here because paths end up in a blip state. The final coupling period then lasts from $t_{2n-1}$ to the final time $t$. But providing that we formally set $t_{2n}=t$ and $\Xi_{2n}=-\sum_{j=1}^{2n-1}\Xi_{j}$, we get
\begin{equation}
J_{k,k,f,f'}=\sum_{n=1}^{\infty} \frac{(i\Delta)^{\small{2n-1}}}{2^{\small{2n-1}}} \int_{t_0}^{t} dt_{\small{2n-1}} ... \int_{t_0}^{t_2} dt_{1}\sum_{\{\Xi_j\}'\{\Upsilon_j\}' }   \mathcal{F}_{n}',
\label{eq:off(t)_appendix}
\end{equation}
with $\mathcal{F}_{n}$ given by Eq. (\ref{Q_appendix}). \\

A similar results holds for $J_{k,\overline{k},f,f'}$ and Eq. (\ref{Fbis_appendix}) is still valid for $\mathcal{F}_{n}'$.

\chapter{Sampling of stochastic variables and convergence}  
\label{appendix:sampling}
\section{Fourier series decomposition}  

For simplicity we take $t_0=0$ and introduce the variable $\tau=t/t_f$ where $t_f$ is the final time of the experiment/simulation. Hence $\tau \mapsto Q_2 (\tau t_f)$ and $\tau \mapsto Q_1 (\tau t_f)\theta (\tau)$ are defined on $[-1,1]$. We extend their definitions by making them 2-periodic functions and it is then possible to expand them in Fourier series. In particular, we have:

\begin{align}
  \frac{Q_2 \left[(\tau_j-\tau_k)t_f \right]}{\pi}&=\frac{g_0}{2}+\sum_{m=1}^{\infty} \frac{g_m}{2}\left[\phi_m(\tau_j)\phi_m^*(\tau_k)+h.c. \right],\\
   \frac{Q_1 \left[(\tau_j-\tau_k)t_f \right]}{\pi}&\theta(\tau_j-\tau_k)=\frac{f_0}{2}+\sum_{m=1}^{\infty} \frac{f^s_m}{2}\left[\phi_m(\tau_j)\phi_m^*(\tau_k)+h.c. \right]\notag \\
  & +\sum_{m=1}^{\infty} \frac{f^a_m}{2}\left[\phi_m(\tau_j)\phi_m^*(\tau_k)-h.c. \right],
\end{align}
  where $\phi_m:\tau \mapsto \exp(im\pi\tau)$, and we have for $m>1$, $\left\{g_m=\int_{-1}^1 d\tau \frac{Q_2 (\tau t_f)}{\pi} \cos m \pi \tau \right\}$, $\left\{f^s_m=\int_{-1}^1 d\tau  
 \frac{Q_1 \left(\tau t_f \right)}{\pi}\theta(\tau) \cos m \pi \tau \right\}$, and $\left\{f_m^{a}=\int_{-1}^1 d\tau  \frac{Q_1(\tau t_f)}{\pi}\theta(\tau) \sin m \pi \tau \right\}$. $g_0$ and $f_0$ are the constant Fourier coefficients. Then we define $h$ and $k$ as
  
  \begin{footnotesize}

  \begin{align}  
 h(\tau t_f) =\sum_{m=1}^{\infty} &\phi_m(\tau t_f)\Big[\left(\frac{g_m}{4}\right)^{\frac{1}{2}}(s_{1,m}+is_{2,m})+\left(\frac{f_m^s}{4}\right)^{\frac{1}{2}}(u_{1,m}+iu_{2,m})+\left(\frac{f^{a}_m}{4}\right)^{\frac{1}{2}}(v_{1,m}+iv_{2,m})      \Big] \notag \\
+&\phi_m^*(\tau t_f)\Big[\left(\frac{g_m}{4}\right)^{\frac{1}{2}}(s_{1,m}-is_{2,m})+\left(\frac{f_m^s}{4}\right)^{\frac{1}{2}}(u_{3,m}+iu_{4,m})+\left(\frac{f^{a}_m}{4}\right)^{\frac{1}{2}}(v_{3,m}+iv_{4,m})      \Big],\\
~\notag \\
 k(\tau t_f) =\sum_{m=1}^{\infty} &\phi_m(\tau t_f)\Big[\left(\frac{f_m^s}{4}\right)^{\frac{1}{2}}(u_{1,m}+iu_{2,m})+\left(\frac{f^{a}_m}{4}\right)^{\frac{1}{2}}(v_{1,m}+iv_{2,m})      \Big] \notag \\
+&\phi_m^*(\tau t_f)\Big[\left(\frac{f_m^s}{4}\right)^{\frac{1}{2}}(u_{3,m}+iu_{4,m})+\left(\frac{f^{a}_m}{4}\right)^{\frac{1}{2}}(v_{3,m}+iv_{4,m})      \Big],
\label{fields_explicit_construction}
\end{align}

  \end{footnotesize}

where $\{s_{i,m}\}$, $\{u_{i,m}\}$ and $\{v_{i,m}\}$ are standard normal variables. One can check that $h$ and $k$ verify the correlations given by Eqs. (\ref{height_1}), (\ref{height_2}) and (\ref{height_3}). \\

 In the numerics, we use Fast Fourier Transform in order to increase the speed of the numerical procedure. As a result, discretization in time and Fourier domain are linked: we use $N=2^p$ steps and increase progressively $p$ until results remain unchanged. 
 
 %In Fig.~\ref{discretization_dynamics_appendix}, we illustrate this procedure for the dynamics of $p_{|T_+\rangle} (t)$ for the dimer problem (see Chapter \ref{dimer}) with initial condition $|T_+\rangle$, with $\alpha=0.02$, $\omega_c=100$, $K=0$, for $p$ from $6$ to $11$. For $N>11$, all the curves give the same result (superposed to the full black curve).
 % \begin{figure}[h!]
  %\center
  %\includegraphics[scale=0.4]{figureAppendice2.pdf}
 % \caption{Time evolution of $p_{|T_+\rangle} (t)$ for the dimer being initially in the state $|T_+\rangle$, for $p=6$ (dashed blue line), $p=7$ (dotted green line), $p=8$ (full yellow line), $p=9$ (dotted red line), $p=10$ (dashed purple line), and $p=11$ (full black line). Parameters are $\alpha=0.02$, $\omega_c=100$, and $K=0$.}
 % \label{discretization_dynamics_appendix}
%\end{figure}\\

%From Equations (\ref{fields_explicit_construction}), we verify that $\overline{k_j -k_{j-1}}=0$ for all $j \in [1,N]$ by construction, which ensures the trace conservation property of the SSE on average.

\section{Definition of the stochastic fields in relation with autoregressive processes}  
Let us consider the discrete random process $h_n$ satisfying
 \begin{align}  
 \overline{h_n h_m}=\frac{1}{\pi} Q_2(t_n-t_m),
 \label{correlation_yule_walker}
\end{align}    
for $n,m>0$. Such a process is not Markovian, and correlations build up at each time step. We can build such a process by considering that $h_n$ depends on all the $h_m$ for $0 \leq m<n$. We write
 \begin{align}  
 h_n=\sum_{p=1}^n \gamma^{(n)}_p h_{n-p}+\sigma_n \epsilon_n,
\end{align}  
where $\epsilon$ is a standard random variable. Parameters $\gamma^{(n)}_p$ and $\sigma_n$ are determined according to Eq. (\ref{correlation_yule_walker}), leading to the following set of equations 
 \begin{align}  
&\sum_{p=1}^n \gamma^{(n)}_p Q_2(t_{n-p}-t_k)=Q_2(t_{n}-t_k), ~ \forall k \in [1,n]\\
&\sum_{p=1}^n \gamma^{(n)}_p Q_2(t_{p})+\sigma_n^2=Q(0).
\end{align} \\

These equations are the equivalent of the standard Yule-Walker \cite{Yule,Walker} equations for Autoregressive processes. Here the order of the process changes at each time step. This notably implies that $\sigma_n$ depends on $n$: the variance of the additional random movement at time $n$ decreases with $n$. \\

This sampling procedure gives comparable results.

\section{Cutoffs and control parameters in the SSE framework}

Discretization in time and fourier space are linked in the case of Fourier decomposition, because we use Fast Fourier Transform algorithm to optimize the sampling. The parameter p introduced in Sec. I is then the only control parameter in our framework.\\

In Time-Dependent Numerical Renormalization group for example, one introduces two control parameters. The first one is related to the time discretization, while another cut-off is used at each time step to restrict the dynamics to a given number of environmental states.\\

Interestingly, there are also two control parameters in the SSE when one uses autoregressive sampling. The first one is the time discretization, and the second one corresponds to the standard deviation of the first random variable (the one used to determine the first move at $t_0$).

\chapter{Additional developments on Blips and Sojourns re-writing}  
\label{appendix_M1_M2}
\section{Expressions of $\mathcal{M}_1^p$ and $\mathcal{M}_2^p$}   

\begin{figure*}[h!]
\center
  \includegraphics[scale=0.2]{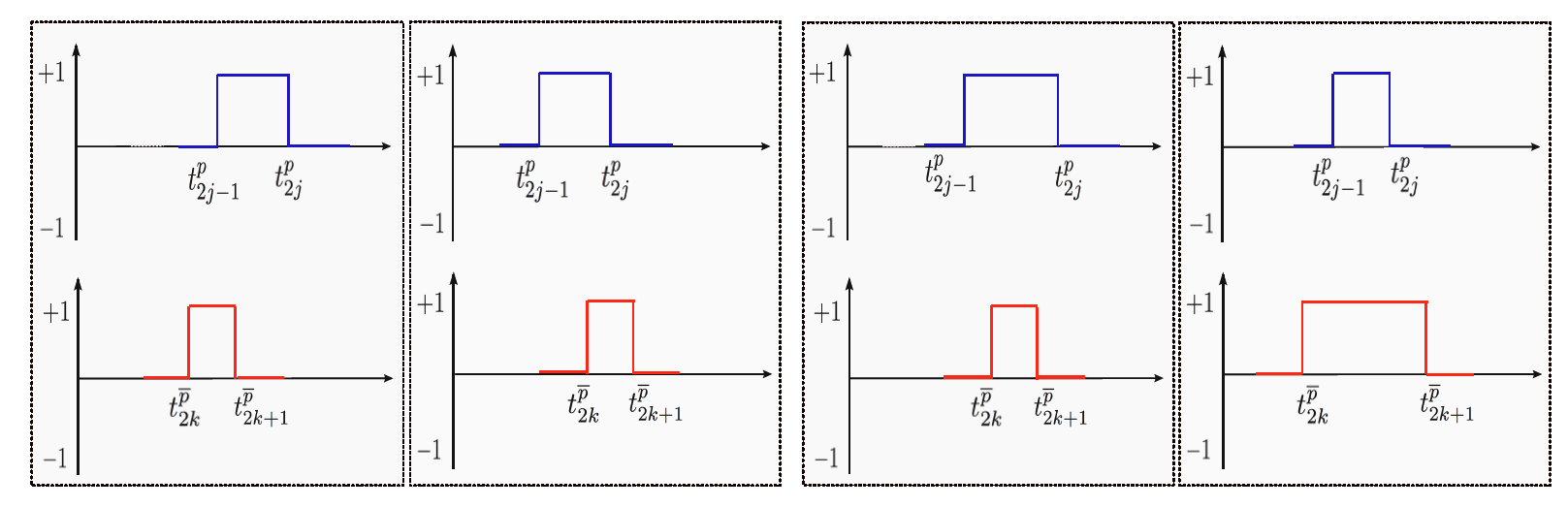}
  \caption{Coupling of a blip of the spin $p$ with a simultaneous sojourn of the spin $\overline{p}$. There are four distinct configurations. }
  \label{blip_sojourn_appendix}
\end{figure*}

For $\mathcal{Q}_1^p$ and $\mathcal{Q}_2^p$, the blip and sojourn variables cannot be simultaneously both non-zero. For $\mathcal{M}_1^p$ and $\mathcal{M}_2^p$, the situation is different as the state of the first spin does not constrain the state of the second one. More explicitly, for $\mathcal{M}_1^p$ for example, one of the spins may be in a blip state while the second one is in a sojourn state, as illustrated in Fig.~\ref{blip_sojourn_appendix}. In the following, we will compute the contribution of these particuliar blip-sojourn configurations.\\

% \begin{widetext} 
%\begin{figure}[h!]
%\center
%\includegraphics[scale=0.6]{blip_sojourn_appendix_1.pdf} \includegraphics[scale=0.6]{blip_sojourn_appendix_2.pdf} \includegraphics[scale=0.6]{blip_sojourn_appendix_3.pdf}  \includegraphics[scale=0.6]{blip_sojourn_appendix_4.pdf} 
%\caption{Coupling of a blip of the spin $p$ with a simultaneous sojourn of the spin $\overline{p}$. }
%\label{blip_sojourn_appendix}
%\end{figure}

The first case (left panel) yields,
\begin{footnotesize}
\begin{align}
-\frac{i}{\pi} \int_{t_{2j-1}^p}^{t_{2j}^p} ds \int_{t_{2k}^{\overline{p}}}^{s} ds' \xi^p(s) \eta^{\overline{p}}(s') L_1(s-s') =-\frac{i}{\pi}& \Xi^p_{2j-1}\Upsilon^{\overline{p}}_{2k} \Big[\int_{t_{2j-1}^p}^{t_{2k+1}^{\overline{p}}} ds \int_{t_{2k}^{\overline{p}}}^{s} ds' L_1(s-s') \notag \\
&+ \int_{t_{2k+1}^{\overline{p}}}^{t_{2j}^p} ds \int_{t_{2k}^{\overline{p}}}^{t_{2k+1}^{\overline{p}}} ds' L_1(s-s') \Big]  \notag \\
=\frac{i}{\pi}& \Big[\Xi^p_{2j-1}\Upsilon^{\overline{p}}_{2k} Q_1(t^p_{2j-1}-t^{\overline{p}}_{2k})+\Xi^p_{2j}\Upsilon^{\overline{p}}_{2k} Q_1(t^p_{2j}-t^{\overline{p}}_{2k}) \notag \\
&+\Xi^p_{2j}\Upsilon^{\overline{p}}_{2k+1} Q_1(t^p_{2j}-t^{\overline{p}}_{2k+1}) \Big].
\end{align}
\end{footnotesize}
The second configuration gives,
\begin{align}
-\frac{i}{\pi} \int_{t_{2j-1}^p}^{t_{2j}^p} ds \int_{t_{2k}^{\overline{p}}}^{s} ds' \xi^p(s) \eta^{\overline{p}}(s') L_1(s-s') &=-\frac{i}{\pi} \Xi^p_{2j-1}\Upsilon^{\overline{p}}_{2k} \left[\int_{t_{2k}^{\overline{p}}}^{t_{2j}^{p}} ds \int_{t_{2k}^{\overline{p}}}^{s} ds' L_1(s-s') \right]  \notag \\
&= \frac{i}{\pi} \left[\Xi^p_{2j}\Upsilon^{\overline{p}}_{2k} Q_1(t^p_{2j}-t^{\overline{p}}_{2k})\right].
\end{align}

The third configuration gives,
\begin{footnotesize}
\begin{align}
-\frac{i}{\pi} \int_{t_{2j-1}^p}^{t_{2j}^p} ds \int_{t_{2k}^{\overline{p}}}^{s} ds' \xi^p(s) \eta^{\overline{p}}(s') L_1(s-s')=-\frac{i}{\pi}& \Xi^p_{2j-1}\Upsilon^{\overline{p}}_{2k} \Big[\int_{t_{2k}^{\overline{p}}}^{t_{2k+1}^{\overline{p}}} ds \int_{t_{2k}^{\overline{p}}}^{s} ds' L_1(s-s')\notag\\
&+\int_{t_{2k+1}^{\overline{p}}}^{t_{2j}^{p}} ds \int_{t_{2k}^{\overline{p}}}^{t_{2k+1}^{\overline{p}}} ds' L_1(s-s') \Big] \\
=\frac{i}{\pi}& \left[\Xi^p_{2j}\Upsilon^{\overline{p}}_{2k} Q_1(t^p_{2j}-t^{\overline{p}}_{2k})+\Xi^p_{2j}\Upsilon^{\overline{p}}_{2k+1} Q_1(t^p_{2j}-t^{\overline{p}}_{2k+1})\right].
\end{align}
\end{footnotesize}

The  fourth configuration gives,
\begin{footnotesize}
\begin{align}
-\frac{i}{\pi} \int_{t_{2j-1}^p}^{t_{2j}^p} ds \int_{t_{2k}^{\overline{p}}}^{s} ds' \xi^p(s) \eta^{\overline{p}}(s') L_1(s-s') =-\frac{i}{\pi}& \Xi^p_{2j-1}\Upsilon^{\overline{p}}_{2k} \left[\int_{t_{2j-1}^{p}}^{t_{2j}^{p}} ds \int_{t_{2k}^{\overline{p}}}^{s} ds' L_1(s-s') \right]  \notag \\
=\frac{i}{\pi} &\Big[\Xi^p_{2j-1}\Upsilon^{\overline{p}}_{2k} Q_1(t^p_{2j-1}-t^{\overline{p}}_{2k})+\Xi^p_{2j}\Upsilon^{\overline{p}}_{2k} Q_1(t^p_{2j}-t^{\overline{p}}_{2k})\Big].
\end{align}
\end{footnotesize}

% \end{widetext} 
  
We finally recover the expression of Chapter 3.
\section{Scaling regime}  

In the scaling regime $\Delta/\omega_c \ll 1$, it is possible to overcome the sign problem naturally arising in our method as shown in Ref. \cite{Peter}. Simplifications occur in Eqs. (\ref{Q_1_2}) and (\ref{Q_1_m_2}) as we can consider that  $ Q_1(t_j-t_k)=2 \pi  \alpha \tan^{-1} \left[\omega_c (t_j-t_k)\right]\simeq \pi^2  \alpha$. Then we have

\begin{align}
\sum_{k=0}^{2 n_p -1} \sum_{j : t_j^q>t_k^p} \Xi_j^q \Upsilon_k^p   Q_1(t_j^q-t_k^p) &=i\pi \alpha  \left[\sum_{j=1}^{2 n_q} \xi_j^q \eta^p_l \right] ,
\end{align}
for $q=p$ or $p=\overline{p}$. $\xi_j^q$ is the value of $\xi^q (t)$ in the interval $[t_j^q,t_{j+1}^q]$ and $\eta^p_l $ is the value of  $\eta^p (t)$ in the interval $[t_l^p,t_{l+1}^p]$. The integer $l$ is defined by $t_l^p<t_j^q \leq t_{l+1}^p$. In the case of $p=q$, we just have $l=j-1$. 

This expression does not depend on intermediate times, but only on the path taken. As a result, there is no need to introduce the time-dependent field $k$. After having introduced the field $h$ as in the main text, we finally recover Eqs. (\ref{eq:p1:two_spins}) and (\ref{eq:SSE:two_spins}) of the main text, with
\begin{figure*}[h!]
\center
  \includegraphics[scale=0.35]{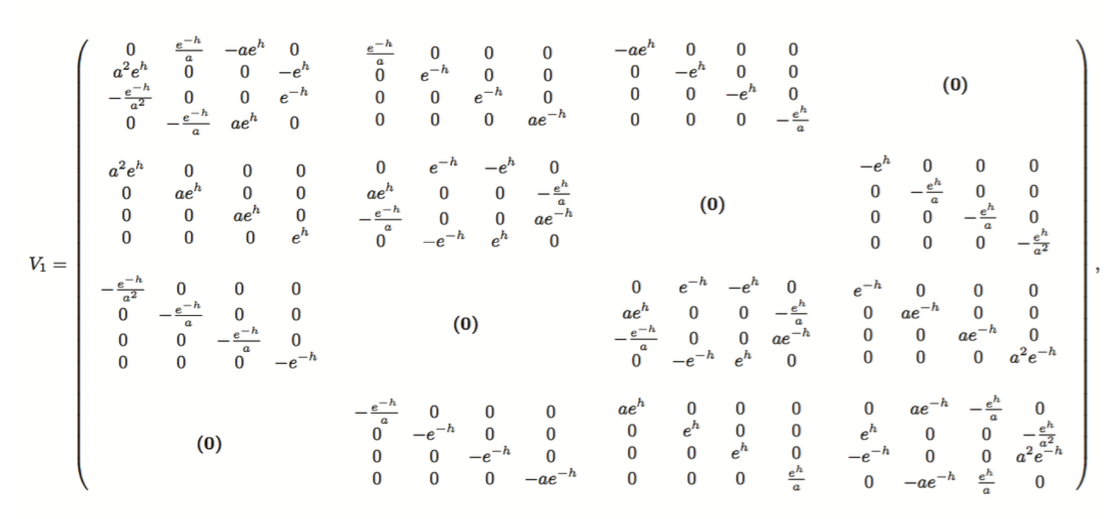}
\end{figure*}

where $a=\exp(i\pi \alpha)$, $|\phi_i\rangle^T=|\phi_f\rangle^T=(1,0,0,0,0,0,0,0,0,0,0,0,0,0,0,0)$.

\chapter{Non-adiabatic response and topology}
 \label{appendix:TDperturbation_theory_topo}
 \section{Chern number and evolution of $\langle \sigma^z \rangle$ with $\theta$}
\label{appendix_index}
One can first use the Poincar\'{e}-Hopf theorem to show that the Chern number is equal to the degree (introduced below) of the mapping $(\theta,\phi)\to \vec{d}/|\vec{d}|$, as used for example in Ref. \cite{fuchs_simon_brouwer}. The degree $deg$ of a smooth map $f: M\to N$ between two connected, oriented and closed $n$-dimensional manifolds $M$ and $N$, is an integer defined by \cite{degree1,degree2}
\begin{align}
deg=\sum_{x \in f^{-1}(y)} \textrm{sign} \det\left(\mathcal{J}\right),
\label{degree}
\end{align}
where $\mathcal{J}$ is the Jacobian matrix of $f$ and $y\in N$ is a regular point with a finite number of preimages. $deg$ is an integer which do not depend on the point $y$. In our precise case, we work with $2$-dimensional manifolds, that fulfill the requirements of the above definition. When $H_0<H$, any regular point $y$ in $N$ has only one pre-image. The sum in Eq. (\ref{degree}) reduces then to one term and we have in general $deg=\pm 1$. When $H_0>H$, the situation is different as there are always two preimages of any regular point in $N$. A computation of the Jacobian for a particular choice of $y$ shows that these two terms compensate and one gets $deg=0$. We recover then Eq. (\ref{C_sigma_z}), when we consider the limit $y\to (0,0,1)$ .

\section{Proof of Eq. (\ref{curvature_sigma_y}) using time-dependent perturbation theory}
 Let us call $|e_t\rangle$ and $|g_t\rangle$ the excited and ground state of the system at time $t$, associated with the eigenenergies $E_e (t)$ and $E_g (t)$. We project the wavefunction of the system at time $t$ on this instantaneous basis,
\begin{align}
|\psi(t)\rangle=a_g (t) |g_t\rangle+a_e (t) |e_t\rangle.
\label{wavefunction}
\end{align} 

We first show that the non adiabatic response of the system will lead to a non-zero expectation value for $\langle \sigma^y(t) \rangle$. Then we compute $a_e (t)$ and $a_g (t)$.
\begin{itemize}
\item Let us define $A(\phi)=\langle \partial_{\phi} \mathcal{H}_0 \rangle$. We have 
\begin{align}
A(\phi)=&\langle \psi(t)| \partial_{\phi} \mathcal{H}_0| \psi(t) \rangle \notag \\
=&|a_g(t)|^2 \underbrace{\langle g_t | \partial_{\phi} \mathcal{H}_0| g_t \rangle}_{=0}+ |a_e(t)|^2 \underbrace{\langle e_t | \partial_{\phi} \mathcal{H}_0| e_t \rangle}_{=0}+ a_g(t) a_e^*(t) \langle e_t |\partial_{\phi} \mathcal{H}_0 | g_t \rangle \notag \\
&+  a_g^*(t) a_e(t) \langle g_t |\partial_{\phi} \mathcal{H}_0 | e_t \rangle.
\end{align}\\
Then 
\begin{align}
A(0)&= \frac{H}{2} \sin \theta \langle \sigma^y \rangle= \left[ a_g(t) a_e^*(t) \langle e_t |\partial_{\phi} \mathcal{H}_0 | g_t \rangle +  a_g^*(t) a_e(t) \langle g_t |\partial_{\phi} \mathcal{H}_0 | e_t \rangle\right] (\phi=0).
\label{A_0}
\end{align}\\
It is clear from Eq. (\ref{A_0}) that $\langle \sigma^y (t) \rangle$ is linked to the non-adiabatic response of the system.\\

\item  To find the time evolution of $a_e(t)$ and $a_g(t)$, we use time dependent perturbation theory, following Refs. \cite{polkovnikov:course,polkovnikov:PNAS}. Inserting expression (\ref{wavefunction}) into the Schr\"{o}dinger equation and projecting on the state $|e_t \rangle$, we get:
\begin{align}
i \dot{a_e}(t) + a_g(t) \langle e_t | \partial_t | g_t \rangle=a_e(t) E_e(t).
\end{align}
Next, we define $\alpha_i (t)=a_i (t) e^{ i \Theta_i (t)}$, with $\Theta_i (t)=\int_{t_0}^t E_i (\tau) d\tau$ for $i=(g,e)$. At first order in $v$, we get\cite{polkovnikov_course},
\begin{align}
\alpha_e (t)=i \int_{t_0}^t d\tau \langle e_{\tau} | \partial_{\tau} | g_{\tau} \rangle \exp \left[i \left(\Theta_e (\tau)-\Theta_g (\tau) \right) \right]+o(v).
\end{align}
Using integration rules on fast oscillating functions \cite{polkovnikov:course}, we finaly reach
\begin{align}
a_e (t)=i\frac{\langle e_t | \partial_t | g_t \rangle}{E_e(t)-E_g(t)}+o(v)=-i v \frac{\langle e_t | \partial_{\theta} \mathcal{H}_0 | g_t \rangle}{\left[E_e(t)-E_g(t)\right]^2}+o(v).
\label{excited_coeff}
\end{align}\\
Inserting the expression (\ref{excited_coeff}) into Eq. (\ref{A_0}), we get:
\begin{align}
\langle \psi(t)| \partial_{\phi} \mathcal{H}_0| \psi(t) \rangle=-iv\frac{\langle g_t | \partial_{\phi}\mathcal{H}_0 | e_t \rangle\langle e_t | \partial_{\theta}\mathcal{H}_0 | g_t \rangle-\langle g_t | \partial_{\theta}\mathcal{H}_0 | e_t \rangle\langle e_t | \partial_{\phi}\mathcal{H}_0 | g_t \rangle}{\left[E_e(t)-E_g(t)\right]^2}+o(v).
\label{sigma_y_curvature}
\end{align}\\
We recognize on the right hand side of Eq. (\ref{sigma_y_curvature}) the expression of the Berry curvature, and we find back Eq. (\ref{curvature_sigma_y}). For $H_0=0$, we have $\mathcal{F}_{\theta \phi }^0=1/2 \sin \theta$.\\

\end{itemize}
\chapter{Other heat engine setup}   

\label{appendix_characteristics}

As stated in the main text, no current is rectified when the system is put in state of overall zero temperature. Here, we consider the case when the conduction electrons are at zero temperature so that the charging energy of the quantum dot forbids electrons to tunnel onto it. However, if the quantum dot is capacitively coupled to an electrical environment via capacitance $C_d$ with impedance $Z_d$, and temperature $T$, the electrons have a new way of gaining energy: photons from the hot impedance can be absorbed, allowing the electrons to tunnel onto the dot. \\

\begin{figure}[h!]
\center
\includegraphics[width=9.9cm]{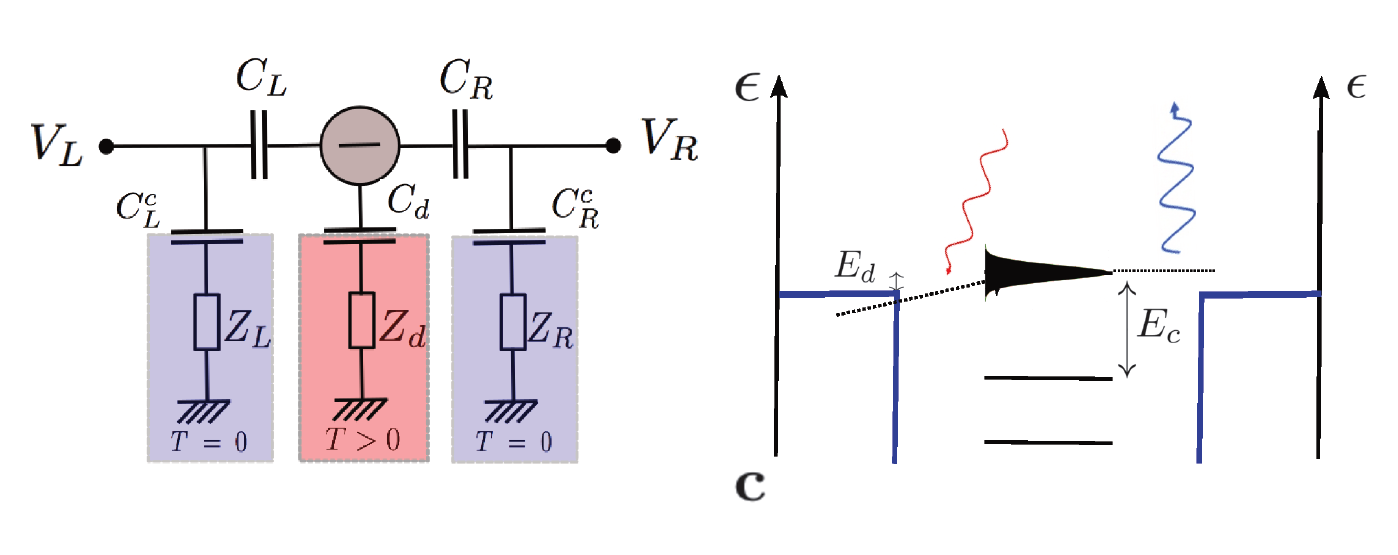}%\includegraphics[width=3.9cm]{cold_electrons_24_04_2.pdf}
\caption{The lateral parts correspond to the energy diagram of the electrons in the corresponding lead at zero temperature and the central part represents the energy levels of the quantum dot. The straight dotted arrows represent  a tunneling event from left to right, associated with curvy arrows which correspond to photon emission/absorption. Photons from a hot impedance $Z_d$ excite zero-temperature electrons tunneling onto the quantum dot. These electrons emit photons in the cold impedances to tunnel off of the quantum dot. The asymmetry in the quantum vacuum fluctuations from the zero temperature latteral impedances leads to rectified current.
}
\label{figsetup_bis}
\end{figure}

We present in Fig. \ref{figsetup_bis} the corresponding circuit and pictorial representation.

In this case we find that $E_c =e^2/2C_\Sigma - e \sum_j (C_j/C_\Sigma) \delta V_j$, where the sum is over $L, R, d$. Here $C_\Sigma = C_L + C_R + C_d$ and $\delta V_d$ corresponds to the voltage fluctuations linked to the central impedance $d$, which is only present in the case of Fig.~\ref{figsetup}(a). The effect of the fluctuating voltages can be taken into account by a unitary transformation $U = \exp ( i \phi_{L} \sum_k c_k^\dagger c_k + i \phi_{R} \sum_q c_q^\dagger c_q + i \phi_{d} d^\dagger d)$, leading to
\begin{eqnarray}
t_{kd} \rightarrow t_{kd} \exp(i \Phi_{L}),
t_{dq} \rightarrow t_{dq} \exp(- i\Phi_{R}).
%t_{kd} \rightarrow t_{kd} e^{i \phi_{L,0} - i\phi_{d,0}},
%t_{dq} \rightarrow t_{kd} e^{i \phi_{d,0} - i\phi_{R,0}}.
\end{eqnarray}
Here, we have introduced the total phase $\Phi_{l}=\phi_{l} - \phi_{d}$, where $\phi_{l} = (e/\hbar) \int_0^t dt' \delta V_l(t')$ for  $l \in \{L, R\}$, and $\phi_{d} = (e/\hbar) \sum_j (C_j/C_{\Sigma}) \int_0^t dt' \delta V_j(t')$ where the sum is over $j\in \{L, R,d\}$. The voltage fluctuations $\delta V_j$ are linked to the environmental degrees of freedom whose dynamics is governed by $H_{env}$. Within $P(E)$ theory, the probability of the electrical circuit $l \in \{L,R\}$  to either absorb or emit a photon is still given by
\begin{align}
P_l(E) = (1/h) \int_{-\infty}^\infty dt e^{i E t/\hbar} e^{K_l(t)},
\end{align}
where $K_l(t) = \langle (\Phi_l(t) - \Phi_l(0)) \Phi_l(0)\rangle$.  \\

We have $K_l = (\eta_{ll}^2 + \eta_{lm}^2) k_l + \eta_d^2 k_{d}$ for $(l,m)=(L,R),(R,L)$ where we have introduced the coupling constants to the environmental baths, $\eta_{ll} = (C_{m} + C_d)/C_\Sigma, \eta_{lm} = C_{m}/C_\Sigma, \eta_d = C_d/C_\Sigma$ ($m=R$ when $l=L$ and vice versa). $k_l(t)$ and $k_d(t)$ are the lead and dot correlation functions taken in isolation $k_j(t) = \langle (\phi_j(t) - \phi_j(0)) \phi_j(0)\rangle$ and they read 
\begin{align}
k_l(t)&=\int_{0}^{\infty} \frac{d \omega}{\omega} 2 Re  \left[ \frac{Z_l}{ R_q}\right]  (e^{-i\omega t}-1), \notag \\
k_d(t)&=\int_{0}^{\infty} \frac{d \omega}{\omega} 2 Re  \left[ \frac{Z_d}{ R_q} \right] \Big\{\coth\left(\frac{\beta \hbar \omega}{2}\right)\left[\cos(\omega t)-1\right]-i\sin(\omega t)         \Big\},
\label{k}
\end{align}
where $Z_l=\sqrt{L_l/C^c_l}$ is the impedance coupled to the lead $l$. For the sake of simplicity, we have specified the case for which both left and right leads have the same correlation function $k_{L}(t)=k_{R}(t)=k_{0}(t)$ in isolation at zero temperature, and the asymmetry of the system comes from different values of $C_L$ and $C_R$. This is one particular choice to break the Left/Right symmetry, which leads to the simple expression above for $K_l$. There are other ways to break this symmetry such as having different impedances $Z_L$ and $Z_R$. The important point is that the capacitive coupling to the hot impedance now enters the $P_l(E)$ functions for the tunneling electrons.\\

The effects of $k_{0}(t)$ and $k_d(t)$ are quite different and we will consider their effects independently. We consider for $k_{0}(t)$ the case where the zero frequency external impedance of the $T=0$ transmission lines $Z=R=\sqrt{L/C^c}$ is small compared to the resistance quantum, $R \ll R_q = h/e^2$, and define the large $\alpha = R_q/R \gg1$ (where $L=L_L=L_R$ and $C^c=C^c_L=C^c_R$). In this case, we have for long time, $k_{0}(t) = -(2/\alpha) [\ln(\alpha E_c t/\pi \hbar) + i\pi/2 + \gamma_e]$, where $\gamma_e$ is the Euler constant \cite{IN}. The isolated effect of this cold impedance would lead to a $P_{0}(E)=(1/h) \int_{-\infty}^\infty dt \exp\left(i E t/\hbar\right) \exp \left[ k_0(t)\right] $ function that vanishes for negative energies, and has a power-law divergence for small positive energies, as in the main text. In contrast for the hot impedance, still in the small resistance limit, the correlation function becomes $k_d(t) = - (Re Z_d/R_q) k_B T t/\hbar$ in the long time limit. The isolated effect of this cold impedance would lead to a normalized Lorentzian for $P_d(E)=(1/h) \int_{-\infty}^\infty dt \exp\left(i E t/\hbar\right) \exp \left[ k_d(t)\right] $ of width $\Delta = (Re Z_d/R_q) k_B T$ \cite{IN}.\\

The combined $P_l(E)$ functions from both effects is found by taking a convolution of the two $P_0(E)$ and $P_d(E)$ functions (convolution theorem), provided the coupling constants $\eta_l$ to the various circuits are properly accounted for. This turns out to be straightforward because they can be absorbed into effective impedances $\tilde \alpha_l=\alpha /(\eta_{ll}^2 + \eta_{lm}^2) $, charging energies ${\tilde E}_{c,l}=(\eta_{ll}^2 + \eta_{lm}^2) E_c$ and Lorentzian widths, $\tilde \Delta_l=\Delta \eta_d^2$, giving
\begin{align}
P_l(E) = \frac{\pi^{2/\tilde{\alpha}_l} e^{-2 \gamma_e/\tilde{\alpha}_l}
}{\Gamma(2/\tilde{\alpha}_l)  \sin(2\pi/\tilde{\alpha}_l ) (\tilde{\alpha}_l \tilde{E}_{c,l})^{2/\tilde{\alpha}_l}}
{\rm Im}( -i \tilde{\Delta}_l - E)^{\frac{2}{\tilde{\alpha}_l}-1}.
\end{align}

The Left/Right asymmetry is now solely contained in the effective parameters $\tilde \alpha_L \neq \tilde \alpha_R$. These $P_l(E)$ functions permit us to calculate the tunneling rates $T_{\pm,l}$ between the leads and the dot, from left to right (+) and right to left (-),  

\begin{eqnarray}
T_{\pm, L} (\Omega) = T_{0,L} \int d\epsilon f(\epsilon \pm \Omega \mp \mu/2) P_L(\epsilon), \label{Gamma_L_appendix}\\
T_{\pm, R} (\Omega) = T_{0,R} \int d\epsilon f(\epsilon \mp \Omega \mp \mu/2)  P_R(\epsilon)\label{Gamma_R_appendix}.
\end{eqnarray}

At zero temperature the Fermi distribution $f$ becomes a step function and equations (\ref{Gamma_L}) and (\ref{Gamma_R}) have a simple interpretation: the electrons in the lead have no way to gain energy without absorbing a photon from the bosonic bath if $\mu=0$. This is why the convolution with the hot impedance is important - it permits the $P(E)$ function to have some probability in the negative energy range, so a photon can be absorbed from that bath. It may then be given back to a cold bath as it tunnels left or right, as can be seen in Fig.~\ref{figsetup_bis}. The asymmetry between the capacitances breaks the left-right symmetry. Equivalently we repeat that one may have taken the same value for $C_L$ and $C_R$ but with different impedances $Z_L$ and $Z_R$. In this case, the level is dephased by a fluctuating phase $\phi_d$, so that we assume a sequential non-resonant tunneling case and the rectified current is given by \cite{RO},
\begin{align}
I =e ~\frac{T_{+,L}(E_d) T_{+,R}(E_d) - T_{-,L}(E_d) T_{-,R}(E_d)}{ T_{+,L}(E_d) + T_{+,R}(E_d) +  T_{-,L}(E_d) + T_{-,R}(E_d) }.
\label{appendix_rectified_current}
\end{align}\\

 This effect leads to a weaker rectified current compared to the setup presented in the main text (it compares to the results obtained for the non resonant case). This is due to the fact that one cannot consider resonant sequential tunneling, because the interaction of the electron on the dot with the external hot environment prevents phase coherence. \\
 
For this setup, the efficiency reads $\eta=IV/J_H$, where $J_H$ is the amount of heat received from the environment. We find a smaller efficiency in this case.
\end{appendices}
\bibliographystyle{numeric-comp}
\bibliography{references}

\end{document}